\RequirePackage{fix-cm}
\documentclass[natbib,smallextended]{svjour3}       % onecolumn (second format)
\smartqed  % flush right qed marks, e.g. at end of proof
\usepackage{graphicx}
\usepackage{amsmath,amssymb}
\usepackage{dcolumn}
\usepackage{fancy}
\usepackage{footnote}
\usepackage[margin=2.0cm]{geometry}
\usepackage{booktabs}
\usepackage{color,transparent}
\usepackage{extpfeil}
\usepackage[dvipsnames,table]{xcolor}
\usepackage[most]{tcolorbox}
\usepackage{lipsum}
\usepackage{empheq}
\usepackage{mathtools}

\usepackage[colorlinks=true,linkcolor=blue,citecolor=blue,urlcolor=blue]{hyperref}

\newtheorem{theorem*}{Theorem}

\newcommand{\subparagraph}{}
\newcommand*\widefbox[1]{{\setlength\fboxsep{10pt}\fbox{\hspace{0.5em}#1\hspace{0.5em}}}}

\graphicspath{{Figures/}} %Setting the graphicspath

\begin{document}

\title{Ultra-light dark matter}


\author{Elisa G. M. Ferreira}

\institute{Elisa G. M. Ferreira \at Max-Planck-Institut f\"ur Astrophysik, Karl-Schwarzschild-Str.~1, 85748 Garching, Germany \\
              \email{elisagmf@mpa-garching.mpg.de}           %  \\
}


\maketitle

\begin{abstract}
Ultra-light dark matter is a class of dark matter models (DM)  where DM is composed by bosons with masses ranging from $10^{-24}\, \mathrm{eV} < m < \mathrm{eV}$. These models have been receiving a lot of attention in the past few years given their interesting property of forming a Bose--Einstein condensate (BEC) or a superfluid on galactic scales. BEC and superfluidity are some of the most striking quantum mechanical phenomena manifest on macroscopic scales, and upon condensation the particles behave as a single coherent state, described by the wavefunction of the condensate.
The idea is that condensation takes place inside galaxies while outside, on large scales, it recovers the successes of $\Lambda$CDM.
This wave nature of DM on galactic scales that arise upon condensation can address some of the curiosities of the behaviour of DM on small scales.
There are many models in the literature that describe a DM component that condenses in galaxies. In this review, we are going to describe those models, and classify them into three classes, according to the different non-linear evolution and structures they form in galaxies: the fuzzy dark matter (FDM), the self-interacting fuzzy dark matter (SIFDM), and the DM superfluid. Each of these classes comprises many models, each presenting a similar phenomenology in galaxies. They also include some microscopic models like the axions and axion-like particles. 
To understand and describe this phenomenology in galaxies, we are going to review the phenomena of BEC and superfluidity that arise in condensed matter physics, and apply this knowledge to DM. We describe how ULDM can potentially reconcile the cold DM picture with the small scale behaviour.  These models present a rich phenomenology that is manifest in different astrophysical consequences. We review here the astrophysical and cosmological tests used to constrain those models, together with new and future observations that promise to test these models in different regimes. For the case of the FDM class, the mass where this model has an interesting phenomenology on small scales $ \sim 10^{-22}\, \mathrm{eV}$, is strongly challenged by current observations.
The parameter space for the other two classes remains weakly constrained. We finalize by showing some predictions that are a consequence of the wave nature of this component, like the creation of vortices and interference patterns, that could represent a smoking gun in the search of these rich and interesting alternative class of DM models.

 %\subclass{MSC code1 \and MSC code2 \and more}
\end{abstract}

\setcounter{tocdepth}{3}
\tableofcontents

%%%%%%%%%%%%%%%%%%%%%%%%%%%%%%%%%
% Introduction and Motivation
\section{Introduction and motivation}

An overwhelming amount of observational data provides clear and compelling evidence for the presence of dark matter (DM) on a wide range of scales.
This component, which is believed to be responsible for the ``missing'' mass in our universe, is the main ingredient for all the structures we have in our universe.  
This is one of the oldest unsolved problems in cosmology, being traced back to the  1930s \citep{Zwicky,Bertone}, and also one of the best measured ones.
The evidence for dark matter first emerged from the study of the rotation curves of galaxies.
From pioneering studies \citep{Rubin}, it was already evident  that the amount of matter necessary to fit the flat observed rotation curves did not match the theoretical curves predicted, assuming Newtonian mechanics and taking into account only the visible matter present in those galaxies. Dark matter was proposed as an additional (non-luminous) component to explain this discrepancy. 

Nowadays, the evidence for dark matter comes from precise measurements on a wide range of scales. From sub-galactic and galactic scales, to clusters, going up to the large scale structure (LSS). On cosmological scales, the observed anisotropies of the Cosmic Microwave Background  (CMB) \citep{Planck}, together with data from Type Ia Supernovae, determine the total energy density of matter with high precision. This together with the bounds on the abundance of the light chemical elements from  Big Bang Nucleosynthesis, which constrains the amount of baryonic matter in the universe, strongly shows the need for a clustering component of non-baryonic\footnote{We are going to see later that there are some ``baryonic'' candidates for DM.} origin, that does not interact (strongly) with photons, and that dominates the matter content of the universe, accounting for approximately $85\%$ of all matter.  The same non-luminous and clustering component is necessary to explain the structures we see in our universe today, as is evident in observations of the large scale structure of our universe \citep{Anderson:2013zyy,LSS}.

With all this evidence coming from precise astrophysical and cosmological observations, cosmologists have converged to a phenomenological model to describe our universe, the $\Lambda$CDM model. This model is currently the concordance model of cosmology and it accumulates a number of observational successes. It exhibits outstanding agreement with current cosmological observations \citep{Anderson:2013zyy}, which is manifested in the parameters of this model being constrained at the percent and sub-percent level.
This incredibly simple model is described by only six parameters and parametrizes a large amount of the universe's history. It describes a universe that is flat and seeded by nearly scale invariant perturbations, composed of baryons, which amount to approximately $5\%$ of the energy density of the universe, a small radiation component, but in its majority is composed of two unknown ingredients. The energy budget of the universe is dominated ($\sim70\%$) by a component responsible for the current accelerated expansion of the universe called dark energy, and a clustering component, the dark matter, making up to $\sim 25\%$ of our universe.
These large-scale observations give a coarse-grained description of these non-baryonic components in the hydrodynamical limit where dark matter is described as a perfect fluid with very small pressure ($w\approx 0$) and sound speed, $c_\mathrm{s} \approx 0$, that does not interact, at least strongly, with baryonic matter. Dark energy is parametrized by a cosmological constant, the simplest model for the present accelerated expansion of our universe.

Therefore, within $\Lambda$CDM, the Cold  Dark Matter (CDM) paradigm emerged from the large scale observations and describes the component responsible for the formation of the structures of our universe through gravitational clustering. In the CDM model, DM is described by a perfect fluid that must be massive, sufficiently cold, which means non-relativistic at the time of structure formation, and collisionless in order to explain the observational data on large linear scales.
This coarse-grained description of a CDM is very successful in fitting the linear, large scales observations from the CMB, LSS, to clusters, and general properties of galaxies. 

However, even though we know the hydrodynamical properties of DM on large scales to a very high precision, the microphysics of the DM component remains unknown. 
This allows for the creation of a plethora of possible models of DM. Those models recover the large scale properties of CDM, but invoke very different objects and phenomena to play the role of DM.

\begin{figure}
\centering
\includegraphics[width=\textwidth]{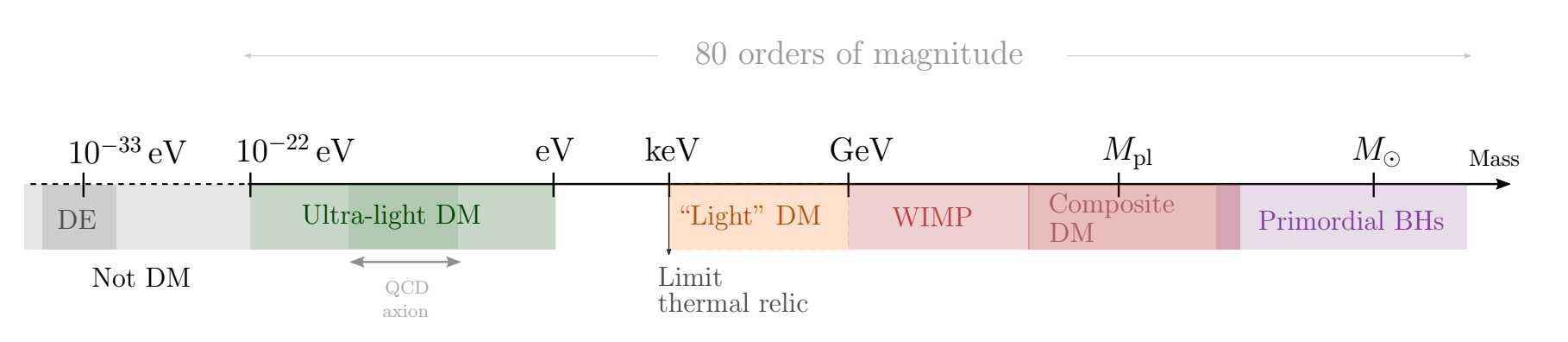}
 \caption{Sketch (not to scale) of the huge range of possible DM models that have been conceived. They span many orders of magnitude in mass, with  DM represented by very distinct phenomena, ranging from new elementary particles to black holes.}
\label{Fig.:Mass_scale}
\end{figure}

This incredible variety of viable models of DM can be seen in the huge range of masses those models cover, as shown in Fig.~\ref{Fig.:Mass_scale}. This figure shows many different broad classes of DM models, and each of which might contain many different specific models. It spans more than 80 orders of magnitude and shows very different hypothesis for DM, from new elementary particles, to composite objects~\citep{Jacobs:2014yca,Khlopov:2019qcr}, up to astrophysical size primordial black holes (for a review on recent bounds see~\citep{Carr:2020gox,Carr:2020xqk}). This shows us that although we have gathered a lot of knowledge about the gravitational properties of DM, the nature of DM is still elusive, with the current data still allowing a huge amount of highly different models. 

The possibility that dark matter could be a long lived particle is very appealing. Specially if these candidates are expected candidates from extensions of the standard model of particle physics.  One class of models that became the preferred candidates for the DM particles are WIMP, weakly interacting massive particles, which represent new elementary particles that interact with baryons not only gravitationally but also through the weak force or a new force of comparable strength \citep{Review_WIMP,DM_review}. The strong motivation for this candidate is because if it is thermally produced in the early universe, the relic abundance of particles that have mass of the order of the electroweak scale, and a coupling of order one, corresponds precisely to the abundance of DM in our universe. The possibility that WIMP could also be discovered by direct detection experiments is also an important motivation to search for this candidate. There is a great experimental effort to constrain the properties of WIMP DM with the parameter space being very restricted over the past few years.  Given the complex phenomenology from the possible models of WIMP DM and their interaction with the standard model particles, the translation of those bounds to the exclusion of WIMP models is not straightforward. The cosmological and astrophysical behaviour of all the classes of WIMP models is similar to CDM, so the avenue to probe this scenario is through direct, indirect and collider experiments (for a complete review of all the searches, current and projected limits on WIMP detection for both spin dependent and independent models, together with indirect detection and collider searches, see~\citep{Arcadi:2017kky}.) 

Another candidate that comes from extensions of the standard model of particle physics is the QCD axion. The axion was introduced to address the strong CP problem of quantum chromodynamics (QCD)~\citep{QCD1,QCD2,QCD3}. The axion can be used in many different contexts in cosmology, including as a candidate for DM. The QCD non-perturbative effects induce a potential for the axion. During the radiation dominated period, the QCD axion starts to oscillate at the bottom of its potential and the axions behaves like dust, contributing to the energy density of the universe as non-relativistic matter. The QCD axion couples weakly to the standard model, which motivated an experimental effort for its direct detection (see these references for a review of axion direct detection searches~\citep{DiLuzio:2020wdo,Sikivie:2020zpn,Graham:2015ouw,Battesti:2007um}). 

Although we have these very well motivated candidates from particle physics, we still have no conclusive evidence for electroweak or other non-gravitational interactions for dark matter. All the knowledge we have about dark matter is gravitational. 
We know that CDM describes the behaviour of DM very well on large scales. However, this beautiful and simple coarse grained description of DM as the CDM is challenged by some curiosities that appear on small scales.

As the observations and simulations of the small non-linear scales and galactic scales improve, a number of challenges have emerged for this coarse grained description from $\Lambda$CDM. 
These discrepancies have been around for decades, such as the cusp-core problem, the missing satellite problem and the too big to fail problem. A particularly curious challenge is the regularity/diversity of rotation curves.  One thing that is surprising about galaxies is that they are extremely diverse, but at the same time they are incredibly regular. This fact is manifest in several empirical scaling relations, such as the well known Baryonic Tully--Fisher relation (BTFR; \citep{McGaugh:2005qe,McGaugh:2007fj}). The BTFR shows the correlation between the total baryon mass (including stars and gas) of the galaxy with the the asymptotic rotation velocity in galaxies. The measured BTFR follows a scaling relation different from the one predicted by $\Lambda$CDM, and it holds for a range of 6 orders of magnitude in mass, with very small scatter. The significance of these discrepancies is disputed and addressing these challenges is an active field of research. Those challenges emerge on scales where baryonic physics is relevant and simulations including several baryonic effects have been perfected pointing in the direction that baryons could possibly explain some of these observations within $\Lambda$CDM.

As the physics of these baryonic processes is complex and as there is no final consensus about the status of theses discrepancies, an alternative explanation for these discrepancies on small scales could be that DM is not the usual CDM, but a component that has different phenomenology on small scales. Even setting aside the small scale problems, given that the observational constraints on these scales are less strong than on cosmological scales, the dynamics on small scales can offer a chance to probe the properties of DM in the hope to help find the microphysics of this component.
Therefore, the small scales are a good laboratory to probe the nature of DM models that have impacts on astrophysical scales.

The simplest modifications of the DM paradigm that have a different phenomenology on small scales, and can potentially address some of the small scales discrepancies is the warm dark matter (WDM) model \citep{Colin:2000dn}). In this model, DM has a small mass leading to a thermal velocity dispersion, modifying  its behaviour on astrophysical scales while maintaining the large scale predictions of CDM.  Even with a small velocity, DM free streams out of potential wells and is enough to suppress the formation of small scale structures addressing some of the small scale problems. Another popular model inspired by those discrepancies is the self interacting DM (SIDM) \citep{Spergel:1999mh}), where the DM particles have a self-interaction in a way to also suppress the formation of structures on small scales.

In the past few years, another class of alternative models has (re)emerged as an appealing class of DM models given their  rich phenomenolgy on small scales. These are the ultra-light dark matter models (ULDM), where DM is composed by ultra-light bosons with masses in the range $10^{-24}~\mathrm{eV} < m < \mathrm{eV}$. Given the small masses of these bosons,  DM forms a condensate or a superfluid on galactic scales. The idea is that the wave nature of DM on galactic scales provides a non-CDM behavior which leads to a different and rich phenomenology for DM on those scales. On large scales DM behaves as CDM, although with different initial conditions for the ULDM in comparison to CDM, maintaining the observable successes of CDM on those scales. 

There are many different realizations of this interesting non-CDM phenomenology on small scales. Depending on the modelling of the ULDM, these produce distinct condensate structures and lead to a different phenomenology. 
There are many different models in the literature describing these possibilities. 
We classify them in this review into three categories, according to the different condensate structure they form. These three classes are the \emph{fuzzy dark matter}, when the ultra-light scalar field system is only subjected to gravity; the \emph{self-interacting Fuzzy DM} when the system also presents (weakly) self-interaction, and \emph{superfluid DM}, where DM forms a superfluid on galactic scales. 
%where these last two classes describe a superfluid that forms on galactic scales, but with distinct dynamics.

This classification is general and based only on the non-linear structure it forms in the halo of galaxies, which is a consequence of the non-relativistic theory they describe. These can be purely phenomenological models of ULDM on small scales, in the absence of a microscopic description. These classes also contain microscopic scalar field theories like QCD axion, axions (cominig from other origins) or axion like particles (ALP), which are part of the FDM class. 
Each of these categories have different properties which lead to different astrophysical consequences, that can be probed by current and future astrophysical observations. For this reason, ULDM models have regained interest in the community in the past few years, with new and exciting experimental  effort to probe many aspects of the small astrophysical scales, opening the avenue to test these models and answer some questions about the nature of DM.

%%%%%%%%%%%%%%%%%%%%%%%%%%%%%%%%%%%%%%%%%%%%%%%

\paragraph*{\textbf{Motivation for this review and detailed plan:} \qquad} 
%\mbox{}\\

This review has the goal of giving an overview of the ultra-light dark matter (ULDM) candidates, focusing in their gravitational effects and mostly in their different phenomenology on small scales.

There are many very good and complete reviews in the literature describing specific models of ULDM or the microscopic models that can be part of the ULDM class like. There are many reviews of axions in cosmology~\citep{Sikivie:2006ni,Arvanitaki:2009fg,Wantz:2009it,Kim:2008hd,Kawasaki:2013ae,Axions_1} and ALPs~\citep{Ringwald:2014vqa,Arias:2012az,Axions_2,Marsh:2017hbv,Niemeyer:2019aqm,Powell:2016tfs}. Axions and ALPs have a whole rich phenomenology of its interactions to the standard model that will not be explored here, but that can be seen in the following reviews~\citep{Axions_1,Axions_2}.
The FDM model oalso has a huge body of literature with many excellent reviews like~\citep{Hui:2016ltb,Suarez:2013iw,Review_SFDM_Arturo}.

We propose to do something different in this review. Instead of studying one single model or a specific microsocpic theory, we study many ULDM models interested in the gravitational phenomenology that these models present. We study the ULDM models by dividing them into classes according to their dynamics on small scales. The three classes proposed in this review encompass many of the models cited above, with the inclusion of (weakly) self-interacting models and the DM superfluid model. We believe this classification is instrumental and shows the general behaviour and phenomenology that each of these model have inside each class.
Therefore, we hope to bring not only a new classification that encompasses many of the models present in the literature, but also to include new models, trying to make a big overview of the entire class of ULDM models.

Another new feature this review brings is a brief review of BEC and superfluidity, and the different descriptions of these phenomena.  Condensation in each of these classes might arise in a different way, given their different descriptions. Bose Einstein condensation and superfluidity are very well understood and well studied macroscopic quantum phenomena in condensed matter physics, being largely studied theoretically and experimentally. However, these phenomena are not so well understood in gravity.
Therefore, understanding their definition, description and differences is particularly important in order to understand if condensation arises in theses models, and the difference in the condensate structure that is expected to form in each of the classes of ULDM models. 

I take this opportunity to also discuss briefly the different views in the literature about the formation of a condensate and the scales where this effect takes place.

With that, we aim to give a general picture of the state of the field to date, trying to describe all the classes of ULDM present in the literature. We hope this review can be a resource to researcher entering this exciting field.
%%%%%%

\vspace{0.5cm}

The review is organized as follows. First, in Sect.~\ref{Sec.:Small_scales}, we start by describing the small scale challenges of $\Lambda$CDM, as a motivation to show the discrepancies these alternative models of DM aim to address. The goal of this section is not only to show  the problems that some of the ULDM models might solve, but also to introduce the reader into some of the concepts of galactic astrophysics. In this way, the reader can understand some of the interesting phenomenology that the ULDM models have on small scales that differ from the ones predicted by the CDM paradigm.
Next, in Sect.~\ref{Sec.:BEC_superfluids}, we introduce the basic concepts of the quantum phenomena of BEC and superfluids. In this section we describe these phenomena, describe approximations and and the structures formed in those system with and without rotation, all in the context of condensed matter physics where they are well defined, understood and tested. The goal is to give a sound basis to the reader so they can understand with a critical view how these concepts can be applied to the case of DM in the next section, given the analogies, approximations and generalizations done in the literature of ULDM.
Following this we are ready to describe the main topic of the review, the ULDM models in Sect.~\ref{Sec.:ULDM}. We start by describing the three classes that we propose to classify the models of ULDM based on the type of non-linear they describe. We then talk about the \emph{fuzzy DM} and the \emph{self-interacting BEC DM} models, showing the conditions for them to condense on galactic scales. We then focus on the \emph{fuzzy DM} model, showing how and  in which conditions the model attempts to solve the small scale challenges, and the interesting astrophysical consequences this class presents. We then talk about the \emph{superfluid DM} model describing its condensation on galactic scales, the formation of the superfluid core and its observational consequences. We also discuss the stability of this construction, and its possible extension to cosmology. The constraints in these models and new windows of observations of the effects of these models are discussed in Section~\ref{Sec.:Observations}. We will constraints from different observations. In the case of the FDM the current bounds show that the mass range where an interesting phenomenology is expected on small scales is strongly constrained. We conclude the review summarizing our discussion.

Since there is no unique literature this review is based on, but a series of reviews and articles referring to specific topics, the main references used are cited in the corresponding sections. The only exception is Section 2 that is based mainly in the following reviews \citep{Bullock:2017xww,DelPopolo:2016emo,MOND7}. 

\paragraph*{\textbf{Conventions:} \qquad}
In the entire review natural units are used, where $c=\hbar =1$, unless stated otherwise. The exception is section 3 where all the $\hbar$ factors are present. With that, the reduced Planck mass is given by $M^2_{\mathrm{pl}}=1/8 \pi G$, where $G$ is the Newtonian gravitational constant. Unless stated otherwise, the metric signature used is $\left( +, -, -, - \right)$, and Greek letters are indices going from $\mu, \nu = 0, \dots, 3$. Sometimes for simplicity we describe partial derivatives as $\partial_{\mu}=\partial / \partial x^{\mu}$.
In the text gray boxes bring definitions necessary for the understanding of  the topics in the section or following sections. Frames text and equations refer to important results or discussions that we would like to highlight.

%%%%%%%%%%%%%%%%%%%%%%%%%%%%%%%%%
%Small Scale Challenges of Cold Dark Matter
\section{Small-scale challenges of cold dark matter}
\label{Sec.:Small_scales}

\begin{figure}[htb]
 \centering
\includegraphics[height=0.5\textwidth , width=0.9\textwidth]{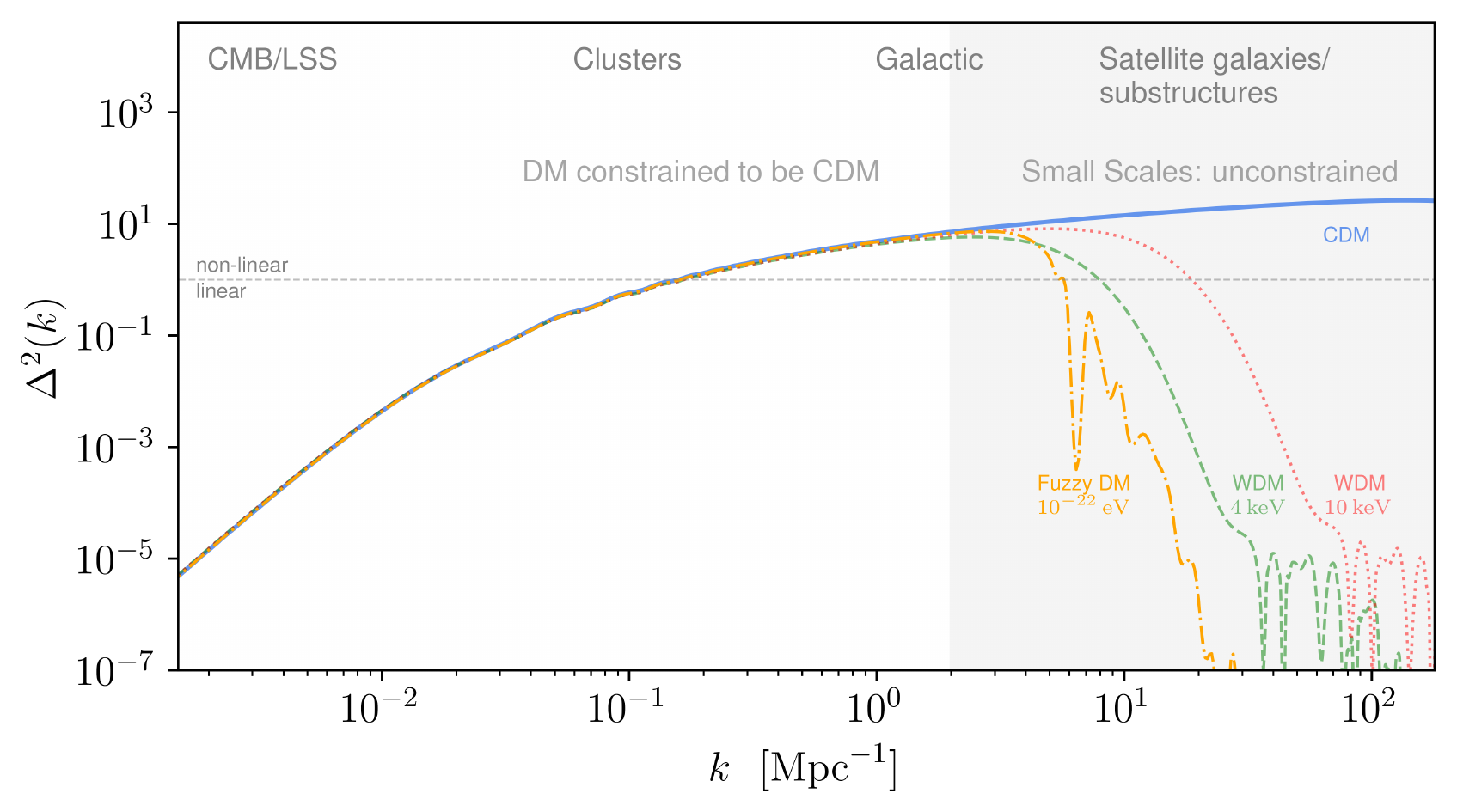}
 \caption{In this figure, inspired from \citep{Kuhlen:2012ft}, we show how the dimensionless power spectrum can be probed by many large scale and small scale observables, which can be seen as a function of  the wavenumber $k$. The solid line shows the linear dimensionless power spectrum coming from a $\Lambda$CDM universe. To show how the small scales might reveal different behaviour for different DM components, we show the linear power spectrum of warm DM (WDM) with mass of $10 \, \mathrm{keV}$ (red dotted line),  WDM with mass of $4 \, \mathrm{keV}$ (green dashed line), and for fuzzy DM with mass $10^{-22} \, \mathrm{eV}$ (orange dash-dotted line). The gray dotted horizontal line represents the limit from linear to non-linear regime, where $\mathrm{\Delta} \sim 1$. The power spectrum for $\Lambda$CDM and for WDM were generated using the Boltzmann code \textsc{\tt{CLASS}} \citep{class,class_wdm}, and for the fuzzy DM using Axion\textsc{\tt{CAMB}} \citep{camb,Hlozek:2014lca}\protect\footnotemark .}
\label{Fig.:Small_scale_power_spectrum}
\end{figure}

In the concordance model of cosmology, DM is described by the CDM paradigm. This hydrodynamical description for DM is in very good agreement with observations from large scales. This can be seen in the power spectrum ($P(k)$), which is the Fourier transform of the two-point correlation function of the matter density perturbations,
\begin{equation}
\mathrm{\Delta} ^2 (k) = \frac{1}{2 \pi^2} k^3 \, P(k)\,,
\label{Eq.:dimensionless_power_spectrum}
\end{equation}
represented here by the dimensionless power spectrum where $k$ is the wavenumber of the fluctuation, shown in Fig.~\ref{Fig.:Small_scale_power_spectrum}. The large scales (around $k \lesssim 0.1 \, \mathrm{Mpc}^{-1}$), measured by the CMB and LSS galaxy surveys, show a good compatibility with the CDM model. This agreement is also robust for the non-linear intermediary scales ($k \sim [10^{-1} - 10] \, \mathrm{Mpc}^{-1}$) with constrains from clusters, weak lensing and Ly-$\alpha$ forest. As we go to smaller and highly non-linear scales ($k \gtrsim 10 \mathrm{Mpc}^{-1}$ equivalent to $M \lesssim 10^{10} M_{\odot}$), these constraints are less strong, and might retain important information about possible deviations from the CDM paradigm. 
We can see on the right side of Fig.~\ref{Fig.:Small_scale_power_spectrum} on galatic and subgalatic scales,  different models of DM would behave very differently from the expected linear behaviour of CDM and this could be probed by the observations on those scales \citep{Zavala:2019gpq,Kuhlen:2012ft}.
%%%%%%%%%%%%%%%%
\footnotetext{The parameters used to generate these power spectra were: $\Omega _b h^2 = 0.022$, $\Omega _c  h^2 = 0.12$, $h=0.67$, $n_s=0.96$, $A_s=2.2 \times 10^{-9}$, and $\tau=0.09$. }
%%%%%%%%%%%%%

On small scales, the formation of structures is highly non-linear and the evolution of structures is studied using large numerical simulations.  In the past few years, those simulations improved in size and precision, simulating the cosmological and small scales.  But when compared to the observations of galaxies, a number of discrepancies emerged, revealing some curious behaviour on small scales. Given the enormous success of the concordance model, these discrepancies attract a lot of interest of the community. They might represent that we need to better take into account the astrophysical processes that happen inside those regions, which indeed have a complex dynamics. 
%The presence of baryons on those scales is known to modify the behaviour o
%; including those processes into the simulations, accompanied with even further  improvements in the observations.  
Or this might indicate that the CDM model is not good to describe the physics on small scales and the coarse grained CDM paradigm needs to be revised. 
An even a more radical approach would be to modify gravity on smaller scales.

In this section we present very concisely the theory of non-linear structure evolution. We show how the numerical predictions assuming the concordance model might be in tension with the current observations of galaxies. These tensions are seen in the counts and density of low-mass objects, and in the scaling relations that show the tight regularity that galaxies present.
We highlight in this section some of the  concepts that are going to be used in the ULDM section and that might not be too familiar for researchers from fields of dark matter phenomenology and cosmology.
%%SHOULD I LEAVE THIS?

\begin{tcolorbox}[colback=gray!10,enhanced,breakable,frame hidden,halign=justify]
%\textbf{Summary of scales and galaxies}\footnotemark
%\vspace{0.0.8cm}
\subsubsection*{Summary of scales and galaxies\footnotemark}
\paragraph*{Galaxy clusters: \qquad} Largest gravitationally bound systems in the universe, with masses $ \sim 10^{14} - 10^{15} M_{\odot}$ (equivalent to $k \sim [1.5 - 6] \times 10^{-1} \, \mathrm{Mpc}^{-1}$), containing  hundreds of galaxies, hot gas and mostly DM.

\paragraph*{Milky-Way (MW) galaxy: \qquad} MW is a barred spiral galaxy and part of the Local Group of galaxies with mass $ \sim 10^{12}\, M_{\odot}$. It has a stellar disk of approximately $30 \, \mathrm{kpc}$ in diameter and $0.3 \, \mathrm{kpc}$ thick, and  $v_{\mathrm{vir}} \sim 100 \, \mathrm{km/s}$ (virial velocity, defined below), with the halo of the MW being hundreds of $\mathrm{kpc}$ in size.

\paragraph*{Dwarf galaxies: \qquad} 
 Dwarf galaxies are low luminosity, small size galaxies, with masses smaller than $10^9\, M_\odot$. Regarding their mass, they can be further divided into:
Bright dwarfs ($M \sim 10^{7-9} M_\odot$), classical dwarfs ($M \sim 10^{5-7}\, M_\odot$), and ultra-faint dwarfs ($M \sim 10^{2-5}\, M_\odot$). Regarding their characteristics, they can be divided into ellipticals, spheroidal and irregulars, that contain gas and star formation.
\vspace{0.1cm}

\begin{changemargin}{0.5cm}{0.55cm} 
\textit{Dwarf Spheroidals (dSphs):} Type of dwarf galaxy with a close to spheroidal shape, they have low-luminosity with a very small quantity of gas and dust, and no recent star formation. They present a large amount of DM and are usually the satellites.
\end{changemargin}
\vspace{0.05cm}

\end{tcolorbox}
\footnotetext{The masses are indicated in terms of the solar mass $M_{\odot}$ which is equivalent to $2 \times 10^{30} \, \mathrm{kg}$ in SI units. Distances are denoted in parsec (pc), where 1 parsec corresponds to 1 arcsecond of measured parallax, and it corresponds in SI units to $3.1 \times 10^{16} \, \mathrm{m}$.}

\subsection{Dark matter halos and substructures}
\label{Sec.:halos_theory}

A halo can be described as a virialized spherical mass concentration of dark matter. Halos are  formed by gravitational collapse of a non-linear overdense regions that stopped expanding to collapse into a sphere in virial equilibrium\footnote{Virial equilibrium means that it obeys the virial theorem $E_{\mathrm{kin}}=-2\, E_{\mathrm{pot}}$ and conservation of energy. So we can describe the system only in terms of the radius $R$ and thee mass, $M$ (or velocity $V$) of the spherical mass concentration.}. The virialization of the halo happens through violent relaxation, where the DM particles scatter on small fluctuations of the gravitational field present in this distribution, taking a time $t_{\mathrm{dyn}}$, the dynamical time, to fully cross the sphere. Once this process is completed, at $t_{\mathrm{coll}}$, the dark matter halo has a radius approximately $1/6$ of the radius of the region it collapsed from, and average density \citep{Schneider}
\begin{equation}
\langle \rho \rangle = \left( 1 + \delta_{\mathrm{vir}} \right) \bar{\rho}(t_{\mathrm{coll}})\,,
\end{equation}
where $\bar{\rho}$ is the mean density, and $ \left( 1 + \delta_{\mathrm{vir}} \right) \approx 178 \, \Omega_{\mathrm{m}}^{-0.6}$. Given this, the dark matter halo is defined  as the spherical region where the density is approximately 200 times the critical density of the universe at a given redshift, with mass given by:
\begin{equation}
M_{200}=\frac{4\pi}{3} R_{200}^3 \, 200 \rho_{\mathrm{cr}}\,,
\label{M200}
\end{equation}
where $\rho_{\mathrm{cr}}=3 H^2(z)/ (8\pi G)$. The virial velocity is given by the mean circular velocity at the virial radius, $V^2_{200} \equiv G M_{200} / R_{200}$. With that, one can express the evolution of the mass and virial radius with respect to $V_{200}$:  $M_{200}={V^3_{200}/10\,G H(z)}$ and $R_{200}=V_{200}/ 10 H(z) $. We can see from these expressions that halos that form early in the evolution of the universe are less massive, while late-forming halos are more massive and larger.

This definition is not unique and depends on the choice of the virial overdensity parameter\footnote{Not to be confused with the dimensionless power spectrum defined in (\ref{Eq.:dimensionless_power_spectrum}).}, $\Delta$, which above was taken to be $\Delta=200 \rho_{\mathrm{cr}}/\bar{\rho}$. More generally, (\ref{M200}) can be written as $M_{\mathrm{vir}}=\left( 4\pi/3 \right) R^3_{\mathrm{vir}} \, \Delta \, \bar{\rho}$. The values of $\Delta$ can vary in the literature, with some common definitions being $\Delta = 333$ at $z=0$ for a fiducial cosmology given by \citep{Planck}, which asymptotes to $\Delta=178$ at high-$z$ \citep{Bryan:1997dn}; or a fixed $\Delta=200$ at all redshifts, usually denoted by $M_{200\mathrm{m}}$.  
 
We identify the DM halos from numerical simulations, the N-body code P$^3$M~\citep{P3M}, and can extract from them the abundance of halos as a function of their mass for a given redshift.
The individual halos can also be analyzed in those simulation and the radial mass profile can be determined. A surprising feature encountered in those simulations is that halos appear to have a universal density profile, averaged over spherical shells. Their functional form is characterized by the Navarro, Frenk and White (NFW) profile \citep{NFW},
\begin{align}
\boxed{\rho_{\mathrm{NFW}} (r)= \frac{\rho_{\mathrm{s}}}{\left( r/r_{\mathrm{s}} \right) \, \left( 1+r/r_{\mathrm{s}} \right)^2} \rightarrow \left\{
  \begin{array}{lr}
    1/r \,, & \,  \,\,\mathrm{for }\,\,\,  r \ll r_{\mathrm{s}} \\
    1/r^3\,, & \, \,\, \mathrm{for }\,\,\,  r \gg r_{\mathrm{s}}
  \end{array}
\right. }
\end{align} 
where $r_{\mathrm{s}}$  is the radius where the slope of the profile changes and $\rho_{\mathrm{s}}=\rho(r_{\mathrm{s}})$. We can see that this profile diverges towards the center of the halo, presenting a cusp. The amplitude of the density profile can be written in terms of $R_{200}$, as we can see from (\ref{M200}),
\begin{equation}
\bar{\rho}=\frac{3}{4\pi R_{200}} \int_0^{R_{200}} 4 \pi r^2 \rho(r) \, dr = 3 \rho_{\mathrm{s}} \int_0^1 \frac{x^2}{cx \left( 1+cx \right)^2} \, dx\,, 
\end{equation}
where $x=r/R_{200}$, and $c := R_{200}/r_{\mathrm{s}}$  is the concentration index and describes the shape of the distribution. With that, the NFW profile can be determined completely by $R_{200}$ (or $M_{200}$ or any other halo radius definition), and the parameter $c$. The shape of the concentration can be inferred from the same P$^3$M simulation, where $c \propto \left( M/M^{*} \right)^{-1/9} \left( 1+z  \right)^{-1}$. We can see that, early-forming halos have a smaller radius, and they are denser than the larger ones, given the higher concentration. The NFW profile can be generalized for a three-parameters profile that  better fits the DM profile of halos for all ranges in mass  \citep{Einasto,Navarro:2003ew,Gao:2007gh}.

Above we presented the spherically averaged density profiles of DM halos, described by the NFW profile. Although this presents a good fit to DM N-body simulations (that assume spherical symmetry and use shells that are distributed radially) and some observations, halos are not spherical. From halo and large cosmological simulations~\citep{triaxial,triaxial_0,triaxial_1,triaxial_2,triaxial_3} we can see, however, that the majority of the DM halos are elliptical or triaxial, with their axis aligned with the cosmic web structure. This non-spherical structure and intrinsic alignment might come at formation of the halos from the tidal field. The halo triaxiality plays a crucial hole in the interpretation of lensing data, cluster morphology and Sunyaev-Zeldovich measurements (for reviews on this topic, see~\citep{review_intrinsic_1,review_intrinsic_2}, and needs to be taken into consideration.

\vspace{0.3cm}
\begin{tcolorbox}[colback=gray!10,enhanced,breakable,frame hidden,halign=justify]
\paragraph*{Surface brightness of galaxies \qquad} \mbox{}\\
Our capacity of observing galaxies is limited by the brightness of the sky. They can only be observed if their surface brightness, which is the brightness per area, is higher than the sky surface brightness\footnotemark  where the area $A$ is the area of the survey ($\mu _{\mathrm{B}} = 23 \, \mathrm{mag}/\mathrm{arcsec}^2$).  This can limit our understanding of the distribution of galaxies, making us miss the fainter ones. 
%The effective surface brightness is defined as~\citep{Binney_Tremaine}, in units of $\mathrm{mag/arcsec}^2$:
%\begin{equation*}
%S_p = M_{\odot} + 21.572 - 2.5 \log_{10} S\,,
%\end{equation*}
%where $S=m+2.5 \, \log_{10} A$ is the surface brightness in units of $L_{\odot}/\mathrm{pc}^2$, with $m$ the total magnitude in the area A, and $M_{\odot}$ is the absolute magnitude and $L_{\odot}$ the luminosity of the Sun. 
The surface brightness of a galaxy is described with respect to the radius $R$~\citep{Binney_Tremaine} as $S_p(R)=S_d \, \exp(R/R_d) \propto \exp(-kR^{1/m}) $, where $R_d$ is the disk scale length, and in the second equality we have the empirical S \'{e}rsic law. 
% that have lower contrast with respect to the background. 
Therefore, we have the following nomenclature for the galaxies with respect to their surface brightness.

\vspace{0.3cm}
    \textit{Low surface brightness (LSB) galaxies:} There is no formal definition for LSB galaxies, but in general they are disk galaxies that have surface-brightness smaller than $\mu_{\mathrm{B}}$. They are believed to make the majority of the galaxies in our universe, and most of the LSB galaxies are dwarf galaxies. However, this is not necessarily the case, with LSB galaxies being galaxies in a broad range of masses, and very diverse morphologies.  This low luminosity is likely associated to a small star formation rate in those galaxies. So those galaxies are believed to be DM dominated. Their rotation curves\footnotemark usually reach much smaller speeds than the ones from high surface brightness galaxies (see below), with a very slow rise before reaching the plateau region given their lower density, but broaden mass distribution.

 \vspace{0.3cm}
	\textit{High surface brightness (HSB) galaxies:} They are usually defined as galaxies that are brighter than $\mu_{\mathrm{B}}$. They are the usual galaxies we study. The rotation curves are known to reach high velocities with a steep rise, coming from the inner region that has a higher density of baryons with narrower mass distributions than LSBs, which is described by the Newtonian baryonic acceleration. This is followed by a Kleplerian fallout to the flat part of the rotation curve. 
\end{tcolorbox}
\footnotetext{The brightness of an object is a measure of the amount of light (luminosity) that we detect: $B=\mathrm{Luminosity}/4 \pi d^2$, where $d$ is the distance to the object. We use magnitude to measure the brightness of an object in a scale without units, and represented by $\mathrm{mag}$.}
\footnotetext{A rotation curve of a galaxy shows the change in the orbital circular velocity of stars or gas clouds  with respect to the distance from the center of the galaxy. An example of a rotation curves can be seen in~Figure \ref{cusp_core} for dwarf galaxies. Different types of galaxies present very distinct rotation curves, such as low or high surface brightness galaxies. This can be seen in Figure~\ref{Fig.:rotation_curves}. }

\subsection{Discrepancies in comparison with observations}

In this section we will show how some of the theoretical predictions from simulations of the small scales considering the $\Lambda$CDM model compare with respect to astrophysical observations. However, this comparison is not straightforward, since we indirectly probe the dark matter inferring it from the visible matter that traces the gravitational potential of galaxies and clusters. There are a few approaches to connect the information of galaxies and the dark matter halos like forward modelling, abundance matching and kinetic measurements, and each of those methods has its difficulties and limitations. The result of this comparison is a series of discrepancies that challenges the results of the simulations, and in some cases limitations in observations. We will present some of these challenges in this section.
Some of those challenges might have complementary origin and solution, and are indeed connected, as we will discuss bellow.  

\subsubsection{Cusp-core}

As we saw above, the expected density profile from colisionless simulations is the NFW profile which is cuspy towards the central region of the halo. Given the complex dynamics of baryonic matter in some galaxies, good laboratories to probe the halo structure are low surface brightness (LSB) galaxies and late-time dwarfs. Those systems are dominated by DM throughout their halo up until the central regions. 
Measuring the rotation curves of dwarf galaxies, \citep{Flores:1994gz,Moore:1994yx}  found that those measurements preferred cored isothermal profiles.  Many other measurements of the rotation curves of those systems \citep{RC1,RC2,RC3,RC4,RC5,RC6,RC7,RC8,RC9,RC10,RC11,RC12} have confirmed this discrepancy, showing that a constant density core with a profile with a slope $\gamma = 0 - 5$ (considering the profile at small radius given by $\rho \sim 1/r^\gamma$). The smallest values for this slope from dissipationless simulations  are too large in comparison to the ones obtained by observations.

\begin{figure}[htb]
\centering
\includegraphics[scale=0.16]{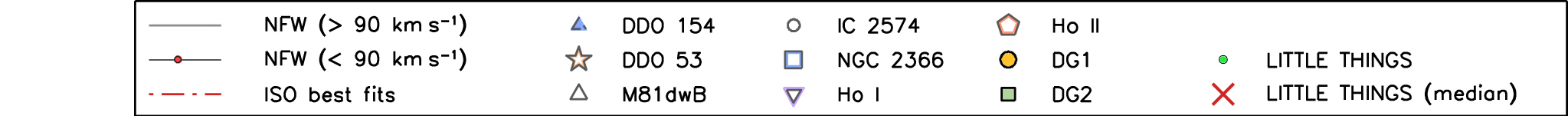}
\includegraphics[scale=0.16]{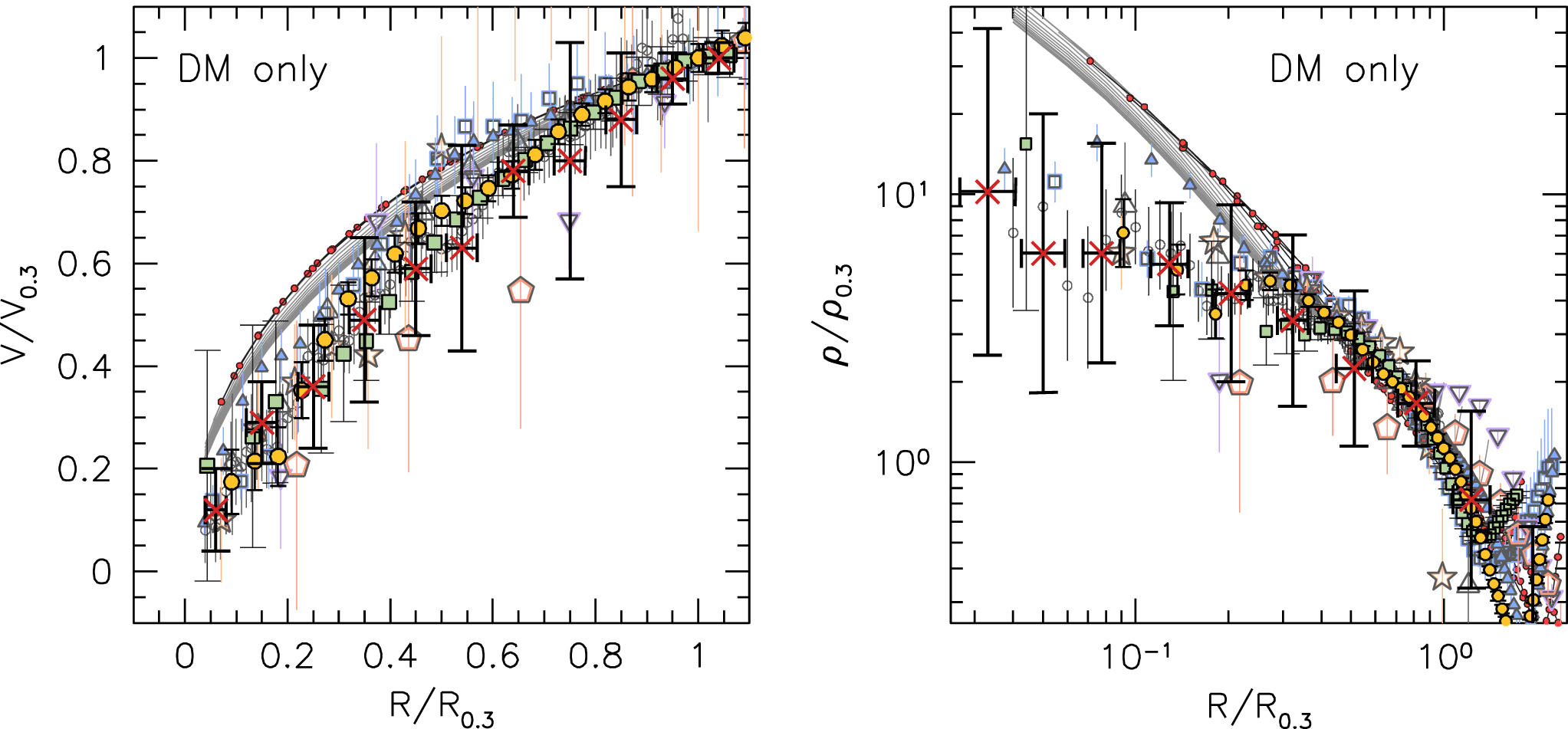}
 \caption{Figure adapted from~\citep{things}, showing the results from the THINGS and LITTLE THINGS surveys. The plot shows a comparison of the velocity versus radius (rotation curve) (\textit{left panel}) and density versus the radius (\textit{right panel}), normalized by $R_{0.3}$, and $V_{0.3}$ and$\rho_{0.3}$, from the theoretical parametrizations of the NFW potential (solid line) and the pseudo-isothermal (dashed line), with the simulated galaxies~\citep{Governato}, represented in the plots by the legend DG1 and DG2.  In the plots observational data from 7 dwarf galaxies measured by THINGS is represented by the other points. The crosses represent the median values of the LITTLE THINGS rotation curves and density profile.  We can see that the galaxies seem to follow a cored profile, while NFW predicts a cusp.}
\label{cusp_core}
\end{figure}

The  recent measurement of nearby dwarf galaxies from the survey THINGS (HI Near Galaxy Survey) \citep{things} and LITTLE THINGS \citep{little_things} confirmed this discrepancy. Measuring the rotation curves from 7 and 26 nearby dwarfs, they found that the inner slope is much smaller than the NFW one ($\gamma=-1$), with $\gamma = 0.29 \pm 0.07$ for the LITTLE THINGS survey, as we can see in Fig.~\ref{cusp_core}.

The situation is more complex for high surface brightness (HSB) objects given its complex inner density structure; or for galaxies with large mass, like spiral galaxies, where at small radii is dominated by baryonic matter. Even in the case of dwarf galaxies, it was pointed out that some systems present cuspy profiles, while others cored ones, presenting an unexpected diversity in the rotation curves \citep{diversity_dwarfs}. Since different results were obtained by different techniques for the same system, this shows that determining the inner slope of galaxies is a hard task. 

The origin for these discrepancies can come from the fact that the simulations take into account only DM, while the properties of galaxies are also influenced by the presence of baryons.  The newest hydrodynamical simulations obtained by many independent groups have shown that baryonic feedback can in fact soften the inner cusps in the profile and generate core-like profiles like the ones observed for dwarf galaxies. The main effects are supernova feedback flattening and dynamical friction from baryonic clumps (for a more detailed list of these and other baryonic processes, see \citep{DelPopolo:2016emo}).  These simulations show a threshold mass of $M_{\mathrm{vir}}\sim 10^{10}\, M_\odot$ below which the simulation predict profiles that are cusped \citep{Feedback1,Feedback2,Feedback3,Feedback4,Feedback5,Feedback6}.

However, not all simulations agree with this result. Additionally, modelling those baryonic feedback effects is challenging, and introduce many new parameters and uncertainties in modelling  assumptions. Finally, not all baryonic processes that might influence the formation and dynamics of galaxies  were included in the simulations, and that might reveal to be important for the result. It is clear that the inclusion of baryonic effects is hinting in the right direction, but until consensus is achieved, alternatives need to be considered. As mentioned before, a modification of the properties of DM might in a simple way account for that, as we will show for the case of Bose--Einstein condensate DM. An early solution to the cusp-core problem, and that explains the rotation curves with exquisite precision is a modification of the dynamics of gravity on small scales, the MOdified Newtonian Dynamics (MOND). This is also a solution for the regularity versus diversity challenge, and  its main points and shortcomings will be presented  at the end of this section.

\subsubsection{Missing satellites}

Structure formation is hierarchical in nature and it is expected that the DM halos are also populated by small subhalos. This is confirmed in $\Lambda$CDM simulations of Milky-Way sized halos, which show that the subhalo mass function diverges toward low masses, limited only by the numerical limit. Those simulations then predict several hundreds of subhalos with $v_{\mathrm{max}} \sim 10 - 30 \, \mathrm{km/s}$, that are large enough to host a galaxy ($M_{\mathrm{peak}} \gtrsim 10^7 \, M_\odot)$, where $M_{\mathrm{peak}}$ is the maximum virial mass the halos had when they formed. On the other side, until 2005 only 12 MW classical satellites were known, with 15 more confirmed ultra-faint satellite galaxies until 2014, with the data from Sloan Digital Sky Survey (SDSS) \citep{Drlica-Wagner:2015ufc}. To date, with the inclusion of Dark Energy Survey (DES) data, a few more ultra-faint candidates were discovered, with the known count of satellites of more than 50. However, the number of MW galaxies satellites is still much smaller than the number predicted from simulations. This is known as the missing satellites problem, and not only appears in the MW, but also in the Local Group.

DES and future observations are expected to discover more of those ultra-faint galaxies, which can alleviate this discrepancy, but there is still a debate if this will solve the problem. Another possibility is that low-mass subhalos are there, but we just cannot see them, since they have very low baryonic content.  One can expect that for low mass subhalos, galaxy formation is suppressed since the photoionizing background heats the gas, reducing its cooling rate and inhibiting gas accretion for $M_{\mathrm{vir}} \sim 10^9\, M_\odot$ \citep{UV1,UV2,UV3,UV4,UV5}.  Star formation is also suppressed since supernova-driven winds could strip the gas out of these halos \citep{Dekel}. Other mechanisms can also suppress the baryon content in the low-mass galaxies, see \citep{DelPopolo:2016emo}, like reionization suppression.  So, the visible subhalos are only a set of the entire distribution of halos that contains the non-visible faint end.  It was shown recently  in the hydrodynamical simulations APOSTLE~\citep{UV5,Solve1,Solve3} that apparently this mechanism can solve the difference in the number of predicted and observed satellite galaxies, thus solving the missing satellite problem. But the question remains if this process needs to be too finely tuned to solve the problem.
%However, there are still some missing  low-mass satellites.

\paragraph{Too big to fail} \mbox{} \\

\begin{figure}[htb]
\centering
\includegraphics[scale=0.13]{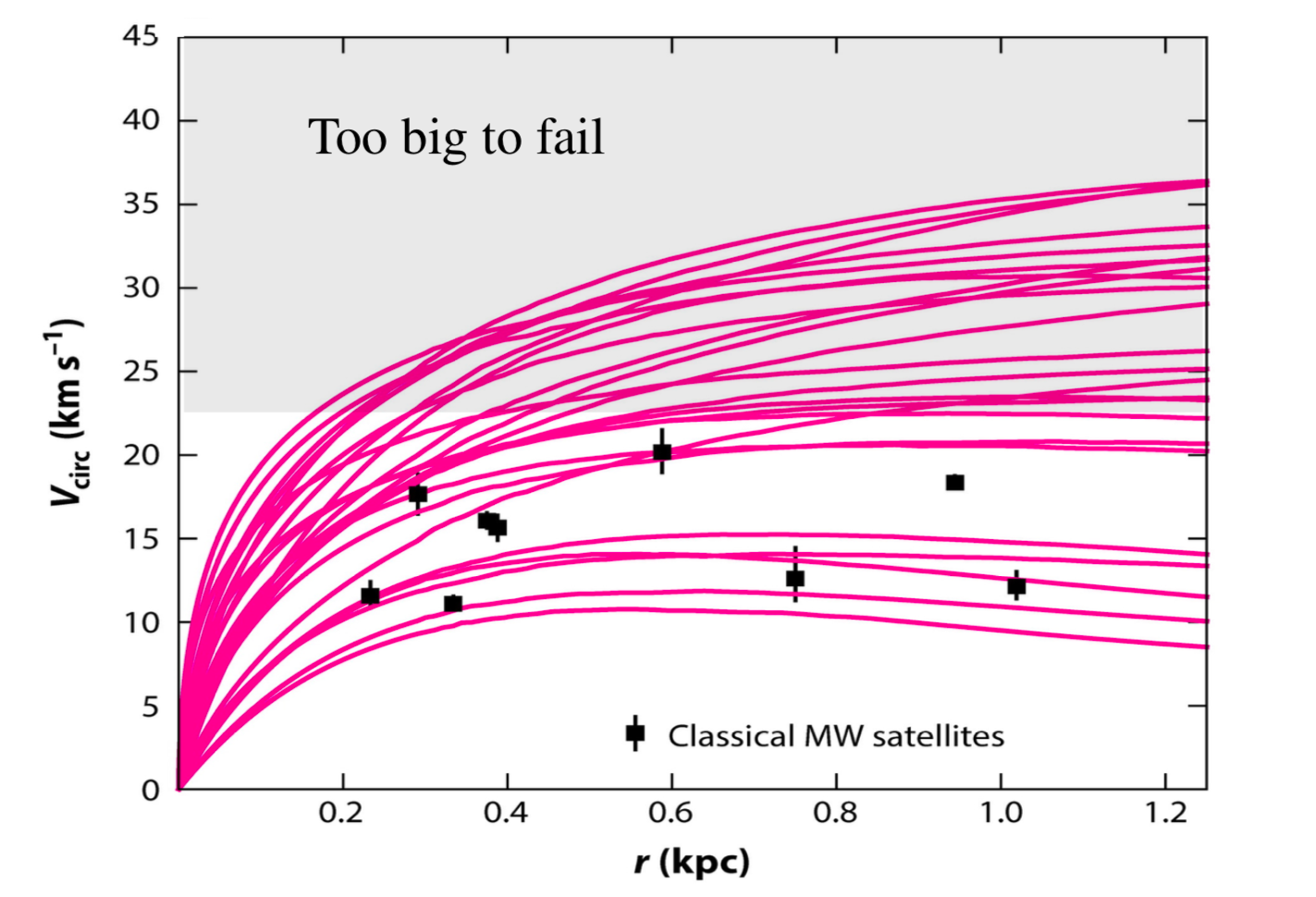}
 \caption{Figure adapted from~\citep{Bullock:2017xww}, showing the circular velocity versus the radius of specific subhalos from the Aquarius simulation that have $V_{\mathrm{max}} > 30 \mathrm{km/s}$ (magenta lines). Those are known to have halos that are very massive and expected to host the formation of starts. However, as we can see from the data points corresponding to classical MW satellites with masses $M \sim 10^{5} - 10^{7} \, M_{\odot}$, in the most massive of those subhalos, with $M > 10^8 \, M_{\odot}$ shown in the gray region of the plot, we do not observe satellites. This means that the more massive subhalos predicted by the simulations are too big to fail to form stars and galaxies.}
\label{Fig.:too big to fail}
\end{figure}

The above mechanism that could solve the missing satellite problem leads to another challenge: the too big to fail problem. When we say that the visible subhalos of the MW are only a set representing the most massive subhalos in the total distribution of subhalos, to have agreement with $\Lambda$CDM simulations these visible MW subhalos need to correspond to the most massive subhalos predicted by the simulations. But, the most massive subhalos predicted by those simulations have central masses\footnote{The central mass is equivalent to quoting $V_{\mathrm{max}}$ since $V_{\mathrm{circ}}^2=GM/R$, where the maximum circular velocity is defined as the peak of the rotation curve and it is a quantity less affected by tidal stripping \citep{Penarrubia:2007zx}.}($V_{\mathrm{max}} > 30 \mathrm{km/s}$) that are too large to host the observed satellite galaxies \citep{too_big1,too_big2}, and the ones that have central mass like the expected by the MW (with $12 < V_{\mathrm{max}} < 25 \mathrm{km/s}$) are not the most massive ones. 
%There is a mismatch between the central masses from simulations and the observed ones. 
So, the puzzle is why should the most massive subhalos, where the gravitational potential is the strongest and the striping gas mechanisms cited above are not important, be too big to fail to form stars and galaxies? This is illustrated in Figure~\ref{Fig.:too big to fail}.
This problem also appears in the galaxies in the Local Group and Local Volume \citep{too_big4,too_big5}, so it is not a specific property of the MW.

This higher central mass from the most massive sub-halos predicted by simulations in comparison to the MW dSphs, seems to be a more general feature that appears in simulations. This discrepancy might indicate a more delicate issue  related to the internal structure of the sub-halos. In this way, the too big to fail problem is more than just the problem related to the missing satellites problem, as stated above.

Like for the other problems, it was proposed that some astrophysical processes driven by baryons could be important on those scales and solve the too big to fail problem. However, these solutions seem to only work for the MW and for very efficient feedback, like the supernova feedback that only solves the too big to fail problem if very efficient. This is an intense topic of debate and no consensus appears to have been reached.  As these notes were being written, there has been claims that the the too big to fail problem has been solved \citep{too_big_solve}.

As for the cusp core problem, different DM physics could solve those problems by having a mechanism that suppresses the formation of small scale subhalos, and that reduces the central densities of massive subhalos (or modifies the dynamics of the central regions). We are going to show how the models with Bose--Einstein condensation address some of those problems.

\subsubsection{Diversity vs. regularity: scaling relations}

Although our universe came from very smooth initial conditions, nowadays the diversity of galaxies that we find in the universe is extraordinary. This incredible diversity of galaxies, though, presents a surprising regularity. This fact is manifest in several scaling relations, that are shown to hold very tightly for a diverse range of galaxies.
These relations relate the dynamical and baryonic properties of galaxies, and hold even for DM dominated systems, and they are one of the most tantalizing aspects of galaxy phenomenology, representing the most pressing challenge for $\Lambda$CDM on small scales.

\begin{figure}
\centering
\includegraphics[height=0.55\textwidth , width=0.4\textwidth]{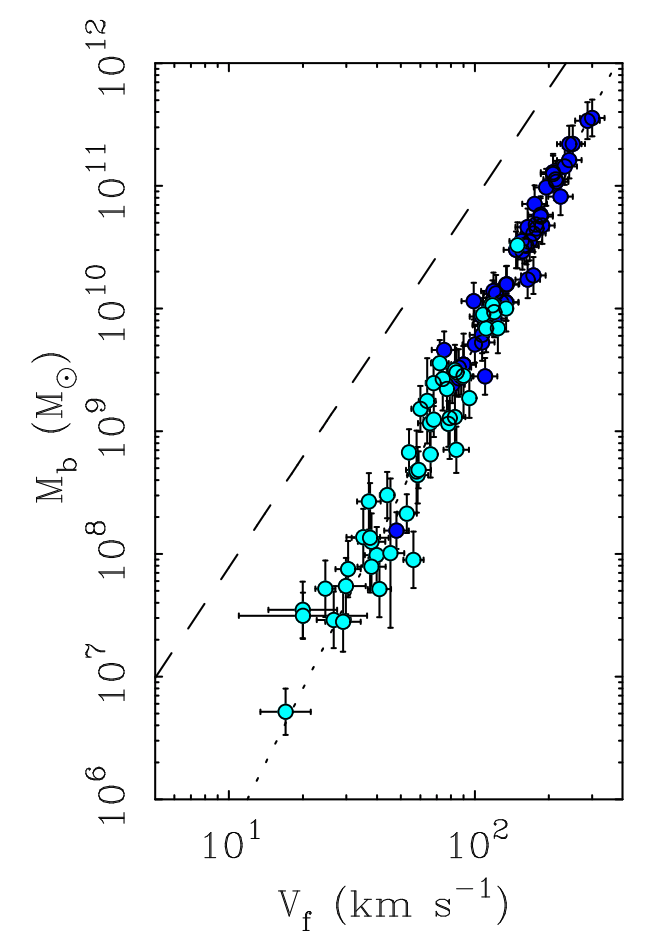}
\hspace{0.6cm}
\includegraphics[height=0.525\textwidth, width=0.5\textwidth]{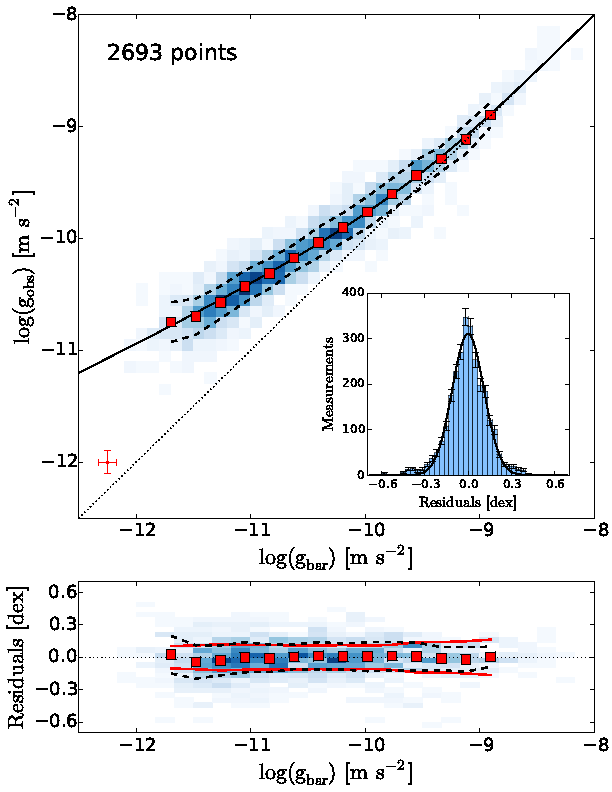}
 \caption{\textit{Left panel:} The figure shows the Baryonic Tully--Fisher Relation (BTFR) from \citep{MOND7}, which shows the relation between the baryonic total mass ($M_{\mathrm{b}}$) and the asymptotic circular velocity. 
Dark and light blue points represent star and gas dominated stars, respectively. The dashed line represents the relation expected for $\Lambda$CDM, with slope equal to 3; while the dotted line which better fits the data, has slope 4. \textit{Right panel:} Plot of the Radial Acceleration Relation from \citep{McGaugh:2016leg}, for 153 SPARC galaxies. The fit to the data is given by the solid line while the dotted line is the unit line. The insert is a histogram of the residuals. The red uncertainty bars represent the uncertainty in each individual point. The lower panel shows the residuals, and the red uncertainty bar shows the mean uncertainty on individual points. The dashed lines represent the rms value in each bin and the solid red lines represent the observational uncertainties and variation between the stellar mass-to-light-ratio from galaxies.}
\label{Fig.BTFR_RAR}
\end{figure}

The most famous of those relations is the Baryonic Tully--Fisher relation (BTFR) \citep{McGaugh:2005qe,McGaugh:2007fj}, which relates the total baryon mass (including stars and gas) of the galaxy to the asymptotic circular velocity in galaxies, $V_{\mathrm{f}}$ (this is the velocity measured at the flat portion of rotation curves):
\begin{equation}
V_{\mathrm{f}}^4=a_0 G M_{\mathrm{b}}\,,
\end{equation}
where $a_0$ is the critical acceleration, a scale that appears in  observations. Its value can be obtained from the data and given by $a_0 \sim 1.2 \times 10^{-8} \, \mathrm{cm/s}$. The BTFR expands the regime of validity of the Tully Fisher relation which relates the luminosity, instead of the total mass, to the circular velocity. Luminosity is a probe of the stellar mass, and in the BTFR the observed gas mass is also considered on top of the stellar mass.  This extends the validity of the scaling relation by many decades in mass. This empirical scaling relation is shown to hold for large ranges of masses, $6$ generations, with a very small scatter, compatible to the size of the error bars.  The left panel of Fig.~\ref{Fig.BTFR_RAR} presents the BTFR. As we can see, the slope of the BTFR is different from the one predicted by $\Lambda$CDM, $V_{\mathrm{f}}^3 \propto M_{\mathrm{b}}$, shown by the dashed line. 

%The mean uncertainty on individual points is illustrated in the lower left corner. Large squares show the mean of binned data. Dashed lines show the width of the ridge as measured by the rms in each bin. This extrinsic scatter closely follows the observed rms scatter (dashed lines): the data are consistent with negligible intrinsic scatter. The solid red lines show the scatter expected from observational uncertainties and galaxy to galaxy variation in the stellar mass-to-light ratio.

There is another general scaling relation that also displays the interesting behaviour of galaxies: 
the mass discrepancy acceleration relation (MDAR). This is more general since the BTFR can be obtained from the MDAR at large distances in the disk. The MDAR is a relation between the gravitational acceleration from baryons alone ($g_{\mathrm{bar}}$), from the distribution of gas and stars in galaxies \citep{McGaugh:2016leg,Lelli:2017vgz}, and acceleration inferred from rotation curves ($g_{\mathrm{obs}}=V^2/r$). As it can be seen in the right panel of Fig.~\ref{Fig.BTFR_RAR}, this scaling relation shows a remarkably tight correlation between these quantities for very diverse and large number of galaxies
% which can be seen by the comparison with the uncertainty of each point in the lower panel.
This can be seen by comparing the interval determined by the solid red lines and the uncertainty in each individual point, represented by the red uncertainty bars on the top figure, with the dashed lines show that the data is compatible with negligible scatter.

This relation shows us that in regions of high acceleration, where $g_{\mathrm{obs}} > a_0$ and baryons dominate, one has $g_{\mathrm{obs}} \sim g_{\mathrm{bar}}$. For low accelerations, in the central regions where it is expected to be DM dominated, this relation deviates from the unit line.  This suggests a very curious behaviour:  the baryon mass distribution dictates the behaviour of the rotation curve at all radii, even for the regions expected to have less baryons. And this behaviour holds even for galaxies that are DM dominated.

%In plain DM parlance, the MDAR implies that by looking only at the baryon mass distribution, one can infer the DM density profile at every radius in the galaxy, even in galaxies where baryons are everywhere subdominant.  

These empirical relations, coming directly from observations, show the surprising feature that in galaxies the dynamics is dictated by the baryon content, even when DM dominates. Even more unexpected these relations are very tight, showing very little spread, even if they come from very diverse types of galaxies. 
As pointed out in \citep{Bullock:2017xww}, in these correlations what dictates the dynamics is the baryon mass, which is the sum of gas and stars, and not only the stellar mass, which is the one that is expected to correlate more with the total feedback energy.  

Recently it was shown by many groups that these relations can be explained within the $\Lambda$CDM paradigm \citep{Navarro:2016bfs,BTFR_EAGLE,Garaldi:2017fse,Dutton:2019gor,Navarro:2018yuk} using the latest hydrodynamical simulations like EAGLE \citep{EAGLE1,EAGLE2}, APOSTLE \citep{UV5}, Illustris \citep{Illustris}, ZOMG \citep{ZOMG1,ZOMG2,ZOMG3}, and NIHAO \citep{NIHAO}. Those simulations include several baryonic effects (like star formation, stellar evolution, metal enrichment, gas cooling/heating, galactic outflows and BH feedback, among others) to their $\Lambda$CDM simulations\footnote{This review will not enter in the details of such baryonic effects that are taken into account in those simulations. This is a field of its own, very rich and fast developing, and discussing those effects is not the scope of this review.}.  Those new large volume and high resolution simulations, like Illustris and EAGLE, have also been able to reproduce the features of the rotation curves of galaxies within $\Lambda$CDM. This large amount of progress in the simulation side is very encouraging.

However, some questions still remain. While the BTFR and the MDAR trends can indeed be reproduced by those simulations, it is pointed out by most of the authors that the  scatter obtained in the scaling relations is larger than the one expected from data (some authors claim that this spread is correlated with the errors in the stellar feedback). The question remains though whether this is a matter of improving the feedback models and/or resolution of the simulation, or if given the stochastic nature of the feedback effects, they will ever be able to give such tight correlations. Another point that is important to be answered is about the importance of these baryonic feedbacks, since these groups do not agree on how very sensitive to the feedback model the simulations are, which is intriguing. 
%A question remains then if this is a matter of improving the feedback models and/or resolution of the simulation, or if given the stochastic nature of the feedback effects, will they ever be able to give such tight correlations. Another point is that is important to be answered is about the importance of these baryonic feedbacks, since some of those authors claim that the results are not very sensitive to the feedback model, which is intriguing. 
The way those effects are introduced in the simulations is by parametrizing their effects, instead of introducing all of these feedback mechanisms from first principles, which is understandable given the complexity of each of those phenomena. This includes many new parameters to the simulations. And different simulations might use different parametrizations.
Those simulations also still do not go all the way until dwarf galaxies\footnote{To my knowledge. But as I said, it is a fast moving field.}, which are DM dominated and where most of the tension is. In summary, this is a very challenging and exciting field and a lot of progress has been done on the simulation side with results that are very encouraging to explain the formation and dynamics of galaxies. But there are some uncertainties in those results and the simulations still do not fully reproduce the observations. 

%%%%%%

\subsubsection{What the small scales tell us}

As we saw above, the small scales hold precious information that can help us understand astrophysical processes, or even the nature of DM. This is revealed by the challenges presented above, which show  rich dynamics on galactic and sub-galactic scales. There are a number of ways that these discrepancies can be addressed. Within $\Lambda$CDM, this can done by including baryonic effects, which as we saw in the previous sections seem to address partially or completely some of those puzzles.  Another proposal for solving some of the puzzles of galactic evolution is more radical and proposes a universe without DM that has a modified force law for small accelerations, the MOdified Newtonian Dynamics (MOND). See box bellow for a discussion of MOND.

A third avenue is to modify the DM paradigm. 
Different models of dark matter can affect the formation of structures in distinct ways, both in the linear and in the non-linear regimes. Therefore, the small scales offer an opportunity to probe the microphyics of DM, beyond the hydrodynamical large scale CDM paradigm. The non-linear regime can be specially changed by modifications of this paradigm, as we can see in Fig.~\ref{Fig.:Small_scale_power_spectrum}. This regime can be probed using galaxies, and for even smaller scales satellite galaxies and studying substructures. This could help find new properties of DM, that could help elucidate its nature. 

%%%%%%
\vspace{0.3cm}
\begin{tcolorbox}[width=\textwidth, enhanced, breakable,  valign=center, colback=white, colframe=black, sharp corners, shadow={0pt}{0pt}{0mm}{black},boxrule=0.5pt,halign=justify,overlay first={
        \draw[line width=.5pt] (frame.south west)--(frame.south east);},
    overlay middle={
        \draw[line width=.5pt] (frame.south west)--(frame.south east);
        \draw[line width=.5pt] (frame.north west)--(frame.north east);},
    overlay last={
        \draw[line width=.5pt] (frame.north west)--(frame.north east);
}]
\paragraph*{\textit{MOND empirical law} - \hspace{0.5cm}} \label{Sec.:MOND}

Milgrom, in~\citep{MOND1,MOND2,MOND3}, motivated by the scaling relations and rotation curves of galaxies, made a remarkable observation about the mass discrepancy in galaxies.  He observed that the mass discrepancy can be determined by the observed baryonic matter, and can be described by the simple empirical law,
\begin{equation}  \label{MOND}
a= \left\{ \begin{array}{lr}
    a_{\mathrm{N,b}} \,, & \,  \,\,\mathrm{for }\,\,\,  a_{\mathrm{N,b}} \gg a_0\,, \\
    \sqrt{a_0\,a_{\mathrm{N,b}}}\,, & \, \,\, \mathrm{for }\,\,\,  a_{\mathrm{N,b}} \ll a_0 \,,
  \end{array} \right.
\end{equation}
where $a_{\mathrm{N,b}} = G M_{\mathrm{b}}(r)/r^2$  is the Newtonian acceleration due to baryons. The scale $a_0$ appears naturally from observations, like we saw in the previous subsection, and its value can be fitted by the data\footnotemark giving $a_0 \sim 1.2 \times 10^{-8} \, \mathrm{cm/s}^2$. This scale separates the regimes where the centripetal acceleration experienced by a particle is given purely in terms of the Newtonian (baryonic) acceleration at large acceleration, and at small acceleration, by the geometric mean of $a_{\mathrm{N,b}}$ and $a_0$. 

The relation works very well fitting the rotation curves of galaxies, both HSB and LSB galaxies.
LSB galaxies (which were predicted by Milgrom), are DM dominated, or in the language of MOND, have low accelerations, given their low stellar surface density. 
It is also remarkably successful in explaining the empirical scaling relations (for a review see \citep{MOND4,MOND5,MOND6,MOND7}). 

More importantly, this empirical relation reveals a very interesting and curious fact. It seems that the dynamics in galaxies is driven by the baryons, even for galaxies that are DM dominated. This seems to indicate a long range correlation between baryons on galaxies. 

 \vspace{0.3cm}
This fact made Milgrom think that a fifth force was responsible for this correlation, instead of DM, and that this relation could come from a modification of gravity at those scales. 
In order to try to get the empirical law (\ref{MOND}) as a modified gravity theory, \citep{Bekenstein:1984tv} described an effective theory for MOND, which we will call full MOND. This can be accomplished by having a scalar field coupled to gravity with effective Lagrangian,
\begin{equation}
\mathcal{L}_{\mathrm{MOND}} = - \frac{2M_{\mathrm{pl}}^2}{3 a_0} \left[ \left( \partial \phi \right)^2 \right]^{3/2}+\frac{\phi}{M_{\mathrm{pl}}} \rho_{\mathrm{b}}\,,
\label{L_MOND}
\end{equation}
which represents a scalar field with a non-canonical kinetic term that is conformally coupled to matter. This Lagrangian, for static and spherically symmetric source, results in a modified Poisson equation
\begin{equation}
\nabla \cdot \left( \frac{|\nabla \phi|}{a_0} \nabla \phi \right) = 4\pi G \rho \,, \qquad \phi '=\sqrt{a_0 \frac{G M(r)}{r^2}}= \sqrt{a_0 a_{\mathrm{N,b}}}\,
\end{equation}
where in the second equation the spherical symmetry was assumed with ``$'$'' denoting derivative with respect to the radial coordinate. This theory describes that on top of the Newtonian force, there is a scalar field mediated force, which is given by the MONDian acceleration. This is a simplified version of their theory, since in their theory they have a way of making an interpolation between the different regimes. This theory also has a fractional power kinetic term, which might be problematic. 

With the current precise observations on large scales, specially from the CMB anisotropies and lensing observations, any theory that does not have DM is not compelling. Indeed, this is a problem for theories without DM, in particular MOND, since the measurement of the third peak of the CMB anisotropies~\citep{Skordis:2005xk,Skordis_2}.
%The consensus in the community is that DM must be present.
This full MOND theory cannot explain galaxy clusters, since it does not predict an isothermal profile. Many attempts were made to extend MOND, by including DM, to try to explain the observation on scales larger than galactic, or extending it to relativistic regimes (see reviews cited above). 

However, the empirical relation (\ref{MOND}) is incredibly successful. That alone, without the assumptions of full MOND (no DM and modified gravity), even in the context of $\Lambda$CDM, is a powerful statement about how DM is distributed in galaxies: in regions where baryons dominate, the theory behaves like Newtonian theory, and in regions where the DM dominates, the DM mass is uniquely determined by the baryonic distribution, $G M(r)/r = \sqrt{G M_{\mathrm{b}}(r) a_0}$.

Given the shortcomings of the full MOND, but the great successes of the empirical law, instead of trying to obtain this theory from a fundamental Lorentz invariant theory, the idea is to obtain the MOND dynamics from a theory of DM. In this way, MOND dynamics emerges only at galactic scales while maintaining the CDM behaviour on large scales. This is achieved in the theory of DM Superfluid that will be presented in Section 4.
\end{tcolorbox}
\footnotetext{A funny numerical coincidence is that the measured scale $a_0$ is related to the Hubble parameter today,  $a_0 \approx c H_0/2\pi$, which is natural units yields $a_0 \sim H_0$. Does this indicate something?}

\vspace{0.5cm}

The goal of this section was two-fold. First, we wanted to give an overview of the so called small scale problems of the CDM, which was the motivation for some ULDM models to be proposed. We wanted to introduce the problems in a way that the reader can understand why the mechanisms proposed by the models in Section 4 address and solve each of those small scale controversies. For example, we are going to see that the FDM model has to have a certain range of mass in order to solve the cusp-core problem and the satellite problem; or that the DM superfluid model has a modified dynamics of small scales, reproducing the MOND behaviour, which explains the rotation curves of galaxies and the scaling relations. 

The second goal, and perhaps the most important was to give an brief overview of the rich astrophysics that takes place in galaxies, introducing important concepts and observations available on these scales. This is important since the main feature of the ULDM models is to present a new phenomenology on small scales coming from the non-CDM behaviour of ULDM. In this way, the  small scale observables offer an important window to test the nature of DM.

%%%%%%%%%%%%%%%%%%%%%%%%%%%%%%%%%
% Bose-Einstein Condensation and Superfluidity
\section{Bose-Einstein Condensation and Superfluidity} \label{Sec.:BEC_superfluids}

%%%%%%%%%%%%%%%%%%%%%%%%%%%%%%%%%%%%
% New text

In this section, we present a short review of Bose--Einstein condensation (BEC) and superfluidity. The goal of this section is to give an introduction to the basic theory, properties and the methods used to describe those systems, so they can be applied for the case of DM in the next section. The different description of those systems and their limit of validity are very important to be able to understand the construction and validity of the DM models presented next and why they present different phenomenologies and astrophysical consequences.

\vspace{0.5cm}

Bose--Einstein condensation is one of the most fascinating phenomenon of quantum mechanics. Since it was theorized in the year of 1920s, by Satyendra Nath Bose and Albert Einstein, its experimental realization opened the door for many advances in the physics of many-body systems, and even to the application of this phenomenon in other fields like cosmology. Its first experimental realization was done in 1995 by two independent groups using  laser and magnetic cooling device to cool down rubidium atoms gas \citep{Anderson:1995gf,Davis:1995pg}. Nowadays, BECs are observed in helium, ultra-cold atomic gases, quasi-particles in solids, multi-component (mixtures) of BECs, among other systems.

Following on the works of \citep{Bose}, which described the quantum statistical properties of photons, Einstein extended this concept to a gas of non-interacting particles of integer spin, later called bosons as a tribute to Bose, that follows a Bose--Einstein statistics~\citep{Einstein}. This Bose gas has the property that at low temperatures a large number of these bosons, described then as quantum oscillators, condense into the lowest momentum state, exhibiting long range coherence. This physical phenomenon initiated the idea of Bose--Einstein condensation. 

A BEC is defined as a system where  at very low temperatures a large fraction of the bosons of the system occupy the lowest energy state of a configuration. This \emph{macroscopic occupancy} of the ground state is an inherently quantum mechanical phenomenon. Physically we can interpret it as a consequence of the wavelike nature of these particles at low temperatures, where the de Broglie wavelength of these bosons is larger than the inter-particle separation, and their wavepackets superpose and form a coherent macroscopic wavefunction describing the entire system. The BEC is then described by a single wavefunction of the system,  linking to the long range coherence property of a condensate.

A few years after BEC was theorized by Einstein, another intriguing macroscopic quantum mechanical phenomenon was discovered: superfluidity. In 1937, \citep{Sf1} and independently \citep{Sf2}, conducting experiments with helium-4 realized that after cooling down this liquid to a certain temperature, the fluid starts flowing without friction, even climbing the walls of the container where it was stored. Fluids that exhibit this behaviour, characterized by a zero viscosity,  are called superfluids. Landau provided a phenomenological description of this effect which rendered him the Nobel prize in 1962. It was proposed by \citep{London}, after the development of laser cooling techniques for atomic gases, that the properties of He$^4$ superfluid are related to BEC. This was not obvious given that the (textbook) description of BEC as an ideal non-interacting Bose gas, contrary to $^4$He that is a strongly interacting fluid. This gave relevance to the, until then, only theoretical ideas of Einstein, and BEC became a rich topic of research.
The relation between superfluids and BECs was confirmed years latter in ultracold atomic gases where almost the entire fluid at low temperatures is condensed and exhibits superfluidity. 

It is very challenging to describe the strongly interacting helium system. A weakly interacting Bose gas was then proposed by Bogoliubov, as a modification of the non-interacting Bose gas model, in order to study the Bose--Einstein condensation and superfluidity. 
In this way, superfluids can be modelled by a Bose--Einstein condensate that has self-interaction, and superfluidity is described as being achieved through interactions in a BEC. 
Notice that BEC can happen even in the absence of self-interaction, as seen above, since it is a statistical property of a gas of bosons in low temperature, but this system does not exhibit superfluidity. 
The weakly interacting theory is used to describe many superfluid systems at certain limits. This description, tough, evolved in the last few years in order to extend and generalize this framework to finite temperature systems, mixtures and even stronger interacting system corrections. New frameworks also emerged in order to describe different systems that cannot be modelled by the weakly-interacting theory. 
One of those ideas based on the hydrodynamical description is to write these systems as an effective field theory (EFT) in order to describe the system macroscopically using symmetry alone without the need of working its microscopic description. This EFT, depending on the symmetries of the system, can recover at some limits the weakly interacting superfluids, but also can be used to describe more general superfluids, superconductors and even systems like the unitary Fermi gas \citep{UFG1,UFG2}, which is a gas of fermions that interact through a strong 2-body coupling that is a superfluid in the ground state.

The theoretical description of those condensed matter systems, together with the experimental efforts, is a field of research that is in fast development. In this review we are going to describe the basic concepts on BEC and superfluid, and detail the different descriptions and properties of these systems. We start by describing the non-interacting ideal gas, where condensation was first conceptualized in order to present   in more detail the definition and the conditions for condensation. We then start to describe superfluidity.  We show first the definition of superfluidity as defined in Landau's theory of superfluidity.  We then describe a more concrete model for a BEC where superfluidity is present, the weakly interacting Bose gas. This is the simplest example of superfluidity. We show how this model describes condensation and superfluidity. We then follow to show the field theory description of the superfluid, where the system is described as a system which undergoes spontaneous symmetry breaking caused by the condensate. This description brings advantages and makes clear the study of many features in BECs and superfluids. As a low energy description of the superfluid, we present the EFT of superfluids as another description of more general superfluid systems. We finalize describing what happens when we rotate a superfluid, showing  the nucleation of vortices upon rotation.

Not linked to what we will discuss in the review, but an important  fact. Nowadays, it is known that superfluidity is not necessarily linked to condensation. Recent investigations seem to point that there are states where you can have superfluidity for the majority or all the particles in the system, while only a small fraction is condensed. This happens for example for liquid helium below a certain temperature.

%%%%%%%%%%%%%%%%%%%%%%%%%%%%%%%%%%
\subsection{Non-interacting ideal gas} \label{Sec.:ideal_BEC}
%%%%%%%%%%%%%%%%%%%%%%%%%%%%%%%%%%

We start our discussion with the non-interacting Bose gas. The properties of this system are a  consequence purely from the quantum statistic of indistinguishable bosons. We will see that in the grand canonical ensemble we can write the Bose distribution function in which we can see the conditions for Bose--Einstein condensation.

We want to describe here a theoretical gas of many non-interacting bosons in a box. In a system with  a large number of particles ($N$), it is impractical to try to determine the state of each particle or even the collective many-body wavefunction that describes this system $\Psi(\mathbf{r}_1, \cdots, \,\mathbf{r}_N,t)$. In this sense, to describe a system with many particles that can occupy many different states, we represent the system using a statistical ensemble description. To describe the state of this collective system, one does not need to label the state of each particle, but to determine the number of particles in each state of the system. The ensemble that is specially convenient for this task of deriving the probability of microscopic states is the \emph{grand canonical ensemble} (GCE). Since our system is composed of bosons, which are indistinguishable particles $\Psi(\mathbf{r}_1,\, \mathbf{r}_2, \cdots, \, \mathbf{r}_N,t) = \Psi(\mathbf{r}_2,\,\mathbf{r}_1, \cdots, \, \mathbf{r}_N,t)$, called Bose symmetry, this ensemble is useful to describe the system where many particles can occupy the same state.

The GCE is a statistical ensemble that describes a system that is in contact in thermal and chemical equilibrium with a large reservoir, in a way that there is exchange of energy and particles with the reservoir. This exchange of particles with the reservoir makes the number of particles in the system to fluctuate, although the number of particle of the system plus reservoir is constant. As the system is in equilibrium the energy and particle number fluctuate around an average.
%The total number of particles and energy of the system and reservoir is conserved, but the number of particles $N$ of the system can change.  
This ensemble can be described by the following constants: the chemical potential ($\mu$) and the temperature ($T$), which hold for a system (of volume $V$). If the GCE is applied to small systems, an additional condition is necessary: that the gas is diluted.  In principle, the probability of finding the system in a state $s$ with energy $\epsilon_s$ and $n_s$ particles, or occupation number $n_s$, is given by:
\begin{equation}
P_\mathrm{s} =\frac{1}{Z^{\tiny{\mathrm{GC}}}_s} e^{\beta \left( \mu \, n_s -  \epsilon_s \right)}\,, \qquad \qquad \mathrm{with} \qquad Z^{\tiny{\mathrm{GC}}}_s = \sum_{s} e^{\beta \left( \mu \, n_s -  \epsilon_s \right)}\,,
\label{Bose}
\end{equation}
where $\beta=1/(k_{\mathrm{B}} T)$, with $k_{\mathrm{B}}$ the Boltzmann constant. The chemical potential $\mu=\left( \partial E / \partial N  \right)_{S,V}$ is the energy required to add one particle to the isolated system, fully determined by $N$, the total number of particles, and $T$. The chemical potential is defined to be negative (so no unphysical negative occupation occurs). The total energy of the system is given by $E=\sum_s  n_s \epsilon_s$.
The normalization $Z^{\tiny{\mathrm{GC}}}_s$ is the \emph{grand canonical partition function}.

With the GC distribution function, we can then evaluate the average occupation number,
\begin{equation}
 \langle n_s \rangle =\sum_{n_s} n_s P_s =  \frac{1}{e^{\beta \left( \epsilon_s -\mu \right) } - 1} \,,
 \label{Eq.:Bose_Einstein_statistics}
\end{equation}
where the sum converges for $\mu < \epsilon_s$. This is the \emph{Bose--Einstein distribution}. This gives us the total number of particles in the system:
\begin{equation}
N= \sum_s n_s =\sum_s \frac{1}{e^{\beta(\epsilon_s -\mu)}-1} \,.
\end{equation}

We can separate the total number of particles into two contributions,
\begin{equation}
N=N_0 + N_{\mathrm{T}}\,, 
\label{number_BEC}
\end{equation}
where $N_0=1/e^{\beta(\epsilon_0 -\mu)}-1$ is the number of particles with $\mathrm{s}=0$, which is the number of particles in  the condensate, with $\epsilon_0$ indicating the lowest energy of the single particle state.  The number of particles that are not in the ground state, not in the condensate,  also called the thermal component of the gas, $N_{\mathrm{T}}={\textstyle\sum}_{s \neq 0} n_s$. We can replace the sum for an integral and, from the partition function, the thermal component is given by, 
\begin{equation}
N_{\mathrm{T}}=\frac{V}{\lambda_{\mathrm{T}}} \int_0^\infty d\epsilon \,  \frac{\epsilon^{1/2}}{e^{ \beta (\epsilon - \mu)}}\,,
\label{Eq.:NT}
\end{equation}
where $\lambda_{\mathrm{T}} = \sqrt{2\pi \hbar^2 \beta}$. For a fixed temperature $N_{\mathrm{T}}$ reaches a maximum when $\mu=\epsilon_0$. So, $N_{\mathrm{T}}$ is limited, meaning that in this limit  there is a finite number of particles not in the ground state. At this same point in this limit, $N_0$ can diverge showing that the number of particles in the ground state grows becoming macroscopically occupied. This macroscopic occupation of the ground state is seen as a condensation and this phenomenon is called \emph{Bose--Einstein condensation}. 

\begin{figure}
 \centering
\includegraphics[width=0.8\textwidth]{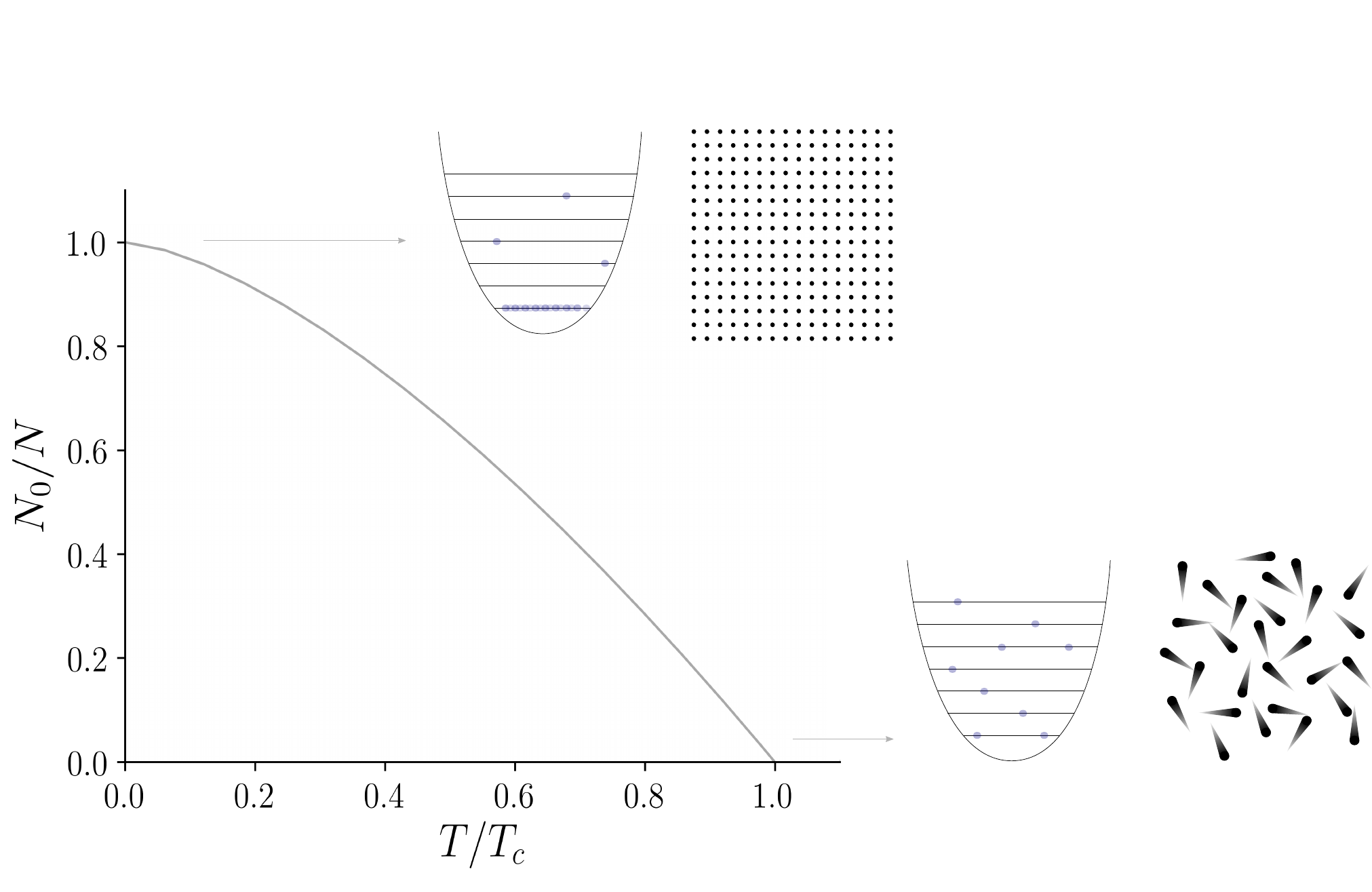}
 \caption{In this figure we plot the number of particles in the ground state, normalized by the total number of particles, with respect to the temperature for the non-interacting Bose gas. We show schematically that for higher temperatures $T\geq T_{\mathrm{c}}$ the system is in the normal state where the particles behave as free particles and occupy all energy levels. As the temperature is lowered,  when $T<T_{\mathrm{c}}$ we have the formation of a condensate  described by a macroscopic wavefunction. When $T=0$, all particles of the system are in the ground state and we have a pure BEC, described by a single wavefunction. }
\label{Fig.:BEC}
\end{figure}

The critical  temperature $T_{\mathrm{c}}$ defines the temperature below which there is the formation of the BEC. We can define it as the temperature above which all the particles of the system are not going to be in the condensate: $N_{\mathrm{T}} (T_{\mathrm{c}}, \mu=\epsilon_0)=N$. The chemical potential can be zero at $T_{\mathrm{c}}$, and from (\ref{Eq.:NT}), for the maximum $\mu=\epsilon_0=0$, we can get that
\begin{equation}
T_{\mathrm{c}} = \left( \frac{2 \pi \hbar^2}{m k_{\mathrm{B}}} \right) \left( \frac{n}{\zeta(3/2)} \right)^{2/3}\,,
\end{equation}
where $n$ is the total number density and $\zeta(3/2)\approx 2.612$ is the Riemann's zeta function. With that, for $T<T_{\mathrm{c}}$, we expect that most of the particles are going to be in the condensate,  and the number of particles in the condensate is:
\begin{equation}
\boxed{N_0/N=  1-\left( \frac{T}{T_{\mathrm{c}}} \right)^{3/2} \,.}
\end{equation}
Complementary, the number of particle in the thermal component is $N_{\mathrm{T}} = N \left( T/T_{\mathrm{c}} \right)^{3/2}$.
From that expression we can see that the occupation becomes macroscopic towards small temperatures, as we can see in Fig.~\ref{Fig.:BEC}. This indicates the formation of a BEC. This condition that a BEC  can form for $T<T_{\mathrm{c}}$ can be translated into the condition: $n \lambda_{\mathrm{dB}}^3 \gg 1$,  where $\lambda_{\mathrm{dB}} = \sqrt{2\pi \hbar^2/(m k_{\mathrm{B}} T)}$ is the thermal de Broglie wavelength that gives the coherence length of the gas. This condition indicates that the gas needs to be dilute in order for condensation to happen. This condition is also equivalent to having the de Broglie wavelength of the particles overlap and the system being described by a macroscopic wave-function.
When $T=0$, all the particles of the system will be in the ground state and the condensate is described by a single macroscopic wavefunction. As we see from Fig.~\ref{Fig.:BEC}, at high temperatures, condensation is broken and the system behaves as a gas of individual massive particles.

\vspace{0.3cm}
\begin{tcolorbox}[colback=gray!10,enhanced,breakable,frame hidden,halign=justify]
%\textbf{Summary of scales and galaxies}\footnotemark
%\vspace{0.0.8cm}
\paragraph*{The de Broglie wavelength  $\lambda_{\mathrm{dB}}$: \qquad} associated wavelength of a massive particle given by $\lambda_{\mathrm{dB}} = h/p =h/mv$, where $p$ and $v$ are the momentum and velocity of the particle with mass $m$, respectively.  For an ideal gas of temperature $T$ in a volume $V$, we have the thermal de Broglie wavelength which determines the coherent length of the gas.
The thermal de Broglie wavelength can be defined then, where the  characteristic thermal momentum is $p_{\mathrm{T}}=\sqrt{2 \pi m k_{\mathrm{B}} T}$, by:
\begin{equation*}
\lambda_{\mathrm{dB}} = \sqrt{\frac{2 \pi \hbar^2}{m k_{\mathrm{B}} T}}\,.
\end{equation*}
When the thermal de Broglie wavelength is much smaller than the interparticle distance ($d$), we have a gas of free particles. Otherwise, we have condensation as studied here.
\vspace{0.05cm} 
 
\end{tcolorbox}

%%%%%%%%%%%%%%%%
%%%%%%%%%%%%%

In summary, Bose--Einstein condensation can happen for ideal gases. The condition for condensation is that the occupation number of the ground state is so large that becomes macroscopic when $T<T_{\mathrm{c}}$. This can be translated in a \emph{condition for condensation}: if $n \lambda_{\mathrm{dB}}^3 \gg 1$, there is the formation of a BEC.
%; and if $n \lambda_{\mathrm{dB}}^3 \ll 1$, there is not. 
%For such large occupation, the quantum corrections are suppressed, and the BEC can be written, as a good approximation, in terms of classical theory. 
With that we can see that this very simple theoretical model already shows this intriguing macroscopic quantum phenomenon  that would be confirmed experimentally many years later. 
%We are going to see that for an interacting theory, the a classical field theory with spontaneous symmetry breaking is a good description for a BEC system.

\vspace{0.3cm}
\begin{tcolorbox}[width=\textwidth, enhanced, breakable,  valign=center, colback=white, colframe=black, sharp corners, shadow={0pt}{0pt}{0mm}{black},boxrule=0.5pt,halign=justify,overlay first={
        \draw[line width=.5pt] (frame.south west)--(frame.south east);},
    overlay middle={
        \draw[line width=.5pt] (frame.south west)--(frame.south east);
        \draw[line width=.5pt] (frame.north west)--(frame.north east);},
    overlay last={
        \draw[line width=.5pt] (frame.north west)--(frame.north east);
}]
\subsubsection*{Condition for condensation of a non-interacting ideal gas }

The condition for condensation of an ideal gas of $N$ bosons in thermal equilibrium with volume $V$ and temperature $T$ is:
\begin{equation*}
T < T_{\mathrm{c}} \quad    \Longleftrightarrow   \quad n \, \lambda_{\mathrm{dB}} \gg 1  \quad \Longleftrightarrow   \quad \lambda_{\mathrm{dB}} \gg d \sim \left( \frac{V}{N} \right)^{1/3} =  \left( \frac{1}{n} \right)^{1/3}  
\end{equation*}
These conditions state that the temperature must be smaller than the critical temperature; or that we have a macroscopic occupation number of the ground state $N_0$; or that the de Broglie wavelength needs to be bigger than the mean space between particles in order to have quantum degeneracy.  It is easy to see that these conditions are equivalent. 

\end{tcolorbox}

\subsection{Landau's superfluid model and criteria for superfluidity} \label{Landau}

A few years after BEC was theorized, another striking macroscopic quantum phenomenon was observed, superfluidity.
Landau constructed a phenomenological theory to explain the results of  superfluidity in helium, which was observed to flow in thin capillaries. This phenomenological theory, however, is quite general to describe superfluids and gives general conditions for the appearance of superfluidity. 

This theory has the goal of explaining why in superfluids charge is transported without friction. As we described above, according to London's ideas, in order to have superfluidity one needs to have a BEC.  The condensate has the role of transporting charge. So we consider a superfluid as the condensate that transports charge without losing energy. Dissipation of the condensate, which is equivalent to friction in the fluid, is caused by exciting particles out of the condensate. We have a supeffluid in the limit of no or low dissipation, and the superfluid is lost  in the limit of high dissipation. We present now the conditions for that to happen.
%We need to set the condition for this excitations to allow superfluidity. 

Consider a superfluid moving through a capillary with velocity $v_{\mathrm{s}}$. The energy of elementary excitations is given by \citep{Pitaevskii},
 \begin{equation}
E=E_{\mathrm{kin}}+\epsilon _p+\mathbf{p} \cdot \mathbf{v}_{\mathrm{s}}\,,
\end{equation}
in the rest frame of the capillary.  The kinetic energy of the fluid is given by $E_{\mathrm{kin}}$, and $\epsilon_p>0$ and $\mathbf{p}$ are the energy of the excitation and momentum in the frame of the fluid, and translated to the frame of the capillary. Dissipation happens when $\epsilon_p+\mathbf{p} \cdot \mathbf{v}_{\mathrm{s}}<0$. This can only be negative if its minimum, when $\epsilon_p+p v_{\mathrm{s}} \cos(\theta)$ where $\theta=n\pi$ for $n$ integers, is smaller than zero: $\epsilon_p-p \, v_{\mathrm{s}}<0$. With that we can determine the critical velocity:
\begin{equation}
v_{\mathrm{c}} = \min_p \frac{\epsilon_p}{p}\,. 
\end{equation}
For $v_{\mathrm{s}}<v_{\mathrm{c}}$, with $v_{\mathrm{c}} \neq 0$, the system transports charge without dissipation and the coherence of the BEC is maintained. 
This is the first criteria for superfluidity. The second necessary criteria is that $v_{\mathrm{c}}$ cannot be zero, so we need to have a condensate that transports the charge. A non-interacting (pure) Bose gas like we saw in the previous section has $v_{\mathrm{c}}=0$, so it cannot be a superfluid. A weakly interacting Bose gas has $v_{\mathrm{c}} \neq 0$ and it is a good representation of a superfluid.
%The complete criteria for superfluidity is the existence of a condensate, otherwise there is nothing to transport the charge without friction; $v_c$ cannot be zero (like it is the case for the non-interacting Bose gas, where $v_c=0$); and the above criteria, $v<v_c$.

As we are going to see in the next section in the case of the weakly interacting BEC because of the spontaneous breaking of the U(1) symmetry, a Goldstone mode appears, the phonon. This mode is gapless $\epsilon_{p=0} = 0$ and it is an elementary excitation of the superfluid. Even for that mode, the critical velocity is not zero, so there is some cost for producing the gapless excitation. For this weakly interacting Bose gas,  the critical velocity is the fluid sound speed, and Landau's criteria for superfluidity becomes:
\begin{equation}
v_{\mathrm{s}} < v_{\mathrm{c}} = c_{\mathrm{s}}\,.
\end{equation}
In summary, given that $v_{\mathrm{c}}$ is nonzero, and that we have a condensate (by construction) in this system, if $v<v_{\mathrm{c}}$, we can say that there is superfluidity. This results, however, is only valid for zero temperature.

Landau also developed the theory for a superfluid at finite temperature, the \emph{two fluid model}. At finite temperatures the fluid has two components: the \emph{superfluid} component that flows without friction, and a \emph{normal fluid} which describes the excitations. In this theory then there are two sounds speeds, for each degree of freedom. In the case of weakly interacting Bose gas, the first sound is $c_{\mathrm{s}}$ associated with the oscillation in density, and the second sound is $c_{\mathrm{s}}/\sqrt{3}$ that corresponds to the speed of propagation of the temperature oscillations.

This phenomenological theory is still an important topic of research as a condition for superfluidity. 
From simulations to experiments it is interesting to ask if the Landau criteria is fulfilled as a criteria for supefluidity. This criteria seems to be valid only in the regime of linear perturbations. This is the case since in Landau's theory the superfluid dissipates only into elementary excitations. However, we know that it is also possible to exist quantum vortices, topological defects present in rotating superfluids. This phenomenological theory does not take that into account, which can change the critical velocity of the superfluid to smaller values, reaching the dissipative regime of the superfluid before than if using Landau's critical velocity \citep{beyond_Landau}.

\vspace{0.3cm}
\begin{tcolorbox}[width=\textwidth, enhanced, breakable,  valign=center, colback=white, colframe=black, sharp corners, shadow={0pt}{0pt}{0mm}{black},boxrule=0.5pt,halign=justify,overlay first={
        \draw[line width=.5pt] (frame.south west)--(frame.south east);},
    overlay middle={
        \draw[line width=.5pt] (frame.south west)--(frame.south east);
        \draw[line width=.5pt] (frame.north west)--(frame.north east);},
    overlay last={
        \draw[line width=.5pt] (frame.north west)--(frame.north east);
}]
\paragraph*{\textit{Landau criteria for superflduidity}} \mbox{}\\ 

Phenomenological conditions for superfluidity (at $T=0$):
\vspace{0.1cm}
\begin{enumerate}
\item Existence of a condensate;
\item $v_{\mathrm{c}} \neq 0$ \qquad {\color{gray} (Non-interacting Bose gas has $v_{\mathrm{c}} =0$ - not a superfluid! Interaction is crucial for super-\\ \hspace*{1.66cm} fluidity.)}
\item  $v < v_{\mathrm{c}}$  \qquad {\color{gray} - system transports charge without dissipation and the coherence of the BEC is maintained.} 
\item[] \vspace{0.05cm}$v_{\mathrm{c}}=$ velocity above which excitations can leave the condensate  {\color{gray}($v_{\mathrm{c}}=c_{\mathrm{s}}$ - interacting BEC)} 
\end{enumerate}

\vspace{0.1cm}
At finite temperatures, two fluid model: the \emph{superfluid} component that flows without friction, and a \emph{normal} fluid which describes the excitations.
\end{tcolorbox}

%%%%%%%%%%%%%%%%%%%%%%%%%%%%%%%%%%
\subsection{Weakly interacting Bose gas - superfluid}
\label{Sec.:Interacting BEC}
%%%%%%%%%%%%%%%%%%%%%%%%%%%%%%%%%%

We now turn to the discussion of interacting systems. 
Inspired by Landau's phenomenological theory, the weakly interacting Bose gas was proposed as the simplest system to study superfluidity, and as a realistic model to understand condensation.
%A weakly interacting Bose gas was then proposed by Bogoliubov, as a modification of the non-interacting Bose gas model, in order to study the Bose--Einstein condensation and superfluidity. 
In this section we are going to model a superfluid by a Bose--Einstein condensate that has self-interaction, and show that, although a BEC can be formed both in the case of the non-interacting and interacting Bose gas, the presence of interaction is crucial for superfluidity \citep{Pitaevskii,quantum_fluid,condensation}.

We present here a microscopic description of superfluid which arises upon condensation. The microscopic system where this happens is a weakly-interacting gas of bosonic particles. 
To describe this interacting gas, first we need to understand how to describe the excitations in this system.

One of the conditions for condensation is that the gas is dilute. However, even in a dilute gas the interaction can play an important role. 
 The way we describe interactions or collisions in a Bose gas is somewhat different than in a classical fluid. Since now we describe it using their wavefunction, we need to have a interatomic potential $V_{\mathrm{int}}$ to enable these collisions. 
In a dilute system at low temperatures, three-body interactions are suppressed, so we are going to describe this system with binary collisions. In such a system, the two-body collisions depend only on one parameter $a$, the $s$-wave (or coherent) scattering length \citep{Harko}, which is the zero energy limit of the scattering amplitude $a= \lim_{T \rightarrow 0} f_{\mathrm{scat}}$. This is valid only for low energies when the other length scales of the problem $ d \gg a$. In this limit, the elastic scattering cross-section becomes constant $\sigma=8\pi a^2$, and the two-body interatomic potential can be written as $V_{\mathrm{int}}(\mathbf{r}-\mathbf{r}')=(4\pi a \hbar^2 /m) \, \delta(r-r') \equiv g \, \delta (r-r')$, which is short-ranged and present only when the atoms interact. The s-wave scattering length $a$ can be positive or negative depending on the system described, representing a repulsive or an attractive interaction. With this we can model the effective interaction Hamiltonian as $\hat{H}_{\mathrm{int}} = \int d \mathbf{r} d \mathbf{r}' \, \hat{\Psi}^{\dagger}(\mathbf{r})  \hat{\Psi}^{\dagger}(\mathbf{r}) \,  V_{\mathrm{int}}(\mathbf{r-r'}) \,  \hat{\Psi}(\mathbf{r'})  \hat{\Psi}(\mathbf{r'}) \approx \int d^3 r \, \hat{\Psi}^{\dagger}(\mathbf{r})  \hat{\Psi}^{\dagger}(\mathbf{r}) \hat{\Psi}(\mathbf{r})  \hat{\Psi}(\mathbf{r})$, where $\Psi$ is the Bose operator.

The dynamics of this many-body interacting system is given by the second-quantized $N$-body Hamiltonian,
\begin{equation}
\hat{H}=\int d^3 r \, \hat{\Psi}^{\dagger}(\mathbf{r}) \left[ -\frac{\hbar^2 \nabla^2}{2m} + V_{\mathrm{trap}}(\mathbf{r}) \right] \hat{\Psi} + \frac{g}{2} \int d^3 r \, \hat{\Psi}^{\dagger}(\mathbf{r})  \hat{\Psi}^{\dagger}(\mathbf{r}) \hat{\Psi}(\mathbf{r})  \hat{\Psi}(\mathbf{r})  \,,
\label{Eq.:Many_body_H}
\end{equation}
where the brackets is the single particle Hamiltonian, and $V_{\mathrm{trap}}(\mathbf{r})$ is the trapping potential, an external potential applied to the system (that in the next section could be the gravitational potential). In the Heisenberg description, we can then write the Heisenberg equations of motion,
\begin{equation}
i \hbar \frac{\partial \hat{\Psi}(\mathbf{r},t) }{\partial t}= [\hat{\Psi}(\mathbf{r},t), \, \hat{H}]=\left( -\frac{\hbar^2 }{2m} \nabla^2 + V_{\mathrm{trap}}(\mathbf{r}) \right) \hat{\Psi}(\mathbf{r},t)+ g \, \hat{\Psi}^{\dagger}(\mathbf{r},t) \hat{\Psi}(\mathbf{r},t) \hat{\Psi}(\mathbf{r},t)\,,
\label{Eq.:full_condensate_Schrodinger}
\end{equation}
with the brackets indicating the commutator. This is the Schr{\"o}dinger equation for the Bose field operator $\hat{\Psi} (\mathbf{r},t)$.

The Bose field operator $\hat{\Psi}^{\dagger}(\mathbf{r})$ and $\hat{\Psi}(\mathbf{r})$ create and annihilate a particle at position $\mathbf{r}$, and obeys the canonical commutation relations with only non-zero commutator given by $[\hat{\Psi}(\mathbf{r}), \, \hat{\Psi}^{\dagger}(\mathbf{r})]=\delta(\mathbf{r}- \mathbf{r}')$.
The Bose field operator describes a continuum spectrum of single particle position eigenstates, and can be re-written in the single-particle basis as
\begin{equation}
\hat{\Psi}^\dagger = \sum_i \hat{a}_i^\dagger \phi_i^{*} (\mathbf{r})\,, \qquad  \hat{\Psi} = \sum_i \hat{a}_i \phi_i (\mathbf{r})\,,
\end{equation}
where $\phi_i$ is the states wavefunction, and the creation and annihilation operators, $\hat{a}_i^\dagger$ and $\hat{a}_i$, create and annihilate a particle from the state $\phi_i$.  They obey the Bose commutation relations\footnotemark, with only non-zero component given by $[\hat{a}_i, \hat{a}_j^\dagger]=\delta_{ij}$. 
\footnotetext{In terms of the creation and annihilation operator, the Hamiltonian of the many-body system is given by, 
\begin{equation}
\hat{H} = \sum_{ij} H_{ij}^{\mathrm{sp}}  \hat{a}_i^{\dagger} \hat{a}_j + \frac{1}{2} \sum_{ijkm} \langle ij | \hat{V} | km \rangle \hat{a}_i^\dagger \hat{a}_j^\dagger \hat{a}_k \hat{a}_m \,,
\end{equation}
where $\langle ij | \hat{V} | km \rangle$ denotes the matrix element for the interaction, and $H_{ij}^{\mathrm{sp}}=\int d^3 \mathbf{r} \tilde{\Phi}_i^{*} (\mathbf{r}) \hat{H}^{\mathrm{sp}} \tilde{\Phi}_j (\mathbf{r})$, where $\tilde{\Phi}_j$ is the states wavefunctions.}

Many-particle systems described by the Hamiltonian (\ref{Eq.:Many_body_H}) are very difficult to be solved. With the exception of a few simple models, in order to find solutions to this problem and be able to study its properties we need to make simplifications. For that we use Bogoliubov's prescription or \emph{mean-field} approximation.
For the general case of a non-uniform gas, the mean-field approximation can be written, in the Heisenberg picture as,
\begin{equation}
\hat{\Psi} (\mathbf{r}, t) = \psi (\mathbf{r}, t) +\delta \hat{ \Psi} (\mathbf{r}, t)\,,
\end{equation}
where $\psi (\mathbf{r}, t) \equiv \langle \hat{\Psi} (\mathbf{r}, t) \rangle$ is classical field called the wavefunction of the condensate.
% and  has the meaning of an order parameter\footnote{In a theory where phase transition happens, which is the case here, the order parameter determines the order of the system: when the symmetry is broken and the system is "ordered" is non-vanishing; and it vanishes in the restored phase, called "disordored".}. 
The density of the condensate is fixed by: $n_0 = | \psi (\mathbf{r}, t) |^2 =n$. 
Like we described for Landau's theory, $\delta \hat{\Psi} (\mathbf{r}, t) $ is a small perturbation of the system with $\langle \delta \hat{ \Psi} (\mathbf{r}, t) \rangle =0$ and describes depletion of the condensate. 

Effectively this approximation leads the many-body problem to be reduced to a single body problem by describing the averaging the effects of all other particles. \textit{Given that the interactions are weak, and that the gas is diluted, quantum fluctuations on the condensate are suppressed. The mean field approximation is valid for dilute system with  $n\, a^3 \ll 1$.}  When this conditions is not met there are deviations  of the mean-field approximation. We  can treat these deviations in perturbation theory, where we invoke non-vanishing moments for the fluctuation operator, like the Hartree--Fock--Bogoluibov, which considers a non-zero $\langle \delta \hat{\Psi} \rangle$, or the Hartree--Fock--Bogoluibov--Popov, for terms up to second order in the perturbation. 

With this approximation, we can write the generalized Gross--Pitaevskii (GP) equation:
\begin{equation}
\boxed{
  i \hbar \frac{\partial \psi(\mathbf{r},t) }{\partial t}= \left( -\frac{\hbar^2 }{2m} \nabla^2 + V_{\mathrm{trap}} (\mathbf{r})  + g \, |\psi (\mathbf{r,t}) |^2 \right) \psi(\mathbf{r},t) \, .}
\qquad \parbox{9em}{Gross--Pitaevskii \\ equation}
\label{Eq.:GP_equation}
\end{equation}
The GP equation is a non-linear Schr{\"o}dinger equation, with non-linearity arising from the self-interaction term. This equation describes the dynamics of the zero-temperature dilute weakly-interacting Bose system by allowing us to determine the shape of the single particle wave function, the condensate.
\vspace{0.2cm}

We can study the case of stationary solutions. The stationary solution can be taken as the solution that provides us with the condensate, the ground state wavefunction. The ground state is the lowest energy state of a quantum mechanical system, with the excited states being the states with higher energy than the ground state. The stationary states $\Psi $  are the eigenfunctions of the Hamiltonian operator, with eigenvalues $\mu$ related to the energy of the system, in a way that for the wavefunction we have
\begin{equation}
i \hbar \frac{\partial \Psi}{\partial t} = \hat{H} \Psi = \mu \Psi \,.
\end{equation}
With that we can write the stationary solution as,
\begin{equation}
\psi (\mathbf{r}, t)  = \psi_{\mathrm{s}} (\mathbf{r} ) e^{-\frac{i}{\hbar} \mu t}\,,
\label{Eq.:stationary_solution}
\end{equation}
where the eigenvalue of the Hamiltonian $\mu$ is also called the chemical potential, and  $\phi_{\mathrm{s}}$ is real field with $\int d\mathbf{r} \, \psi^2 = N_0 = N$. The Gross--Pitaevskii equation becomes:
\begin{equation}
\left(-\frac{\hbar^2}{2m} \nabla^2 + V_{\mathrm{trap}} (\mathbf{r}) + g \,  | \psi_{\mathrm{s}}(\mathbf{r}) |^2 \right) \psi_{\mathrm{s}} (\mathbf{r}) = \mu \psi_{\mathrm{s}} (\mathbf{r})\,.
\end{equation}
In the Thomas-Fermi limit, which is the approximation where the interaction energy is bigger than the kinetic energy for a large number of particles, the kinetic energy can be neglected so we have $\mu \psi=(g \, n+ V_{\mathrm{trap}}) \psi$. As a solution of this equation, we get that in the Thomas-Fermi limit,
\begin{equation}
n(\mathbf{r}) = |\psi (\mathbf{r}) |^2 = \left\{ \begin{array}{ll}
 ( \mu -  V_{\mathrm{trap}} (\mathbf{r})) / g  \,, & \,  \,\,\mathrm{for }\,\,\, \mathbf{r} \,\,\, \mathrm{where}\,\,\, ( \mu -  V_{\mathrm{trap}} (\mathbf{r})) / g > 0 \,, \\
    0\,,  & \, \,\, \mathrm{otherwise} \,.
  \end{array} \right.
\end{equation}

\paragraph*{Fluid description} \mbox{}\\

We can decompose the complex macroscopic condensate wavefunction into,
\begin{equation}
\psi(\mathbf{r},t)=|\psi(\mathbf{r},t)| \, e^{i \theta (\mathbf{r},t)}\,,
\label{Eq.:Mandelung_1}
\end{equation}
where, as we saw above , $\psi$ is normalized to the total number of particles,  $|\psi(\mathbf{r},t)| = \sqrt{n(\mathbf{r},t)}=\sqrt{\rho(\mathbf{r},t)/m}$, and $\theta(\mathbf{r},t)$ is the phase distribution. Inserting  this  into the GP equation we get two equations. We make the following redefinition
\begin{equation}
\mathbf{v}(\mathbf{r},t) \equiv \frac{\hbar}{m} \, \nabla \theta(\mathbf{r},t)\,.
\label{Eq.:Mandelung_2}
\end{equation}
This, together with (\ref{Eq.:Mandelung_1}) are called Madelung transformation. With those new variables, the GP equation results in two equations,  the \emph{Madelung equations} \citep{Mandelung}:
\begin{empheq}[box=\fbox]{align}
& \frac{\partial \rho}{\partial t} + \nabla \cdot (\rho \mathbf{v}) = 0\,,  \label{Eq.:continuity}  \\ 
& \rho \frac{\partial \mathbf{v}}{\partial t} + \rho \, (\mathbf{v} \cdot \nabla)\mathbf{v} = -\nabla \left( P_{\mathrm{int}} + P_{\mathrm{QP}} \right) - n \, \nabla V_{trap} \,. \label{Eq.:Euler}
\end{empheq}
%\begin{align}
%& \frac{\partial \rho}{\partial t} + \nabla \cdot (\rho \mathbf{v}) = 0\,, \\
%& m \frac{\partial \mathbf{v}}{\partial t} + m (\mathbf{v} \cdot \nabla)\mathbf{v} = -\nabla \left( V_{trap} + g \, n - \frac{\hbar^2}{2m} \frac{\nabla^2 \sqrt{n}}{\sqrt{n}} \right)\,.
%\end{align}
They are a representation of  the GP equation into ``hydrodynamical'' equations since they have a similar  form  as the continuity equation and Euler equation for a perfect fluid. However, the second Madelung equation describes a fluid with a potential flow, given the definition of the velocity,
with \emph{zero vorticity} $\mathbf{\nabla \times \mathbf{v} =0}$. This corresponds  to the  main characteristic of the superfluid that it flows without friction, has irrotational flow. 
This equation also differs from the perfect fluid Euler equation by the presence of the quantum pressure term.

The second Madelung equation, the Euler-like equation, reveals more interesting properties of the superfluid. This equation has two pressure terms, $P_{\mathrm{int}} $ and $P_{\mathrm{QP}} $ that are respectively the pressure term, and the quantum pressure:
\begin{align}
  P_{\mathrm{int}}&= K \, \rho^{(j+1)/j} \Bigr|_{j=1} = \frac{g}{2} n^2 = \frac{g}{2m^2} \rho^2 \,,\nonumber\\
  \nabla P_{\mathrm{QP}}&=-n \nabla Q=- n \nabla \left[ \frac{\hbar^2}{2m^2} \frac{\nabla^2 \sqrt{n}}{\sqrt{n}} \right]\,.
\end{align}
The pressure term comes from the self-interaction which gives a polytropic type of pressure. For the two-body interaction, which is the case we show here,  $j=1$. If we have a three-body interaction, for example, the pressure would have polytropic index $j=1/2$, giving $P_{\mathrm{int}} \propto \rho^3$. The constant $K$ depends on the interaction constant.
The \emph{quantum pressure} is defined in terms of the quantum potential $Q=-(\hbar^2/2m) \nabla^2 \sqrt{n}/\sqrt{n}$ (see definition bellow). 
%The quantum pressure is a pure quantum mechanical pressure that arises due to the uncertainty principle (see more details below), and is characteristic of this quantum system. 
%The quantum pressure comes from the quantum mechanical kinetic term that arises due to the uncertainty principle (see more details below), and is characteristic of this quantum system. 

\vspace{0.3cm}
\begin{tcolorbox}[colback=gray!10,enhanced,breakable,frame hidden,halign=justify]
  	\paragraph*{Quantum pressure: \qquad} Quantum pressure (QP) is the name given to the term\footnotemark
     \begin{equation}
      \nabla P_{\mathrm{QP}} =  -n \nabla \left[ \frac{\hbar ^2}{2m} \frac{\nabla^2 \sqrt{n}}{\sqrt{n}}\right] \,, \qquad\qquad P_{ij,\,\mathrm{QP}}= - \left( \frac{\hbar}{2m} \right)^2 \rho \, \partial_i \partial_j \ln \rho\,,
	\end{equation}  
where $P_{ij,\,\mathrm{QP}}$ is the quantum pressure tensor. Together with the second term on the left hand side of  equation (\ref{Eq.:Euler}) this term comes from the spatial part of the kinetic term. However, those two components are very different. The classical component describes the kinetic pressure due to the motion of the particle.  The quantum pressure comes from the quantum part of the  kinetic term that arises due to the Heisenberg uncertainty principle or can be seen as the curvature of the amplitude of the wave function. This is an additional force term that appear in the Madelung equation due to the zero point motion of particles.  In the mean field approximation, this term is still present, given the classical wave function describing the system, so the term quantum is also misleading in this context.
    This term is repulsive and counter-acts attraction from a potential or attractive interaction, supporting the system against collapse. With that, the system cannot have vanishing size.
 This term modifies the dispersion relations of the excitations of the condensate and it is important at small scales, for scales smaller than the healing length. This term is negligible for large scales, the Thomas-Fermi approximation, and for a uniform superfluid, since $n=\mathrm{const.}$.
      \\

    \paragraph*{Healing length: \qquad} defined as the length for which $P_{\mathrm{int}} = P_{\mathrm{PQ}}$, given by,
    \begin{equation}
    \xi = \frac{\hbar}{\sqrt{2mgn}}\,.
    \end{equation}
    It is the length for which the interactions "heal" (coarse-grain) any density of phase perturbations in the condensate.
\end{tcolorbox}
\footnotetext{One can notice that the form of the quantum potential coming from the Madelung equations, and consequently the quantum pressure tensor, are very similar to the Bohm quantum potential \citep{Bohm:1951xw,Bohm:1951xx}. Some authors point that  Bohm rediscovered the quantum pressure in his new interpretation of quantum mechanics. However, some authors claim this equivalence is not so clear. Some authors also claim that since the quantum pressure tensor has non-diagonal components, and it is not isotropic, so it cannot be called pressure \citep{Hui:2020hbq}.}

\paragraph*{Condensate solution} \mbox{} \\

Having established how we describe the weakly interacting system in the mean-field approximation, we can now describe what is the ground state, the condensate. We can solve analytically the GP equation for a few simple cases and obtain the condensate solution as for the cases with no interactions, in harmonic potential, and other simple systems \citep{quantum_fluid}. We are going to present some interesting cases here.

In the case of a uniform gas, the wavefunction $\Phi_0 (\mathbf{r})= \sqrt{N} \phi_0/\sqrt{V}$ describes the condensate formed, while the remaining functions form a complete set of functions orthogonal to the condensate. 

\textit{Solitons - \,\,}   Solitons are a localized solution of the one-dimensional GP equation with $V_{\mathrm{trap}}=0$, which is integrable in this limit. They are also called solitary waves, and are solutions described by a permanent and localized wave, that maintains its shape and velocity upon collisions. They are obtained when the dispersion term, the kinetic spatial term, and the non-linear term from the interaction cancel out.
If the interaction is repulsive ($g>0$), a dark soliton is formed,which is given by the solution $\psi_{\mathrm{d}}(x) = \psi_0 \tanh(x/\sqrt{2 \,\xi})$, where $\psi_0=\lim_{x\rightarrow \infty} \psi$. A bright soliton is the solution for attractive interactions ($g<0$), with $\psi_{\mathrm{b}}(x) = \psi_0 e^{-i\mu t/\hbar} \cosh(\sqrt{2 m |\mu |/\hbar} \, x)^{-1}$. They are called dark and bright since they represent a decrease and a concentration of condensate, respectively.

%In reality, “soliton-like” objects are more interesting representing solutions in 3-dimensional condensates in a trapping potential of near to or non-integrable systems.

\paragraph*{Collective excitations, dispersion relation and sound speed} \mbox{}\\

The excitations in the superfluid which represent  perturbations of the condensate, are an important part of this system. They represent sound waves, called the \emph{phonons} that propagate through the condensate. We are going to show how they arise.

Here we are going to work in the case $V_{\mathrm{trap}}=0$, for simplicity. The case where a trapping potential is present, like for example a condensate in the presence of a gravitational potential, is studied in Sect.~\ref{Sec.:stability}. For a homogeneous condensate, to study the perturbations around the condensate, we perturb the classical wavefunction $\psi (\mathrm{r},t)$,
\begin{equation}
\psi(x,t)=\psi_0 + \delta \psi^{(1)} + \delta \psi^{(2)}+ \cdots \,,
\end{equation}
where we assumed that the motion is only in the $x$-directions without  lost of generalization. The perturbations are small and $\delta \psi^{(i)}$ indicate the $i^{\mathrm{th}}$ order in perturbation. To  linear order, we can re-write the linearized GP equation: $i \hbar \partial_t \delta \psi^{(1)}=-(\hbar^2/2m) \partial^2_x \delta \psi^{(1)} + \mu ( \delta \psi^{(1)}+\delta \psi^{* \, (1)})$.  We make an ansatz for the solution as travelling waves, $\delta \psi^{(1)} = A \, e^{i(kx-\omega_k t)}+ B \, e^{-i(kx-\omega_k t)}$. The parameters $A,\, B$ are determined by the initial conditions. Substituting this ansatz into the linearized GP equation,  we can see that the dispersion relation is given by,
\begin{equation}
\omega_k^2 = c_s^2 k^2 + \frac{\hbar^2}{(2m)^2} k^4\,,
\label{Eq.:dispersion_relation_QM}
\end{equation}
where the sound speed is defined as the term coming from the linear part of the dispersion relation,
\begin{equation}
c_s^2=\frac{g \,  n_0}{m}\,.
\end{equation}
The sound  speed appears because of the presence of the interaction, therefore since for a superfluid the presence of an interaction is crucial, we can say that a superfluid is a fluid that has a sound speed. With this we can easily see the definition and properties of a superfluid.
\begin{itemize}
\item[] \textbf{Superfluid:} In the presence of interactions,  a sound speed is present which determines the behaviour of the excitations on large scales. 

- For \textit{large wavelengths} (small $k$), the higher $k$ terms do not contribute and the dispersion relation is given by:
\begin{equation}
\omega_k = c_s k\,, 
\end{equation}
which is the dispersion relation of a \emph{sound wave}. The superfluid is characterized by excitations that propagate  as waves, the phonons. Because of this property phonons can mediate a long range forces ($F \sim 1/r^2$). This long-range force  is the responsible for the effective dynamics of a superfluid: flowing without friction. We are going to see later (in the field theory description) that this gapless mode can also be viewed as the Nambu-Goldstone modes coming from the spontaneous symmetry breaking of the U(1) symmetry of the system caused by the formation of the BEC.

This limit where the quantum pressure term can be ignored, or more specifically when $P_{\mathrm{int}} \gg P_{\mathrm{QP}}$, is the Thomas-Fermi approximation we saw above, and can alternatively be defined by wavelengths bigger than the healing length. 

This is the case for a repulsive interaction ($g>0$). The situation is different for an attractive interaction ($g<0$), where $w_k$ is imaginary and the solutions is unstable given by exponentially growing or decaying functions. This means that it is not possible to form a stable condensate in these cases. 

- For \emph{small wavelengths} (large $k$), the quantum pressure term dominates, and the dispersion relation is given by  $\omega_k = \hbar k^2 /2m$, which describes a free particle. In this limit, the system stops exhibiting superfluidity. 

In general, for intermediary frequencies, the full  dispersion relation (\ref{Eq.:dispersion_relation_QM}) does not propagate as a wave and shows two degrees of freedom: a gapless mode (the phonon), that propagates as a wave, and a massive mode, related to particle creation. We are going to discuss these again in the field theory section.

\item[]
\item[] \textbf{No interactions - BEC:} In the limit where $g \rightarrow 0$, the BEC stops exhibiting superfluidity. The phonon becomes gapped, the dispersion relation is given by: $\omega_k = \hbar k^2 /2m$, which is the dispersion relation of a free particle. Since $\omega_k^2>0$ in this case, the solution of the linearized GP equation (without a external potential) is a stable oscillatory solution.
\end{itemize}

We showed above that the superfluid can decay into collective elementary excitation, the  long-wavelength sound-wave quanta, the phonon.  These are excitations with linear dispersion relation that behave as periodic fluctuations in density in the superfluid.
When studying linear perturbations around a classical condensate background, only the phonon excitation is expected.
However, a superfluid can also deplete into other excitations called rotons and maxons. 
The linear part of the dispersion relation is only a part of the dispersion relation at small momenta.
For higher momenta, the dispersion relation presents a maximum and then a minimum. Near the maximum we have the maxons excitations, and near the minimum, for even  higher momenta, we have the roton excitations.
%Rotons are massive quasiparticle  that appears in high density superfluids and are absent in dilute superfluid. 
So, phonons, rotons and maxons are excitations described by different parts of the same dispersion relation. 
%However, the phonon is the only intrinsic excitation of a superfluid, present in every superfluid (with its definition overlapping tot he one of a superfluid).
And lastly, for a rotating superfluid, there is also the vortex nucleation of the condensate, as we will see in Sect.~\ref{Sec.:vortices}.

\vspace{0.3cm}
\begin{tcolorbox}[colback=gray!10,enhanced,breakable,frame hidden,halign=justify]
\paragraph*{U(1) symmetry group: \qquad} The unitary group of degree $n=1$  is the group associated with $n\times n$ unitary matrices ($U^{*}U=UU^{*}=1$) under matrix multiplication, $U=e^{i\alpha}$, with $\alpha$ the parameter of the group. For $n=1$, this is the unitary transformations complex numbers, which corresponds to phase rotations. The $U(1)$ group is isomorphic\footnotemark  ~to the $SO(2)$, the group of the $2\times 2$ orthogonal rotations in $\mathbb{R}^2$ with unit determinant.
	The U(1) symmetry can be global, when it acts in the same way in all points in space-time, or local, acts differently at each place in space and time.
\\ 

\paragraph*{Noether's theorem: \qquad}   One of the most important results theoretical physics, Noether's theorem definition can be found in basically every textbook in field theory and even classical mechanics. The theorem states that for each continuous symmetry of a action there is an associated  conserved current $j^{\nu}(\mathbf{x},\, t)$, $\partial_{\nu} j^{\nu}(\mathbf{x},\, t) =0$ . This has profound implications since the conservation laws that describe the dynamics of a systems are then associated with the symmetries present in the problem.
\vspace{0.1cm}  	
  	
% We can illustrate this in terms of the global U(1) symmetry cited above\footnotemark. Consider a complex scalar field ($\Psi (\mathbf{x},t)$) theory with action:
We can illustrate this for a system with the global U(1) symmetry that, according to Noether's theorem has a conserved particle number density\footnotemark. Consider a complex scalar field ($\Psi (\mathbf{x},t)$) theory with action:
  	\begin{equation}
  	S=\int d^4x \, \mathcal{L} = \int d^4x \, \left[ (\partial_\mu \Psi )^{*} (\partial_\mu \Psi ) + V(\mathbf{x}) \, \Psi^{*} \Psi  \right]\,,
  	\end{equation}
  	where $V(\mathbf{x})$ is a potential. On top of being invariant under Lorentz transformations, this system has U(1) symmetry, with the action invariant under continuous rotations of the phase of the complex scalar field: $\Psi \rightarrow e^{i\alpha} \Psi $ (and c.c.). Since this is a global symmetry, $\alpha$ is independent of $(\mathbf{x},\, t)$. With that, we can evaluate the  Noether current and charge of this system:
  	\begin{equation}
  	j^\nu = i (\Psi^\dagger \partial^\nu  \Psi - \Psi \partial^\nu \Psi^\dagger)\,, \qquad  Q = \int d^3 x \, j^0 \,,
  	\end{equation}
  	which are conserved: $\partial_\nu j^\nu = 0$ and $dQ/dt=0$. From $j^{\nu}=(m n, \, \mathbf{j})$, the conserved charge can be written in terms of the local density of Q, $Q=\int d^3 x \, \rho(\mathbf{x},\, t)$, and the conserved current implies a continuity equation for $\rho$. In this way, $Q \propto \Psi^\dagger \Psi$, which implies the conservation of the norm and it is associated with number conservation. 
  \vspace{0.3cm}
  	
\end{tcolorbox}
\footnotetext{Isomorphism is a one on one mapping between the elements of groups preserving its group operations. In the case of $U(1)$ and $SO(2)$, the isomorphism takes a complex number of unit norm into a rotation, $e^{i \theta} \mapsto \left( \begin{array}{cc} \cos\theta & -\sin\theta \\
\sin\theta & \cos \theta \end{array} \right) $, where the rotation angle is the argument of $z$.}
\footnotetext{For a more general derivation of Noether's theorem in field theory for any continuous symmetry and the connection with the generators of the symmetry group see \citep{qft}}

Some comments are in order. When defining $\psi$ as the condensate wavefunction, this quantity  is actually a mean-field value of the wavefunction, the degree of freedom that defines the condensate. This description of averaging selecting the condensate is consistent with the theory of critical phenomena, like phase transitions. This Bose system can be seen as a system with spontaneous breaking of a symmetry of the description. In our case the U(1) symmetry which is the symmetry of the Hamiltonian. This is analogous to the the spontaneous symmetry breaking in a ferromagnet. The difference is that, since we have a Bose system, the idea of spontaneous symmetry breaking to the thermodynamical limit of a finite size Bose gas defines the number of particles of the system, and this is only consistent with the picture of having a condensate that can change the number of particles in the ground state, if the number of particles is conserved.

We can also understand this argument for the symmetry breaking of the system by analysing another approach to the weakly interacting system initially developed by \citep{Penrose:1956zz}, and \citep{Beliaev}. 
In this approach the condensate wavefunction is identified, using the density matrix $\bar{\rho}(\mathbf{r},\, \mathbf{r}')=\langle \hat{\Psi} (\mathbf{r}') \hat{\Psi} (\mathbf{r})  \rangle$, to the number density of particles $n(\mathbf{r})=\bar{\rho}(\mathbf{r},\, \mathbf{r})=\sum_i n_i |\phi(\mathbf{r})|^2$, where we are working in the stationary case. The formation of a BEC, which means that the ground state has a macroscopic occupation, leads to the factorization $N=n_0+\sum_{i \neq 0} n_i$, which means  $\bar{\rho}(\mathbf{r},\, \mathbf{r}')= \Phi_0 (\mathbf{r}') \Phi_0 (\mathbf{r}) + \sum_{i \neq 0} n_i \phi_i^{*} (\mathbf{r}' )\phi_i (\mathbf{r})$.  The field operator then can be factorized in the presence of a condensate into $\hat{\Psi} (\mathbf{r}) = \Phi_0 (\mathbf{r}) + \delta \hat{\Phi}  (\mathbf{r})$, with $\langle  \delta \hat{\Psi}  (\mathbf{r})  \rangle =0$. This means when a condensate form,
\begin{equation}
\langle  \hat{\Psi}  (\mathbf{r})  \rangle = \Phi_0  (\mathbf{r}) \neq 0\,. 
\end{equation}
A state that has conserved particle number has $\langle  \hat{\Psi} \rangle =0$. So this condition above is seen as describing a symmetry breaking, more specifically \emph{Bose symmetry breaking}, and the consequence is a system  where particle number is not conserved in the condensate. In the absence of a condensate this goes back to the particle conserving condition:
\begin{equation}
\Phi_0  \left\{ \begin{array}{lr}
    = 0 \,, & \,  \,\,\mathrm{for }\,\,\, T >T_{\mathrm{c}}\,, \\
    \neq 0 & \, \,\, \mathrm{for }\,\,\,  T <T_{\mathrm{c}}  \,.
  \end{array} \right.
\end{equation}
This suggests that condensation comes from a spontaneous symmetry breaking theory.

%%%%%%%%%%%%%%%%%%

With that, the interacting condensate system above can be understood as a particle conserving system of bosons with U(1) symmetry, described by the classical field $\psi$, the wavefunction of the condensate, where the formation of a Bose--Einstein condensate is a phase transition, coming from a spontaneous breaking of the symmetry (that can be seen as spontaneous coherence). This description of the condensate makes us see a parallel with the formalism used in \emph{field theory}.

%%%%%%%%%%%%%%%%%%%%%%%%%%%%%%%%%%
\subsubsection{Field theory description}
%%%%%%%%%%%%%%%%%%%%%%%%%%%%%%%%%%

% NEW (follow more our notes)
Given the suggestion that we can describe condensation as spontaneous symmetry breaking process, we turn now to the description of this system using a field theory language.
The methods from field theory are very appropriate to describe this type of system where spontaneous symmetry breaking is present. Given the description we had above for the superfluid and the properties of this system, we represent this system as a massive complex scalar field\footnotemark \, with self-interactions with a global $U(1)$ symmetry  that is spontaneously broken by the presence of the ground state, the condensate, with superfluidity arising upon condensation. We are going to work here, in order to best illustrate this description, in the homogeneous case, where there is no trapping or external potential applied to the system.
% In this way we represent this system as a complex scalar field with self-interactions with a global $U(1)$ symmetry that is spontaneously broken by the presence of the ground state, the condensate
%Therefore, we can rewrite the weakly interacting Bose system in a field theory formalism as a system carrying a conserved U(1) charge (the number of particles)
\footnotetext{We have used here a complex scalar field that has a continuous symmetry since, as we saw before according to Noether's theorem, this system has a conserved charge. The $U(1)$ symmetry gives us a system with conserved particle number. For this reason it is essential to use a complex scalar field. }

Given that, we describe this system by the Lagrangian density for a two-body interaction,
  \begin{align}
  \mathcal{L} &= (\partial_\mu \Psi )^{*} (\partial_\mu \Psi ) - m^2 \Psi^{*} \Psi - \frac{g}{2} (\Psi^{*} \Psi )^2 
\label{L_2body}
  \end{align}
where we consider $g > 0$ in order to get a stable condensate with long range coherence, as discussed above.
As we saw above, this system has a $U(1)$ global symmetry, which means that it is invariant under continuous rotations of the phase of the complex scalar field,
\begin{equation}
\Psi \rightarrow \Psi e^{i \alpha}\,, \qquad \Psi^{*} \rightarrow \Psi e^{-i \alpha}\,,
\end{equation}
where $\alpha$ is a constant\footnote{It is equivalent to re-write this, in a language of $SO(2)$ symmetry, where the complex field can be written as $\Psi = (1/\sqrt{2}) (\Psi_1 + i \Psi_2)$, where the system as an invariance under rotations:  $\left\{ \begin{array}{r}
\Psi_1 \rightarrow \cos \alpha \, \Psi_1 - \sin \alpha \, \Psi_2  \\ 
\Psi_2 \rightarrow \sin \alpha \, \Psi_1 + \cos \alpha \, \Psi_2 
\end{array} \right. $.}.
The equation of motion is given by,
\begin{equation}
\partial^{\mu} \partial_{\mu} \Psi + m^2 \, \Psi + g | \Psi |^2 \Psi =0 \,.
\label{Eq.:EOM_section3}
\end{equation}

In order to describe the condensate, like we did previously, we separate the condensate contribution to the perturbations, and assume the mean-field approximation,
\begin{equation}
\Psi (\mathbf{x}, t) = \psi (\mathbf{x}, t) + \delta \Psi (\mathbf{x}, t) \,,
\label{Eq.:mean_field_field_theory}
\end{equation}
where $\psi(\mathbf{x}, t) $ is the background that gives the condensate field, and $\delta \Psi (\mathbf{x}, t) $ are the excitations of the condensate that are considered small with respect to the background solution.

As we saw in (\ref{Eq.:stationary_solution}), the stationary solution determines the ground state of the system. Using that, the condensate wavefunction can be written as
\begin{equation}
\psi = v \, e^{i \mu t}\, ,
\label{Eq.:Psi_bg}
\end{equation}
where $\theta_{\mathrm{bg}} =\mu t$ is the phase of the ground state where $\mu$ is the chemical potential\footnotemark .
This is the background solution for the the equation of motion (\ref{Eq.:EOM_section3}) as long as,
\begin{equation}
\mu^2=m^2+g\, v^2\,.
\end{equation}
\footnotetext{We identified the chemical potential in our superfluid  as the time derivative of the ground state phase, $\mu=\partial_t \theta_{\mathrm{bg}}$. This is only valid if $\partial_i \theta_{\mathrm{bg}}=0$, which is the chemical potential in the frame where the superfluid moves with velocity $\mathbf{v}$.  If this is not true, the chemical potential is given by $\bar{\mu}=\sqrt{\partial_\nu \theta_{\mathrm{bg}} \partial^\nu \theta_{\mathrm{bg}}}$, which is the chemical potential in the rest frame of the fluid.}
%%%%%%%%%%%%%%%%%%%%%%%%%%%
%%%%% CHECK THIS!!!

When we have a spontaneous symmetry breaking, only the ground state, the state that broke the symmetry, is not invariant under $U(1)$ anymore. This means that we still have a conserved current given by $j^\nu = i (\Psi^\dagger \partial^\nu  \Psi - \Psi \partial^\nu \Psi^\dagger)$, as we saw above in the definition of SSB. Using the equations of motion one can see that this current is conserved. For the ground state (\ref{Eq.:Psi_bg}), this current gives the number density: $n=2\mu v^2$. In the field theory description the conservation of the norm ($\int |\Psi|^2$), related to the number density in the condensate and conservation of the number of particles in the condensate seen in the quantum mechanical approach above, is interpreted as a consequence of the global symmetry of the system which leads to a conserved Noether charge, $Q = \int j^0 \propto \int n \propto \int |\Psi|^2 $ (conservation of the number density of field quanta).

%%%%%%%%%%%%%%%%%%%%%%%%%%
\vspace{0.5cm}
\begin{tcolorbox}[colback=gray!10,enhanced,breakable,frame hidden,halign=justify]
\paragraph*{Spontaneous symmetry breaking (SSB): \qquad} 
SSB is the phenomenon where the state of the theory is not invariant (not symmetric) under the symmetry transformations  ($U$) of the Hamiltonian (or action) that describes the system. This stable state $| \psi \rangle$ spontaneously broke the symmetry of the system.  This mechanism offers an explanation for why there exist stable states, like a condensate, that do not respect the symmetries of a system. This allow for the existence of different symmetry related state $U| \psi \rangle$, of same energy but different phases defining a set of symmetry-broken states. To distinguish them, or their phases, we have the \emph{order parameter} of the system, defined as $\mathcal{O}=[Q,\, \Phi]$, where a state breaks a symmetry $U=e{^i\alpha Q}$ if $\Phi$ exists.
The order parameter can be used to identify if a symmetry was broken $\langle \mathcal{O} \rangle \neq 0$, where the system is said to have \emph{long range order}, given by its two-point function ($C(\mathbf{x}$) being proportional to a constant. For unbroken systems,  $\langle \mathcal{O} \rangle = 0$, where the $C(\mathbf{x},\, \mathbf{x}') \propto \exp(-| \mathbf{x} - \mathbf{x}'| / l)$. The coherent length, $l$, is infinite when there is long range order \footnotemark.
\end{tcolorbox}
\footnotetext{For a very complete and extensive review on SSB, see \citep{Beekman:2019pmi}.}
%%%%%%%%%%%%%%%%%%%%%%%%%%%

\begin{figure}
 \centering
\includegraphics[width=\textwidth]{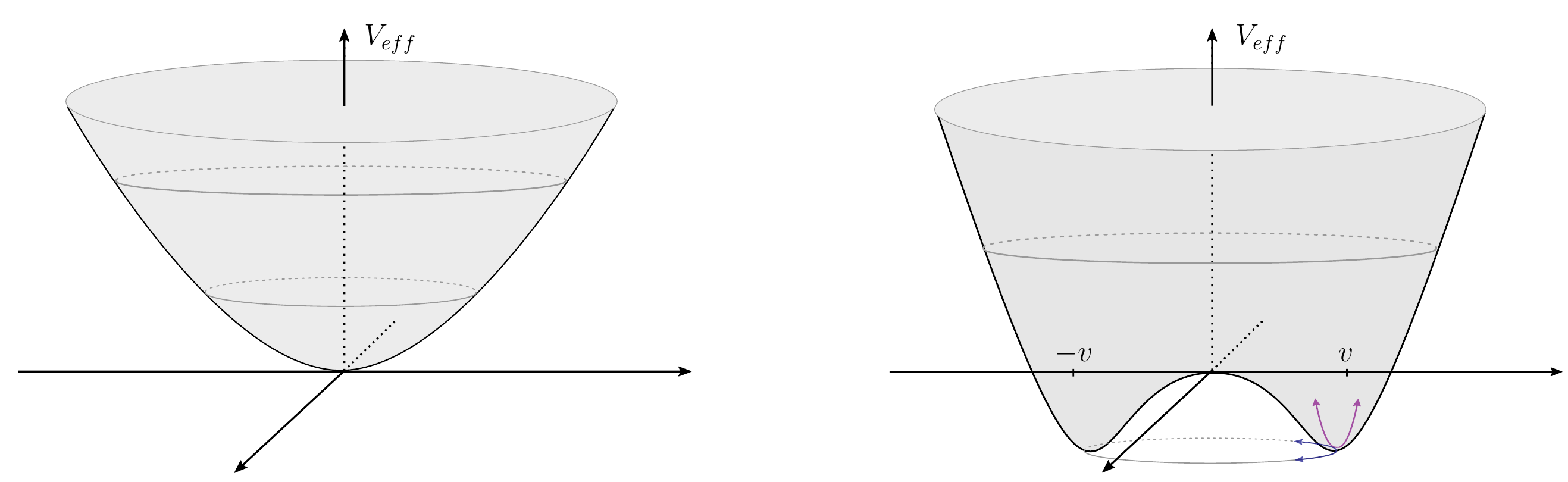}
 \caption{Potential of the weakly self-interacting system. \textit{Left panel:} Potential of the symmetry restored phase, $m^2>0$ which has a minimum at $\psi_{\mathrm{s}}=0$. This ground state is symmetric, respects the symmetry of the system. \textit{Right panel:} Symmetry breaking potential, for $m^2 <0$. This phase has a degenerate minima not invariant under the symmetry of the system. It represents the condensate state.}
\label{Fig.:SSB_sf}
\end{figure}

This is the background solution corresponding to the condensate. This ground state is responsible for  spontaneously breaking the $U(1)$ global symmetry.  And we can see that explicitly. Since this solution represents the ground state, we can compute the energy functional for this system,
\begin{equation}
E=\int d^3 x \,\,  \mathcal{U}= \int d^3 x \, (\partial_0 \Psi \partial^0 \Psi + \partial_i \Psi \partial^i \Psi + V_{eff})\,,
\label{Eq.:energy}
\end{equation}
where $V_{eff}=m^2 \Psi^{*} \Psi + g \Psi^{*} \Psi $. The ground state is given by the stationary, minimum energy state. %This can be found by finding the minimum of the energy (\ref{Eq.:energy}), which amounts to finding the minimum of $\mathcal{U}=v^2 (\mu^2 - m^2)+ (g/2)\, v^4$. This has a minimum at $v=0$, when $\mu^2 < m^2$, and $v_{0,\, \mathrm{s}}=\pm \sqrt{|\mu^2-m^2| / g}$, when $\mu^2 > m^2$. The value of the minima $v_{0,\, \mathrm{ssb}}=\pm \sqrt{|\mu^2-m^2| / g}$ are called vacuum expectation value and are the value that the field assumes at the ground sate, apart from a phase.
% UP: Minimizing E
%Down: Minimizing U
This can be found by finding the minimum of the energy (\ref{Eq.:energy}), which amounts to finding the minimum of the potential energy. The set of solutions for the minimum of this potential energy is:
%This can be found by finding the minimum of the energy (\ref{Eq.:energy}), which amounts to finding the minimum of $\mathcal{U}=v^2 (\mu^2 - m^2)+ (g/2)\, v^4$. The set of solutions for the minimum of this potential energy is:
%%%%%%%%%%%%%%% CHOOSE HOW TO REPRESENT THE GROUND STATE. EQUATION OR TEXT.
% FOR NOW V
\begin{equation}
\left\{ \begin{array}{lcl}
    v_{0,\, \mathrm{s}}=0\,, & \qquad \mathrm{for }\,\,\, m^2 > 0 \,, & \qquad \qquad \mathrm{Symmetry \, restaured} \\
   v_{0,\, \mathrm{ssb}}=\pm \sqrt{\frac{|m^2|}{g}}\,, & \qquad  \mathrm{for }\,\,\, m^2 < 0\,, & \qquad \qquad \mathrm{Symmetry \, broken - condensate} \,.
  \end{array} \right.
\end{equation}
%This has a minimum at $v_{0, \, \mathrm{s}}=0$, when $m^2 > 0$, and $v_{0,\, \mathrm{s}}=\pm m/\sqrt{g}$, when $m^2<0$. 
The value of the minima $v_{0,\, \mathrm{ssb}}=\pm m/\sqrt{g}$ are called vacuum expectation value and are the value that the field assumes at the ground sate, apart from a phase.
As we can see from on the left panel of Fig.~\ref{Fig.:SSB_sf}, the symmetry restored phase when $m^2 > 0$ has a well defined minimum at $v_{0,\, \mathrm{s}}=0$, and this is the normal phase, with no condensation. This ground state $\psi_{\mathrm{s}}=0$ is preserved under the symmetry of the system ($U(1)$ symmetry), the rotations of the phase of the complex field.

The symmetry breaking phase when $m^2 < 0$, there is a continuous set of minima with the ground state given by $ \psi_{\mathrm{ssb}} = v_{0,\, \mathrm{ssb}} e^{i \alpha }$, corresponding to all the possible phases in the circle $\alpha \in [0,\, 2\pi )$, as seen in the right panel of Fig.~\ref{Fig.:SSB_sf}. All of these classical backgrounds are not invariant under $U(1)$ symmetry, the symmetry of the system, which means that $\phi_{i,\, \mathrm{ssb}} \rightarrow \phi_{j,\, \mathrm{ssb}}= v_{0,\, \mathrm{ssb}} e^{i \alpha } \neq \phi_{i,\, \mathrm{ssb}}$. In this way, we say that the symmetry is spontaneously broken by this condensate ground state. 
%A consequence of that is that the number of particles in the condensate is undetermined. 
From the SSB we can see that the condensate has long-range order, with the field having the role of the order parameter of the system.
%As we can see from on the left panel of Fig.~\ref{Fig.:SSB_sf}, the symmetry restored phase when $\mu^2 > m^2$ has a well defined minimum at $v_{0,\, \mathrm{s}}=0$, and this is the normal phase, with no condensation. This ground state $\psi_{\mathrm{s}}=0$ is preserved under the symmetry of the system, rotations of the phase of the complex field.The symmetry breaking phase when $\mu^2 < m^2$, we have now a continuous set of minima with the ground state given by $ \psi_{\mathrm{ssb}} = v_{0,\, \mathrm{ssb}} e^{i \alpha }$, corresponding to all the possible phases in the circle $\alpha \in [0,\, 2\pi )$, as seen in the right panel of Fig.~\ref{Fig.:SSB_sf}. All of these classical backgrounds are not invariant under the symmetry of the system, of the action, which means that $\phi_{i,\, \mathrm{ssb}} \rightarrow \phi_{j,\, \mathrm{ssb}}= v_{0,\, \mathrm{ssb}} e^{i \alpha } \neq \phi_{i,\, \mathrm{ssb}}$. In this way we say that the symmetry is spontaneously broken by the condensate. A consequence of that is that the number of particles in the condensate is undetermined. From the SSB we can see that the condensate has long-range order, with the field having the role of the order parameter of the system.
 %%%%%%%%%%%%%%%%
%%%%%%%%%%%%%

%%%%%% ALSO NEED TO UNDERSTAND IF MINIMIZES THE POTENTIAL ENERGY OR THE TOTAL ENERGY
\footnotetext{An alternative way of writing this Lagrangian that leave explicitly the spontaneous breaking of the symmetry by the finite charge: 
\begin{equation}
\mathcal{L} = |(\partial_\mu - i \mu)\Psi|^2 - m^2 |\Psi|^2 - g |\Psi|^4\,.
\label{Eq.:L_weakly_int}
\end{equation} 
This is equivalent to the usual way we introduce the chemical potential in the Hamiltonian: $\mathcal{H}-\mu \mathcal{N}$, where $\mathcal{N}=j^0$ is the conserved charge. For this modified Lagrangian, the condensate would have trivial phase $\theta_{\mathrm{bg}}=0$. This is equivalent to what we did in the text where the Lagrangian had the canonical kinetic term, but the phase of the condensate had a time dependency with the chemical potential $\theta_{\mathrm{bg}} = \mu t$.}

%%%%%%%%%%%%% EXCITATIONS %%%%%%%%%%%%%%%%
\paragraph*{Excitations} \mbox{}\\

We consider now fluctuations around the classical condensate configuration in order to study the spectrum of this system. Considering small fluctuations means perturbing each degree of freedom of the field around the condensate (\ref{Eq.:mean_field_field_theory}). This is equivalent, in the polar notation to:
\begin{equation}
\psi(\mathbf{x}, t) = (v+\rho) \, e^{i(\mu t+\pi)}\,,
\end{equation}
where $\rho(\mathbf{x},t)$ can be interpreted as a perturbation in the radial direction, and $\pi(\mathbf{x}, t)$ a perturbation in the angular/phase direction.
Plugging this into the Lagrangian, we have,
\begin{equation}
\mathcal{L}=-(\partial_\mu \rho)^2+(v+\rho)^2 \left[ g\,v^2 + 2\mu \dot{\pi}+\dot{\pi}^2 -(\partial_i \pi)^2  \right] - \frac{g}{2} (v+\rho)^4\,.
\label{Eq.:L_pert}
\end{equation}
With this expansion is already easy to see that $\rho$ has a mass term (the term accompanying the $\rho^2$ term), so the perturbation in the radial direction is massive. The perturbation in the phase has no mass term and it is going to be massless. This massless excitation was already expected from the \emph{Goldstone theorem} (see below), where a SSB leads to the appearance of a massless excitation, the Nambu-Goldstone (NG) boson $\pi$. In the context of a superfluid this gapless excitation is the \emph{phonon}. 

For low energy theories, the phonon is the only degree of freedom that is excited in the theory, as the massive mode can be integrated out. Therefore, at low energies a superfluid is completely described by the phonon excitations. 

\emph{Low energies} here mean energies lower than the mass gap of the massive mode $\rho$. We are going to work on this limit here to obtain the spectrum of this NG boson. In this limit, the kinetic term of $\rho$ can be neglected and the equation of motion becomes:
\begin{equation}
g(v+\rho)^2 = g\, v^2 + 2\mu \dot{\pi} + \dot{\pi}^2 - (\partial_j \pi)^2 \equiv X \,.
\end{equation}
Using this we can rewrite the on-shell Lagrangian in terms of $X$ as,
\begin{equation}
\mathcal{L}=\frac{X^2}{2g} = \frac{1}{2g} \left[  g\, v^2 + 2\mu \dot{\pi} + \dot{\pi}^2 - (\partial_j \pi)^2  \right]^2 \,.
\label{Eq.:EFT_weakly}
\end{equation}
This Lagrangian depends only on the phonon $\pi$, as $\rho$ was integrated out, and is invariant under \emph{shift symmetry}, $\pi \rightarrow \pi + c$, inherited from the $U(1)$ symmetry of the complex scalar field. This is the effective Lagrangian at leading order in derivative expansion for the phonon. To obtain the dispersion relation we can expand this Lagrangian, using 
%the canonically normalized field 
$\pi_{\mathrm{c}}=\mu \pi$,
\begin{align}
\mathcal{L} &=  \frac{1}{2g} \frac{1}{4 \mu^2}  \left\{ g^2 n^4+ 4 g n \mu \dot{\pi}_{\mathrm{c}}+  \left( \frac{n g}{\mu} + 4 \mu^2 \right) \dot{\pi}_{\mathrm{c}}^2 - {\color{CornflowerBlue} \frac{n g}{\mu} (\partial_j \pi_{\mathrm{c}} )^2 } 
+ 2 \left[ \dot{\pi}_{\mathrm{c}}^3 - \dot{\pi}_{\mathrm{c}} (\partial_j \pi_{\mathrm{c}})^2 \right] 
+ {\color{BurntOrange} \frac{1 }{4\mu^2} }\left[ \dot{\pi}_{\mathrm{c}}^2 - {\color{BurntOrange} (\partial_j \pi_{\mathrm{c}} )^2} \right]^2 \right\} \,,
\label{Eq.:L_expanded}
\end{align}
where we used that $n=2\mu v^2$.
%%%  ABOUT THE LINEAR TERM
%The terms linear in the fluctuations reduce to a total derivative term when we use theequations of motion (3.8). Thus, they yield no contribution to the action, assuming that the fields vanish at infinity.
Taking the Fourier transform of the field, we compute the equation of motion and can see that the dispersion relation of the phonon is given by,
\begin{equation}
\omega_k^2 = {\color{CornflowerBlue} \frac{g\, n}{\mu} k^2 }+{\color{BurntOrange}  \frac{1}{(2\mu)^2} k^4}=c_{\mathrm{s}}^2 k^2 + \frac{1}{(2\mu)^2} k^4\,,
\end{equation}
where $c_{\mathrm{s}}$ is the sound speed of the superfluid. 
This expression shows us the general behaviour we already seen in the excitations calculated in the QM approach. For long wavelengths (small k), the higher order term in $k$ is suppressed, and the dispersion relation is $w_{k,\, \mathrm{long}}^2=c_{\mathrm{s}} k^2$, which is the dispersion relation characteristic of a sound wave. Therefore, the phonon propagates as a wave with sound speed $c_{\mathrm{s}}$ in the long wavelength regime. 
%This dispersion relation means that the energy necessary to m
%One might think that the Goldstone mode can very easily be excited. And this is true in some sense. For instance, due to the gaplessness, such a mode becomes populated for arbitrarily small temperatures. Landau’s argument
%objects moving through the fluid with velocity less than cs do so frictionlessly.

The second term of the dispersion relation is characteristic of a massive particle. This means that for short wavelengths, the system behaves as a system of normal particles propagating, and not a superfluid.

We can re-write these expressions in the \emph{non-relativistic (NR) regime}, which is the regime we are interested in comparing to the previous approach (and also the limit where we describe, in the next section, the behaviour of DM in galaxies). In this limit, $g \, v^2 \ll m^2$, which implies $\mu^2 \approx m^2$. The dispersion relation in this case is given by 
\begin{equation}
w_{k,\, \mathrm{NR}}^2=c_{\mathrm{s}}^2 k^2 + \frac{1}{4m^2} k^4\,, \qquad \qquad \mathrm{with}\,\,\, c^2_{\mathrm{s}}= \frac{g\, n}{m}\,, 
\end{equation}
which is equivalent to the one found in (\ref{Eq.:dispersion_relation_QM}).

In the absence of interactions, we recover the ideal Bose gas from Sect.~\ref{Sec.:ideal_BEC}, with background solution is given by $ \Psi_{\mathrm{ideal}}=ve^{imt}$, and dispersion relation of a massive particle $w_k^2= (1/4m^2) k^4$, showing again that although this system condenses in to a BEC, in the absence of interactions there is no superfluidity.
It is easy to see from (\ref{Eq.:L_expanded}) that in the relativistic regime, $g \, v^2 \gg m^2$, $c^2_{\mathrm{s}}=1/3$.

In this way we showed that the field theory description of the superfluid is very good to describe the general properties of the superfluid. In order to properly compare with the results obtained above in the QM approach, we are going to show that in the non-relativistic limit, the field theory description gives us the GP and Madelung equations obtained above.

Before doing that, one comment is in order. In (\ref{Eq.:EFT_weakly}), we showed that in the low energy regime we can re-write the microscopic theory of a superfluid as an effective theory only of the phonon with a non-canonical kinetic term. We did this here in the case of a weakly self interacting system with two body interaction, but in section~\ref{Sec.:EFT_superfluid} we are going to extend this idea to general superfluid systems and show the construction of the EFT of superfluids. 

%%%%%%%%%%%%%

\vspace{0.3cm}
\begin{tcolorbox}[colback=gray!10,enhanced,breakable,frame hidden,halign=justify]
  	\textit{Goldstone theorem:} The Goldstone theorem \citep{Goldstone:1961eq,Goldstone:1962es,Nambu:1961tp} or Nambu--Goldstone theorem, states when a spontaneous symmetry breaking occurs a mode with energy that vanishes as $k \rightarrow 0$ is present in the spectrum of excitations of the system. This mode is called Nambu--Goldstone (NG) boson and is a massless particle, in the case of relativistic systems or collective excitations with zero energy gap for non-relativistic systems\footnotemark .
 %The Goldstone theorem \citep{Goldstone:1961eq,Goldstone:1962es,Nambu:1961tp} or Nambu-Goldstone theorem, states that in  a system with SSB a massless particle, in the case of relativistic systems or collective excitations with zero energy gap for non-relativistic systems,  called Nambu--Goldstone (NG) boson is present in a system with SSB. 
 When a symmetry is spontaneously broken, the Noether theorem still applies so there is still conserved currents. The stable state responsible for the SSB  is not invariant under this conserved charge, $Q |\psi \rangle \neq 0$ (or $\langle \mathcal{O} \rangle \neq 0$). This condition implies that there must be a state, the NG mode, with $E_{\mathbf{k}} \rightarrow 0$ as $\mathbf{k} \rightarrow 0$, whose quanta  is a massless boson.
% , exciting a gapless mode $|s \langle = Q |\psi \rangle$.
The NG boson still exists if the symmetry is not exact or broken by an external potential $\mu$, but in this case the mode has a gap $\mu$ at $k \rightarrow 0$. For ordinary NG bosons (type A), the number of NG bosons created  is equal to the number of broken symmetry generators $Q$, $n_{\mathrm{BG}}$.

 The Goldstone theorem described here is valid for a system that is invariant under Lorentz transformation, with the appearance of $n_{\mathrm{BG}}$ NB bosons.
 %For this system we can determine, given the symmetry broken, the number of Goldstone bosons that will appear in the system SSB, $n_{\mathrm{BG}}$. 
 However, SSB is an important phenomena in many systems that are not Lorentz invariant, like the BEC or supefluid, with a number of bosons that will appear in the theory, $n_{\mathrm{NG}}$, called Nambu--Goldstone (NG) bosons, which might not be equal to the number of broken generators, like in the Lorentz invariant case.
A  generalization of the Goldstone theorem, which includes systems  that do not have Lorentz symmetry, exists and can be found in \citep{Watanabe:2011ec,Watanabe:2012hr}. In these works they classify and generalize the Nambu--Goldstone theorem for any symmetry, including non-relativistic systems invariant under Galilean symmetry, showing how to compute the number of NG bosons created by the breaking of such symmetry.
\vspace{0.3cm}

\end{tcolorbox}
\footnotetext{Some translational symmetry needs to be maintained in the system, like Lorentz symmetry for relativistic systems, so that momentum is still a well defined quantum number. }

\paragraph*{Recovering the other approaches: \qquad }  

We want to show that we can recover the GP theory presented in the previous subsection, and emphasize that the field theory description is compatible to describe the superfluid. The field theory presented above is a fully relativistic theory, which means that the action is invariant under Lorentz transformations on top of the global U(1). However, GP description shown above is non-relativistic. Therefore, in order recover the GP and Madelung equations we need to take the non-relativistic limit of the relativistic field theory above\footnote{One can also already start from a non-relativistic action for a field, called Schr{\"o}dinger field, which directly yields the Schr{\"o}dinger equation. See \citep{Beekman:2019pmi} for this derivation.}.

Starting from our field theory for the weakly interacting bosons (\ref{L_2body}), we take the non-relativistic limit of the Lagrangian. We do that by talking the limit $c \rightarrow \infty$
%\footnote{In this part of the section, we are working in natural units. Keeping track of all the $c$s present, the Lagragian is given by $S=-\int d^4 x \, \sqrt{-g} (g^{\mu\nu} \partial_{\mu} \Psi^{*} \, \partial_{\nu} \Psi - m^2 c^2 \Psi^{*} \Psi )$} 
  and assume that in this limit the field has a very fast phase rotation in time, which allows us to rewrite the fields as,
\begin{equation}
\Psi  (\mathbf{x}, t)  = \frac{1}{2m} \psi (\mathbf{x}, t)   e^{-imt}\,.
\end{equation}
With that, the Lagrangian can be written as:
\begin{equation}
\mathcal{L} = \frac{i}{2} \left( \dot{\psi} \psi^{*}-\psi \dot{\psi^{*}}  \right) -\frac{1}{2m} |\nabla \psi|^2 - \frac{g}{16m^2} (\psi^{* } \psi)^2\,.
\end{equation}
From this non-relativistic Lagrangian we can evaluate the equation of motion for the scalar field $\psi$,
\begin{equation}
i \dot{\psi} = \left( - \frac{1}{2m} \nabla^2 + \frac{g}{8m^2} |\psi|^2 \right) \psi\,,
\end{equation}
which is exactly the Gross--Piatevskii equation like shown above, in the absence of a trapping/external potential. 

From that, we can also derive the Madelung hydrodynamical equations. If we make the following substitution:
\begin{equation}
\psi \equiv \sqrt{\rho/m} e^{i \theta}\,, \qquad \mathbf{v} \equiv \frac{1}{m} \nabla \theta\,.
\end{equation}
The vorticity of the superfluid is zero and the momentum density has a non-zero curl. Plugging this in the equations of motion we recover the Madelung equations (\ref{Eq.:continuity}) and (\ref{Eq.:Euler})  in the absence of an external potential. This shows again that we can recover the equations that describe the interacting BEC using the field theory approach.

\vspace{0.3cm}

In this section we showed how to describe a weakly interacting BEC. We showed the standard treatment of the theory, where the many-body quantum system is described by the GP equation. We showed that condensation can be thought as a spontaneous symmetry breaking process and showed that the system can be described in a equivalent way using the field theory approach. We specialized in both cases in the mean-field theory, which is valid for dilute systems, and simplifies the significantly the study of the system.
 %in the mean-field approximation using the field theory approach, the Gross--Pitaevskii approach and the hydrodynamical one. We can see that they are equivalent in some regimes. The choice of which description to use depends on the observables you want to verify from the theory. 
 
It is important to comment on the validity of this theory and the approximations made. The theory presented above is only valid for zero temperatures and in the mean-field approximation, that holds for $na^3 \ll 1$, where we can describe the condensate as a classical wavefunction and the limit where quantum corrections are not important. As we cited above, there are correction to the mean field and other approximations where one can study this model (see \citep{Pitaevskii} for some examples). For finite temperature, one has to describe the superfluid as a two-fluid model.

In the cases we are going to study, we will extend a bit the validity of the zero temperature description, as a first approximation, since in galaxies the temperature is obviously not zero. However, since the occupation number will be very high in our problem, the classical description is safe.

\paragraph*{Two fluid model} \qquad

The description presented above for the superfluid  is valid for a zero temperature dilute weakly interacting Bose gas. However, as already described in Landau's phenomenological theory, for finite temperature system  the superfluid has to  be described as a two-fluid model: a mixture of a superfluid and a normal fluid. As we saw previously in this description, for $T<T_c$, the fluid is a mixture of  normal and superfluid with most of the fluid in the  superfluid phase, while for temperatures higher than $T_c$, the coherence of  the superfluid is broken and the  fluid is mostly  in the normal phase.

Connecting Landau's theory and the microscopic description of a superfluid by a weakly interacting  theory, a two-fluid relativistic theory of superfluid was developed. This can be linked to  the non-relativistic one described above, and a "hydrodynamical" and field theory approaches are developed. We are not going to describe this  in  details in this review, since this  is not  used to describe the current ULDM models in the literature. However, the two-fluid description should be used for finite temperature systems to describe a superfluid and we believe it is important to describe realistic DM superfluid models. For a review of the two-field model, see \citep{Schmitt:2014eka}. In Sect.~\ref{Sec.:ULDM}, we present one work where the  two-fluid formalism is used to describe the self-interacting BEC DM.
\vspace{-0.3cm}
\begin{center}
\rule{0.2\textwidth}{.4pt}
\end{center}

The weakly interacting Bose system studied in this section is the prototypical description of a BEC system that presents superfluidity. It is a microscopical description that shows the behaviour of the condensate and its excitations. But this description is limited and cannot be used to describe all the models of possible superfluids and  realistic experimental systems.
The interesting point  is  that this theory  can be recast as a spontaneous symmetry breaking theory of the U(1) symmetry of the many-body system. BEC and superfluidity are a consequence solely of the spontaneous symmetry breaking, independent on the specific model chosen. Therefore, this hints us to describing  these symmetry breaking systems using the effective field theory approach, where the Lagrangian of the system is described by symmetry alone.  The hope is to be able to describe more complicated superfluid systems. We are going to describe this approach then in the next subsection.

%%%%%%%%%%%%%%%%%%%%%%%%%%%%%%%%%%
\subsection{Effective field theory of a superfluid}
\label{Sec.:EFT_superfluid}
%%%%%%%%%%%%%%%%%%%%%%%%%%%%%%%%%%

We described in the previous section the construction of a microscopic effective theory for the weakly interacting BEC that can be used to describe superfluids. This description is based on London's idea~\citep{London}, that has its roots in the superfluid/superconductor hydrodynamics, that the BEC can be described by a theory with spontaneous symmetry breaking caused be the condensate, with superfluidity arising upon condensation and being described by the Goldstone boson, the phonon, at low energies. 
We saw that we can write an effective Lagrangian for the phonon that describes the behaviour of the superfluid, matching many observations, in the low energy and momentum regimes.

This procedure shows us that we can describe the hydrodynamical degrees of freedom of a theory by the Goldstone modes created by the SSB of a symmetry, the global $U(1)$ symmetry in our case. This is more general than the simple weakly interacting two-body interaction case showed above. 
This is already the case of hydrodynamics that describes macroscopically the behaviour of low energy variables and interactions of system given a symmetry, coarse-graining over the smallest scales. 
This is the perfect playground for the use of effective field theories (EFT), and  EFT techniques are very appropriate for this task. An EFT describes the low energy (long distances) behaviour of a system, without having to refer to its underlying microscopic theory, by parametrizing our ignorance of those short scales.

The idea is to use EFT methods in order to describe the dynamics of a superfluid. This was developed in \citep{Son:2002zn,Dubovsky:2005xd,Dubovsky:2011sj,Son}, where they develop the general formalism to describe fluid hydrodynamics without dissipation as a EFT. In this approach the Lagrangian that describes the system is constrained by symmetry alone.\footnote{There are many other references that also develop EFTs for different condensed matter systems and under different conditions, as it can be seen in these references~\citep{Kolmogorov,Burgess:1998ku,Hofmann:1998pp}, including discussions on dissipation, generalizations and modelling the UV physics that affects the EFT. This is not an extensive list, but showing just some examples of these constructions. For an EFT of pions, one can check the following references~\citep{Gell-Mann_1,Weinberg_1,Gell-Mann_2,Weinberg_2,Weinberg_3}, as some examples.}  
This approach is very powerful since not only can be applied to many different systems, but describes the system without the need of its microscopic understanding. It is also very powerful since it is an expansion over momentum. At leading order in the  expansion, we describe the low energy theory. But this description allows us to go beyond leading order in the long-wavelength expansion.

In our case, we want to construct the EFT that reproduces, in the long wavelength regime (low energies), the superfluid hydrodynamics, as presented above.
This theory is a theory of the phonon, which is the only degree of freedom that is excited at low energies. This is the Goldstone mode produced by the SSB of the $U(1)$ global symmetry by the ground state. We are going to work here in the non-relativistic regime, but one can see the references above for the relativistic case. Restricting to the non-relativistic regime does not imply any loss of generality of the argument, with the system only subjected to different symmetries than the relativistic case.

\vspace{0.3cm}

Inheriting the knowledge of a superfluid from previous sections, at low energies, the only dynamical degree of freedom that describes a superfluid is the phase of the condensate, the phonon. Therefore, in the non-relativistic regime, we need to construct the EFT of this phase $\pi$. The superfluid is described by a theory where the symmetry is spontaneously broken by the ground state,
\begin{equation}
\theta = \mu t -\pi\,.
\end{equation}
The theory that is described by this phase $\mathcal{L}_{\mathrm{eff}} (\theta)$ is invariant under \emph{shift-symmetry} $\theta \rightarrow \theta +c$ which is inherited from the $U(1)$ symmetry of the complex scalar field. 

EFT states that to construct the effective Lagrangian for the phonon, we have to write all the terms that are compatible with the symmetries of the problem. This system has \emph{shift symmetry} and \emph{Galilean symmetry}. For the shift symmetry in order for the Lagrangian to be invariant under this symmetry, only terms that are acted by a derivative  can appear in the theory: $\mathcal{L} = \mathcal{L} \left( \dot{\theta}, \partial_i \theta \right)$. This may contain terms with any power of the derivative of the field. 

However, this theory has more symmetries. In a generic space-time and adding a gauge field, which is a natural extension of the simple scalar field model, we require that this Lagrangian is invariant with respect to three-dimensional general coordinate transformations and gauge invariance. The most general Lagrangian $\mathcal{L}=\mathcal{L} \left( D_t \theta, g^{ij}D_iD_j \theta \right)$ that is invariant under these symmetries, shift symmetry, gauge invariance and general coordinate invariance, is given by,
\begin{equation}
\mathcal{L}=P\left( X \right)\,, \qquad \qquad  \mathrm{with} \,\,\,\,\,  X=D_t \theta - \frac{g_{ij}}{2m} D_i \theta D_j \theta\,,
\label{P(X)_general}
\end{equation}
where $D_t \theta = \dot{\theta}+A_0$ and $D_i \theta = \partial \theta - A_i$. In flat space, $g_{ij}=\delta_{ij}$, the general coordinate invariance corresponds to a Galilean symmetry.

As it can be seen in \citep{Son}, Galilean symmetry is not enough to constrain the NLO terms and one needs to consider the full general coordinate invariance. This is equivalent to considering an additional constraint for the theory, which is known from fluid hydrodynamics,
\begin{equation}
T_{0i}=m\, j_i\,,
\label{Eq.:Greiter}
\end{equation}
where $T_{0i}$ is the off-diagonal component of the energy-momentum tensor. As stated in \citep{Greiter,Dubovsky:2011sj} this additional constraint (with an analogous constraint in the case of relativistic systems \citep{Dubovsky:2011sj})
states that only one degree of freedom carries all the current and momentum. Introducing a new symmetry, full general coordinate invariance, is equivalent to assuming this relation.

With the two symmetries of the system, in the absence of gauge fields $A_0=A_i =0$, the Lagrangian that describes this system is given by:
\begin{equation}
\mathcal{L}=P\left( \dot{\theta} - \frac{(\partial_i \theta)^2}{2m} \right)\,.
\label{P(X)_simple}
\end{equation}
This is a Lagrangian that has a non-canonical kinetic term and it obeys (\ref{Eq.:Greiter}).
%\footnote{To describe a relativistic superfluid one has to impose Lorentz symmetry, which will give an effective Lagrangian, $\mathcal{L}=\mathcal{L}(T)$ where $T=-(1/2) g^{\mu \nu} \Theta_\mu \Theta_\nu$, and $\Theta$ is the relativistic phase coming from a relativistic complex field.}.

At the background $\theta = \mu t$, and $T=0$, this Lagrangian density is equal to the pressure $P=P(\mu)$. With that we can evaluate the particle number density,
\begin{equation}
n=P' (X)\,,
\label{number_density}
\end{equation}
where $'$ indicates the derivative with respect to $A_0$. For the condensate then the equilibrium number density at chemical potential $\mu$ is $n(\mu)=P' (\mu)$, where $P(\mu)$ is the thermodynamical pressure, defined up to a constant.

Given the Lagrangian (\ref{P(X)_simple}) with $\theta=\mu t + \phi$ we can write the Lagrangian as:
\begin{equation}
L=P(\mu)-n\dot{\phi} + \frac{1}{2} \frac{\partial n}{\partial \mu} \dot{\pi}^2- \frac{n}{2m} (\partial_i \pi)^2\,.
\end{equation}
We can see from that the phonon speed of sound
\begin{equation}
c_{\mathrm{s}} = \sqrt{\frac{n}{m} \frac{\partial \mu}{\partial n}} = \sqrt{\frac{\partial P}{\partial \rho}}\,,
\end{equation}
where $\rho = mn$ is the mass density.

One limit of this EFT Lagrangian is the quadratic Lagrangian for the weakly interacting BEC, shown in (\ref{Eq.:EFT_weakly}).  We can see that by considering the special case where $P(X)$ is written as a polynomial, $P(X) \propto (\dot{\theta}/m)^n$. Depending on the power chosen we will have a superlfuid with a different equation of state:
\begin{equation}
\left\{ \begin{array}{lll}
   n=2: & \qquad P \sim \rho^2  & \qquad \qquad \mathrm{BEC/Sf \, (2-body)} \\
   n=3/2: & \qquad P \sim \rho^3 &  \\
   n=5/2:, & \qquad P \sim \rho^{5/3}  & \qquad \qquad \mathrm{Unitary \, Fermi \, gas} 
  \end{array} \right.
\end{equation}
These represent different systems. The case $n=2$ is equivalent to the previous case, in section~\ref{Sec.:Interacting BEC}, where we had a superfluid with a two-body interaction described by the microscopic weakly interacting theory, where we obtained $P(X)=X^2$ for the low-energy (\ref{Eq.:EFT_weakly}). The case with $n=3/2$  can correspond to the same theory as the previous case, a weakly interacting theory, but with a three-body interaction, with this effective Lagrangian obtained by integrating out the massive radial mode, like done in the previous section. These two equivalences shows us an interesting aspect of this EFT and from the hydrodynamics of superfluids: the interaction is linked to the equation of state of the superfluid, and this can be seen by a different choice of $P(X)$ in the EFT. The case $n=3/2$ can also represent another completely different superfluid system, like we will see in the case of the superfluid DM in the next section, where the this case does not come from a  weakly interacting microscopic theory with three-body interaction. This is also the case for the other example we show here,  the unitary Fermi gas, which cannot be described by a microscopic theory like we did in the previous section, and being described with this EFT if $n=5/2$.

One comment is in order here. Usually in quantum field theory having fractional exponents can be problematic, leading to caustics or superluminal propagation. However, in the case of the superfluid this is not a problem. Before reaching these regimes (like the formation of caustics), the superfluid coherence is broken, and the EFT description of the superfluid is no longer valid.

If one wants to add an external or  trapping potential $V_{\mathrm{ext}}$, like for example if the gas is under the influence of a gravitational potential, this corresponds to making $A_0=V_{\mathrm{ext}}$. With that, the Lagrangian is given by (\ref{P(X)_general}), with $X=\dot{\theta} - \frac{(\partial_i \theta)^2}{2m}-V_{\mathrm{ext}}$. In the case of the a condensate in a gravitational potential this is given by 
\begin{equation}
\boxed{\mathcal{L}=P\left( X \right)\,, \qquad \qquad  \mathrm{with} \,\,\,\,\, X=\dot{\theta} - \frac{(\partial_i \theta)^2}{2m}-m \Phi \,,}
\end{equation}
where $\Phi$ is the gravitational potential. This is going to be the case studied in the next section for the DM superfluid.

With that Lagrangian we are able to describe the theory as the other approaches we used to describe the BEC and superfluid theories.

\vspace{0.3cm}
\begin{tcolorbox}[width=\textwidth, enhanced, breakable,  valign=center, colback=white, colframe=black, sharp corners, shadow={0pt}{0pt}{0mm}{black},boxrule=0.5pt,halign=justify,overlay first={
        \draw[line width=.5pt] (frame.south west)--(frame.south east);},
    overlay middle={
        \draw[line width=.5pt] (frame.south west)--(frame.south east);
        \draw[line width=.5pt] (frame.north west)--(frame.north east);},
    overlay last={
        \draw[line width=.5pt] (frame.north west)--(frame.north east);
}]
\paragraph*{\textit{Equivalence of the EFT description: \,\, \qquad}} 

For low energies, and in the non-relativistic case, the EFT of superfluids is equivalent to the microscopic description presented above. Considering $P(X)\propto (\dot{\theta}/m)^n$.

\begin{align*}
&\mathrm{2- body} \quad \mathcal{L}= - |\partial \Psi|^2 -m^2 |\Psi|^2 - \frac{g}{2} |\Psi|^4 \,\, \quad \Longleftrightarrow \qquad \mathcal{L}=P(X) \propto X^2  \,\,\,\,\, \longrightarrow \,\, p \propto \rho^2 \\
&\mathrm{3- body} \quad \mathcal{L}= - |\partial \Psi|^2 -m^2 |\Psi|^2 -  \frac{g_3}{2} |\Psi|^6 \quad \Longleftrightarrow \quad \mathcal{L}=P(X) \propto X^{3/2} \,\, \longrightarrow \,\, p \propto \rho^3
\end{align*}

\end{tcolorbox}

\paragraph*{Superfluid hydrodynamics: \qquad}

 From this formalism we can also describe the superfluid hydrodynamics. From (\ref{number_density}) we re-write the field equation with respect to the number density to obtain the continuity equation:
\begin{equation}
\dot{n}+ \frac{1}{m}	\partial_i \left( n \partial_i \pi  \right) = 0\,.
\end{equation}
The gradient of the field $\pi$ can define the velocity of the superfluid, $\mathbf{v}_{\mathrm{s}} = -\mathbf{\nabla} \theta /m = \mathbf{\nabla} \pi /m$, we can derive the second equation of superfluid hydrodynamics:
\begin{equation}
\dot{\pi}= -\mu(n) - \frac{m v_{\mathrm{s}}^2}{2}\,.
\end{equation}

\paragraph*{Validity of the EFT: \qquad} 

As we saw above, this theory is valid for low energies, or long-wavelength, and breaks for high energies.  The Lagragian shown here is valid in leading order in derivative expansion. In this regime it reproduces the results from hydrodynamics of superfluids. But this framework also allows us to go beyond leading order in this momentum expansion, the next-to-leading order (NLO) Lagrangian.  In \citep{Son}, they show a prescription to take into considerations next-to-leading order terms. This can be done in this framework at arbitrarily order, only requiring that the NLO Lagrangian respects the symmetry of the system, and at the cost of introducing new free parameters. This might allow the study of those systems in a regime that is challenging for the microscopic perturbative description.

However, the validity of this effective theory constructed here still needs to be checked as higher order terms in the Lagragian can only be neglected if the sound speed of the theory is not too small.  In Sect.~\ref{Sec.:4_validity}, we describe the validity of the EFT for that concrete example of superfluid and show that the theory is valid for the parameters of the model.

This theory is also only valid in the absence of dissipation. For a discussion of how to describe dissipative phenomenon in this EFT approach can be seen in \citep{Berezhiani:2020umi}.

\vspace{0.3cm}

%%%% Final words of the section

The formalism of the EFT presented here is used to describe the low-energy dynamics of the superfluid Goldstone mode, the phonon. This formalism is more general, tough, and translates into a EFT language the hydrodynamics of fluids at zero temperature and without dissipation, so it can be generalized to describe superfluids with different equation of state, supercondutivity, unitary Fermi gas, among other systems. This low energy EFT approach is very useful to describe the dynamics of various physical systems and writing the superfluid in this macroscopic effective Lagrangian offers us the chance to study the dynamics of this system without having to work out the details coming from the microscopic short distances physics. With that this formalism allows us to study the behaviour of more complicated superfluids, with different equations of state that might come from different and more complicated interactions. This will be useful to describe the DM superfluid model, in Sect.~\ref{Sec.:ULDM}.

%%%%%%%%%%%%%%%%%%%%%%%%%%%%%%%%%%
\subsection{Rotating superfluid - quantum vortices} \label{Sec.:vortices}
%%%%%%%%%%%%%%%%%%%%%%%%%%%%%%%%%%

When we rotate a normal fluid, the fluid rotates together with the recipient in a homogeneous way, like a rigid body, with vorticity $\nabla \times \mathbf{v_n} =  \nabla \times \mathbf{\mathbf{\Omega} \times \mathbf{r}} \neq 0$, where the normal fluid velocity $\mathbf{v_n}$ of a rotational fluid is given by the cross product of the angular velocity and the position.

As we saw above, a superfluid, described by a weakly-interacting BEC, has irrotational flow: $\mathbf{v_s}=(\hbar/m) \nabla \theta$ which gives $\nabla \times \mathbf{v} = 0$, where viscosity is absent. This is the defining property of a \emph{superfluid}. In another language, this means that the circulation, around a closed contour $\mathcal{C}$ in a superfluid is given by:
\begin{equation}
\Gamma = \oint_\mathcal{C} \mathbf{v}\cdot d\mathbf{l}=\frac{\hbar}{m} \oint_\mathcal{C}  \nabla \theta \cdot d\mathbf{l} = \frac{\hbar}{m} \int \nabla 
\times \nabla \theta =0\,,
\end{equation}
where $d\mathbf{l}$ is a length element on the path $\mathcal{C}$ and $A$ is the area enclosed by this contour. When a superfluid is rotated, this property says that the superfluid would not rotate, but would remain stationary. So, how can we rotate a superfluid and maintain the irrotational flow? This is possible if the superfluid phase presents a singularity.

A superfluid is a state where the system is described by one macroscopic wavefunction. In the presence of this singularity, this wavefunction is single-valued, $\psi(\theta) =\sqrt{\rho /m} \, e^{i\theta}= \psi(\theta+2 \pi n)$, which leads to the quantization of the circulation:
\begin{equation}
\Gamma = \oint _{\mathcal{C}} \nabla \theta \cdot d\mathbf{l}= \frac{\hbar}{m} \Delta \theta = 2\pi n \frac{\hbar}{m}\,.
\end{equation}
This properties  above describe a vortex. The way a superfluid rotates is  inhomogeneous by forming quantized vortices  \citep{vortices1,vortices2,vortices3}.

From that we can see that the azimuthal velocity is of a irrotational fluid is given by $v_{\phi} = n (\hbar /m) (1/r)$, where $r$ is the distance to the center of the closed loop $\mathcal{C}$ and $\phi$ is the azimuthal angle. At the center we have the \emph{vortex core}, as $\lim_{r \rightarrow 0} \psi \rightarrow 0$, of size equal to the healing length, where the density $\rho$ vanishes and $\theta$, the phase, rotates by $2\pi$ around the core. The flow in the center is given by the \emph{vortex line}.  The vorticity of the rotating superfluid is given by:
\begin{equation}
\nabla \times \mathbf{v}_s =  \frac{\hbar}{m} \sum_{i=1}^{N_v} 2\pi \delta (\mathbf{r}_i) \mathbf{\hat{z}}\,.
 \end{equation}
where $\mathbf{r}_i$ is the location of the $N_v$ vortices, and we considered that the vortex lines are in the $z$-direction.  The vorticity is non-zero only at each vortex.
 So the flow is irrotational  in most of the superfluid, except in the vortices. Given that the vortices are line singularities, they form a lattice of uniformly distributed vortices in the superfluid and carry the angular momentum of the rotation $L_z=n \, L_{qm}$, where $L_{\mathrm{qm}} = \hbar N $ is the the minimum angular momentum necessary to have  one quantized vortex and $N=\int |\psi|^2$ is the number of particles in the condensate. This configuration is energetically preferable (instead of for example concentric sheets around the superfluid). With that, the spatially averaged vorticity is given by:
 \begin{equation}
\langle \nabla \times  \mathbf{v}_s \rangle = \frac{\hbar}{m} n_v \,  \mathbf{\hat{z}}=2 \Omega \,  \mathbf{\hat{z}} \,,
 \end{equation}
where $n_v=2\Omega/(\hbar/m)$ is the density of vortices, which is related to the angular velocity \citep{Feynman}. Although most of the superfluid has irrotational flow, the rotational flow is given by the vortices in a way that the whole superfluid then \emph{effectively} flows as a normal fluid $\langle \mathbf{v}_s \rangle=\mathbf{v}_n$, allowing the superfluid to rotate. 
 
 %%%%%%%%%%%%%%%%
 \begin{figure}
 \centering
\includegraphics[width=0.5\textwidth]{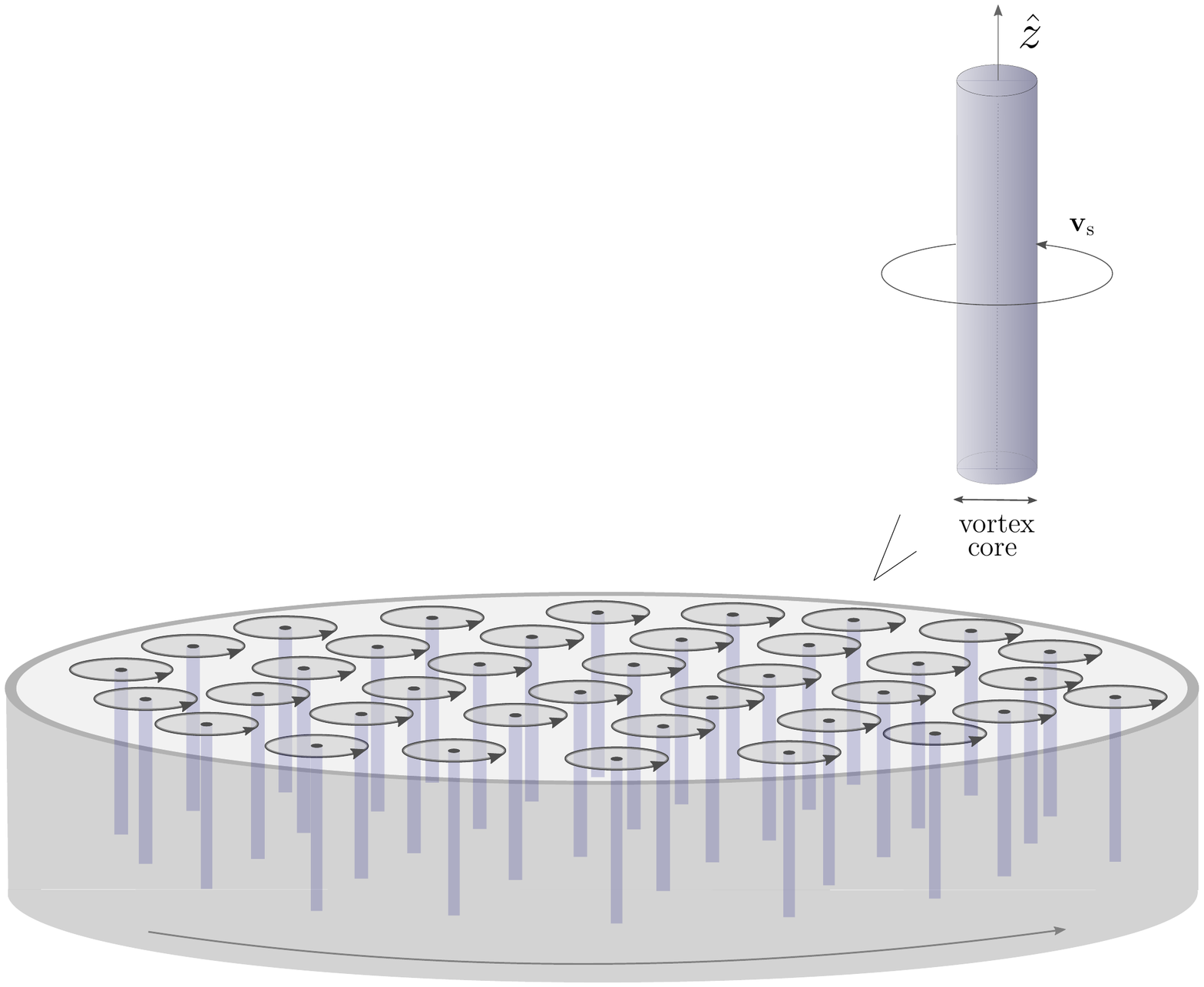}
\includegraphics[width=0.35\textwidth]{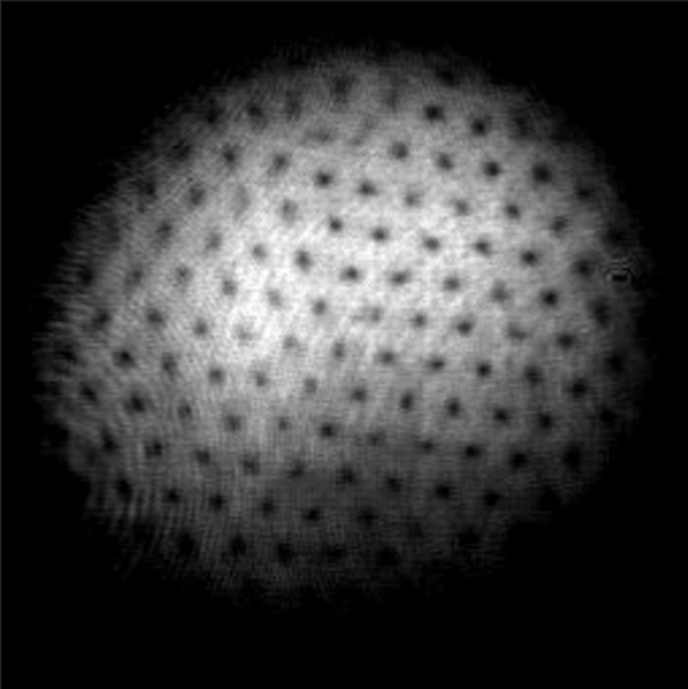}
 \caption{\textit{Left panel:} Schematic of a superfluid in a rotating box. A lattice of vortices is homogeneously formed in the superfluid. \textit{Right panel:} Figure from \citep{vortex_exp} showing the experimental observation of a highly ordered vortex lattice.}
\label{Fig.:Vortices}
\end{figure}
%%%%%%%%%%%%%

Given that the superfluid has irrotational flow, if it is rotated by a small $\Omega$, the superfluid will remain stationary. There is a critical angular velocity, $\Omega_{\mathrm{c}}$, above which the nucleation of vortices in the superfluid occurs. The critical velocity is given by \citep{Guadagnini:2017gtz}:
\begin{equation}
\Omega_c=\frac{1}{mR^2} \ln \left( \frac{R}{\xi} \right)\,,
\label{Eq.:critical_angular_velocity}
\end{equation}
where $R$ is the radius of the cylinder where the superfluid is contained. Through the dependence in the healing length we can see that the critical angular velocity depends on the interaction, for the case of a weakly interacting BEC that behaves as a superfluid. The case where the interaction is attractive, $g < 0$, it is known not to produce vortices. This is also the case where the BEC has a finite size coherence, producing only smaller soliton cores. This condition for  the formation of cores can also be re-written as $L \gg L_{\mathrm{qm}}$, where $L$ angular momentum of the applied rotation to the system.

If there is the formation of vortices, the area density of vortices is given by $n_{\mathrm{v}}=n/A_{\mathrm{BEC}}=m \Omega/\pi \hbar$. From that, we can calculate the size of the vortices created ($\pi R_{\mathrm{v}}^2=1/n_{\mathrm{v}}$) which depends on the mass and the angular velocity of the fluid, $R_{\mathrm{v}}= \sqrt{\hbar/m \Omega}$.

But what exactly is the structure of these vortices and how are they described? As we saw before, we defined the vortex as a singularity, where the wavefunction is zero at the vortex line, with a quantized circular irrotational flow around the vortex line.  Given this, we represent the vortex as an object in cylindrical coordinates: $\psi(\mathbf{r})=f(r,z) e^{in\phi}$. To describe the density distribution, their structure and size, one has to solve the Gross--Pitaevskii equation for this object, which in the case of the weakly interacting Bose gas is,
\begin{equation}
-\frac{\hbar^2}{2m} \left[ \frac{1}{r} \frac{\partial}{\partial r} \left( r \frac{\partial f}{\partial r} \right) + \frac{\partial ^2 f}{\partial r^2}  \right] + \frac{\hbar^2}{2m} \frac{n^2}{r^2} f + V(r,z) f + \frac{g}{8m^2} f^3 = \mu f \,.
\end{equation}
In the limit $n=0$ we recover the standard Gross--Pitaevskii equation. So the term that contains $n$ is called the centrifugal barrier and it is the kinetic energy term from the azymutal velocity of the vortex. Solving this equation gives the density profile of the vortex, from which we can determine the size, density and general structure of the vortex. 

In the presence of a trapping or external potential, changes the density of the condensate which also changes the dynamics of the vortices. For details of the formation of vortices in the presence of a trapping potential see also \citep{Fetter:2009zz}.

%Nevertheless, the presence of a nonuniform trapping potential Vtr(r) and the resulting nonuniform density in Eq. (2.26) significantly affect the dynamics of vortices, so that qualitative classical pictures based on uniform incompressible fluids may not always apply

Since its discovery in Helium 4 superfluid, vortices have been one of the central topics in the research of supefluidity \citep{Madison:2000zz}. They have been experimentally observed in many systems like Helium 4 and Helium 3, superconductors and atomic BECs, including multi-component BECs. These new observational advances allowed us to vizualize and study those vortices, as it can be seen in left panel of Fig.~\ref{Fig.:Vortices}.

%%%%%%%%%%%%%%%%%%%%%%%%%%%%%%%%%%%%%%%%%%%%%%%%%%%%%%%%%%%%%%%

\subsection{BEC in wave turbulence - kinetic theory}
\label{Sec.:wave_turbulence}

We studied in this section the BEC and the superfluid, and showed that their dynamics is described by the Gross--Pitaevskii equation, which is a non-linear Schr\"{o}dinger equation. This equation expresses the evolution of the condensate, which is described by the collective wavefunction formed by the macroscopic occupation of the ground state by the bosons, in the mean field self-potential. 
This theory is capable of describing many  different condensate systems and describe the properties of BECs that can be tested with experiments.
There is, however, an alternative way to describe the physics of a BEC, using the theory of wave turbulence. 

Wave turbulence is the theory  that describes  non-equilibrium statistical systems using random non-linear interacting waves. These random interacting waves are the fundamental constituent of the theory.  Waves in a non-linear medium interact and behave very differently than waves in vacuum and wave turbulence describes these systems of interacting non-linear waves. This approach can be used to describe many different physical system from quantum fluids to astrophysics, which is our main interest.  Wave turbulence arises in classical context in systems like non-linear optics, surface waves in water, magnetic turbulence in interstellar gases, among many others (see~\citep{Nazarenko2011_intro} for a list of examples and references).  Quantum fluids are also physical systems where wave turbulence occurs, and this approach can be used to explain exotic superfluids, atomic BEC, second sound waves in $^4$He superfluid, and other systems as it can be seen in~\citep{Kolmakov4727}. Wave turbulence is a formalism specially useful to model these systems numerically and for experimental studies.

Wave turbulence can arise either in strong or weak interacting system of random non-linear waves. Weak turbulence is a limit where it is considered that all the waves are weak and have random phases. It provides a theoretical framework for wave turbulence theory to describe many physical system, representing  those systems by a wave kinetic equation. This equation describes the evolution of the wave spectrum, two point correlation function or the probability distribution function, via averaging over random phases, of a system. This kinetic equation is different for each system and can have a  three wave non-linear interactions, four-wave interactions  or even higher order depending on the system described. Higher moments of this equation allow the study of deviations of gaussianity or to explore the limits of validity of the wave turbulence. 

%The kinetic equation presents only the first moment, but higher moments of this equation can also be studied to verify deviiatiion This equation contains not only the kinetic equation as its first moment, but also all higher moments that allow to study deviations from gaussianity and intermittency, as well as the validity of the underlying WT assumptions. 

Wave turbulence can also be used to describe Bose Einstein condensates and superfluids. It can offer a way of describing different stages of a BEC systems in a nearly equivalent way of its proper description through the non-linear Schr\"{o}dinger equation presented above. We are going to show now how and for which limits a BEC can be described using wave turbulence approach. We are going to show that wave turbulence with four wave interaction, which reproduces the non-linear Schr\"{o}dinger equation, can only describe the classical initial evolution of a BEC, breaking as the condensate evolves. We also show that the quantum evolution of a condensate can be described by this theory.
Wave turbulence also offers a description for a BEC and superfluids outside the weak interaction limit, but we are not going to discuss this case here.

\vspace{0.3cm}

We want to develop a statistical description of the wavefunction and of the non-linear Schr\"{o}dinger equation, and determine the kinetic equation which is the main equation of the wave turbulence theory. This is done by describing the evolution of an ensemble of waves (in contrast with the Boltzmann equation that is the equation that describes the evolution of a distribution of particles). For that we introduce the Wigner distribution~\citep{wigner}
\begin{equation}
f_{\mathbf{p}}(\mathbf{x},t) \equiv \int d\mathbf{y}\, e^{-i\mathbf{p.y}} \langle \psi(\mathbf{x}+\frac{\mathbf{y}}{2}) \psi^{*}(\mathbf{x}-\frac{\mathbf{y}}{2}) \rangle \,,
\label{Eq.:Wigner}
\end{equation} 
where the average is the average with respect to the random initial phases of the field. The Wigner distribution is a quasi-probability distribution, it behaves like a probability distribution but it does not obey all the axioms of probability theory by Kolmogorov~\citep{Kolmogorov33book}. It can be used to represent this ensemble of waves with random initial phases, or of random classical fields. As a quasi-probability it can acquire negative values indicating interference of waves in phase-space~\citep{interference_waves} describing the ondulatory behaviour that can be captured in this description.

We want now to describe the evolution of this ensemble. We want to have a kinetic theory that is nearly equivalent to the GP equation (\ref{Eq.:GP_equation}) for a weakly interacting system. For that, as we can see from the interacting Hamiltonian of this theory (\ref{Eq.:Many_body_H}), we need a four wave interaction. With that we can rewrite (\ref{Eq.:GP_equation}) using (\ref{Eq.:Wigner}) as
\begin{equation}
\frac{\partial f}{\partial t}+\frac{\mathbf{p}}{m} \cdot \nabla_{x} f = 2 \, \mathrm{Im} \int d\mathbf{y}\, e^{-i\mathbf{p.y}} \langle \psi(\mathbf{x}+\frac{\mathbf{y}}{2}) \psi^{*}(\mathbf{x}-\frac{\mathbf{y}}{2}) U_{\mathrm{tot}}(\mathbf{x}+\frac{\mathbf{y}}{2}) \rangle \,,
\label{NLS_kinetic_field}
\end{equation}
where we wrote $f_{\mathbf{p}}(\mathbf{x},t)$ as $f$ for simplicity of notation. Here we are considering that $U_{\mathrm{tot}} = g |\phi(x)|^2$. This is the kinetic equation or the wave kinetic equation.
% In the case we are interested in the next section of the gravitational interaction this is the case and then the total potential can be the self-interaction plus the gravitational interaction. There are some subtleties in the case of the gravitational potential and we will discuss this in Sect.~\ref{Sec.:description_condensate}.
 
This equation gives the evolution of the distribution that describes the an ensemble of waves that is the two point correlation function of the field.
This distribution can be interpreted as the wave action or the particle number density or particle occupation number, where this second interpretation leads us to the idea we had before that $\psi$ can be thought as the classical limit for the quantum wavefunction or field of a weakly interaction Bose gas     \citep{optical_turbulence,kinetic_theory_BEC_1}.  

We want to obtain the closed form of the kinetic equation, expressed in terms of $f$.
The above kinetic equation presents, in the case of the four wave interaction, a four point correlation function. For the case of weak coupling that we are studying here,  the Wick theorem is approximately valid and we can express this four-point function into the sum of two-point functions, reducing the higher order problem into a lower order one:
\begin{equation}
\langle \psi_1 \psi^{*}_2 \psi_3 \psi^{*}_4 \rangle = \langle  \psi_1 \psi^{*}_2 \rangle  \langle  \psi_3 \psi^{*}_4 \rangle +  \langle  \psi_1 \psi^{*}_4 \rangle  \langle  \psi_3 \psi^{*}_2 \rangle + \langle \psi_1 \psi^{*}_2 \psi_3 \psi^{*}_4 \rangle_{\mathrm{conn}}\,,
\label{Eq.:Wick_theorem}
\end{equation}
where we simplified the notation writing  $\psi_i = \psi(x_i)$, and the last term represents the connected part of the correlation function, the non-diagonal part of the operator. The connected part is non-zero if the distribution is non-Gaussian, and it is also higher order in the interaction.

In the regime of small non-linearities and taking the random phase approximation~\citep{random_app}, which allows us to ignore the higher order correlations, we can write the closed form for this equation. We wish to work in Fourier space, so we take the Fourier transform of the field. The above approximations correspond to ignoring higher Fourier moments and writing equation (\ref{NLS_kinetic_field}) in terms of the two-point correlation function $\langle \psi_{k_i} \psi_{k_j} \rangle = n(k_i) \, \delta(k_i - k_j)$, where $\delta(k_i - k_j)$ is the Dirac delta~\citep{Nazarenko_BEC_vortices,Nazarenko2011,Kolmogorov}. 
Given these assumptions, we can re-write the kinetic equation in closed form:
\begin{equation}
\dot{n}_k = 4\pi \int \, n_1 n_2 n_3 n_4 \, \left[ \frac{1}{n_k} + \frac{1}{n_3} - \frac{1}{n_1} \frac{1}{n_2}  \right] \delta(\mathbf{k}+ \mathbf{k}_3- \mathbf{k}_1 -\mathbf{k}_2 )\, \delta (k^2 + k^2_3 -k_1^2 -k_2^2)\, d\mathbf{k}_1 d\mathbf{k}_2 d\mathbf{k}_3 \,. 
\label{kinetic_equation_closed}
\end{equation}
The above equation is the equation for the evolution of the wave spectrum. 
One can also also write the equation for the  higher moments if interested in investigating deviations from gaussianity or the limits where this description breaks, among other phenomena.

This equation reproduces the non-linear Schr\"{o}dinger equation, which is the equation that governs the evolution of a wavefunction. Any wavefunction. It can be thought as the quasi-classical limit at high occupation number of the quantum kinetic equation (we discuss this later in this section). 
This equation describes the evolution of any ensemble of classical waves that is described by the Schr\"{o}dinger equation. It describes a non-condensed system, since in this description condensate density, which is the density of particles in the ground state which in the language of waves translates into waves with $k=0$, is relatively small (in comparison to a strong BEC deep into the region $T\ll T_{\mathrm{cr}}$ where the number of particles in the ground state is almost equal to the total number of particles). When the condensate density is large or the population of lowest momentum states is large, the system is not weakly non-linear anymore and the above description breaks.
However, this description can be used to the initial stages of a condensate, when the condensate density is still small. We call this limit ``weak'' condensate. When we have self-interactions the system can form a superfluid, and for this four wave system this equation describes the vicinity of superfluid transition. \textit{Therefore, the kinetic equation with four-wave interaction can describe the classical initial evolution of a BEC, when we have a ``weak'' condensate.} 

As temperatures drop and the condensate density becomes larger, we have a ``strong'' condensate.  In this stage of the evolution of the condensate the above description with a four wave interaction is not a descriptiion of the system, and the evolution of the occupation number of the condensate needs to be described by the wave kinetic equation with \textit{three-wave} interaction\citep{Nazarenko2011,Kolmogorov,optical_turbulence,WT_BEC}. This three wave interaction representation describes the later phase in the evolution when the condensate is strong and fluctuations on top of the condensate are only given by phonons. Only at this stage the theory can again be described by weak turbulence, but now involving three-waves.  The intermediary regime between these two description is more complicated.  After the four-wave description breaks down, the system is highly non-linear composed by a gas of hydrodynamical vortices. Only after these vortices annihilate and most of the systems is in the condensed phase, one can use the three-wave description.

One possible solution of (\ref{kinetic_equation_closed}) is the Kolmogorov--Zakharov (KZ) spectra, which is a non-equilibrium steady-state solution. Within this solution we can have turbulent cascade processes, with a dual cascade for different direction of the energy flux. The interpretation of these dual cascade processes in BEC is interesting and  corresponds to techniques used in experimental realizations of BEC. Inverse cascading, which is the non-equilibrium transfer of particle to the lowest energy momentum can be though as condensation. The the initial process of  BEC formation can be achieved in this process as a non-equilibrium condensation. The forward cascading is a processes is the energy transfer to higher momentum states, higher energy level. When the condensate is in a trap, these particles are going to leave the trap, and this is called evaporative cooling.

We studied above the case without a trapping potential. When in a trapping potential, the condensate density is now coordinate dependent and the behaviour in this trap will depend on the relation between the characteristic mean free path of the excitation wave packets and the size of the trap or the range of the force that produces this potential.

We are gong to see in Sect.~\ref{Sec.:description_condensate} another solution of the above four-wave kinetic equation in the case of long-range interactions, like the gravitational interaction. The potential is present and it is the gravitational potential. This is described by the Landau kinetic theory, and can describe the initial stages of a condensate in the presence of gravitational interactions. Different than in the KZ description of condensation discussed above, in this case condensation does not arise from a cascading process but from a dissipation process.

\vspace{0.3cm}

We described above the four-wave classical kinetic equation. This is a good \textit{classical} limit representation of the initial stages of a BEC. However, this is not a description of strong condensates, as a three wave kinetic equation needs to be adopted instead. This is also not a description of the quantum condensate. Wave turbulence theory can also be used to describe the full quantum regime of condensate, and not only the classical limit described until now. To be able to do that we first need to discuss the statistical distributions that the kinetic equations describe. 
%With this we can also see that the four wave classical kinetic equation to have a relevant description needs to be in a region where the ultraviolet (UV) catastrophe does not take place.

One of the distributions that is a solution of the classical equation (\ref{kinetic_equation_closed}) is the Rayleigh-Jeans (RJ) distribution, $n_k = T/(k^2+mu)$, with $T>0$ and $\mu > 0$.  
This is similar to what we find when we are studying classical limit of the systems described by the Gross--Pitaevskii equation. Thu full quantum system described by (\ref{Eq.:full_condensate_Schrodinger}) has a occupation number that describes a Bose-Einstein statistics. The classical limit of this system, described by the Gross--Pitaevskii equation has a mean occupation number in equilibrium described by the RJ distribution, the classical limit of the Bose-Einstein distribution: $\langle n_k \rangle = k_B T/(\epsilon_k - \mu)$, where $\mu <0$. This system describes a condensate only when the occupation of the ground state is macroscopic, $\langle n_k \rangle \rightarrow \langle N_0 \rangle $.
 
The temperature $T$  in this RJ distribution is related to the initial energy of the system $E_0 = \int d\mathbf{k} \, \omega_k n_k = T   \int d\mathbf{k} $, where $\omega_k = k^2 + \mu$, in connection to thermodynamics. From this we can see that since each degree of freedom of the theory has the same energy $T$, for a continuous and infinite system, the this and the energy diverge. This is the classical RJ catastrophe or the UV catastrophe.

Thus, the UV catastrophe is inherent to the ensemble of classical nonlinear waves. It is argued then that for the RJ solution to be a relevant solution of the kinetic equation a cutoff needs to be introduced to regularize the UV catastrophe. This is as a phenomenological way of making the classical description of the system valid. 
Therefore, describe a BEC in such a classical theory one needs to have a momentum cutoff in the theory. This truncated system then can be used to describe the evolution of a classical ensemble of waves via the kinetic theory derived above. In realistic systems, this cutoff sometimes comes naturally from dissipation or limits of the simulation or experiment. If one is working in a description where the UV catastrophe does not take place, where the evolution period of the system is not threatened by this divergence, then the above description for the classical condensate also holds. This will be the case of Sect.~\ref{Sec.:description_condensate}.

Another solution to make the description valid (avoiding the UV catastrophe) is to go to a quantum statistics. We can modify the kinetic theory in order to obey a Bose Einstein statistics.
We generalize the kinetic equation to:
\begin{equation}
\dot{n}_k = 4\pi \int \, n_1 n_2 n_3 n_4 \, \left[ \left( \frac{1}{n_k} +1 \right)  \left( \frac{1}{n_3} +1 \right) -  \left( \frac{1}{n_1} +1 \right)   \left( \frac{1}{n_2} +1 \right)   \right] \delta(\mathbf{k}+ \mathbf{k}_3- \mathbf{k}_1 -\mathbf{k}_2 )\, \delta (k^2 + k^2_3 -k_1^2 -k_2^2)\, d\mathbf{k}_1 d\mathbf{k}_2 d\mathbf{k}_3 \,. 
\end{equation}
The Bose-Einstein statistics is now a solution of this equation. This is the \textit{quantum kinetic equation} and can be used to describe the full quantum BEC.

\vspace{0.3cm}

We showed above that we can use wave turbulence as an approximate description of some regimes of a BEC. The classical four wave kinetic equation can be used to describe the initial stages of a BEC in the classical limit. Different description will arise for different stages of the evolution of the condensate.Wave turbulence can also describe both classical condensate evolution and quantum evolution. 
%Therefore this theory can only approximate the BEC evolution, and
%Therefore, this theory is not a microscopic theory of a BEC or a superfluid, but it gives a description of a condensate at some stages and limits  in terms of random non-linear interacting waves. 
% it offers a nearly equivalent description of this system at different regimes: classical, quantum, weak condensate, strong condensate, ... each of these regimes being represented by a different kinetic equation. 
 A BEC is a quantum phenomena, but its classical evolution can approximately be described by wave turbulence which is a convenient description, in particular for simulations.

We are going to use kinetic theory to describe the formation of the BEC in Sect.~\ref{Sec.:description_condensate}. 

%%%%%%%%%%%%%%%%%%%%%%%%%%%%%%%%%%%%%%%%%%%%%%%%%%%%%%%%%%%%%%%
\subsection{Summary and discussion: what is a condensate?}
\label{Sec.:discussion_sec3}

We saw in this section an introduction to two of the most interesting phenomena in quantum mechanics, the BEC and superfluids. We went through all these concepts in detail with the goal to give a proper definition of Bose Einstein condensation in the context it is well understood and measured.
 
BEC is the phenomena of macroscopic occupation of the ground state that happens at low temperatures. BEC is a consequence of the quantum statistics of bosons, and it is an inherently a quantum phenomena.
Equivalent to having a macroscopic occupation of the ground state, condensation can be though a the regime where the interparticle distance is smaller than the de Broglie wavelength of the bosons, which leads to a superposition of these wavefuctions, creating a macroscopic wavefunction that describes the condensate, which is a macroscopic quantum object. One of the main properties of a BEC is that it presents macroscopic (long range) quantum coherence.

We also saw that we can describe a more realistic condensation processes by using a weakly interacting Bose gas, which exhibits superfluidity upon condensation. This theory is described by the fully quantum many-body Hamiltonian. For a large number of particles ($N \gg 1$), this Hamiltonian is very complicated to be studied.  But, when the interactions in the BEC are weak, the BEC is dilute\footnote{The BEC has 3 scales in the absence of a trapping potential: the de Broglie wavelength, the $s$-wave scattering length $a$ and the inter-particle distance $d$.  In order to describe the scattering of two particles that have large $\lambda_{dB}$ as the scattering of two bodies, we have to have $  d \gg a$.  The mean field approximation is applicable in the limit where we have many particles $N$ large and $na^3 \ll 1$, meaning that the interactions in the condensate are weak, which translates to $a \ll \lambda_{db}$. A condensate that follows this condition is said to be dilute, which means that for fixed $n$, the bosons almost don't interact, $a$ must be very small.} $na^3 \ll 1$, and for large N,  we can take the \textit{mean field approximation}. In the mean field approximation, we can make the huge simplification that the many-body wavefunction can be approximated by an effective single-particle wavefunction. This means that the wavefunction of the condensate can be written as:
\begin{equation}
\hat{\Psi} (\mathbf{r}, t) = \psi (\mathbf{r}, t) +\delta \hat{ \Psi} (\mathbf{r}, t)\,,
\label{classical_separation}
\end{equation}
where $\psi (\mathbf{r}, t) \equiv \langle \hat{\Psi} (\mathbf{r}, t) \rangle$ (this is a classical quantity, because if it was quantum this average would be zero).   The field $\psi (\mathbf{r}, t)$ is the classical field or classical wavenumber. Quantum effects are suppressed in this limit (the depend on $1/N$), and the BEC is well approximated by a classical theory\footnote{The mean field approximation, $N$ large and weak coupling, and the classical limit ($\hbar \rightarrow 0$) not always coincide. The mean field approximation is called semi-classical in some places of the literature. In some instances the mean field $N \rightarrow \infty$ can be recast as a classical limit\citep{mean_field_vs_classical}. The mean field approximation is usually concerned to systems that preserve the number of particles, like condensates.}. 
The classical field that represents the condensate satisfies the (classical) non-linear Schr{\"o}dinger equation or Gross--Pitaevskii equation.
Therefore, the classical Gross--Pitaevskii or non-linear Schr{\"o}dinger description of a condensate is a mean field description of condensate. 

An important detail about the mean field approximation.
In the classical limit of a scalar field we also have (\ref{classical_separation}), where $\psi = \langle \hat{\Psi} \rangle$ is the scalar field and $\delta \hat{\Psi}$ are suppressed quantum corrections.  Now, when a condensate can be treated as classical we are also in the limit where we can expand (\ref{classical_separation}), but we have  $|\psi(\mathbf{r},t)|^2 = n_0 =n$, which means that there is a macroscopic occupation of the ground state. 
For the classical limit of the condensate, as the term $\delta \hat{ \Psi} (\mathbf{r}, t)$ becomes more and more important and the mean field approximation break this can be seen as depletion of the condensate. This limit can be broken if temperature or interactions are increased, then quantum and thermal fluctuations deplete the condensate\footnote{It is also possible to describe a non-condensate quantum system in a classical limit. This was seen in the wave turbulence theory. In this case, there is no macroscopic occupation of the ground state but the entire system is still in the high occupation classical limit. In this case $\delta \hat{\Psi}$ is not the depletion from the condensate but it represents the quantum correction to the system. }. At this point the mean field approximation breaks and this classical description of the condensate cannot be used anymore.

Although the condensate to be formed depends on the quantum statistics of bosons and it is a quantum phenomena, it can be treated as classical in the mean field approximation.
In an equivalent way, in the field description of BEC,  the condensate can be described in the mean field limit by a classical field\footnote{There is only one example in the literature of condensed matter physics where there is classical ``condensation''. This happens for electromagnetic light waves in nonlinear optics~\citep{condensation_light_1,condensation_light_2}. Kinetic condensation is achieved when the light beam goes from a disordered to a coherent state. However, as it was emphasized in~\citep{condensation_light_theory} this classical condensation is a process analogous to the (genuine) Bose Einstein condensate, having similar properties and obeying the non-linear Schr{\"o}dinger equation. This is a very new and active field of study and it is going to be very interested to see the development of this field.}.

\textit{Summarizing}: A BEC can be described by a coherent classical scalar field that satisfies the Gross--Pitaevskii equation, \textbf{in the mean field approximation} where most of the particles are in the ground state, $\mathbf{|\psi(\mathbf{r},t)|^2 = n_0 =n}$.  
%Notice that this last condition is very important for the validity of the approximation.

\vspace{0.5cm}

 The term ``condensate'' is used in the literature loosely meaning different things for different authors. I will use throughout this review the definition presented here. Therefore, every time I am using the term condensate, I am referring to the definition described here.

%%%%%%%%%%%%%%%%%%%%%%%%%%%%%%%%%
% Ultra-Light Dark Matter
%% NEW SECTION 4 

%%%%%%%%%%%%%%%%%%%%%%%%%%%%%%%%%%%
\section{Ultra-light dark matter}
\label{Sec.:ULDM}
%%%%%%%%%%%%%%%%%%%%%%%%%%%%%%%%%%%

After the introduction in the previous sections of the concepts that are going to be applied in this part of the review, we are finally ready to discuss the ULDM models.

Ultra light  DM denotes a class of models where DM is composed by ultra-light bosons. These models were introduced as a new class of DM models that can address the small scale challenges of $\Lambda$CDM, but mainly as models that offer a novel and rich phenomenology in galaxies that can be tested with small scale observations.
The general idea of those models is that inside a virizalized DM halos, ULDM thermalizes and forms gravitationally bounded cores that can be described as a BEC or a superfluid. In this way these models behave like CDM on large scales, with modified initial conditions, recovering the incredible observational successes of this description, while inside galaxies they present a wave-like behaviour. 

In order to have this behaviour inside galaxies, the mass of this bosonic DM has to be very small. There are many models in the literature of ULDM that present this wave-like behaviour in galaxies, and the specific range of masses where this wave-like behaviour happens in galaxies depends on the specifics of the models. However, we can estimate in a model independent way the range of masses of the ULDM particles in order to present this behaviour in galaxies. The mass of the ULDM has to be:
%There are many models in the literature that present specific mechanisms for this process. However, we can estimate in a model independent way the bound on the mass of the ULDM where there is the formation of this condensed core inside the galaxy. 
\begin{equation}
10^{-25}\, \mathrm{eV} \lesssim m \lesssim 2 \, \mathrm{eV}\,.
\label{Eq.:Up_low_bound_m}
\end{equation}
The lower bound on the mass is very general and comes from the fact that the size of the condensate core cannot be larger than the halo, since we want the condensate only on galactic scales and normal DM on larger scales. The maximum case we can have for the formation of a condensate is where the de Broglie wavelength of the ULDM particle is of the order of the size of the halo. Taking this bound at virialization, there is a maximum value on how large the de Broglie wavelength can be $\lambda_{\mathrm{dB}} < R_{200}$. Taking $z_{\mathrm{vir}} \sim 2$ and for halos with mass of order of $10^{12} M_\odot$, we can see that for a spherical halo \citep{9} this imposes a lower bound on the mass
\begin{equation}
m > m_{\mathrm{H}} \equiv \frac{2 \hbar}{\sqrt{3G}} (R\, M)^{-1/2} \approx 10^{-25} \left( \frac{M}{10^{12} \, M_\odot} \right)^{-1/2} \left( \frac{R_{200}}{100\, \mathrm{kpc}} \right)^{-1/2} \, \mathrm{eV}\,.
\end{equation}

We can also impose an upper bound on the mass asking the question: what is the biggest mass I can have that ULDM forms a core inside the galaxy? Again, to answer this question one needs to work with a specific ULDM model to study the Jeans theory of this model and the solutions of the equations in this region, which would give a bound within this model for the creation of these cores. However, one can try to be more general. The non-CDM behaviour happens in the regions where the wave behaviour takes place. So, the interesting non-CDM behaviour occurs on scales of the order or smaller than the scale that characterizes the wave which given by the de Broglie wavelength. This is the maximal case where the de Broglie wavelength of the ULDM particle is of the size of the galaxy. But one can also think that we can obtain a galaxy size wave as the superposition of the de Broglie wavelength of each of the ULDM particles (which are themselves smaller than the galaxy in this hyphothesis). Then we can calculate the biggest mass, which means smaller de Broglie wavelength if each particle, for which this superposition yields a galaxy size wave. This translates to the condition that de Broglie wavelength of the boson DM is larger than the inter-particle distance between each boson,
\begin{equation}
\lambda_{\mathrm{dB}} \sim \frac{1}{mv} > l= \left( \frac{m}{\rho} \right)^{1/3}\qquad \Longrightarrow \qquad m < \left( \frac{\rho}{v^3} \right)^{1/4}\,,
\label{cond1}
\end{equation}
where assuming a spherical halo, the interparticle distance is defined as the radius of a sphere with density $\rho$. This gives a bound on the mass of the DM particle. 
This condition is the same as the condition that a gas need to have in order to condensate into a BEC, as we can see in the box in Section 3.1 since it is equivalent to the condition of having macroscopic occupation number of the ground state for temperatures bellow the critical temperature of the system. This condition does not determine condensation of DM in the halo, since showing that condensation happens in the halo is much more complicated than this ideal gas condition that is not realistic for the halo, but it is a $0^{th}$ order condition for this phenomena together with the assumption of thermalization. But here we use it only as a condition to form a galaxy size macroscopic wave from the superposition of the individual particles' waves.

We use the density and velocity of the dark matter halo like described in Sect.~\ref{Sec.:Small_scales} from standard spherical collapse~\cite{Berezhiani:2015pia}, and take this bound at virialization:
\begin{align}
  \rho_{200} &= 200 \rho_{\mathrm{cr}} \sim1.95 \times 10^{-27} \, \left( 1+z_{\mathrm{vir}} \right)^3  \, \mathrm{g/cm}^3\,, \nonumber\\
      V_{200} &\sim 85 \left( \frac{M}{10^{12} M_\odot}\right)^{1/3} \sqrt{1+z_{\mathrm{vir}}} \,\, \mathrm{km/s}\,, \label{virial_MW}
\end{align}
where we derived these expressions assuming $H_0 \sim 70\,  \mathrm{km \, s}^{-1} \, \mathrm{Mpc}^{-1}$ and a halo mass of order of the MW. This gives the bound:
\begin{equation}
m \lesssim 2.3 \, (1+z_{\mathrm{vir}})^{3/8} \left( \frac{M}{10^{12} M_\odot}\right)^{-1/4} \, \mathrm{eV}\,.
\end{equation}
Taking $z_{\mathrm{vir}} \sim 2$ and for halos with mass of order of $10^{12} M_\odot$, we have an upper bound for the mass of the ultra-light DM particle in order to have galaxy-sized wave coming from the superposition of the wave of each particle: $m \sim 2 \, \mathrm{eV}$\footnote{Remember here what we said before that it is much harder to show condensation for ULDM under the influence of gravity in the halo of a galaxy. Condensation has been shown to happen in the case of the FDM in~\citep{Levkov:2018kau}, and we discuss this in more detail in Section 4.1.}.

I would like to bring the readers attention to following. The issue of condensation of ULDM in the halo is one of huge debate in the literature and we are going to discuss this in Sect.~\ref{Sec.:discussion}.

\vspace{0.3cm}

The range of masses showed above is just an estimate of the maximal higher and lower masses that the ULDM can have. Each specific ULDM model has a mass range where this behaviour in galaxies takes place and that it is in agreement with observations which depends on the specifics of the models. However, those bounds have to always fall within this general range (assuming ULDM is all the DM in the universe - for studies where this is not true see~\citep{Hlozek:2014lca,Hlozek:2017zzf}). We will see in Section 5 that for the FDM model, using CMB and LSS observations and assuming again the FDM is all the DM, that the lower bound in the mass is very close and within this theoretical estimation. Now for the DM superfluid model, as we will see in Section 4.3, respects the upper bound presented here with the mass of its particle very close to this limit.

The mass range for the ULDM presents masses that are much smaller than the ones usually considered for DM candidates and cannot be produced thermally in the early universe. Therefore, ULDM is a \emph{non-thermal relic}, having to be created by a non-thermal mechanism in order to be cold today and behave as DM. There are many non-thermal production mechanism (see 'DM relics' panel bellow) but since in this review we are being agnostic on the type of particle that consists our ULDM, we are not assuming any creation mechanism, unless we are talking about a specific microscopic candidate. 

With that, for a ULDM candidate with a mass in the range (\ref{Eq.:Up_low_bound_m}) we are going to have a non-CDM like behaviour coming from the presence of this core with wave-like behaviour inside the galaxy, while having a CDM like behaviour on large scales, with different initial conditions resulting from the different mass and dynamics of the specific ULDM model. ULDM is the name used to the collection of models that have the characteristic stated above. There are many realizations of this behaviour which are present in each of the specific ULDM model, that yield a different description of DM in galaxies, and a distinct and rich phenomenology in galaxies. We are now going to see that those models can be classified into three main classes according to their descriptions in galaxies.

%%%
\paragraph*{Classification of ULDM models} \mbox{}\\
%%%

The idea of having  DM condensation on small scales is not new and has been around for 30 years \citep{Sin:1992bg,Ji:1994xh,Old_3}. For this reason there are many models in the literature that were developed to describe a DM component with that behaviour on galactic scales. These models receive many names in literature. They are either models that have a microscopic description or phenomenological models, which allow for the inclusion of different interactions and for a different dynamics to describe the evolution and non-linear structures of DM in the halo, which in turn can lead to distinct and rich astrophysical consequences on small scales.

One possible model of ULDM is the axion. This model helped to bring a lot of attention in the literature to this class of models. In the case of the axion, we have a model that has a microscopic description and a well defined cosmology.  The QCD axions and general axions can behave like DM in a large range of parameters. This is also the case for axion like particles (ALPs), which is another microscopic scalar theory that can describe DM. These microscopic theories behave like DM, but only present interesting phenomenology in small scales for a more limited range of masses. On small scales, the behaviour of these microscopic scalar theories can coincide with the behaviour of other phenomenological models of ULDM. For all of those the non-relativistic action that yield a non-linear Schrödinger-Poisson equation.

In this review we am going to classify the ULDM models according to the description they present on the small scales, given by their non-relativistic dynamics on those scales. Each of these classes can contain both phenomenological and microscopic models that yield the same non-relativistic model. This classification is instrumental  since it elucidates the physics responsible for the non-relativistic evolution and non-linear structures that are formed, and separates the different phenomenology each of those classes present. 

These different descriptions also yield different conditions for condensation (or if it condenses of not) and formation of the condensate core. Each of these non-relativistic descriptions is going to describe a different phenomena upon condensation, being possible to have either a BEC or a superfluid, the latter in the presence of interactions.

According to this criteria, we classify the ULDM models into three categories (which somehow agrees with what was suggested in \cite{Sharma:2018ydn}).:

\textbf{Fuzzy dark matter (FDM):} The first category is given by a gravitationally bounded scalar field model. It described by a non-linear Shrödinger equation subjected to a gravitational potential, coupled to the Poisson equation (see 'ULDM classes' box). In this model condensation under the influence of the gravitational potential is achieved in galaxies where the gravitational attraction is counteracted by the quantum pressure. This class of model can be called \emph{fuzzy DM}, since this name is already very well established for these gravitationally bounded BECs. One of its main realizations, which coined the name \emph{fuzzy dark matter} is presented in \cite{Hu:2000ke,Hui:2016ltb}, where the DM is given by a light particle with $m \sim 10^{-22} \, \mathrm{eV}$. The FDM model is the ULDM model that was studied the most in the literature both theoretically and numerically. With a particle with this mass, the FDM model is known to be able to solve some of the challenges from small scales presented above, and to be in agreement with large scale observations. The mass of this model does not have to necessarily have this value and need to be determined by observations, although this is the value that gives the most appealing modifications on small scales.  \textit{Therefore, this model has one free parameter, the mass of the FDM particle $m$} (we are considering in the review the case where all DM is composed of ULDM). Some interesting phenomenology also emerges from this model, as we will discuss in detail in the next subsections, that can be probed by current and future astrophysical observations.
This model has also been called in the literature by wave DM, $\psi$DM, among other names \citep{Sin:1992bg,Ji:1994xh,psiDM1,psiDM2,psiDM3,Matos1,Matos:1998vk,Matos:2008ag}. 

Axions and ALPs, can also be thought to be in this class, since they yield exactly the same physics on small scales. However, these models are more general and can describe DM  for different values of their mass  (like the QCD axion that does not produce such structures). 
%In the case of the axion, we have only one scales in the problem which can be taken as $m_a$. In the case of the ALPs we have two free parameters $m_a$ and $f_a$.

\textbf{Self Interacting FDM (SIFDM):} The second category is called \emph{self-interacting FDM} (SIFDM), but it also receives the names repulsive DM, scalar field DM, fluid dark matter, among others in the literature \citep{1,2,3,4,5,6,7,8,10,11,12,13,14,Chanda,Chavanis:2017loo,Chavanis:2016dab,15,16,17,18,9}. In these models, DM is described by a scalar field model, in the presence of gravity, with a (usually) 2-body self-interaction. The presence of the interaction makes this model present superfluidity upon condensation. This case is described by the  interacting BEC presented in Section 3.3, which is the simplest example of a superfluid. The presence of the interaction controls the stability of the core and for this reason this model presents a different  phenomenology  depending not only on the mass of the particle, as for FDM, but given the strength and sign of the interaction. For a repulsive interaction, the  condensate has a long range coherence and presents superfluidity. 
The 2-body case is characterized by having an equation of state (EoS),  $P \sim n^2$, as we saw in the previous section. Higher order interactions, describe SIFDM with different equations of state.

%The case of the QCD axion was studied both in the contexts of SIFDM and FDM \citep{Sikivie:2009qn,Erken:2011dz,Axions_1} (although the axion presents a self-interaction, sometimes they use the QCD axion masses in the context of the FDM model). The validity of this description as a DM model was disputed in \cite{Chanda}, since long-range coherence in the condensate can only be achieved in the presence of a repulsive interaction, and in the case of attractive interaction, like it is the case of the axion, a long-range coherence and superfluidity are not achieved.
%We will describe this in details in the next subsection.

%As a very general description, both of these models have the goal of producing this condensed stable state in galaxies that suppresses the growth of structures. Below, we will show in more details the consequences that each of those models presents. 
%%% I am not sure about this paragraph

\textbf{DM superfluid:} The third category is called \emph{DM Superfluid} \citep{Berezhiani:2015bqa,Berezhiani:2015pia,Khoury:2016ehj,Hodson:2016rck,Berezhiani:2017tth}. 
%This model is also a model of an interacting BEC, and it is described as a strong self-interacting theory of light bosons that condenses into a BEC and exhibits superfluidity. 
This theory was proposed with the goal of reproducing the MOND empirical law on small scales, presenting a natural framework for the emergence of this theory. Different than in the case of SIFDM, in order to reproduce MOND on small scales it requires that the EoS is given by $P \sim n^3$, like what is expected by MOND, with a more intricate dynamics describing the small scales. To accomplish that, this model is described using the EFT of superfluids which allows us to describe superfluids with a more general dynamics and EoS.
%Given this EoS, this could be though as coming from an effective field theory, like the usual interacting BEC, but with a  3-body interaction. As we will see bellow, this description does not work to give the proper Lagrangian of MOND.

%%%%%%%%%%%%%%%%%%%%%%%%%%%%%%%%%%%%%%%%%%%%%%%%%%%%%%%%%%%%%%%%%%

\vspace{0.3cm}
\begin{tcolorbox}[width=\textwidth, enhanced, breakable,  valign=center, colback=white, colframe=black, sharp corners, shadow={0pt}{0pt}{0mm}{black},boxrule=0.5pt,halign=justify,overlay first={
        \draw[line width=.5pt] (frame.south west)--(frame.south east);},
    overlay middle={
        \draw[line width=.5pt] (frame.south west)--(frame.south east);
        \draw[line width=.5pt] (frame.north west)--(frame.north east);},
    overlay last={
        \draw[line width=.5pt] (frame.north west)--(frame.north east);
}]
\subsubsection*{ULDM classes}

Classification is based on the different ways they achieve condensation.  
 
\begin{itemize}

\item[] {\color{darkgray}Fuzzy DM (FDM)} \qquad described by a ultra-light scalar field under the influence of gravitational  \hspace*{3.cm} \qquad potential. Forms a BEC on galactic scales.

\begin{minipage}[c]{0.4 \textwidth}
\end{minipage}
\begin{minipage}[l]{0.65 \textwidth}
\begin{equation*}
\hspace*{3.5cm} i \dot{\psi} = -\frac{1}{2m} \nabla^2 \psi +  V_{\mathrm{grav}}  
\end{equation*}
\end{minipage}
\begin{minipage}[r]{0.1 \textwidth}
\end{minipage}

\item[]
\item[] {\color{darkgray}Self-Interacting~FDM} \,\,\,\,\, described by a self-interacting scalar field with 2-body (or higher) interaction.  \\ \hspace*{1cm}  {\color{darkgray}(SIFDM)}

\vspace{-0.37cm}
\begin{minipage}[l]{0.85 \textwidth}
\begin{equation*}
\hspace*{2.5cm} i \dot{\psi} = -\frac{1}{2m} \nabla^2 \psi +  V_{\mathrm{grav}} +  g \, |\psi|^2 \psi + g_3 \, |\psi|^4 \psi + \cdots  
\end{equation*}
\end{minipage}
\begin{minipage}[r]{0.1 \textwidth}
{\color{gray}(Superfluid)}
\end{minipage}

\item[]
\item[] {\color{darkgray}DM Superfluid} \qquad \qquad described by a superfluid with specific EoS to reproduce MOND in galaxies.

\begin{minipage}[l]{0.85 \textwidth}
\begin{equation*}
\hspace*{2cm}\mathcal{L}=P(X) 
\end{equation*}
\end{minipage}
\begin{minipage}[r]{0.1 \textwidth}
{\color{gray}(Superfluid)}
\end{minipage}

\end{itemize}

\end{tcolorbox}
%%%%%%%%%%%%%%%%%%%%%%%%%%%%%%%%%%%%%%%%%%%%%%%%%%%%%%%%%%%%%%%%%%

There are many amazing reviews in the literature that focus in different parts of the ULDM class of model, either focusing in microscopic models, like describing axions~\cite{Sikivie:2006ni,Arvanitaki:2009fg,Wantz:2009it,Kim:2008hd,Kawasaki:2013ae,Axions_1} or ALPs~\citep{Ringwald:2014vqa,Arias:2012az,Axions_2,Marsh:2017hbv,Niemeyer:2019aqm,Powell:2016tfs}, or focusing in one of the classes like the FDM~\cite{Hui:2016ltb,Suarez:2013iw,Review_SFDM_Arturo}, for which the axions and ALPs describe the same non-linear theory (given by a Schrödinger-Poisson system in the absence of self interactions). This review follows a classification between the different classes according to their non-relativistic description and includes not only the FDM, but also the other classes of ULDM. 

Given that, in this review we are going to be describing the dynamics each of these classes of models present on small scales, and the different cosmological and astrophysical consequences this new phenomenology brings. The exception is in the case of the for the FDM where we also are going to talk a little bit about the cosmology of the axions and ALPs. However, in general, we are going to remain agnostic about the origin of this field and we are going to work out only the gravitational consequences they have\footnote{The list of models we present here that compose each of the classes is not completely exhaustive and it only aims to show the diversity of models in the literature, and the different mechanisms they describe. However, all the possible dynamics present on small scales coming from ULDM are described here and can be within one of these three classes. In \cite{Lee:2017qve} one can find a more complete list of references.}. Understanding these consequences will prepare the field for the next section, where we discuss constraints on these models. We briefly discuss in the end of this section the big role that simulations have in studying these models and also the regimes where these ultra-light fields can behave as dark energy.

 %%%%%%%%%%%%%%%%%%%%%%%%%%%%%%%%%%%%%%%%%%
\begin{figure}[htb]
\centering
\includegraphics[scale = 0.3]{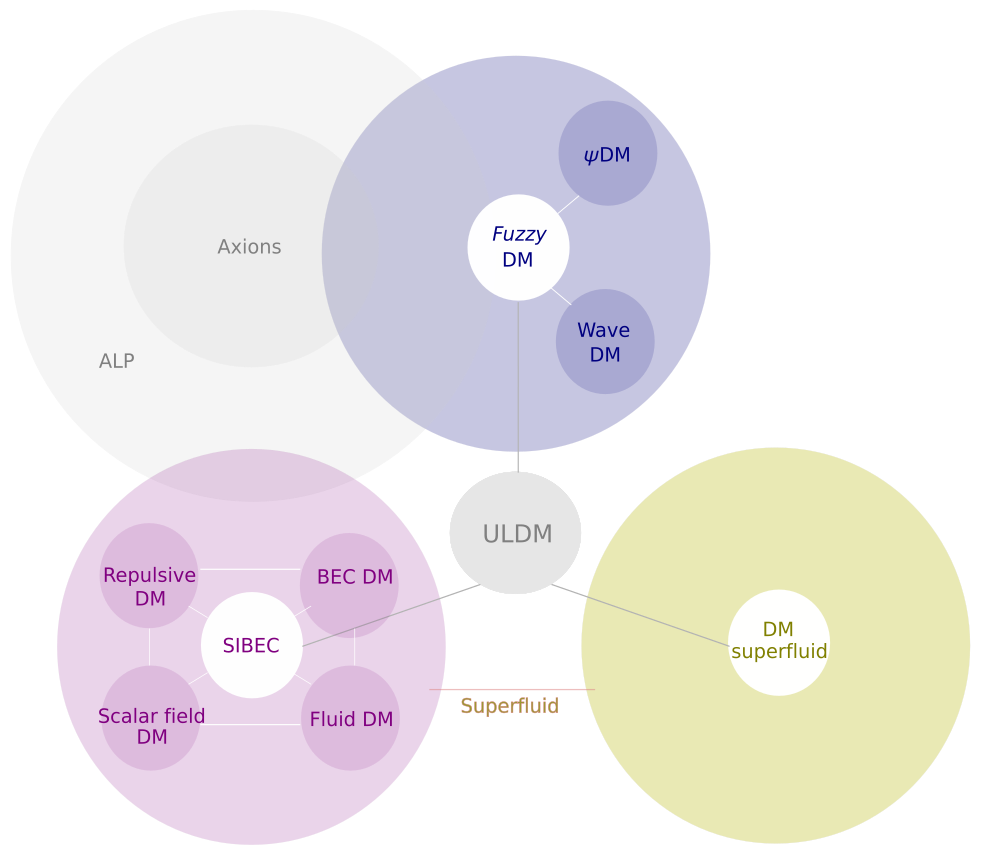}
 \caption{Map of the ULDM classes of models.}
\label{Fig.:models_ULDM}
\end{figure}
%%%%%%%%%%%%%%%%%%%%%%%%%%%%%%%%%%%%%%%%%

\vspace{0.5cm}

%%%%%%%
\vspace{0.3cm}
\begin{tcolorbox}[colback=gray!10,enhanced,breakable,frame hidden,halign=justify]
%\textbf{Summary of scales and galaxies}\footnotemark
%\vspace{0.0.8cm}

\subsubsection*{DM relics}
Dark matter candidates can have distinct formation mechanisms, with the main ones being thermal or non-thermal relics produced in the early universe.
%Dark matter differs in the production mechanisms in the early universe that took place for its formation. 
Depending on this mechanism, different masses and couplings for these are allowed as the correct relic abundance of DM is obtained.
%DM can have different masses and still describe the non-relativistic and cold fluid we know it describes DM at large scales.

\paragraph*{Thermal relics: \qquad} 
This refers to the particles, including DM, that are produced from the hot and high density thermal bath of the photon-baryon plasma in the universe.  
Initially, in the early universe, the universe was in a state where it was hot and dense where particle and photons were very close to thermal and chemical equilibrium. This means that the time scale of the particle interactions ($1/\Gamma$) in the plasma are much bigger than the expansion time of the universe $H \ll \Gamma = n \langle  \sigma v  \rangle$, where  $\langle  \sigma v  \rangle$ is the thermally averaged cross section \citep{Kolb:1990vq}.  As evolution follows and those quantities redshift, at some point $H \sim \Gamma$, and the particle decouples from the thermal bath (at the temperature $T_{\mathrm{fo}}$).  This is a simplified description of the process called freeze-out or decoupling. Depending on the interaction rate of each particles, they decouple at different times. This process is described by the Boltzmann equation and it is how electrons, neutrons and neutrinos are formed. If $T_{\mathrm{fo}} \ll m$, where $m$ is the mass of the particle, the particle decouples as a non-relativistic particle, and it is called a cold relic; otherwise, if $T_{\mathrm{fo}} \gg m$ or $T_{\mathrm{fo}} \sim m$, and we have hot and warm relics.

We can assume that DM is a component that was in contact with the thermal bath and it is a particle produced through decoupling from the thermal bath like described above. 
For the thermal relics, the particle with smaller mass is hotter.
% It is hard to find a robust lower bound for the thermally produced DM. If thermal DM interacts only with the dark sector (WDM) (or of it was never in thermal equilibrium), thermal DM cannot be smaller than $\mathcal{O}({keV})$. Otherwise, if it interacts with the standard model than $M^{\mathrm{th}}_{\mathrm{DM}} > \mathcal{O}({MeV})$. 
%XXXXXXXXXXX
%Indirect detection experiments 
% So, from structure formation there the mass of thermal DM cannot be smaller than $\mathcal{O}({keV})$. However, it is very challenging to 
%This considers that all DM is given by this thermal DM, but if this is not the only component for DM, this bound can be evaded.
For cold relics, in the case of WIMP created through this mechanism, for the WIMP to have the correct abundance of DM today, the averaged cross section is roughly $\sigma_{\mathrm{DM-DM}} \simeq 10^{-8} \, \mathrm{GeV}^{-2}$. This is of the same order of magnitude of the electroweak cross section: $\sigma_{\mathrm{weak}} \simeq \alpha^2/m^2_{\mathrm{weak}}$ with $\alpha \simeq \mathcal{O}(0.01)$ and  $m_{\mathrm{weak}} \simeq \mathcal{O}(100\, \mathrm{GeV})$~\citep{Kolb:1990vq}.
%For cold relics, in the case of WIMPS, created through this mechanism, to obtain the observed DM abundance the averaged cross section needs to be of the order of the one characteristic of weak interactions $\sigma_{\mathrm{DM-DM}} \simeq 10^{-8} \, \mathrm{GeV}^{-2}$. This is known in the literature as the WIMP miracle.
Many models of DM are produced thermally like supersymmetric candidates, more complicated WIMP candidates or particle DM decays\footnotemark.

\paragraph*{Non-thermal relics: \qquad} As we saw above, there is a limit for the mass of the DM particle  that can be created thermally. The only way of having smaller mass candidates of DM is by having a non-thermal mechanism to produce those DM particles. There are many mechanisms that can produce non-thermal DM candidates that include decaying from topological defects (see for example~\citep{Sikivie:2006ni,Hiramatsu:2012gg} where axions are produced from the decay of axion strings or domain walls), decaying from a massive parent particle, vacuum misalignment, among others \citep{Axions_1,Dev:2013yza}. Vacuum misalignment or vacuum displacement is one of the genesis mechanism for these ULDM and axions. The vacuum displacement mechanism \citep{Preskill:1982cy,Abbott:1982af,Dine:1982ah,Carroll:1999iy} can be described, in a concise way in the following way. A massive scalar field in an FRW universe, when $H > m_\varphi$, is overdamped and it behaves nearly  as a constant. So, if we consider that initially this field was displaced from its minimum, $\varphi=\varphi^{*}$, the field has a potential energy given by $\varphi ^{*}$. When $H \sim m_\varphi$, the field starts to evolve and begins to oscillate in its potential, and in turn redshifts like matter. The mass and the initial displacement fix the energy density of this misalignment field. If we consider that the ultra-light particles are created by this mechanism, it imposes a lower bound on the mass $ H(a_{\mathrm{eq}}) \approx 10^{-28}\, \mathrm{eV}$, in order to start behaving like DM around equality. For more detail on this mechanism for axions, see \cite{Axions_1}.

%For DM and DE to have the right abundances, the parameter ranges 10-33 to 10-2 eV. are of particular interest. In these ranges, axions are produced non-thermally, thus evading hot DM bounds
%There may also be a thermal sub-population, tested by measurements of Nef
 
\end{tcolorbox}
%\footnotetext{There may also be other mechanisms for the formation of different candidates of DM as combination of thermal and non-thermal production or production as a result of particle-antiparticle asymmetry, which I will not comment in this review.}
\footnotetext{Even the axion has a thermal production channel, if the axion is in contact with the thermal bath, with this axion being hot and contributing to a fraction of the effective number of neutrinos (see~\citep{Axions_1} for a review on that topic).}

%%%%%%%%%%%%%%
\subsection{FDM and SIFDM} 
\label{Sec.:stability}
%%%%%%%%%%%%%%

In this section we are going to describe the FDM and the SIFDM models. Although they both describe different non-relativistic dynamics and structure formation in the halo, and according to our classification are in different classes, we describe them in this section together since they can be described by a relativistic action.  Even tough we are interested in the non-CDM phenomenology of these models on small scales, these models can also modify the initial conditions for the evolution of the matter perturbations, and depending on the mass, modify the evolution of the model. In this way, the relativistic theory allows us to study the cosmology of this model and with that make predictions that can be tested by large scale observations like CMB and LSS. This helps us to describe the model in different scales and use observations from large and small scales to constraint the parameters of the model. 

The ULDM is described as a very light scalar field minimally coupled  to gravity given by the action,
\begin{equation}
S= S_{\mathrm{EH}}+S_{\phi} = \int d^4 x \sqrt{-\bar{g}} \left[ \frac{R}{16\pi G} + \frac{1}{2} g^{\mu \nu} \partial_\mu \phi \partial_\nu \phi -\frac{1}{2} m^2 \phi^2 -\frac{g}{4!} \phi^4 \right]\,,
\label{axion_action}
\end{equation}
where $S_{\mathrm{EH}}$ is the Einstein Hilbert metric, $R$ is the Ricci scalar, $g_{\mu \nu}$ is the metric, $\bar{g}$ is the determinant of the metric and $g$ is self-interaction coupling.

The axion or ALPs are described by an action like this. In this case, this action has a microscopic theory behind it.  This potential comes from non-perturbative effects in QCD, for the axion, or other concrete string theory models, and gives a small mass for the axion or ALP. Since in these cases this action comes from a well defined microscopic theory, the parameters relate to scales from this theory and the mechanisms that originated this particle. There is only a range in the parameter space where the ALPs behave like DM, and that also gives an interesting modification of structure formation. This happens for masses around $m \sim 10^{-22} - 10^{-20}\, \mathrm{eV}$, which is similar to the range of masses for the FDM model.

But this relativistic action could also be phenomenological that allows the coupling to have a different sign and values, different from the axion case. We are going to see soon that the attractive and repulsive interactions yield different phenomenologies, with the repulsive allowing for much bigger collapse cores.

In this section we are going to explore the cosmological consequences of this relativistic action, and after go to the non-relativistic regime to study the structure formation in the FDM and SIFDM models and describe the condensate formed in the center of the galaxy.
To explore cosmological evolution, we are first going to study the concrete case where this action is the action for the ALPs. We also comment a little on the case of axion. This is very useful since in this case when we have a microscopic theory, we can identify the scale of the parameter and their relations, determine the initial conditions and the DM abundance.

% Notice that I could have done the same analysis of the cosmology but for the case of the phenomenological action with self-interaction. In that case, given the constraints in order for this component to behave like DM, the parameters $m$, $g$  are independent and are free parameters of the theory. The initial conditions are also not determine and depend either on the mechanism of formation in this theory or is a free parameters too.

%%%%
\subsubsection{Formation: ALPs} 
\label{Sec.:cosmology_FDM}
%%%%

We are going to briefly describe here the formation mechanism for ALPs (see~\citep{Axions_1} for a more complete description).

An ALP is a pseudo-Nambu Goldstone resulting from the spontaneous symmetry breaking of a global $U(1)$ symmetry described by the complex scalar field
\begin{equation}
\Psi = v e^{i\phi/f_a}\,,
\end{equation}
where $f_a$ is the scale of the spontaneous symmetry breaking and $\theta_a = \phi/f_a$ is the misalignment angle. In the case of the QCD axion, the symmetry broken is the chiral global $U(1)_{PQ}$, the Peccei-Quinn symmetry, and the introduction of this Goldstone boson solves the strong CP problem. When the symmetry is broken, the massive radial component $v$ is fixed at the vacuum expectation value (vev), $v_{0,ssb}=f_a/\sqrt{2}$, making the radial field non-dynamical while there is a continuous set of minima with the ground state given by $\Psi = v_{0,ssb} e^{i\theta_a},$ corresponding to all the possible phases in the circle (if the reader wants to refresh the memory on SSB, see Section 3.3.1 and the box that refers to SSB in that section). The pseudo-Goldstone boson $\phi$ is invariant under shift-symmetry, inherited from the $U(1)$ symmetry of the complex scalar field. 

However, non-perturbative effects, coming from string theory models or from instatons in the case of the axion, can induce a potential that break the shift symmetry explicitly, although softly, which leads to a residual discrete symmetry. This potential gives a small mass to the ALPs, and has the form:
\begin{equation}
V(\phi) = \Lambda_a^4 \left[ 1-\cos \left( \frac{\phi}{f_a} \right) \right]\,,
\end{equation}
where $\Lambda_a^4$ is the scale of spontaneous symmetry breaking\footnote{A similar mechanism that generates a dynamical DE component at late times in the context of the DM superfluid model can be seen in~\citep{Ferreira:2018wup}, presented also briefly in  Sect.~\ref{Sec.:unified} of this review.}. This potential is not unique and the overall constant added was chosen arbitrarily. For small field values $\phi \ll f_a$, this potential can be expanded into,
\begin{equation}
V(\phi) = \frac{1}{2} m^2 \phi^2 + \frac{g}{4!} \phi^4 + \cdots\,,
\end{equation}
where $m=\Lambda_a^2/f_a$ and $g = -\Lambda_a^4/f_a^4 < 0$.  Since the spontaneous symmetry breaking scale $f_a$ is usually much higher than the explicit symmetry breaking scale $\Lambda_a$, the mass is usually very small, with the self-interaction coupling $g$ even smaller. For the QCD case, $\Lambda_{\mathrm{QCD}} \sim 200 \, \mathrm{MeV}$~\citep{Axions_1},  and $10^9 \, \mathrm{GeV} \lesssim f_a \lesssim 10^{17} \, \mathrm{GeV}$ coming from astrophysical constraints, then we can see that $10^{-10} \, \mathrm{eV} \lesssim m_{\mathrm{QCD}} \lesssim 10^{-2}\, \mathrm{eV}$. For string theory models, there is a variety of cases, but typically $\Lambda_{st} \sim T_{\mathrm{SUSY}} \, e^{-S_{i}}$, where $S_{i}$ is the instaton action which generally is $S_{i} \gg 1$.

Although very small, given the approximate shift-symmetry, the ALPs mass is protected against radiative corrections, and interactions with the standard model are suppressed by powers of $f_a$. 

For the case of the QCD axion, the mass generated by the non-perturbative QCD effects has a time dependence that scales with the power law of the temperature. For a discussion about that, see~\citep{Axions_1}.

Given all that, the action for the ALP is given exactly by the action (\ref{axion_action}) (usually the self-interaction is omitted since this is suppressed by powers of $f_a^3$). 

\subsubsection{Cosmological evolution} 

Having established that there is a microscopic theory where the action (\ref{axion_action}) can come from, we are going to study the cosmology in the general case, which describes any model in the FDM and SIFDM classes.
If we ignore the small interaction, this model corresponds to the FDM class. In the presence of interaction this model corresponds the SIFDM class, where the interaction can be attractive or repulsive, while in the case of ALPs and axions it can only be attractive ($g<0$). 

We are going to focus only in the matter sector now and omit the Einstein Hilbert action (which gives GR, the background theory where our field evolves). The action we are going to work with is:
\begin{equation}
S_{\phi}= \int d^4 x \sqrt{-g} \left[ \frac{1}{2} g^{\mu \nu} \partial_\mu \phi \partial_\nu \phi -V(\phi) \right]\,,
\label{action_field_only}
\end{equation}
where the potential can be given by only the mass term, describing the FDM, or both the mass and the interaction, describing SIFDM.

We can study the evolution of this field  in a flat Friedmann-Robertson-Walker background (FRW) background, given by the metric $ds^2 = dt^2 - a^2(t)dx^2$, where $a(t)$ is the scale factor. The equation of motion for the ALP is given by:
\begin{equation}
\ddot{\phi} + 3H \dot{\phi} + \frac{dV}{d\phi} = 0\,,
\label{eq.:alp_eom}
\end{equation}
where $H=\dot{a}/a$ is the Hubble parameter.
The cosmological evolution is going to depend on the competition between the potential term and the Hubble friction term. 
Lets consider the case of the FDM, where $V(\phi) = (1/2) \, m^2 \phi^2 $ to illustrate the cosmological evolution. The background evolution of this field proceeds in the following way. In the early universe, $H \gg m$, the Hubble friction dominates and the solution is a constant given by the initial conditions $\phi_{\mathrm{early}} = \phi (t_i) $. In the case of the ALP, this initial condition is known and given by the formation of the ALP with $\phi_i= f_a \theta_a (t_i) = f_a \theta_i$. At early times the ULDM is subdominant and has equation of state $w = -1$ behaving like dark energy.
As the universe expands, the Hubble parameter becomes smaller and smaller, until a point where it is smaller than the mass of the field $H \ll m$. The solution of the equation in this case is oscillatory. As the field oscillates, the equation of state also oscillates around zero, giving an averaged out equation of state of a dust like component $w =0$. In this limit the field behaves like DM and the energy density evolves as $\rho= \rho(a(t_{*})) \, (a(t_{*})/a)^3$, where $t_{*}$ is the time when $H(t_{*}) \sim m$.

We can already see that the lower the mass is, more and more will take until the Hubble parameter to become smaller than the mass, prolonging the early period where the field behaves as dark energy. For higher masses, the FDM behaves like DM earlier in the history of the universe. Therefore, the mass completely controls when the period of DM domination starts. 

In the case of ALP, for them to behave like all the DM of the universe, it has to start oscillating before matter-radiation equality, which gives $m > 10^{-28} \, \mathrm{eV} \sim H(a_{\mathrm{eq}})$. However, not all of this regime the ALP as DM presents interesting phenomenological consequences for structures, since for heavier masses, ALP behaves closer and closer to CDM. We are going to see this in more details later, but the sweet spot in mass for the ALP to have this distinct regime on small scales is $m \sim 10^{-22} - 10^{-20}\, \mathrm{eV}$.

To obtain the full solution of (\ref{eq.:alp_eom}), one need to solve this equation coupled to the Friedman equation that describes the evolution of the scalar factor according to the components of the universe:\linebreak $
H^2 = (1/M_{pl})\left( \rho_{\phi} + \rho_r + \rho_b + \rho_{\Lambda} \right)$, where the contributions in the energy density come from the FDM or SIFDM, radiation, baryons and cosmological constant, respectively. The energy density and pressure of this scalar particle is given by:
\begin{equation}
\rho_{\phi} = \frac{1}{2} \dot{\phi}^2 + V(\phi)\,, \qquad p_{\phi} =\frac{1}{2} \dot{\phi}^2 - V(\phi)\,,
\label{presure_FDM}
\end{equation}
where in the case that the field is oscillating, can be averaged over time.
In the case of the FDM,  we can take $a(t) \propto t^p$, which is valid for the period of radiation and matter dominations, and we have:
\begin{equation}
\phi(t) = a^{-3/2} \left( \frac{t}{t_i} \right)^{1/2} \left[ C_1 \, J_n (mt) + C_2 \, Y_n (mt) \right]\,,
\end{equation}
where $C_{1,2}$ are determined by the initial conditions, $n=(3p-2)/2$ and $J_n$ and $Y_n$ are the Bessel functions of first and second kind.

We can then compute the relic density of FDM, which is the energy density of those particles today.  For the case where FDM is the ALP and behaves like DM, and all the DM is made of ALPs, the density fraction is:
\begin{equation}
\Omega_{\mathrm{ALP}} \sim \frac{1}{6} \, \left( 9\, \Omega_r \right)^{3/4} \left( \frac{m}{H0} \right)^{1/2} \left( \frac{\phi_i}{m_{pl}} \right)^2\,.
\end{equation}
The initial value of the field displacement determines the relic density of ALPs, and in order to have the DM density observed today, the initial value of the field must be $\phi_i > 10^{14} \, \mathrm{GeV}$ for the masses of the ALP that correspond to DM behaviour. A similar calculation can be made for the axion and this can be found in~\citep{Axions_1}.

With that we saw that we have 2 scales that are important for the DM ALPs: $f_a$ the spontaneous symmetry breaking scale that determines the initial conditions of the ALPs, and $\Lambda_a$ the scale of explicit breaking that determines the mass given the previous scale. The the temperature $T_{*}$  associated with $t_{*}$, the time when the the field starts to oscillate and behave like DM, is set after the mass is determined.
This is similar in the case of the FDM, where the only one degree of freedom $m$, although the initial condition $\phi_i$ can be unknown. If one assumes the FDM is described by an ALP, then the initial condition is determined.

In the case of the SIFDM, we have an extra parameter in comparison to the FDM, the interaction strength. The interaction term acts as a pressure term in the equation of motion (\ref{eq.:alp_eom}). This pressure can be attractive or repulsive. There is then a competition between the Hubble friction, the mass and the pressure. We are going to study more about these effects in the next section.

\textit{Cosmological perturbations:} \, We have studied the background evolution and now we need to study the cosmological perturbations. This is important in order for us to study the cosmological consequences of this scenario that can be tested with cosmological observations. 
We are not going to present here an extensive description, but a summary of the most important results. Notice that like for the cosmological evolution, the procedure here is general for any FDM or SIFDM.

First, we perturb the scalar field and the metric into small perturbations on top of the background values:
\begin{equation}
g_{\mu\nu} (\mathbf{x}, t) = g_{\mu \nu}^{(0)}(t) + \delta g_{\mu\nu} (\mathbf{x}, t)\,, \qquad \phi (\mathbf{x}, t) = \phi_0(t) + \delta \phi(\mathbf{x}, t)\,,
\end{equation}
where the $0$ indicates the background quantities. We are only going to be interested in the scalar perturbations, in this review. We are going to work on conformal time $\eta$ defined as $d\eta = dt/a$.
The perturbed metric for the scalar metric perturbations is described by $4$ functions (following the convention from~\citep{Mukhanov})
\begin{equation}
ds^2 = a^2(\eta) \left\{ \left( 1+ 2\Phi \right) d\eta^2 +2B_{,i} \, d\eta dx^i + \left[ \left( 1 - 2\Phi \right) \delta_{ij}-2 E_{,ij} \right]  dx^i dx^j \right\} \,.
\label{perturbed_metric}
\end{equation}
When doing perturbation theory in general relativity, due to gauge invariance, new degrees of freedom that are fictitious and not physical might be introduced.One of the procedures to deal with this is to fix a gauge, which fixes these spurious variables.  There are many ways of doing that which leads to the many possible gauges. The final physics given by all these gauges is the same, but each gauge offers a better description of different phenomena. The Newtonian gauge ($E=0$ and $B=0$) is useful in the Newtonian limit. 
One can also choose the comoving gauge ($B=0$ and $v=0$), where $v$ is the velocity of the matter fluid
We are going to use both gauges whenever they are useful.

After a gauge is chosen, one substitutes  the perturbations in the action (\ref{action_field_only}), ignoring the interaction for simplicity, to obtain the second order action for the perturbations. And from that one can obtain the equation of motion for the axion perturbation.

We can also re-write the perturbations in terms of the fluid variables: $\rho_{\phi} = \rho_{\phi,\, 0} + \delta \rho$ and $p_{\phi} = p_{\phi,\, 0} + \delta p$. From (\ref{presure_FDM}), we can identify the perturbations in the fluid variables with the ones of the scalar field and metric~\citep{Hwang:2009js}
\begin{equation}
\delta \rho = \langle \dot{\phi_0} \dot{\delta \phi} - \dot{\phi}_0^2 \Phi +m^2 \phi_0 \delta \phi \rangle \,, \qquad \delta p =  \langle \dot{\phi_0} \dot{\delta \phi} - \dot{\phi}^2_0 \Phi -m^2 \phi_0 \delta \phi \rangle \,, \qquad a (\rho + p) v = k \langle \dot{\phi_0} \dot{\delta \phi}  \rangle\,.
\label{perturb_fluid}
\end{equation}
where we have taken the Fourier transform of the fields with $k$ denotes the wavenumber, and the background pressure and energy density are averaged. 

We are interesting in obtaining the sound speed of the FDM particle. We showed before that the FDM behaves like dark matter at the background level, but we also need to show that the sound speed is small as expected for dust. This different sound speed is going to give a different Jeans scale and give a different structure formation for this model.
Since we are interested in calculating the sound speed, we are going to work in the comoving gauge. 
 In that gauge the equations, assuming the averaged background equation of state (which is zero), for the scalar perturbations of the metric and fluid simplify and can be combined to give
\begin{align}
\ddot{\delta} + 2H \dot{\delta} - 4\pi G \rho \, \delta + \frac{k^2}{a^2} \frac{\delta p}{\rho} = \ddot{\delta} + 2H \dot{\delta} +\left( \frac{k^2}{a^2} c_s^2 - 4\pi G \rho \right) \delta = 0\,, \label{delta_relativistic}
\end{align}
where $\delta=\delta \rho/\rho$ is the density contrast and  $\omega_k^2 = \frac{k^2}{a^2} c_s^2 - 4\pi G \rho$ is the dispersion relation.

By definition, the sound speed is defined as the term that accompanies the gradient, the term with $k^2$. To obtain the expression for the sound speed, we need to compute the perturbation of the pressure $\delta p$. For that, make the simple procedure from~\citep{Hwang:2009js}, where we assume an \textit{ansatz} for the field perturbation, $\delta \phi (x,t) = \delta \phi_{+}(x,t)\, \sin(mt)+\delta \phi_{-}(x,t)\, \cos (mt)$ and substitute that in (\ref{perturb_fluid}). Ths gives us $\delta p$, and the sound speed can be written as:
\begin{equation}
c_s^2 = \frac{k^2}{4m^2a^2} \left( \frac{1}{1+ \frac{k^2}{4m^2a^2}} \right)\,.
\end{equation}
This is the (relativistic) sound speed of our FDM fluid. It is valid outside the Hubble horizon and inside. For sub-Hubble horizon modes, when $k/(ma) \ll 1$ the sound speed becomes:
\begin{equation}
c_s^2   \xrightarrow[{k/ma \ll 1}]{}  c_{s, \, n}^{2} = \frac{k^2}{4m^2 a^2}\,.
\label{cs_newt}
\end{equation}
This corresponds to sound speed that can be obtained in the non-relativistic Newtonian theory, as we will see  in the next section.

There are two competing terms in equation (\ref{delta_relativistic}) which are the terms inside the dispersion relation. The scale $k_{J}$ for which $\omega_k (k_J) = 0$ separates the regimes where each of those terms dominates in the equation. We can also write this in terms of $\lambda_J=2\pi a /k_J$, the Jeans length. 
For modes with $\lambda < \lambda_J$ the dispersion relation is negative, and the solution of (\ref{delta_relativistic}) is that the perturbations oscillate. While when $\lambda > \lambda_J$ perturbations grow. So there is only gravitational instability for the modes that are outside the Jeans length. 
In a theory of DM with a finite Jeans length, the growth of perturbations will be suppressed for scales smaller than $\lambda_{\mathrm{J}}$. That is exactly the effect the small mass if the ULDM models has. This leads to important cosmological consequences for these models.

%%%%%%%%%%%%%%%%%%%%%%%%%%%%%%%%%%%%%%%%%%%%%%%%%%%%%%%%%%%%%%%%%%%%%%%%%%%%%%%%%%%%%%%%%%%%%%%%%%%%%%%%%%%%%%%%%%%%%%%%

%%%%%%%%%%%%%%%%%%%%%%%%%%%%%%%%%%%
\subsubsection{Evolution on small scales} 
\label{Sec.:FDM_SIFDM_NR}
%%%%%%%%%%%%%%%%%%%%%%%%%%%%%%%%%%%

We have finally reached the section where we are going to describe the behaviour of the FDM and of the SIFDM on small scales. It is this different behaviour that s used to classify the models into different classes. 

The action that describes the SIFDM and the FDM is (\ref{axion_action}). We are interested in studying the behaviour of DM in galaxies, so we are on sub-Hubble scales. In this limit, $H \ll m$ and the field is oscillating fast and  behaving as DM. Inside the Hubble horizon and for the small velocities  we have in galaxies ($v_{\mathrm{vir}} \ll c$), we are in the non-relativistic limit of our theory. In the Newtonian gauge, the limit where $B=0$ and $E=0$ in (\ref{perturbed_metric}), and with no anisotropic stress $\Phi = \Psi$, we can write the action for the ULDM field~\citep{Niemeyer:2019aqm}:
\begin{equation}
S_{\phi} = \int d^4 x \, a^3 \left[  \frac{1}{2} (1-4\Phi) \, \dot{\phi}^2 - \frac{1}{a^2} (\partial_i \phi)^2 - (1-2\Phi) V(\phi) \right]\,.
\end{equation}
Since in the non-relativistic limit the field varies slowly, the fast oscillations that we had for the field can be factored and we can re-write the field as:
\begin{equation}
\phi= \frac{1}{\sqrt{2ma^3}} \left( \psi \, e^{-imt} + \psi^{*} \, e^{imt} \right)\,.
\end{equation}
With the field in this form and assuming the $\dot{\psi} \ll m\psi$, we have the total non-relativistic action, nicludnig the Einstein Hilbert action, that describes this theory given by~\citep{16,Chavanis:2017loo}
\begin{equation}
S = \int d^4 x \left[  \frac{i}{2} \left( \psi \partial_{t} \psi^{*} - \psi^{*} \partial_t \psi \right) - \frac{|\nabla\psi|^2}{2m}-\frac{g}{16m^2} |\psi|^4 - m (\psi \psi^{*} - \langle \psi \psi^{*}  \rangle ) \Phi - \frac{a}{8\pi G} (\partial_i \Phi)^2 \right] \,.
\end{equation}
The equations of motion of the action yield the Sch\"{o}dinger- Poisson system of equations:
\begin{align}
&i \dot{\psi} =-\frac{3}{2}  i H\psi -\frac{1}{2m a^2} \nabla^2 \psi + \frac{g}{8m^2} |\psi|^2 \psi +m\Phi  \, \psi\,, \\         &\nabla^2 \Phi = 4 \pi G \left( \rho - \bar{\rho}  \right)\,.
\end{align}
If we consider time scales much smaller than the expansion, we can ignore expansion of the universe and write the equation of our system as:
\begin{equation}
i \dot{\psi} = -\frac{1}{2m} \nabla^2 \psi + \frac{g}{8m^2} |\psi|^2 \psi +m\Phi \, \psi\,.
\label{Eq.:NLS_ULDM}
\end{equation}
This is the non-linear Schr{\"o}dinger equation. The gravitational potential term can be re-written in the form \\ $-Gm^2 \psi \int d^3 x^{'} |\psi (\mathbf{x}^{'})|^2/|\mathbf{x}-\mathbf{x}^{'}|$. This non-linear equation is the Gross--Pitaesvkii equation described in the previous section and describes the evolution of a wavefunction or a field. We can use this equation to analyze the properties of this system, analytically and numerically.

\vspace{0.3cm}
\begin{tcolorbox}[width=\textwidth, enhanced, breakable,  valign=center, colback=white, colframe=black, sharp corners, shadow={0pt}{0pt}{0mm}{black},boxrule=0.5pt,halign=justify,overlay first={
        \draw[line width=.5pt] (frame.south west)--(frame.south east);},
    overlay middle={
        \draw[line width=.5pt] (frame.south west)--(frame.south east);
        \draw[line width=.5pt] (frame.north west)--(frame.north east);},
    overlay last={
        \draw[line width=.5pt] (frame.north west)--(frame.north east);
}]
\subsection*{Fuzzy DM vs SIFDM}

\begin{equation*}
\begin{array}{lc}
\begin{dcases}
\displaystyle i \dot{\psi} = -\frac{1}{2m} \nabla^2 \psi +  m \,\Phi \, \psi + {\color{BlueViolet} \frac{g}{8m^2} \, |\psi|^2 \psi + \frac{g_3}{12m^3}\, |\psi|^4 \psi + \cdots}  \\
\displaystyle  \nabla^2 \Phi = 4 \pi G \left( \rho - \bar{\rho}  \right)
\end{dcases}

&
\qquad\Longrightarrow \qquad
\begin{dcases}
g_i = 0 \,\, & \,\, \text{Fuzzy DM} \\
g_i \neq 0 \,\, & \,\, \text{{\color{BlueViolet}SIFDM}} \\
\end{dcases}

\end{array}
\end{equation*}
\vspace{0.05cm}
\end{tcolorbox}

\vspace{0.5cm}

We can also rewrite the field theory above as a set of hydrodynamical-like equations, in this long wavelength limit. For that, if we identify (using the theory in the presence of expansion):
\begin{equation}
\psi \equiv \sqrt{\frac{\rho}{m}} e^{i\theta}\,, \qquad \mathbf{v} \equiv \frac{1}{a \, m} \mathbf{\nabla} \theta = \frac{1}{2 i m  a} \left( \frac{1}{\psi} \mathbf{\nabla} \psi - \frac{1}{\psi^{*}} \mathbf{\nabla} \psi^{*}  \right)\,.
\end{equation}
The vorticity of the superfluid is zero and the momentum density has non-zero curl. The comoving equations of motion for $\psi$ are:
\begin{align}
& \dot{\rho} + 3H\rho +\frac{1}{a} \mathbf{\nabla} \cdot \left( \rho \mathbf{v} \right) = 0\,, \\ 
& \dot{\mathbf{v}}+H\mathbf{v}+\frac{1}{a} \left( \mathbf{v} \cdot \mathbf{\nabla}  \right) \mathbf{v} = - \frac{1}{a} \mathbf{\nabla} \Phi + \frac{\mathbf{\nabla} P_{\mathrm{int}}}{\rho}+ \frac{1}{2a^3 m^2} \mathbf{\nabla} \left( \frac{\nabla^2 \sqrt{\rho}}{\sqrt{\rho}} \right)\,. \label{hydro2}
\end{align}
These set of  equations are the Mandelung equations, generalized for an expanding universe. The second term in the right hand side of equation (\ref{hydro2}) comes from the self-interaction term, where $P_{\mathrm{int}}$ is the pressure from the interactions.
The last term of the second equation is the quantum pressure. This is present even in the absence of interaction and it is going to be important for the effects and formation of the condensate for the FDM model. The quantum pressure  has the role of not allowing the FDM to cluster and collapse, which also makes the density of this collapsed region to have a finite value. \textit{In this way, this model has naturally a cored profile inside the condensate region, addressing the cusp-core problem.}  This form of the equations is useful for numerical simulations that can reveal some properties of the DM scalar field. However, as we can see from the quantum pressure term, these equations are not defined for $\rho = 0$.

The quantum pressure term is present in the ULDM models and it is not present in other candidates of DM. As we discussed above, the behaviour of FDM or SIFDM in each regime depends on a competition between the gravity, the pressure term and the quantum pressure. We can roughly say that then the non-CDM behaviour expected for our models will take place on scales where the quantum pressure term dominates. The sign and size of the interaction might affect this a lot. A naive estimate of this effect is that the scales where this quantum pressure term matters is for scales smaller than the de Broglie wavelength of the particle, $\lambda < \lambda_{\mathrm{dB}}$. The de Broglie wavelength for a typical MW like galaxy is given by
\begin{equation}
\lambda_{\mathrm{dB}} \simeq 0.2 \, \left( \frac{m}{10^{-22} \, \mathrm{eV}} \right)^{-1} \, \left( \frac{V_{200}}{v} \right) \, \mathrm{kpc}\,,
\end{equation}
where we used the virial velocity (\ref{virial_MW}). For a particle with $m=10^{-22}\, \mathrm{eV}$, this would mean that the wave-like behaviour of the ULDM particles would be relevant in a MW like galaxy on scales smaller than $0.2 \, \mathrm{kpc}$. If we consider dwarf galaxies, for example, where virial velocities are much smaller $v_{\mathrm{vir}}^{\mathrm{dw}} \sim 10 \, \mathrm{km/s}$, these non-CDM effects would take place on scales of order of their halo size. For scales $\lambda > \lambda_{\mathrm{dB}}$, the quantum pressure term is not important anymore and the particles behave like free particles, in a CDM like way\footnote{Here, when I say that is has a CDM like behaviour, I mean it behaves like a free particle and not like a condensate. Therefore, it follows the hydrodynamical description of CDM. They can have the same type of behaviour like CDM and be described by the same equations, but ULDM have different initial conditions and, for small mass describing DE after equality, can modify the expansion.}.

This scale where the condensate behaviour becomes important is called the coherent length. We are going to show now a more precise determination of this coherent length in both the FDM and SIFDM. And together with this analysis, it is going to be possible to better understand what is going on in the center of halos that drives this non-CDM behaviour.

\subsubsection{Description of the condensate} 
\label{Sec.:description_condensate}

Now that we have our description of the FDM and SIFDM on small scales, we want to understand what takes place inside the halos, where the Sch\"{o}dinger-Poisson equations describes the evolution of the system.
We want to describe here what is the picture we have in mind for what happens on those small scales.  

The special feature of the ULDM models is that they present a non-CDM behaviour in galaxies. As we saw in the previous section, the ULDM has a cosmological evolution very close to CDM for large scales. Different than in CDM, the ULDM has a non-zero Jeans length showing that on small scales this component is going to behave differently than CDM. Therefore, outside galaxies, ULDM behaves like CDM but with a suppressed power spectrum, and inside galaxies, in those homogeneous sub-Jeans scales ULDM can have a non-CDM behaviour.
%And this difference is more than just a different initial condition yielding a power spectrum that is suppressed on scales smaller than the Jeans length of the theory.

Inside these homogenous sub-Jeans regions inside galaxies, the ULDM thermalizes and forms gravitationally bound compact objects, called Bose stars or solitons, where a Bose--Einstein condensation or superfluid is formed. This was described in many references in the literature~\citep{Chanda_24,Chanda_25,Chanda_26,Chanda,Chanda_28,Chanda_29,Sikivie:2009qn,Erken:2011dz}, both in the presence and in the absence of interactions. The coherence length of this condensate sets the region where the wave behaviour of the condensate is important and changes the dynamics. Outside the condensate, on scales larger than the coherent length, the ULDM behaves as particles following the particle  description as CDM (decoupled axion following~\citep{Sikivie:2009qn}), with different initial conditions than in CDM.

There have been many studies of the properties of these condensed gravitationally bounded objects, in particular in the context of axions~\citep{96,16,Chanda_34,Chanda_35,Chanda_36,Chanda_37,Chanda_38,Chanda_39,Chanda_40,Chanda_41}.  In this section we are going to study the thermalization and formation of these compact objects both in the context of  FDM, but also in the context of the SIFDM. It is interesting to see that the size and phenomena, BEC or superfluidity, described by each of these models can differ a lot in each case, and in the case of the SIFDM it differs with the sign of the self-interaction.

The formation process of this Bose--Einstein condensate by gravitational interaction in the center of halos or in axion miniclusters is shown to take place in our universe, with relaxation times dominated by faster gravitational relaxation time, which is smaller than the age of the universe~\citep{Chanda_28,Chanda_29,Sikivie:2009qn,numerical_fuzzy_1,Levkov:2018kau,Chanda_2}. We are going to show here we can describe this formation of the condensate and obtain the relaxation times in the case of the FDM and the SIFDM. 

\begin{figure}[htb]
\centering
\includegraphics[scale=0.35]{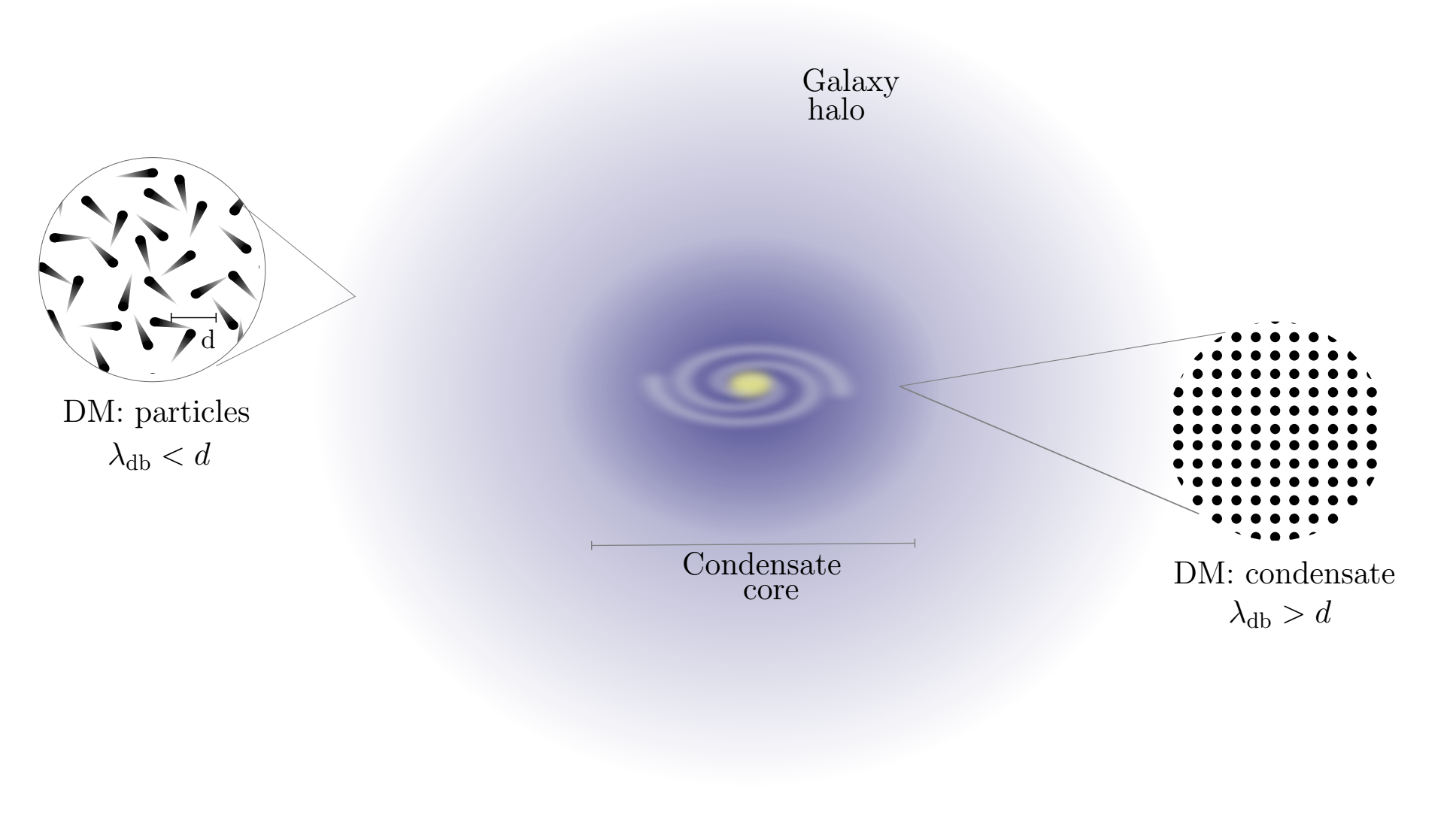}
 \caption{Schematic figure of the behaviour of ULDM in galaxies, where a condensate core is expected to form in the inner parts of the galaxy, while DM behaves like normal DM in the outskirts or outside galaxies.}
\label{Fig.:scheme_core}
\end{figure}

Summarizing, the picture that we have to have in mind is shown in Figure~\ref{Fig.:scheme_core}. Inside the halos of galaxies, a condensate core is formed\footnote{The picture described here is not the same as described in the entire literature. We are going to discuss this in Section 4.1.3.}.  These Bose clumps are called solitons or Bose stars.
This is also the picture that we have for the DM superfluid model, presented in Section 4.2.

%%%%%%%%%%%%%%%%%%%%%%%%%%%%%%%%%%%%%%%%%%
%%%%%%%%%%%%%%%%%%%%%%%%%%%%%%%%%%%%%%%%%%

As we saw from the description of the ULDM given above, we are describing the theory purely classically. It is valid then to ask if our ULDM can be treated as a classical theory or not. Specially since this model has such small masses.  

Classicality is an emergent concept that can be applied for system that are composed by a large number of constituents.  In this limit, the quantum effects of the theory are suppressed by this macroscopic number of particles, and the theory can be described by a classical theory. In this way, there is a limit where this classical approximation breaks and the quantum effects become important.

Lets first talk about what happens in the halo. When taking the classical limit of a theory, we can think that we decompose our quantum field $\hat{\psi}$ into
\begin{equation}
\hat{\psi} = \psi + \delta \hat{\psi}\,,
\end{equation}
where $\langle \hat{\psi} \rangle = \psi$ is the classical field (since if it was quantum this expectation value would be zero) and $\delta \hat{\psi}$ are the quantum corrections on top of the classical field. These quantum correction are suppressed by large occupancy number of the states, and for a coherent state is given by: $\delta \hat{\psi}/ \psi \propto 1/\mathcal{N}$. 
Therefore, if we are in a system with a large number of quantum constituents, the quantum corrections can be neglected and the system can be treated as classical. And this classical field is going to obey classical equations. This is, very roughly since this can be made much more precise mathematically, the definition of a classical field. On top of that it is a condensate, there is a large occupation number of the ground state, so $\mathcal{N}=\mathcal{N}_0$.

We can then estimate the occupancy number that we have in a halo. If we consider a MW-like galaxy, and we take as an example the axion, then the number density of axions in the galaxy is:
\begin{equation}
n_{\mathrm{gal}} =  \frac{\rho_{\mathrm{gal}}}{m} \approx \frac{\mathrm{GeV/cm}^3}{10^{-5} \, \mathrm{eV}} \,.
\end{equation}
Since we can write the occupation number roughly as $\mathcal{N} \sim n \lambda_{\mathrm{dB}}^3$, given the de Broglie wavelength of an axion with mass $m \sim 10^{-5} \, \mathrm{eV}$, we have that in a galaxy $\mathcal{N} \sim 10^{46}$. This shows that in galaxies today, this ultra-light particle has a huge occupancy number and we can consider a classical evolution. We can extend this estimation for earlier times,  if we assume the QCD axion where the potential arises from non-perturbative effects on scales $\Lambda_{\mathrm{QCD}}$. For those axions, the energy density (assuming they are most or all the DM in the universe) is given by $n_a \sim \rho_a/m \sim (T_{\mathrm{eq}}/T_{\mathrm{QCD}}) \, \rho_{\mathrm{tot}} \sim T_{\mathrm{eq}} T_{\mathrm{QCD}}^3 /m$. The de Broglie wavelength can be estimated to be at most of the size of the Hubble horizon at that time $\lambda \sim H_{\mathrm{QCD}}^{-1} \sim T^2_{\mathrm{QCD}}/M_{pl}$. This yields an even larger occupation number $\mathcal{N} \sim 10^{61}$. 
%Therefore, we can generally consider this field, after it is formed, as classical throughout all the evolution of the universe that we are interested in to describe DM, from equivalence until today.

If you start your theory with a classical scalar field described by classical equations of motion, then this description is valid for almost all the evolution of the universe (early times might require quantum treatment tough), specially the cosmological times we are considering. So it is a good description for the evolution of the ULDM\footnote{There is one subtly here that there might be effects on local systems that can deplete the condensate and break coherence. I will discuss that in Section 4.1.3.}.

As we saw in Section 3, in the classical limit, we can describe the properties of the (classical) condensate using a classical theory. And that is what we are going to do here. In the classical point of view, the condensate formed is going to be described by a slowly varying in space, homogeneous and stable field that presents long-range correlation~\citep{Chanda}.  In the case of FDM and SIFDM, condensation takes place in the presence of gravity, and of both gravity and self-interactions in the case of SIFDM. So we need to show how this gravitational thermalization takes place. We are going to describe these condensates, showing their size, given by their correlations length, and the condensate solution using linear theory.

%\subsubsection{Description of condensation?}
\vspace{0.5cm}
\subparagraph{SIFDM}\mbox{}\\
\label{Sec.:SIBEC}

We are first going to treat the SIFDM model, which is described by the presence of a self-interaction~\citep{16,Chanda,Chavanis:2017loo,Chavanis:2016dab}. We specialize here to the two-body interaction, since higher order interaction are usually suppressed in low-energy systems like the one we are interested. But it is easy to generalize this for higher order interactions. From what we saw in the previous section, the theory of a self-interacting condensate describes a superfluid in certain regimes. We are going to see here for which conditions this occurs in the SIFDM model.

We are going to work in the limit where we ignore gravity in order to investigate the effect of the self interaction in the model in a very similar way as described in Sect.~\ref{Sec.:Interacting BEC}. The Schr{\"o}dinger equation that describes model is given by:
\begin{equation}
i \dot{\psi} =  -\frac{1}{2m} \nabla^2 \psi + \frac{g}{8m^2} |\psi|^2 \psi\,. 
\end{equation}
We decompose the field into a homogeneous background solution, which represents the condensate, plus a perturbation part: $\psi (\mathbf{x},t)= \psi_{\mathrm{c}}(t) + \delta \bar{\Psi} (\mathbf{x},t)$.  The condensate part satisfies the Gross--Pitaevskii equation,
\begin{equation}
i \dot{\psi}_{\mathrm{c}} =   \frac{g}{8m^2} |\psi_0|^2 \psi_{\mathrm{c}}\,,
\end{equation}
that has a simple periodic solution $\psi_{\mathrm{c}}(t) = \psi_0 \, e^{-i \mu_{\mathrm{c}} t}$, where $|\psi_0|^2 = n_0$ is the number density of particles that fixes the amplitude of $\psi_0$, and $\mu_{\mathrm{c}} = g n_0/8m^2$. 

The equation describing the evolution of the perturbation, making the field redefinition $\delta \bar{\Psi} = \psi_{\mathrm{c}} \delta \Psi$, is 
\begin{equation}
i \delta\Psi = -\frac{1}{2m} \nabla^2 \delta \Psi + \frac{g n_0}{8m^2} \left( \delta \Psi + \delta \Psi^{*} \right)\,.
\label{Schr_int}
\end{equation}
Since $\delta \Psi$ is a complex scalar field, we can decompose the field into a real and a imaginary parts, $\Psi = A+iB$. We want to determine the dispersion relation of this system, so we write the equation of motion in Fourier space:
\begin{equation}
\frac{d}{dt	} \begin{pmatrix} A_k \\ B_k \end{pmatrix}= \begin{pmatrix} 0 & \frac{k^2}{2m} \\ -\frac{k^2}{2m} - \frac{g n_0}{4m^2} & 0 \end{pmatrix}  \begin{pmatrix} A_k \\ B_k \end{pmatrix}\,,
\end{equation}
where we call $\zeta_k=k^2/2m + gn_0/4m^2$.
The dispersion relation is given by:
\begin{equation}
\omega_k^2= \frac{g n_0}{4m^2} \frac{k^2}{2m} +\frac{k^4}{4m^2}\,.
\label{dipersion_relation}
\end{equation}
We an see that for $\omega^2_k >0$ we have an oscillatory solution:
\begin{equation}
\delta \Psi_k = Z \left( \omega_k+\zeta_k \right) e^{ i \omega_k t} + Z^{*} \left( \omega_k - \zeta_k \right) e^{- i \omega_k t}\,,
\end{equation}
where $Z$ is an arbitrary complex parameter. When $\omega_k^2 < 0$, the solution of the equation for $\delta \Psi_k$ is given by exponentials,
\begin{equation}
\delta \Psi_k = c_1 \left( \gamma_k - i \zeta_k \right) e^{\gamma_k t} + c_2 \left( \gamma_k + i \zeta_k \right) e^{- \gamma_k t} \,,
\end{equation}
where $\gamma_k = (k/ \sqrt{2m}) \sqrt{- \zeta_k}$ are the eigenvalues of the matrix, and $c_1$ and $c_2$ are constants given by the initial conditions. 

The  regimes where the dispersion relation is positive or negative are separated by the modes with wavenumber,
\begin{equation}
\omega_k^2 = 0 \qquad \qquad \Longrightarrow \qquad k^2_{*}=- \frac{g \, n_0}{2m}\,, 
\label{Eq.:Jeans_length_SIFDM}
\end{equation}
from where we can also determine the wavelength $\lambda_{*}=2\pi/k_{*}$ that divides these regimes. This wavelength is proportional to the healing length calculated in the previous sections. However, can see from this scale that what is actually going to determine if we have a stable oscillatory solution or a exponentially growing instability is the sign of the interaction.

For a repulsive interaction, $g>0$, the homogeneous configuration is always stable, and it is always going to be described by an oscillatory solution, either if $\lambda$ is bigger or smaller than $\lambda_{*}$ . This means that for all wavelength we can have a stable solution that can describe a condensate. From the dispersion relation we can also see that, in the \emph{long wavelength} regime
\begin{equation}
\omega_k \simeq c_s k\,,
\end{equation}
which is the dispersion relation of the phonon, that propagates as a wave, mediating long range correlation. This means that the SIFDM with a repulsive interaction is a \textbf{superfluid}, with the theory fully described by this propagating phonon. On the other hand, for very small wavelengths, large $k$, the term $\omega_k= k^2/2m$ dominates, which is the dispersion relation of a free massive particle, and the system stops exhibiting superfluidity. In the intermediary regime, where we should consider both terms of the dispersion relation, and the theory described by two degrees of freedom the phonon and the massive particle associated to particle creation away from the condensate. The scales that determine what we mean by long-wavelength regime is where $\lambda \gg \lambda_{*}$ in a way that the linear term dominates the dispersion relation. Since $\lambda_{*}$ is proportional to the healing length, this condition for the long-wavelengths is equivalent to the condition in Sect.~\ref{Sec.:Interacting BEC}  that the healing length  gives us the scale where quantum pressure (QP) can be neglected, which is what we are describing here.

When we have an attractive interaction, given by $g<0$ we have two regimes of stability. For $k < k_{*}$ ($\lambda > \lambda_{*}$) we have exponential growing solutions, which means that perturbations grow parametrically. Given this instability, the condensate cannot be formed on these scales. For $k > k_{*}$ ($\lambda < \lambda_{*}$), the solution oscillates and is stable, forming a condensate. This stable configuration, however, is different than in the case for repulsive interaction, forming a localized object, with maximum size given by $\lambda_{*}$. This localized stable solutions are called \emph{soliton}. Therefore it makes sense that this is the healing scale is the scale of the stability, since this is the scale below which the interaction ``heals'' perturbations of the condensate.

\begin{empheq}[box=\widefbox]{align}
\begin{array}{ll}
g > 0 & \qquad \longrightarrow \,\, \, \, \forall \, \lambda \qquad \text{Solution oscillates. Condensate {\color{gray}(long range)}} \\
g < 0 & \qquad \longrightarrow \left\{ \begin{array}{ll}
\lambda > \lambda_{*} & \qquad \text{Structures grow. No condensate.} \\
\lambda < \lambda_{*} & \qquad \text{Solution oscillates. Condensate {\color{gray} (finite size)}}
\end{array} \right.
\end{array} \nonumber
\end{empheq}

The case of the attractive interaction is not a superfluid even in the stable localized regions, the solitons. The only stable regions are for $\lambda < \lambda_{*}$ and this is the regime where the linear term is always subdominant in the the dispersion relation. The long-wavelength regions in this case are the regions where instability happens and there is no formation of a condensate.

\vspace{0.3cm}
\textit{Condensate solution - \,}
Having studied the stability of the system and determined the regimes where we have stable and unstable solutions, we can now describe the background solution for each case. As we saw in the ``condensate solution'' part of Sect.~\ref{Sec.:Interacting BEC}, one possible solution for the weakly interacting BEC is given by the solitons. For repulsive interactions the condensate solution is a dark soliton,  while for a attractive interaction one has a bright soliton.

What is important is that  for the repulsive interaction we can create a condensate of any size desired. So the size of the condensate is not limited and it will depend only on the choice of mass and strength of the interaction.
For the case of attractive interaction, we can only have localized stable solutions, giving a finite size bright soliton, with maximum size $\lambda_{*}= 2\pi \sqrt{2m/|g| \, n_0}$. Therefore we can see that we could only have a soliton, for an attractive interaction, that is relevant on galactic scales if the mass is very small but with a not very small coupling. An interplay between these two parameters need to occurs then if one wants galactic sized $\lambda_{*}$.

This condition also shows us that the QCD axion is not a good candidate for a ULDM (at least not to represent  most of the ULDM, but it can still represent a fraction of DM). For the QCD axion the interaction, given by $g_{\mathrm{a}} = -\Lambda^4 / f_{\mathrm{a}}^4$ is negative, and it is extremely small with $g_{\mathrm{a}} \sim -10^{-48}$, because $\Lambda \sim 0.1 \, \mathrm{GeV}$, for the typical QCD scale, and $f_{\mathrm{a}} \sim 10^{11} \, \mathrm{GeV}$, for typical Peccei-Quinn scale. Given that the mass of the axion is approximately $m \sim 10^{-5} \, \mathrm{eV}$, the soliton length, for $n_0 \sim n_{\mathrm{gal}}$, is $\lambda_{\mathrm{s}} \sim 2.8 \times 10^{11} \, \mathrm{km} \sim 9 \times 10^{-6} \, \mathrm{kpc}$. This is much smaller than any galactic scales, which are of the order of tens to hundreds of kpc. The QCD axion, then, produces these small and localized clumps of axions. 
%One thing that is being investigated by some researchers is the possibility of many of those small cores  to be created in a halo, with the possibility of them clumping together to form a bigger condensate.

\vspace{0.3cm}
\textit{Occupancy number evolution -  \,}
With those solutions in hand, we can understand how the evolution of the occupancy number for the condensate will behave for each mode. Determining the evolution of the occupation number is very important since having high occupancy number of the ground state of the system it what is actually the definition of a condensate. And this definition is independent of the system described, representing the best way of showing that condensation took place.
This is given by: $\mathcal{N} = |\psi_k|^2 / V$, where $\psi_k$ is constructed from the exponential and oscillatory solutions described above, with a random phase. The average occupation number evolves as:
\begin{equation}
 \begin{array}{ll}
    \langle \mathcal{N}_k (t) \rangle = \langle \mathcal{N}_k (t_i) \rangle \left\{ 1+ \frac{1}{2 \gamma^2_k} \left( \frac{g n_0}{4m^2} \right) \sinh^2 \left[ \gamma_k \left( t-t_i \right) \right]   \right\}\,, & \qquad \mathrm{for} \,\, \omega_k^2 < 0 \\
   \langle \mathcal{N}_k (t) \rangle = \langle \mathcal{N}_k (t_i) \rangle \left\{ 1+ \frac{1}{2 \omega^2_k} \left( \frac{g n_0}{4m^2} \right) \sin^2 \left[ \omega_k \left( t-t_i \right) \right]   \right\}\,, & \qquad \mathrm{for} \,\, \omega_k^2 > 0 
  \end{array} 
\end{equation}
For $g > 0$, an repulsive interaction, $\gamma_k$ is imaginary, with $\gamma_k = i \omega_k$, so the occupation number oscillates and the oscillations are stable. The ratio  $\langle \mathcal{N}_k (t) \rangle /  \langle \mathcal{N}_k (t_i) \rangle$ which has the largest value is obtained for modes that minimize $\omega_k$, which are the modes with $k \rightarrow 0$. These are the longest wavelengths. Since the long wavelength dominates, this means that  long range correlation is present, and we can have a long range condensate.

For $g < 0$,the occupation number grows exponentially. The fastest growth is given by the modes $k=k_{*}$ that maximize $\gamma_k$. So the modes  $k > k_{*}$ or $\lambda < \lambda_{*}$, where $*$ denotes the characteristic scale where instability sets in, will dominate and the stable configuration of the system will be localized clumps. The size of these clumps will be given by the mass and interaction of the model. 

\vspace{0.5cm}
\subparagraph{FDM: only gravity}\mbox{}\\

We are now going to describe a model without interaction, where the ultra-light particles are under the influence of the gravitational potential. 
We can think that gravitational potential has the same effect as an attractive interaction, in a way that quantum pressure has to counter-act the gravitational collapse. This gives a good picture to what to expect, from the knowledge obtained for the SIFDM. However, one needs to remember that the FDM model is described by a non-interacting theory, which means that it condenses into a BEC, but does \textbf{not} exhibit superfluidity in any regime.

%\paragraph{Evolution around the condensate and stability} \mbox{} \\

The Schr{\"o}dinger equation for DM in a gravitational potential, in the absence of interaction, is,
\begin{equation}
i \dot{\psi} = -\frac{1}{2m} \nabla^2 \psi +  m\Phi \, \psi\,,
\end{equation}
which is coupled to the Poisson equation:
\begin{equation}
\nabla^2 \Phi = 4 \pi G \left( m |\psi|^2 - \bar{\rho}  \right)\,,
\end{equation}
where the average background density, $\bar{\rho}$, was subtracted. Expanding the field as done previously $\psi (\mathbf{x},t)= \psi_{\mathrm{c}}(t) + \delta \Psi (\mathbf{x},t)$, the equation for the condensate is trivial and $\psi_{\mathrm{c}} = \psi_0 =  \mathrm{const.}$. For the  fluctuations we can write the linearized systems of equations that govern the evolution of the perturbation:
\begin{align}
& i \dot{\delta \Psi} = - \frac{1}{2m	} \nabla^2 \delta \Psi + m \Phi\,, \\
& \nabla ^2 \Phi = 4\pi G \, m \, n_0 \left( \delta \Psi + \delta \Psi^{*} \right)\,.
\end{align}
These can be combined into the equation:
\begin{equation}
i \dot{\delta \Psi} = - \frac{1}{2m} \nabla^2 \delta \Psi + 4\pi G m^2 n_0 \nabla^{-2} \left( \delta \Psi + \delta \Psi^{*} \right)\,.
\end{equation}
One can notice that the equation above is very similar to the equation we had for the interacting case (\ref{Schr_int}) for an attractive interaction. With that, we expect that there is an instability for long wavelengths, and that the condensate stable solution is only given for a finite region, forming a localized core.  To determine this lets take the Fourier transform of the fields. Like in the interacting case, the instability is divided by the regimes where the dispersion relation is smaller or bigger than zero. For the parameters of the FDM, we can determine the wavenumber that separates the regimes as:
\begin{equation}
\tilde{\omega}_k^2 = 0 \,, \qquad \Longrightarrow \qquad k_{\mathrm{J}} = \left( 16 \pi G m^3 n_0  \right)^{1/4}\,.
\end{equation}
This scale is the \emph{Jeans scale} and it separates the regimes where gravity dominates and collapse happens ($k < k_{\mathrm{J}}$), and the regime where the quantum pressure dominates and the solution is stable and oscillates ($k > k_{\mathrm{J}}$). In this regime we can have a condensate. The quantum pressure term counteracts the gravitational attraction and  any attempt to localize the particle is accompanied by an increase in energy. So stability below the Jeans scale arises because of the uncertainty principle. 
%The Jeans scale is the geometric mean of the the dynamical scale, defined in Sect. 2, and the Compton scale $\lambda_c = h/(m\,c)$.

\begin{empheq}[box=\widefbox]{align}
\left\{
\begin{array}{ll}
\lambda > \lambda_{\mathrm{J}} & \qquad \text{Structures grow. No condensate.} \\
\lambda < \lambda_{\mathrm{J}} & \qquad \text{QP dominates, solution oscillates. Condensate {\color{gray} (finite size)}}
\end{array} \right. 
\nonumber
\end{empheq}

We can estimate the size of the coherent condensate core. Rewriting the Jeans wavelength as: 
\begin{equation}
\lambda_{\mathrm{J}} = \frac{2\pi }{k_{\mathrm{J}}} = \frac{\pi^{3/4}}{2} \left( G \rho \right)^{-1/4} m^{-1/2} = 94.5 \left( \frac{m}{10^{-22} \mathrm{eV}} \right)^{-1/2}   \left( \frac{\rho}{\rho_{\mathrm{crit}}} \right)^{-1/4} \left( \frac{\Omega_{\mathrm{m}} h^2}{0.12} \right)^{-1/4} \, \mathrm{kpc}\,,
\label{Eq.:Jeans_length_FDM_NR}
\end{equation}
where $\rho_{\mathrm{crit}}$ is the critical density. For \emph{fuzzy DM} in an overdense region $\rho = 10^6 \, \rho_{\mathrm{crit}}$, 
\begin{equation}
\text{Fuzzy DM:} \,\, m \sim 10^{-22} \, \mathrm{eV} \qquad \qquad \longrightarrow \qquad \qquad \lambda_J^{FDM} \sim 3 \, \mathrm{kpc}\,,
\end{equation}
%we can see that the Jeans wavelength is of order $55 \mathrm{kpc}$, 
forming a condensate that is of the order of the scales of the halo of galaxies. As we can see, for a MW-like galaxy, this core formed is smaller than the halo. So we expect that in the outskirts of the halo the DM is not going to be condensed and is going to behave like normal matter, with the profile following the NFW profile. For the QCD axion, if we assume that we can have an axion without interaction, $\lambda_{\mathrm{J}} \sim 1.7 \times 10^{-7} \, \mathrm{kpc} $, which is a very small scale in comparison to galaxies. This stable bound  system is called a Bose star.

\vspace{0.5cm}

Adding the expansion of the universe, we can see that the system of equations becomes
\begin{align}
 &\frac{i}{a^{3/2}} \partial_t \left(a^{3/2} \psi\right) = - \frac{1}{2m} \frac{\nabla^2 \psi}{a^2} + m \Phi \psi \,, \\
 &\nabla^2 \Phi = 4\pi  G a^2 \left( m |\psi|^2 - \bar{\rho} \right)\,.
\end{align}
Like before, we determine the linearized equations for the perturbations $\delta \Psi$ around the coherent homogeneous background that evolves as $\psi_{\mathrm{c}} \propto a^{-3/2}$. The equation that describes the evolution of the perturbation then is
\begin{equation}
i \dot{\delta \Psi_k} = - \frac{k^2}{2ma^2}  \delta \Psi_k - \frac{3}{2} m \Omega_{\mathrm{a}} \frac{H^2 a^2}{k^2} \left( \delta \Psi_k + \delta \Psi^{*}_k \right)\,,
\end{equation}
where $\Omega_{\mathrm{a}} = m n_0/ \rho_{\mathrm{tot}}$ is the density parameter of our FDM particles. The real part of the perturbations obeys the following equation:
\begin{equation}
\ddot{A}_k + 2H \dot{A}_k - \frac{3}{2} \Omega_{\mathrm{a}} H^2 A_k + \left( \frac{k^2}{2ma^2} \right)^2 A_k = 0\,.
\end{equation}
All the terms in this equation would also be present from a massive DM component like CDM, except for the last term, which is related to the quantum pressure. The Jeans length in this case is given by:
\begin{equation}
\frac{k_{\mathrm{J}}}{a} = \left( 6 \Omega_{\mathrm{a}} \right)^{1/4} \sqrt{Hm}\,.
\end{equation}

\vspace{0.3cm}
This stability analysis shows us that if we want to construct a ULDM model with the desired feature of having a condensate in galactic scales, for the FDM model we need to have a very small mass, of the order of $m \sim 10^{-22} \, \mathrm{eV}$. This mass is within the bounds we had obtained earlier for all the ULDM models. For the FDM model the only parameter that controls the size of the condensate is the mass. We need then to determine the mass of the FDM particle for which this model can address the small scale problems, and for which the model is still in accordance with the very precise cosmological observations. This is what we show in the next subsection. We are going to put bounds on this parameter according to the observations in galaxies and on  large scales.

\vspace{0.3cm}
\textit{Condensate solution - \,} 
The ground state solution for the FDM is called Bose star, and it is a gravitational bound stable state in $3$-dimensions. This can be obtained by minimizing the Hamiltonian that is described by system with gravity alone, at fixed particle number ($N$)~\citep{Chanda}.
 In the absence of a exact solution, one can have an ansatz for this solution in analogy to the ground state of the hydrogen atom we assume spherical symmetry and have
\begin{equation}
\psi_{\mathrm{bs}} (r) = \psi_s e^{-i \mu_s t} = \sqrt{\frac{N\, k^3_{\mathrm{bs}}}{\pi}}\, e^{-k_{BS} r} \, e^{-i\, \mu_{s} t}\,,
\end{equation}
which corresponds to the ground state $E=-25G^2m^5N^3/512$, $\mu_s < 0$ and where the characteristic wavenumber of the Bose star is given by:
\begin{equation}
k_{\mathrm{bs}} = 5 G m^3 N/16 \sim k_{\mathrm{J}}\,,
\end{equation}
given that $N \sim n_{\mathrm{bs}}/k^3_{\mathrm{bs}}$. Therefore, the Bose star wavelength coincides with the Jeans length that determines the core, the stable region with no gravitational collapse.

However, the study of the Bose star solution needs to be done numerically. We are going to see next how the formation of these Bose stars takes place in kinetic theory. We see the formation of the BEC in these gravitationally bound structures, having the Bose star, and see how this Bose star grows. We are also going to discuss the cores formed by other simulations in Section 4.1.4 and in Section 4.3. With those we can determine the profile, mass and size of the core.

\vspace{0.5cm}
In the case of the SIFDM, we saw that what determines  the size of the condensate is the mass and of the strength of the interactions. Like for the FDM, in order to construct a SIFDM model that addresses the small scale problems and that are allowed by current observations, we need to put bounds on those parameters. This model was much less studied than the FDM model. There are a few works that present some bounds on those parameters (see references in the definition of this class), but with much less constraints than in the FDM case. This is also understandable since here we have an extra parameter in the model. A more complete analysis for the bounds of the SIFDM is going to be presented in a future publication and it is being considered by the authors while this review is being written.

\paragraph{Initial evolution of the condensate}\mbox{}\\

The picture that we have for the ULDM is that in the interior of galaxies it forms a core where the ULDM condenses and forms a Bose--Einstein condensate, in the case of FDM, and a superfluid, in the case of the SIFDM (and of the DM superfluid which we will see in the next section). Previously, we have determined the region where condensation can happen in the halo and calculated the coherence length of the condensate in the case of FDM and SIFDM,  and calculating what is the ground state solution of the condensate, the soliton or Bose star, for each case.

We want to study if there is the formation of a BEC in the center of galaxies in the presence of gravitational interactions, on top of the self-interaction in the case of SIFDM. We showed before that the ULDM is described by a classical theory.
The reader, then, might be wondering that since we are using a classical theory we cannot fully describe thermalization and the formation of the condensate. Indeed, as we showed in Sect.~\ref{Sec.:wave_turbulence}, and also in~\citep{Chanda}, classical fields follow a Rayleigh-Jeans distribution which present the UV catastrophe, and, thus, cannot describe a condensate. However, in~\citep{Chanda} the authors showed that a classical field theory is capable to describe the phase transition that represents the formation of a condensate and the theory has a well defined thermal equilibrium if one introduces a ultra-violet cut off in the theory.  We saw the same conclusion in wave turbulence, in Sect.~\ref{Sec.:wave_turbulence}, where the four-wave classic kinetic equation can only describe the initial stages of a BEC if a ultra-violet cut off is present in the theory.  Therefore, it is possible to describe the formation of a condensate with a classical theory.

To study describe the formation of a BEC in the center of galaxies here we use the kinetic equation from wave turbulence. We can use the four wave kinetic equation  to describe the initial stages of evolution of a BEC.
This formalism can be used  in the case of ULDM to show that Bose--Einstein condensation caused by gravitational forces indeed happens in the center of halos (and of axion miniclusters). From this theory we can also obtain a prediction for the condensation time and obtain the properties of the condensate distribution. This was done in~\citep{Levkov:2018kau} that made this analysis for the FDM model, and showed numerically the formation of the condensate and growth of the Bose stars, and in~\citep{Chanda_2} where the role of interaction was also studied. This is what we are going to show now.

We are going to proceed in the same fashion as in Sect.~\ref{Sec.:wave_turbulence}, but with the system subjected to a gravitational potential. The difference from the previous procedure is that  the gravitational interaction is long range, while in  Sect.~\ref{Sec.:wave_turbulence} we studied only the role of short-range interactions.

In galaxies, we are in the regime of high occupation number, and the classical description is valid. For the system we are considering, $p^2 \ll 2mT$, therefore for the period where the evolution is going to be described by this kinetic theory the UV catastrophe is not a problem, and the classical kinetic description of the system is valid.

In this regime, we can see that this system described by the non-linear Schr{\"o}dinger equation (\ref{Eq.:NLS_ULDM})  can be described in wave turbulence  by the classical kinetic equation (\ref{NLS_kinetic_field}) where the non-linear process is the  four-wave resonant interaction. 
%This description gives a statistical description for the system as the evolution of a ensemble of waves given by the Wigner function (\ref{Eq.:Wigner}). 
In the case of the ULDM subjected to gravitational potential, the kinetic equation is given by:
\begin{equation}
\frac{\partial f}{\partial t}+\frac{\mathbf{p}}{m} \cdot \nabla_{x} f = 2 \, \mathrm{Im} \int d\mathbf{y}\, e^{-i\mathbf{p.y}} \langle \psi(\mathbf{x}+\frac{\mathbf{y}}{2})\psi^{*}(\mathbf{x}-\frac{\mathbf{y}}{2}) U_{\mathrm{tot}}(\mathbf{x}+\frac{\mathbf{y}}{2})  \rangle \,,
\label{NLS_kinetic_field_ULDM}
\end{equation}
where $f$ is the Wigner distribution described by (\ref{Eq.:Wigner}) and the potential $U_{\mathrm{tot}}$ is given by
\begin{align}
U_{\mathrm{tot}}(\mathbf{x}) &= U_{\mathrm{G}} + g \, |\psi(\mathbf{x})|^2 =  4\pi G m \int d\mathbf{x}' \, \Delta^{-1}_{\mathbf{x}-\mathbf{x}'} \left(  |\psi(\mathbf{x})|^2 - n \right)
+ g\, \int d\mathbf{x}' \,  \delta(\mathbf{x}' - \mathbf{x}) \, |\psi(\mathbf{x})|^2 \,,
\end{align}
and $\Delta^{-1}_{\mathbf{x}-\mathbf{x}'} $ is the Green function coming from the Poisson equation
\begin{equation}
\Delta^{-1}_{\mathbf{x}-\mathbf{x}'}  = \frac{1}{4\pi |\mathbf{x} - \mathbf{x}'|}\,.
\end{equation}
\textit{These two equations are equivalent in this limit to the Schr{\"o}dinger-Poisson system that describes the evolution of the ULDM field in galaxies.} This is the Landau kinetic equation for the gravitating ensemble of random phase classical waves inside a structure of radius $R$ that in our case of interest is the halo of a galaxy, but it can be an axion minicluster or a periodic box, in the case of simulations.

To obtain the closed form of this kinetic equation we need to make some assumptions, as detailed in Sect.~\ref{Sec.:wave_turbulence}. In our case, we are going to assume as a initial distribution Gaussian distributed ULDM particles, described by a Gaussian random field $|\tilde{\psi}_{\mathbf{p}}|^2 = 8\pi^{3/2} \tilde{N} \, e^{-\tilde{p}^2} $ with random phases $\mathrm{arg} \, \tilde{\psi}_{\mathbf{p}}$, where $\psi = \tilde{\psi} v_0^2 \sqrt{m/G}$, $\tilde{\mathbf{p}} \equiv \mathbf{p}/mv_0$ and $v_0$ is the initial velocity. This initial configuration is the Fourier transform of $\tilde{\psi} (t,\tilde{\mathbf{x}})=\tilde{\psi} (0,\tilde{\mathbf{x}})$, with $\tilde{\mathbf{x}} = mv_0 \, \mathbf{x}$, which is an isotropic and homogeneous field. This is an uncorrelated field, a field that has minimal coherence length.
This type of initial configuration reinforces that there is no seed for condensation in the halo or axion minicluster, and that condensation arises simply by the gravitational interaction or gravitational interaction plus self-interaction. This initial condition is also well motivated from axions formed by the Kibble mechanism~\citep{Fairbairn:2017sil}.

%With that, to evaluate the four point correlation function present in (\ref{NLS_kinetic_field_ULDM}), the connected term in the Wick theorem (\ref{Eq.:Wick_theorem}) does not contribute initially. Non-Gaussianities arise later as the system evolves due to the small interactions, both the gravitational and the self-interaction, and the connected part becomes more and more relevant. 

Using Wick theorem, we can write our kinetic equation in closed form for the SIFDM as~\citep{Chanda_2}
\begin{equation}
\frac{\partial f}{\partial t}+\frac{\mathbf{p}}{m} \cdot \nabla_{x} f = F_1 + F_2 + I(f)\,,
\label{NLS_kinetic_regime}
\end{equation}
where the $F_1$ and $F_2$ correspond to the two point correlation functions in the Wick theorem, the first two terms in (\ref{Eq.:Wick_theorem}), and $I(f)$ comes from the connected part\footnote{Called $\mathrm{St} f$ in reference~\citep{Levkov:2018kau}.}, the last term of (\ref{Eq.:Wick_theorem}). The first two terms are given by:
\begin{align}
F_1 &= 2\, \mathrm{Im} \, \langle U_{\mathrm{tot}}(\mathbf{x}-i \nabla_{\mathbf{p}}/2) \rangle f(\mathbf{x},\mathbf{p}) \,, \\
F_2 &= \frac{2}{(2\pi)^6} \, \mathrm{Im} \, \int d\mathbf{q} d\mathbf{q}' d\mathbf{y} d\mathbf{y}' \, \left[4\pi G m^2  \, \Delta^{-1}_{\mathbf{y}} + g\, \delta(\mathbf{y}) \right]\, e^{i \left( \mathbf{y}'\cdot \mathbf{q} - \mathbf{y} \cdot \mathbf{q}'  \right) } f(\mathbf{x}+\frac{\mathbf{y}}{2},\mathbf{p}+\mathbf{q} ) \, f(\mathbf{x}+\frac{\mathbf{y}'}{2},\mathbf{p}+\mathbf{q}' )\,. 
\end{align}
They are $T$--odd terms.  They are order one in the interactions $\mathcal{O}(g)$ and $\mathcal{O}(G)$. The potential obeys: $\Delta \langle U_{\mathbf{tot}} \rangle (\mathbf{x})=(4 \pi G m)/(2 \pi )^3 \int d\mathbf{p} \, (f(\mathbf{x},\mathbf{p}) - n )$.

We can separate the contribution in $I(f)$ by coming from the the self-interaction and from the gravitational interaction $I(f)=I_g (f) + I_{\mathrm{G}}(f)$, respectively,  where
\begin{align}
I_g = 2 \, \mathrm{Im}\, \int d\mathbf{y} \, e^{-i\mathbf{p} \cdot \mathbf{y}} \, g \langle \psi_{+} \psi^{*}_{-} \psi_{+}  \psi^{*}_{+}  \rangle_{\mathrm{conn}}\,, \qquad I_{\mathrm{G}}= 2 \, \mathrm{Im}\, \int d\mathbf{y} \, e^{-i\mathbf{p} \cdot \mathbf{y}} \,  \langle \psi_{+} \psi^{*}_{-}  U_{\mathrm{G}} \rangle \,.
\end{align}
Here we simplified the notation by writing $\psi_{\pm} = \psi(\mathbf{x} \pm \mathbf{y}/2)$. These terms are second order in the interactions. Both of those interactions act in very different scales. The  self-interaction is short-range and it was the case we studied in Sect.~\ref{Sec.:wave_turbulence}. The case of the gravitational interaction needs to be treated carefully since the gravitational interaction is long-range. Here we follow~\citep{Levkov:2018kau} where the treatment from Landau for long range Coulomb interactions was used for gravity.

For the initial Gaussian distribution,  $F_1$, $F_2$ and $I(f)$ vanish, and the ULDM distribution is initially static. 

We are interested in describing kinetic relaxation of ULDM in the halo of a galaxy (or in an axion minicluster), which is the condensation, and obtain the relaxation time of this process. As we saw before, inside the halo $y = \Delta x \ll R$, the field and the potential are  homogeneous. In the center of a homogeneous spherically symmetric halo where the condensate is formed (or a homogeneous box in the case of simulations), the terms $F_1$ and $F_2$ vanish and we do not need to take them into account when studying these initial stages of the condensate, where the four-wave kinetic equation is valid. Therefore, in our case
\begin{equation}
\frac{\partial f}{\partial t}+\frac{\mathbf{p}}{m} \cdot \nabla_{x} f \approx I(f)\,.
\end{equation}

Lets first treat the gravitational part.
To estimate $I_{\mathrm{G}}(f)$, we can follow Landau's treatment. In the regime where the above equation is valid $\Delta x \ll R$, and in the kinetic regime $(mv)^{-1} \ll R$, we can expand the potential as $U_{\mathrm{G}}(\mathbf{x}+\mathbf{y}/2)=U_{\mathrm{G}}(\mathbf{x})+ (\mathbf{y}/2) \cdot \nabla U_{\mathrm{G}}(\mathbf{x})$. 
With that the gravitational scattering integral we were calculating can be described by a \textit{diffusion process in phase space}: $I_{\mathrm{G}} = - \nabla_{\mathbf{p}} \cdot \mathbf{s}$ where the Landau flux $\mathbf{s}$ is
\begin{equation}
\mathbf{s}= \frac{1}{(2\pi)^3} \int d\mathbf{x}' d\mathbf{p}' \, \mathcal{F}^{\mathbf{x}' \mathbf{p}'}_{\mathbf{x} \mathbf{p}} \, \nabla_{\mathbf{x}}(4\pi G m^2 \Delta^{-1}_{\mathbf{x}-\mathbf{x}'})\,, \qquad \mathcal{F}^{\mathbf{x}' \mathbf{p}'}_{\mathbf{x} \mathbf{p}} = \int d\mathbf{y} d\mathbf{y}' \, e^{-i(\mathbf{p}\cdot \mathbf{y}+\mathbf{p}'\mathbf{y}')} \langle \psi_{+} \psi^{*}_{-} \psi_{+}'  \psi_{+}^{*\prime}  \rangle_{\mathrm{conn}}\,.
\end{equation}
Plugging this new $I_{\mathrm{G}}$ in the kinetic equation, we can rewrite it as a the equation for the evolution of the four-point function $\mathcal{F}$, which involves a six-point correlation function. To be able to write this equation in closed form and solve it, one use Wick theorem. Ignoring the connected part in the limit of small separations $y \ll R$, the Landau flux is then $s_i=\int d\mathbf{p}' \, \Pi_{ij}(\mathbf{u}) (f'^2 \partial_{p_j} f - f^2 \partial_{p_j'} f' )$, where $\mathbf{u}= (\mathbf{p}'- \mathbf{p})/m$ and $f'=f(\mathbf{x}', \mathbf{p}')$. The term $\Pi_{ij}(\mathbf{u})$ is given by the integral 
\begin{equation}
\Pi_{ij}(\mathbf{u}) = \int dt'd\mathbf{y} \,  \partial_i \bar{\Delta}^{-1}_{\mathbf{y}} \, \partial_j \left[ \bar{\Delta}^{-1}_{\mathbf{y}+\mathbf{u}t'} - g \, \delta(\mathbf{y}+\mathbf{u}t')\right] \,,
\end{equation}
where $\partial_i \bar{\Delta }^{-1}_{\mathbf{y}} =  4\pi G m^2 \Delta^{-1}_{\mathbf{y}}$. Then this Landau flux defines $I_{\mathrm{G}}$ through $I_{\mathrm{G}} = - \nabla_{\mathbf{p}} \cdot \mathbf{s}$. However, to take the above integral one needs to introduce a long-time and a short-time cutoffs, given the logarithmic divergence of the Poisson Green's function. Therefore, the integral is taken in the interval $(mv^2)^-1 \ll t' \ll R/v$, which is the time that correspond to the regime where relaxation can happen since it can only happen inside the halo $y  \ll R$ where the field is homogeneous, giving the upper limit, and for distances bigger than the de Broglie wavelength $y \gg (mv)^{-1}$, since diffusion is only sensitive to fluctuations at long distance~\citep{Chanda_2}. The fact that diffusion is not sensitive to short distance scales already tells us that the self-interaction contribution be sub-dominant in the relaxation process. So the condensation in the halo happens much faster because of the gravitational interactions. In this time range, the dominant contribution of this integral yields $\Pi_{ij} \approx \Lambda (u^2 \delta_{ij} - u_i u_j)/u^3$, where $\Lambda \equiv \log (mvR) $ is the Coulomb logarithm. We can compute the relaxation time due to gravity with this, which yields $I_{\mathrm{G}} \sim f/\tau_{\mathrm{G}}$.

The $I_g$ contribution is going to be sub-dominant, but we can still evaluate their contribution to the condensate. Since the self-interaction is short range, we do not need to resort to Landau's treatment. We can just set $U_{\mathrm{tot}} = g |\psi|^2$, and solve (\ref{Eq.:NLS_ULDM}). Considering again that we have a homogeneous distribution inside the halo and ignoring the connected part of the Wick theorem since we want the result in leading order of $g$, then the kinetic equation can be simplified to $d^2f/dt^2 \sim 8g^2 \langle n \rangle f$, which tells us that the relaxation rate from the self-interactions is $dI_g/dt \sim 8g^2 n^2 f = \gamma f$, where $d/dt = \partial_t + \mathbf{p} \cdot \nabla_{\mathbf{x}} /m$. This relaxation rate does not grow with time in this limit, but when the connected correlations start to become relevant the relaxation rate starts varying with time. The relaxation time from self-interactions is then given by $\tau_g =1/\sqrt{\gamma}$.

With that the relaxation time for this weakly interacting Bose gas subjected to gravity in a halo is given by:
\begin{equation}
\tau_{\mathrm{G}} = \frac{\sqrt{2} m v^6}{12 \pi^3 G^2 n^2 \Lambda}\,, \qquad \tau_g =\frac{1}{ \sqrt{8} \, |g| n } \,.
\end{equation}
This gives the condensation time. The relaxation time for the gravitational interaction is inversely proportional to $G^2$ (and the cutoff scale), while the self-interaction one is inversely proportional to $g$. The gravitational relaxation time is much faster than the self-interaction one.
The total condensation time is  $\tau_{\mathrm{tot}} \sim 2 \tau_{\mathrm{G}} \tau_g/(\tau_g + \sqrt{\tau_g^2 + 4 \tau^2_{\mathrm{G}}}) \rightarrow \tau_{\mathrm{G}} \tau_g/(\tau_{\mathrm{G}} + \tau_g) \rightarrow \tau_{\mathrm{G}}$, is dominated by the gravitational one. In~\citep{Chanda_2} this is estimated to be of the order $\tau_g/\tau_{\mathrm{G}} \sim 10^5$ for the QCD axion. Therefore, the formation of Bose stars mainly happens due to gravity, even in the presence of the self-interactions. However, self-interaction is important as we saw above that the presence of the self-interactions lead to a different phenomenology and size of the soliton core. Notice that thermalization from gravitational interactions showed here is different then the one arising from short-range interactions through power-law turbulent cascades~\citep{Chanda_28}, arising in this system through a diffusion process.

In the virialized halo, we can see that the formation of the Bose star, for the FDM and SIFDM models, can happen in the universe with formation time smaller than the age of the universe. For example, in dwarf galaxies, the condensation times for a FDM particle of mass $m_{22}$ would take $\sim 10^6\, \mathrm{yr}$.

This formalism shows us that we can describe the first stages of the formation of a BEC in the halo of galaxies using kinetic theory. This approximate description of the BEC can help us estimate the condensation time and study the properties  of the Bose star. This formalism is particularly useful for numerical simulations which were performed in~\citep{Levkov:2018kau} for the FDM model.  The formation of the condensate can be seen numerically with no seed condensate in the simulation. It is seen that the initial evolution of the condensate also follows (\ref{NLS_kinetic_regime}) for $t<\tau_{\mathrm{G}}$, as expected for the wave turbulence with four-wave interaction.
The Bose stars grow after formation and the condensate becomes stronger (more particles condensate). The first decade if this growth shows a growth in the mass of the Bose star $M_s \simeq c v_0 (t/\tau_{\mathrm{
G}}-1)^{1/2}$, with $c=3\pm 0.7$. Only the first decade of the growth was seen both from computational limitation, but also this description does not hold for when the condensate starts to get stronger, and the description with four-wave interaction is expected to break.

That study shows us that we can indeed have the formation of a Bose--Einstein condensate of the ULDM particles inside the galaxies, forming Bose stars and solitons. This shows that the picture of having ULDM forming condensed cores inside galaxies is indeed valid.

%%%%%%%%%%%%%%%%%%%%%%%%%%%%%%%%%%%%%%%%%%%%%%%%%%%%%%%%%%%%%%%%%%%%%%%%%%%%%%%%%%

\subsubsection{Discussion}
\label{Sec.:discussion}

Before going further and describing the observational consequences of the FDM and SIFDM, we are going to briefly discuss the picture presented here and some different interpretations regarding condensation that are present in the literature.

The picture that we have showed until now for the behaviour of the ULDM is the following. ULDM is described by a classical theory that gives its cosmological evolution and its non-relativistic evolution in galaxies. Inside the halos of galaxies, the ULDM thermalizes and forms gravitationally bound systems, the Bose stars or soliton, where a BEC or a superfluid is formed. This condensed core is smaller than the size of the galaxy and the coherent length gives the region where the wave-like behaviour of the condensate is manifested, being of the order of the de Broglie wavelength in the case of the FDM. The formation of these objects through gravitational interactions can occur inside the halo of galaxies, without any condensate seed, in a time smaller than the age of the universe, as shown from kinetic theory.

Therefore, in this picture, after the halo of the galaxy is formed and virializes, thermalization in the inner regions of the halo happens and a condensate core is formed. This coherent length of the condensate, which changes if we are in the FDM or SIFDM cases, sets the size of the core that is smaller than the radius of the galaxy. This condensed core in the inner region of the galaxy is surrounded by a shell where DM behaves like a free particle (since $\lambda_{\mathrm{dB}} \ll d$) instead of a wave like it is inside the core. Outside the cores, in the outskirts of the halo, DM follows the profile predicted by CDM. A condensate core can also be formed in the center of a more massive system like a cluster. However, this condensate is very small, smaller than the ones in galaxies, in comparison with the size of the cluster.
Therefore, in the picture presented here, a BEC is formed in the interior of galaxies due to the gravitational interactions.

However, many authors in the literature there is a different view. For some authors coherence of the ULDM is established initially, in the initial stages of formation of the ULDM.
This is the case for axions, as pointed out in Sect.~4.7 of~\citep{Axions_1} (where a very good discussion of this topic is presented).  The cosmic axion field is described as a classical coherently oscillating scalar field.  This classical field $\phi$ comes from the expectation value of the quantum axion field, which can be represented in the Heisenberg picture as  $\langle \phi | \hat{\phi}(x) | \phi \rangle = \phi (x)$. The state of the axion field ${\phi}$ is going to be  coherent state~\citep{qft,Itzykson:1980rh}, with this coherence established initially. For the population of axions that is produced through vacuum misalignment, in the case where the Peccei-Quinn symmetry is broken during inflation, the inflationary evolution is responsible to making the axion coherent within the Hubble radius. For the axion population that form via the decay of topological defects, coherence of the field is obtained via thermalization after formation with rate of the order of Hubble rate $H_0$, as presented in~\citep{Axions_1}.

Therefore, as it is claimed in~\citep{Axions_1}, the axions coming from any of the populations can be described cosmologically as a classical coherently oscillating scalar field.  This classical description of the axion is said to hold throughout the evoltuion of the universe. This is discussed in~\citep{Dvali:2017ruz} where the validity of this mean field approximation is studied. They wanted to determined what is the quantum break time, which is defined as when quantum effects become important again for the description of the axion and the classical description of the axions is not valid anymore. The quantum breaking time is determined by the rate of axion scattering, and it is found that it exceeds the age of the universe  for the QCD axion, but this would also be true for an axion with  a smaller mass. Therefore, they conclude that it is safe to treat the axion as a classical coherently oscillating scalar field if coherence was established initially\footnote{In this work, they consider that the axion is already represented by a classical coherent field, with all axions in the zero momentum. Thus, it already started with a classical uniform axion field. They do not discuss the thermalization process or any mechanism that led axions to this state. Only studied the maintenance and validity of the classical approximation.}

This classical description also holds inside the halos of galaxies. This has been shown in~\citep{Allali:2020ttz} for a ultra-light fields that behave like DM. The transition from a quantum to a classical system is called decoherence\footnote{Coherence in this reference is the term used to describe the pure quantum mechanical system. This notation comes from quantum mechanics since coherent states only occur in quantum mechanics. Therefore, it has a slight different meaning than the classical coherent scalar field we have been talking until now. The coherent classical field we have been talking until now is the field that came from the expectation value of the (quantum mechanical) field operator of this ULDM taken in these coherent states $\langle \phi | \hat{\phi}(x) | \phi \rangle = \phi (x)$. So we call this classical field of coherent classical field since the field was in a coherent state during its initial quantum stages. This term is usually used in the literature. Refer to~\citep{Allali:2020ttz} for a proper definition of decoherence.}.   This is an important question since decoherence is known to occur very fast in  macroscopic systems due to interactions of the system with the environment. However, if we have DM, where interactions with the environment are known to be very weak interacting mainly gravitationally, decoherence might proceed less efficiently. 
In this reference they study decoherence of a ULDM, including axions, in the halo (ULDM overdensity) with its environment (the diffuse hydrogen in galaxies in this study). They find that ULDM in the halo can be treated as classical inside galaxies, since decoherence would take place very fast, with a decoherence rate of the order of the Hubble rate $H_0$ for ULDM with $m \sim 10^{-7} \, \mathrm{eV}$. 
The same result holds for a BEC in the halo of galaxies. If we have a Bose star that formed in the halo, the decoherence time is very fast for any ULDM that has $m < \, \mathrm{eV}$, so the condensate, after formed, can be treated as classical and described by classical equations like the Schrodinger-Poisson system of equations.

With that, the picture that some authors have is that since coherence is established from initial conditions, and this classical picture can be maintained throughout the evolution of the universe, then when the Jeans length stable regions are formed, the axion in this region is already a coherent field. Thus, inside these regions in the interior of galaxies, coherence was established by initial conditions, so no thermalization is necessary to happen inside the halo. 

However, the coherence of this classical field could be broken during the evolution of the universe by local processes. This can come from many out of equilibrium processes that occur in the universe. One example of this is the formation of halos, since virialization happens through violent relaxation. In this case the DM particles scatter on small fluctuations of the gravitational field, and coherence can be lost.  The system can then be describe by an ensemble of classical incoherent waves, with very small coherence length.
Also, as pointed out in~\citep{Dvali:2017ruz}, even the classicality of the axion could be challenged in small overdense regions and quantum breaking could occur in these regions.

With all that, we can see that either having coherence from initial conditions or not, the ULDM field has to either thermalize or re-thermalize on galactic scales. We showed above that thermalization of ULDM can take place in the halos of galaxies in the presence of gravity or gravity and self-interaction, without any need of previous coherence of the field. And in these regions, it was shown that a BEC is going to be formed. And after its formation from the results from~\citep{Allali:2020ttz} we can treat the evolution of the Bose star as classical. This picure is valid for the FDM, SIFDM and for the DM superfluid.  So, for the phenomenology of ULDM in galaxies, if the field was already a coherent field or not, does not alter the conclusions we present in this review, since this coherence can be reached at late times in the halo.

We just want to emphasize one last thing. Bose--Einstein condensation is a quantum mechanical phenomena. It can only arise because of quantum mechanics. The definition of a condensate is the one given in Sect.~\ref{Sec.:discussion_sec3}. A BEC can be  described classically in the classical limit of the many--body theory or the field theory description of a condensate, or by an approximate classical theory as an ensemble of waves like wave turbulence (where we can make a parallel with the definition of a condensate as particles in the ground state which in the language of waves translates is waves with $k=0$).

This discussion about thermalization or re-thermalization, classical or not, in the context of ULDM  has been presented in many places in the literature of ULDM with many different point of views and interpretations. For the interested reader some of these discussions are present in the following references~\citep{Sikivie:2009qn,Erken:2011dz,Axions_1,Chanda,Castellanos:2013ena,Davidson:2013aba,Davidson:2014hfa}.
We hope we have presented here a clear view of what happens in the interior of halos and how this can be interpreted, unifying some aspects of these interpretations.

%%%%%%%%%%%%%%
\subsubsection{Cosmological and astrophysical consequences of the FDM}
%%%%%%%%%%%%%%

%This section is dedicated to describing and discussing the main cosmological and astrophysical consequences of having DM described as above. 

Now that we have a description of what happens cosmologically and inside the halos of galaxies for these models, we are going to discuss the rich phenomenology that these models exhibit. 
We are going to study their predictions and discuss their cosmological and astrophysical consequences. These models behave differently than CDM in two ways. They present a CDM like behaviour on large scales, with modified initial conditions, and inside the halos of galaxies, they form a core in the inner part of the halo where a non-CDM behaviour is described, while a NFW behaviour is present in the outskirts of the halo, outside the core (see Figure~\ref{Fig.:scheme_core}).

We need to study the predictions coming from these modified initial conditions ,and from the presence of these cores and their non-CDM behaviour.  We can think about three main groups of consequences of these models. First,  is the suppression of the small scales structure, that is going to affect many observables both cosmological and astrophysical; the second are related to the presence of the cores in the center of galaxies; and the third are related to dynamical effects that arise from the BEC or superfluid formed in the central regions of galaxies. We are going to describe these predictions in this section. We are going to focus mainly in the FDM in this section. Each of those effects can be probed with different observables, which can lead to bounds in the parameters of the models. We are going to talk about how these effects are measured and the constraints obtained in Section 5.  For other observational consequences of FDM, see~\citep{Hui:2016ltb}.

A small comment. Although we showed above the we call solitons the ground state of the SIFDM, and Bose stars the ground state of the FDM, in the literature the term solitons is used for both. Therefore, from this point on we will use the term soliton to also describe the ground state of the FDM.

%%%
\paragraph{Suppression of structure formation}\mbox{}\\
%%%

One of the effects coming from the FDM and SIFDM classes of models is the suppression of small scale structure. This is a consequence of the fact that these models present sizable Jeans scale which cuts off the structure formation for wavelengths smaller than $\lambda_{\mathrm{J}}$. The Jeans length for the FDM model is given by (\ref{Eq.:Jeans_length_FDM_NR}). Therefore, as we saw in the previous section, for modes that are larger than the Jeans length $\lambda >\lambda_{\mathrm{J}}$, gravitational instability takes place and structure formation can happen, while modes smaller than the Jeans length $\lambda <\lambda_{\mathrm{J}}$ have oscillatory solution and no structure formation takes place. So structure formation suffers a cut off on scales of the order of the Jeans length. 
The same happens for the SIFDM model, but the scale of this cut off is different, analogous  to the Jeans length, or the healing length, given by (\ref{Eq.:Jeans_length_SIFDM}).

We can quantify this suppression by computing the power spectrum for the ULDM models. We are working in linear perturbation theory, as showed above, and we can evaluate the linear suppression of the power spectrum. In analogy to what  it was done for the WDM model, the modifications of the power spectrum with respect to the $\Lambda$CDM power spectrum are encoded in a transfer function $T_{\mathrm{FDM}}(k,z)$. We can relate the power spectra as~\citep{transfer_FDM,Axions_1}
\begin{equation}
P_{\mathrm{FDM}} (k,z) = T^2_{\mathrm{FDM}}(k,z) \, P_{\Lambda \mathrm{CDM}} (k,z) = T^2_{\mathrm{FDM}}(k,z) \, \left( \frac{D(z)}{D(0)} \right)^2 \, P_{\Lambda \mathrm{CDM}} (k) \,,
\end{equation}
where $ P_{\Lambda \mathrm{CDM}} (k)$ is the power spectrum of $\Lambda$CDM at $z=0$ which in turn is the primordial power spectrum transformed by an appropriate transfer functions as defined in~\citep{transfer_1,Eisenstein:1997ik}; and $D(z)$ is the growth factor given by~\citep{Peebles_book}:
\begin{equation}
D(z) = \frac{5 \Omega_m}{2H(z)} \int_{0}^{a(z)} \frac{da'}{(a'H(a')/H_0)^3}\,.
\end{equation}
The FDM transfer function is given by:
\begin{equation}
T_{\mathrm{FDM}} = \frac{\cos(x^3_{\mathrm{J} (k)})}{1+x^8_{\mathrm{J}}(k)}\,,
\end{equation}
where 
\begin{equation}
x_{\mathrm{J}} (k) = 1.61 \, \left( \frac{m}{10^{-22}\, \mathrm{eV}} \right)^{1/18} \left( \frac{k}{k_{\mathrm{J,eq}}} \right)\,, \qquad k_{\mathrm{J,eq}} = 9 \, \left( \frac{m}{10^{-22}\, \mathrm{eV}} \right)^{1/2} \mathrm{Mpc}^{-1}\,.
\end{equation}
The scale $k_{\mathrm{J,eq}}$ is the Jeans length at matter-radiation equality, the time when perturbations start to grow.

This transfer function presents a fast decay at $k=k_{\mathrm{J,eq}}$, which leads to a suppression of the power spectrum on those scales. The mode where the  power spectrum decays to half of its value $T(k_{1/2})=1/2$ is the half mode given by~\citep{Li:2018kyk} $k_{1/2} = 5.1 \, (m/m_{22})^{4/9} \, \mathrm{Mpc}^{-1}$, where $m_{22} = 10^{-22}\, \mathrm{eV}$. If $k_{1/2} < k_{nl} \sim 0.1$ where $k_{nl}$ is the non-linear scale which is around $1\, \mathrm{Mpc}^{-1}$, the suppression of the power spectrum can be probed by linear CMB and LSS observables. Otherwise, we need non-linear observables. Therefore, roughly speaking, we can probe the suppression on the power spectrum on linear scales for FDM with masses $m \lesssim 10^{-23}\, \mathrm{eV}$. One interesting fact is that the $k_{1/2}$ of the FDM is the same as the one for WDM, although they are different functions of the wavenumber.

Beyond linear observables, this suppression of the power spectrum on small scales induces a suppression in the formation of FDM halos. This can be estimated by calculating the linear half-radius ($R_{1/2,\mathrm{lin}}$) which is the radius where half of the mass of the spherically symmetric system is contained.  Then for  $R_{1/2,\mathrm{lin}} \sim \lambda_{1/2}/2$, where $\lambda_{1/2}=2\pi/k_{1/2}$, we obtain the mass of the smallest halos that can be formed in this theory~\citep{Bullock:2017xww,Niemeyer:2019aqm}
\begin{equation}
M_{\mathrm{lin}} = \frac{4\pi}{3} \, R_{1/2,\mathrm{lin}} \langle \rho_{\mathrm{FDM}} \rangle = 4\times 10^{10} \, M_{\odot} \, \left( \frac{m}{10^{-22}\, \mathrm{eV}} \right)^3 \left( \frac{\Omega_m}{0.3} \right) \left( \frac{h}{0.7} \right)^2 \,,
\label{mass_lin}
\end{equation}
where $ \langle \rho_{\mathrm{FDM}} \rangle = 3/(8\pi G) H_0^2 \, \Omega_m$. Therefore, FDM predicts a large suppression of halos for $M < 10^{10}\, M_{\odot}$ if the mass is $m_{22}$. 
Bellow, in Section 4.1.5, we are going to see how this can be calculated more specifically given the cores in the halos and see how these predictions help address the small scale problems.

This suppression of the power spectrum on small scales also suppresses the formation of galaxies. It is found in simulations that the number of sub-halos in FDM in comparison to CDM is reduced by a factor of $\sim (3M/M_{1/2})^{2.4}$. This suppression of formation of small galaxies is larger in FDM at higher redshifts, in comparison to CDM. This opens up an important question about FDM being able to produce small scales structures at early times to be probed by Lyman-$\alpha$ forest. 

Therefore, the linear suppression of the power spectrum can affect both in the linear and non-linear part of the theory. This can lead to the following effects. This suppression can be probed by probing the linear power spectrum by the CMB and the matter power spectrum through LSS surveys. Or even better by observables that probe even lower scales, like Lyman-$\alpha$ forest and 21-cm from neutral hydrogen. The suppression of small halos also affects the non-linear scales. The substructure of this model is going to be different than in CDM, with a suppression of substructures on small scales. The linear theory can predict a minimal mass for the structures formed, which can be probed by the population of satellites observed (and can be related to the missing satellites problem, as we will discuss in the next subsection). The substructures can also be probed directly by gravitational sensitive probes like gravitational lensing and streams, which are affected by the substrucutre present in the halo, and will be affected differently if the DM presents this suppression. We are going to see these observables and the bounds they can put in the models in Section 5. The substructures in the FDM model are also going to be changed by details of the presence of the core, which lead to different predictions that can be tested, as we will see bellow.

The entire numerical calculation of the power spectrum for the FDM model can be done using the software Axion\textsc{\tt{CAMB}} \citep{camb,Hlozek:2014lca} or a modification of software \textsc{\tt{CLASS}} madei in~\citep{Urena-Lopez:2015gur}. For our discussion in the next section we used power spectra generated by Axion\textsc{\tt{CAMB}}.

We discuss now the effect of the FDM in two observables that can probe  this suppression, together with other phenomenology of the FDM like the change in the rate of expansion: the CMB and the matter power spectrum.

\vspace{0.3cm}
\textbf{CMB - \,}
We want to review the modifications that FDM can cause in the observables of the CMB so we can understand how we can use this observation to probe the mass of the FDM. It is also possible to probe the fraction of FDM in the universe, if this is not assumed to be all the DM in the universe. We are interested in the case where FDM is all the DM in the universe, but we briefly comment here on the case where it is not.

%\begin{figure}[htb]
%\centering
%\includegraphics[scale = 0.2]{CMBS4.png}
% \caption{Figure adapted from~\citep{CMBS4}, illustrating the window of mass that each of the CMB observables can constrain. In this review we are mainly going to focus in primary CMB (blue) and CMB lensing (red). Although we are not discussing, the other effects can also constrain other effects of ULDM that are model dependent.}
%\label{Fig.:CMBS4}
%\end{figure}

The low mass of the FDM can alter the CMB in many different ways. We are mostly interested here in the effects in the primary CMB and CMB lensing, which are the ones that probe the DM in the ranges of mass we are interested for the ULDM models. There are other CMB observables that can probe other aspects of the microphysics of these models, and of their formation, as we can see in Figure 2 from~\citep{CMBS4}.

ULDM affects the CMB  in two main ways\footnote{Notice here that I am focusing only in the gravitational effects of the ULDM models. The CMB can also be affected by aspects related to the microphysics of some of these models and their formation mechanisms, which is specific to come models. For example, putting bounds in the axion isocurvature contributions can offer contraints in the axion decay constant $f_a$. It can even probe interactions in the dark sector. For more details on these other effects, check~\citep{CMBS4,Hlozek:2017zzf}}. The different expansion rate caused by the ULDM models affects the primary CMB, which is the adiabatic, unlensed without secondary effects CMB spectra. The suppression of the power spectrum that leads to a different clustering present in the ULDM models can be seen in the secondary lensing anisotropies.

%The CMB peaks acoustic peaks constrain many aspects of the universe at recombination. The angular size of the baryon acoustic oscillations (BAO) is responsible for setting the position of the fist peak in the CMB, at $z=z_{\mathrm{rec}}$}. The first peak determines the curvature, and the relation between the acoustic peaks determines the matter to radiation density. 

\begin{figure}[htb]
\centering
\includegraphics[scale = 0.5]{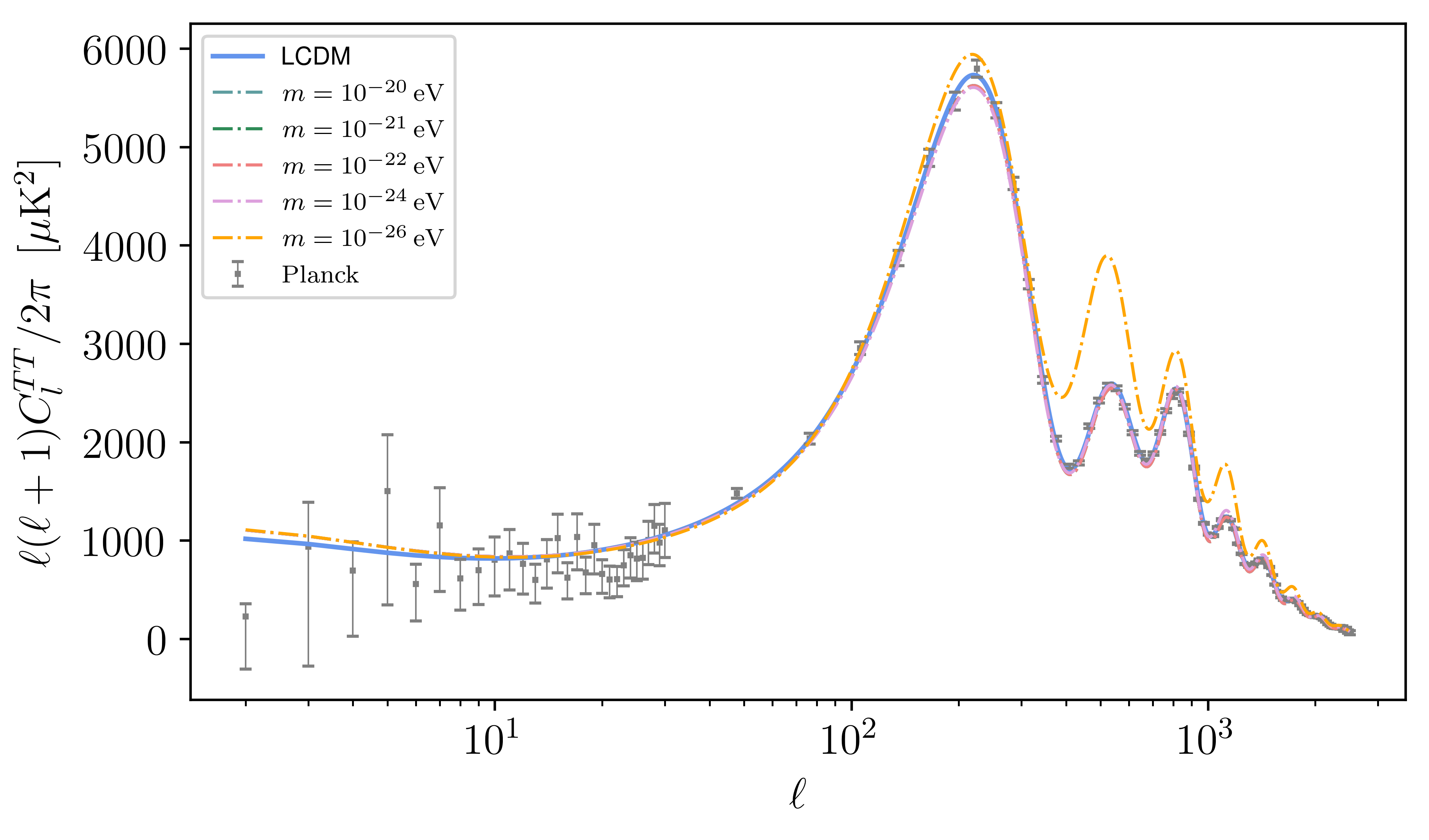}
\includegraphics[scale = 0.45]{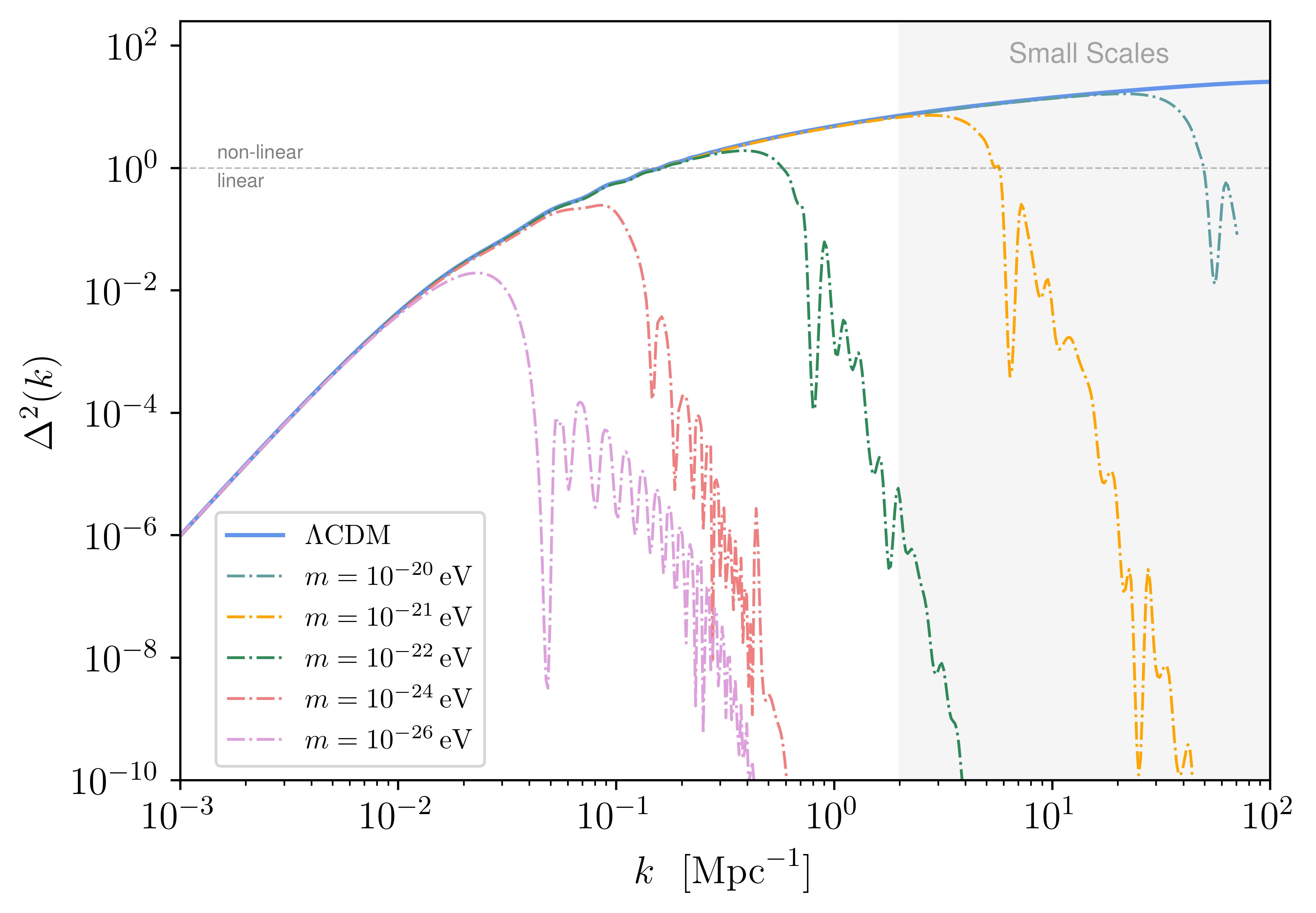}
 \caption{\textit{Left panel:} Angular power spectrum of the temperature CMB anisotropy for the FDM model for different masses where FDM is considered all the DM in the universe. We compare this with the $\Lambda$CDM model best fit to the \textit{Planck} data~\citep{Planck} showed by the gray data points. \textit{Right panel:} Matter power spectrum for the same FDM model as used in the left panel}
\label{Fig.:power_spectrum}
\end{figure}

We assume now that all the DM is given by ULDM. The primary CMB is affected by the expansion rate. Depending on the mass of the axion their oscillations take place at different redshifts.  If $z_{\mathrm{osc}} \lesssim 1100$, then after recombination the ultra-light field is behaving like dark energy and has a very different expansion rate than in the case of $\Lambda$CDM. In this case, the amount of dark energy in the universe will be much bigger, since this will be composed of $\Lambda$ plus the one from ULDM, affecting the first peak and the Sacks Wolfe plateau (the plateau around $\ell \sim 100$ that can be seen in Figure~\ref{Fig.:power_spectrum}).  The amount of the component that behaves as DM will be much smaller in this case, which in turn affects the other peaks. This can be seen in the left panel of Figure~\ref{Fig.:power_spectrum}. If the oscillations take place at $z_{\mathrm{osc}} \gtrsim z_{\mathrm{rec}} \approx 1100$, the ultra-light field behaves as DM before recombination, so they can alter the expansion rate during the radiation time. This affects the Silk damping tale, enhancing the higher acoustic peaks because of the reduction in the angular scale of the diffusion distance. This effect can be degenerate with changes in $N_eff$. As the mass gets heavier, which indicates that the ULDM behaves like DM since very early in the universe, the ULDM behaves more and more like CDM, and the angular  power spectrum is very close to the  $\Lambda$CDM one\footnote{In this section where we are treating the FDM model, when I say heavier FDM masses I mean $m \gtrsim 10^{-25} \, \mathrm{eV}$.}.

In this figure we plot the angular temperature power spectrum of coming from theoretical predictions of the FDM model for different masses, obtained using Axion\textsc{\tt{CAMB}}. We also plot for comparison the data from the \textit{Planck} 2018 $Cl$ TT power spectrum~\citep{Planck} and the $\Lambda$CDM model best fit to this data.  We can see that for masses smaller than $10^{-25}\, \mathrm{eV}$ we can even visually see the deviation of the power spectrum from the data and from the $\Lambda$CDM one, with an enhancement of the size of all the peaks, and changing the relation between the second and third peaks, which indicates less DM with respect to baryons. For higher FDM masses, we cannot visually distinguish it from the $\Lambda$CDM one. Therefore, the primary CMB can put bounds on the mass of the ULDM according to their modified expansion rate at recombination.

Other effects might arise when the density of the axions is not equal to the total DM energy density. There is a degeneracy between the amount of ULDM, $\Omega_{\mathrm{ULDM}}$, and the amount of curvature (or dark energy) and matter.  This can be seen in more detail in~\citep{Axions_1}.

The ULDM can also affect the secondary lensing anisotropies. The small scale suppression of the power spectrum  can be seen as a lensing deflection power on scales $\ell > 1000$. This can be seen in the lensing convergence power spectrum.
This effect from the suppression of clustering can also be seen in the matter power spectrum, as we show bellow. This effect is degenerate to the one coming from massive neutrinos. For smaller mass ULDM, one can use the effects of expansion to break this degeneracy, but this is not the case in the range of masses we are interested for ULDM that has an important effect on small scales.

This shows that CMB is a powerful observable to probe many aspects of ultra-light particles, specially in the low-mass range. We are going to see in Section 5 the constraints obtained by CMB observations in the FDM model and discuss some forecasts.

\vspace{0.3cm}
\textbf{Matter power spectrum - \,}
The matter power spectrum brings information about the matter density contrast in the universe with respect to the scale. The matter power spectrum contains a huge amount of information. Considering the full shape of the power spectrum, we can measured the equality scale ($k_{\mathrm{eq}}^{-1}$) which sets and can be inferred by the peak of the matter power spectrum (a bit hard to see in right panel of Figure~\ref{Fig.:power_spectrum}) and its overall shape. The other feature that is present in the power spectrum are the BAO features, from where we can obtain the BAO frequency and infer the  sound horizon at baryon drag (at $z_{\mathrm{rec}}$).

The ULDM affect the matter power spectrum both by having a different expansion rate and from the suppression of clustering on small scales. A different expansion rate would alter the BAO, presenting a different sound horizon in the power spectrum. This is more relevant for smaller masses of the FDM particle, when the FDM behaves like dark energy for a longer time in the evolution of the universe.

The other effect which is more dramatic on small scales is coming from the different clustering that the FDM presents. As we can see in right panel of Figure~\ref{Fig.:power_spectrum}, for a FDM that represents all the DM in the universe, the matter power spectrum for FDM presents a suppression of the small scale power spectrum. As we saw, this suppression comes from the Jeans length of the FDM that suppresses the power spectrum via the transfer function. We can see that the smaller the mass is, the effect is more dramatic. For masses that affect the linear scales, this can be easily tested using observations. However, we can see that for the heavier FDM particles, this suppression occurs on the small scales, where there is not a lot of observational data. 

We can use galaxy surveys to probe the biased matter power spectrum. This can give us bounds in the mass of the FDM, as we will see in Section 5. However, for heavier masses we can see that the suppression in the power spectrum occur on smaller scales, not probed by galaxy surveys. For that we need new observables, like Lyman-$\alpha$ which allows us to probe scales of order $0.5 \, \mathrm{Mpc}/h \lesssim \lambda \lesssim 100 \, \mathrm{Mpc}/h$. This permits to constrain higher masses for the FDM. Another window of observation that allows us to probe the smaller scales is 21-cm from neutral hydrogen, which can gives the matter power spectrum on scale $k > 10\, \mathrm{Mpc}^{-1}$. We will discuss these observables in Section 5.

One important point about using the power spectrum is the issue of the bias. To infer the matter power spectrum one needs to observe biased tracers of the DM distribution. The bias relates each of those tracers with the underlying DM distribution. This bias for each tracer is unknown, and each probe presents a different bias. Therefore, when obtaining constraints on the mass of the FDM, there is a degeneracy with bias. There is also the possibility of a scale dependence in the bias of ULDM. This is still a not so well studied problem for the ULDM, with a few studies from numerical simulations for the FDM case~\citep{Cooray:2002dia,Hlozek:2014lca}.
One can also obtain the matter power spectrum from shear measurements, coming from gravitational lensing. This are unbiased tracers and can provide measurements of the FDM mass with a complementary approach.

\vspace{0.5cm}
\subparagraph{Halo density profile}\mbox{}\\

We saw above that the suppression of the power spectrum in the FDM model leads to a suppression in formation of small mass halos. Only halos with $M > M_{\mathrm{lin}}$ where $M_{\mathrm{lin}}$ is given by (\ref{mass_lin}) are formed in the FDM. 
This suppression can lead to a different number of low mass halos, modifying the halo mass function. 

The halo mass function (HMF) describes the density of halos per unit of mass. To determine the HMF of the FDM one needs to resort to either simulations or semi-analytic methods. Usually simulations are performed and the HMF can be fitted, as done in~\citep{marsh_20}.  In this reference they have a simulation of colisionless particles with initial conditions coming from the FDM. The fitted HMF obtained is given by:
\begin{equation}
\left( \frac{dn}{dM} \right)_{\mathrm{FDM}} = \left[ 1+\left( \frac{M}{M_0} \right)^{-1.1} \right]^{-2.2}\, \left( \frac{dn}{dM} \right)_{\mathrm{CDM}}\,.
\label{Eq.:HMF_1}
\end{equation}
This HMF presents a suppression for low mass halos, characterized by the scale $M_0 = 1.6 \times 10^{10} \, m_{22}^{-4/3}$.  
The HMF of CDM depends on the redshift and mass of the halo, while the suppression term in brackets  is redshift independent, which is a consequence of the FDM modification only coming from the initial conditions. This means that this also does not take into account the effective sound speed of the FDM, only through the suppression of the initial power spectrum. This HMF is very accurate for higher masses, in agreement with the CDM HMF, but it presents an uncertainty in the low-mass end, showing that this HMF is reliable to show the suppression on those scales, but not so reliable to obtain the slope of the HMF for FDM for low halo masses.

There is also another HMF obtained using different methods.
In~\citep{marsh_6} with the aim of taking into account the scale dependent linear growth from the FDM, they obtain the HMF from a modified Press-Schechter approach. This yields
\begin{equation}
\left( \frac{dn}{dM} \right)_{\mathrm{FDM}} = -\frac{\rho_m}{M} \, f(\nu) \frac{d \ln \sigma^2}{d \ln M}\,,
\label{Eq.:HMF_2}
\end{equation}
where $\nu \equiv \delta_c / \sigma$, $\delta_c$ is the critical collapse overdensity and $\sigma (M)$ is the variance of the power spectrum. The variance is calculated by smoothing the power spectrum with a spherical top-hat window function. The function $f(\nu)$ comes from the Sheth-Thormen model~\citep{STM} and it is given by $f(\nu) = A (q/ 2\pi)^{1/2} \nu [1+(q\nu)^(-2p)]\, exp(-q\nu^2/2)$, with parameters $A=0.3222$, $p=0.3$ and $q=0.707$. The critical overdensity $\delta_c$ brings more details about the evolution of the FDM. At $z=0$, it can be described by comparing $\delta_c^{\mathrm{FDM}} (k) = \mathcal{G}(k)\, \delta_c$, where $\delta_c = \delta_c (z=0,k=k_0)$ and $k_0=0.002$ is the pivot scale. The scale dependent function$ \mathcal{G}(k)$ is the ratio of the scale dependent growth factors from $\Lambda$CDM and FDM~\citep{marsh_6,HMF_2} (in~\citep{HMF_1} a numerical fitting for this function was found).  There are many other modifications of this HMF and alternative formulations that can be seen in~\citep{STM,HMF_2,HMF_3,HMF_4}.

We are now going to see how this prediction can be probed by the luminosity functoin and reionization.

\vspace{0.3cm}
\textbf{Luminosity function and reionization - \,} 

Observations that probe the low-mass end of the halo mass function can be used to test the FDM and put constraints in the mass. A sensitive probes is the luminosity function of galaxies, which can inherit the suppression coming from the halo mass function. Observations at high redshifts of galaxy counts, reionization history can be used to test the suppression of the HMF caused by the FDM. 

The luminosity function $\phi(L)$ is a map between the galaxy luminosity and the dark matter halo. If we want to obtain cumulative galaxy number density so we can compare with observations, we need to relate the UV magnitude of a galaxy with the mass of the halo. If there are less halos formed at early times, this would lead to less galaxy formation. This can be one characteristic of the FDM model to be tested, but this suppression cannot be too severe otherwise this model could lead to less than expected high-z galaxies than seen in observations, like the ones from the Hubble Ultra Deep Field~\citep{HUDF}. This lack of galaxies could also impact the efficiency of the reionization of the intergalactic medium (IGM), which takes place through star formation. This would impact the optical depth of the CMB, which is constrained by CMB observations~\citep{Planck}.

The relation between the HMF and the the luminosity function is usually done by abundance matching~\citep{Kravtsov:2003sg,Vale:2004yt,Conroy:2005aq}. For that, one needs to assume a mapping between the halo mass and the galaxy luminosity or the UV magnitude of the galaxy. This is done by matching the cumulative UV luminosity function $\Phi (L,z)$ for magnitude smaller than $M_{UV}$ (which is the same as luminosities higher than $L_{UV}$) given by the integral of $\phi(L)$  in this interval, with the cumulative HMF for halos masses bigger than a given $M_h$: $\Phi (<M_{UV},z) = n(>M_h,z)$.  That, together with fixing $\phi(L)$ at low-z with observations, fixes the mass to luminosity (or UV magnitude) of galaxies.

Given the modified HMF presented by the FDM model, we can then predict the luminosity function of this model. In~\citep{marsh_20,HMF_2} we can see some examples of the luminosity function for FDM.  The low-mass suppression in the HMF leads to a luminosity function that ends at smaller magnitudes. Therefore, depending on the mass of the FDM particle, we can predict a cumulative luminosity function that ends at different magnitudes. This can be used to put constraints in the FDM by using, for example, observations of dwarf galaxies or measurements of high-z galaxies like the one from the HUDF. We discuss this in Section 5.

Depending on the mass of the FDM particle we can predict less galaxies at high-z, which would alter the reionization history. This can be seen  given that the UV luminosity function is related to the flux of ionizing photons~\citep{Niemeyer:2019aqm}, 
\begin{equation}
\mathcal{F}_{\mathrm{ion}}= f_{\mathrm{esc}} \int \phi_{UV} (L) \, \gamma(L) \, dL\,,
\end{equation}
where $f_{\mathrm{esc}}$ is the scape fraction, related to the fraction of ionizing photons that escapes the galaxy without being absorbed, and $\gamma(L)$ the conversion rate, which describes the conversion between the UV luminosity of galaxies to the luminosity of ionizing photon (for more details see~\citep{marsh_20} and references therein). With that we can write the Thomson optical depth to CMB:
\begin{equation}
\tau = \int_0^z dz'\, \frac{(1+z')^2}{H(z')} Q(z')\, \sigma_T \bar{n}_H \, \left( 1+\eta_{\mathrm{He}} \, Y/4X  \right)\,,
\label{optical_depth}
\end{equation} 
where $\sigma_T$ is the Thomson cross-section, $\bar{n}$ is the mean comoving hydrogen number density, $X$ is the hydrogen fraction, $Y=X-1$ is the helium fraction, and $\eta_{\mathrm{He}}$ is the ionization state of helium. 

We can then use this modified reionization history, coming from the modified luminosity function, to constrain the mass of the FDM. One can use, for example, the CMB where $\tau(z_{\mathrm{rec}})$ is measured.

%%%
\vspace{0.4cm}
\subparagraph{Sub-halo mass function}\mbox{}\\
%%%

As discussed above, the FDM model is characterized by a suppression in the formation of the small structures. This minimal mass of structures formed is going to impact the substructures in the halo. Therefore, a smaller number of sub-halos is present in the FDM in comparison to CDM. This suppression can be seen in the sub-halo mass function, which is given by $dn_{\mathrm{sub}} (m)/d\, \ln M$, where $n_{\mathrm{sub}}$ is the number of sub-halos and $M$ is the halo mass. This can also be obtained from simulations and semi-analytic calculations \citep{marsh_20,marsh_22,marsh_6,marsh_26}, where a fitting form for the FDM sub-halo mass function is obtained \citep{Xu_thesis,Schutz:2020jox}:
\begin{equation}
\left(\frac{dn_{\mathrm{sub}}}{d\ln M}\right )_{\mathrm{FDM}}  = f_1(M) + f_2 (M) \left(\frac{dn_{\mathrm{sub}}}{d\ln M}\right )_{\mathrm{CDM}} \,,
\label{SHMF_FDM}
\end{equation}
where the functions present in this fitting formula are given by:
\begin{equation}
f_1(M) = \beta \exp \left[ - \frac{1}{\sigma} \left( \ln \frac{M}{M_1 \times 10^8 \, M_{\odot}} \right)^2  \right]\,, \qquad 
f_2 (M) = \left[ 1+\left( \frac{M}{M_2 \times 10^8 \, M_{\odot}} \right)^{-\alpha_1} \right]^{-10/\alpha_1} \,.
\end{equation}
The influence of tidal stripping of cores in the sub-halo mass function is studied in~\citep{SHMF_tidal}, showing that core stripping influences this function for sub-halo masses smaller than $10^7 \, M_{\odot}$. Including this effect the parameters in (\ref{SHMF_FDM}) are: $\alpha_1=0.72$, $\sigma=1.4$, $\beta = 0.014 \, m_{22}^{3/2}$, $M_1 = 4.7 \,m_{22}^{3/2} $, and $M_2 = 2.0 \,m_{22}^{1.6} $. For these parameters, this fitting formula agrees with simulations for masses smaller than $5 \times 10^{-21}\, \mathrm{eV}$.
%%%%%%%%%%%%%%%%%%%
\begin{figure}[htb]
\centering
\includegraphics[scale=0.5]{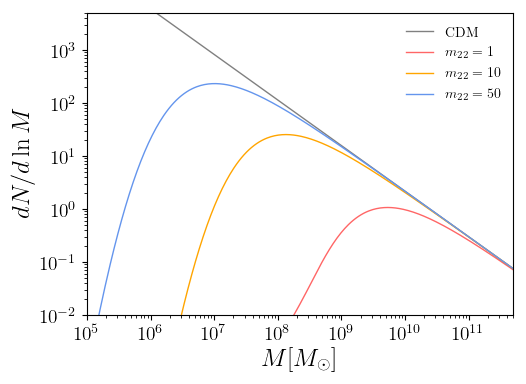}
 \caption{Comparison between the sub-halo mass function from CDM and FDM for different masses of the FDM particle. In the case of the FDM, we consider that all DM is given by the FDM model.}
\label{Fig.:SHMF}
\end{figure}

In Figure~\ref{Fig.:SHMF} we compare the sub-halo mass function for FDM with different masses with the CDM one. Here we used the following CDM sub-halo mass function:
\begin{equation}
\left(\frac{dn_{\mathrm{sub}}}{d\ln M}\right )_{\mathrm{CDM}}  =  a_{\mathrm{CDM}} \, \left( \frac{M}{10^8 \, M_{\odot} } \right)^{-\alpha_0} \,.
\label{SHMF_CDM}
\end{equation}
where $a_{\mathrm{CDM}} = 113.094$ and $\alpha_0 =0.86$, which was obtained as a fit to the sub-halos of the Aquarium simulation~\citep{Springel:2008cc}.
 We can see in the figure the suppression of the small scale structures by the redshift dependent cut in the sub-halos mass function for smaller sub-halo masses, characteristic of the FDM model.

%It is found that the minimum mass of these halos of order $M \sim 3 \times 10^8 \, \left( m / 10^{-22} \, \mathrm{eV} \right)^{-3/2} \, M_{\odot}$, and the abundance largely reduced in comparison to CDM. 
With the sub-halo mass function expected for the FDM model, one can test if the predictions from this model are in accordance with observations. 

%%%%%
\paragraph{The central soliton}\mbox{}\\
%%%%

We have shown in Section 4.1.3 that there is the formation of a soliton (or Bose star) in the central parts of the galaxy, which is the ground state solution of the Schr\"{o}dinger-Poisson system. However, there is no analytical solution for the soliton and we have to obtain this solution numerically. This was done by many authors~\citep{Hui:2016ltb,16,numerical_fuzzy_1,numerical_fuzzy_3,95,soliton_Hui_1,96,soliton_Hui_2,Guzman:2004wj,BarOr}, and here we are going to quote the numerical solutions obtained by~\citep{numerical_fuzzy_1,numerical_fuzzy_3}. 
Here we will restore the $\hbar$ factors for clarity.

An interesting characteristic of the Schr\"{o}dinger-Poisson is that tit has scaling symmetry, which allows to re-scale the quantities of this problem by an arbitrary variable $\beta$ as $\left\{ t, x, \Phi, \phi  \right\} \rightarrow \left\{ \beta^{-2}\,t, \beta^{-1}\, x, \beta^{2} \, \Phi, \beta^{2}\, \phi \right\}$~\citep{Ji:1994xh,Guzman:2006yc} . This is also valid for the SIFDM, although more subtle~\citep{Guzman:2006yc}, and the interaction term scales as $g \rightarrow \beta^{-2}\, g$. The solutions also obeys this scaling transformation, and the physical quantities transform as: the energy density of the soliton transforms as $\rho_s \rightarrow \beta^4 \rho_s$, the radius as $r \rightarrow \beta^{-1} r$, and the mass of the soliton as $M_s \rightarrow \beta \, M_s$. This means that one can  simple re-scale the solution to the equilibrium scale of interested, like the virial scale, if interested in the cores in galaxies, or study the axion star in QCD axion miniclusters~\citep{Niemeyer:2019aqm}.

By assuming spherical symmetry, we need only to determine one of those physical quantities, like the soliton density, and we can derive the other parameters from it. 

The density of the soliton core can be approximated by~\citep{Hui:2016ltb}: $
\rho_c \simeq 4 \times 10^{-3} (G m^2 /\hbar^2)^3 M^4 \simeq (3M/4) R^{-3}_{1/2,c}$
where the half-mass radius is $ R_{1/2,c} \bar{\rho}_{1/2} \simeq 4 \hbar^2 / GMm^2$, the radius where the density drops to one half of its value\footnote{Notice that this is the half-radius of the soliton core, since it is the radius where the core density drops in half. This is different than the half-radius of the halo, for example.}. 
This can be re-written as~\citep{numerical_fuzzy_1,numerical_fuzzy_3}
\begin{equation}
\rho_c \simeq  \frac{ 1.9 \times 10^{-2}}{[1+0.091\, (r/R_{1/2,c})^2]^8} \, \left( \frac{m}{10^{-22} \, \mathrm{eV}} \right)^{-2} \left( \frac{r_c}{\mathrm{kpc}} \right)^{-4}\, M_{\odot} \, \mathrm{pc}^{-3}\,,
\label{rho_c}
\end{equation}
With the density, we can compute the soliton mass
\begin{equation}
M_c = 4\pi \int_0^{\infty} r^2 \, \rho_c(r)\, dr = 2.2 \times 10^8 \,  \left( \frac{m}{10^{-22} \, \mathrm{eV}} \right)^{-2} \left( \frac{R_{1/2,c}}{\mathrm{kpc}} \right)^{-1}\, M_{\odot}\,.
\end{equation}
Here we are approximating these expression by considering that $R_{1/2,c}=r_c$ is the radius of the condensate, and the mass of the soliton core is the mass enclosed in the sphere with this radius. We can see that $95\%$ of the soliton mass is within $r \leq 3r_c$, and this means that (\ref{rho_c}) is accurate to $2\%$ for $0 \leq r \leq r_c$.

We are interested in studying the soliton cores formed inside the galaxies, so we use the scaling to obtain the solution soliton radius in a virial halo of MW-like mass
\begin{equation}
r_c \simeq 0.16 \, \left( \frac{m}{10^{-22} \, \mathrm{eV}} \right)^{-1} \left( \frac{M}{10^{12}\, M_{\odot}} \right)^{-1/3}  \, \mathrm{kpc}\,.
\end{equation}
With that we can see that the FDM soliton core for a FDM with mass $m = 10^{-22} \, \mathrm{eV}$ is going to be smaller than the halo of the MW, with size $r_c \sim 0.16\, \mathrm{kpc}$ in the center region of the halo. In the outskirts of the halo, FDM is expected to behave as free particles, like CDM, so the core is enveloped by shell with in NFW-profile: 
\begin{equation}
\rho_{\mathrm{halo}} \simeq \left\{ \begin{array}{lcl}
         \rho_c & \mbox{for}
         & r \leq r_c \\ \rho_{\mathrm{NFW}}  & \mbox{for} & r > r_c
                \end{array}\right. \,,
\end{equation}
which is the  picture we showed in Figure~\ref{Fig.:scheme_core}. We use this simplified picture to describe what happens in the halo. We can see that the soliton core is inside the Jeans length, as expected, if one compare with (\ref{Eq.:Jeans_length_FDM_NR}). But it is of the order of the de Broglie wavelength if we use the velocity in the soliton $v^2_c = r/(GM_c)$,:
\begin{equation}
\lambda_{dB} (v=v_c) = 3.91\, r_c\,  \left( \frac{r}{r_c} \right)^{1/2}\,.
\end{equation}

In~\citep{numerical_fuzzy_3}  based on the scaling of the Schorinder-Poisson system and using the result of simulations it was also derived the core-halo mass relation for the FDM, showing that $M_c \propto (1+z)^{1/2}\, M^{1/3}$, where $M$ is the halo mass. This shows that MW like halos with $M \sim 10^{12} \, M_{\odot}$ at a redshift around $z\sim 8$ will have soliton cores forming after the halo collapse with mass $M_c \sim 10^{9}\, M_{\odot}$. 
These soliton cores can be seen in FDM simulations, as we will see in Section 4.3.

\vspace{0.5cm}

The presence of these soliton cores in the interior of the galaxies lead to a rich phenomenology. Since no gravitational clustering happens in these regions, the soliton core has a cored profile $\rho_c(r)$, which is different than the one expected from CDM. Therefore, if we can probe the central density of galaxies and obtain a non-cored profile, we can use this to put a bound on the FDM mass. Dwarf galaxies can also be used to do that. Since those galaxies are DM dominated, one can measure their central density and compare with the bound obtained by the expected core to constrain the mass of the FDM. We are going to discuss this in detail in Section 5, and also in a extreme case in Section 4.1.5, where we discuss how the core can solve the cusp-core problem.

%%%%%
\paragraph{Dynamical effects - relaxation}\mbox{}\\
%%%

In  this section we are going to discuss the third class of observational effects of the ULDM, the dynamical effects. These are effects that arise because of the wave-like behaviour that ULDM has inside the soliton cores in galaxies. This can modify the dynamics of objects that are present inside this region where condensation takes place. We are going to see here two of those effects that arise from the relaxation between the FDM and macroscopic objects that move in or through the condensed core. These effects can lead either to heating, which is seen as an energy injection in the orbit of the macroscopic object, or to cooling or dynamical friction, where orbit of the object loses energy to the FDM field. 

These two effects are present in CDM and other DM models, but are modified here by the wave-like behaviour of the FDM. Effects like an enhanced heating can lead to observational effects such as the increase in the velocity dispersion of systems and their expansion. On the other hand, this modified dynamical friction can be used to explain effects like system that were expected to have merged but did not, which represents a challenge for the CDM paradigm.

In the same way that we studied relaxation between the FDM particles leading to the formation of a BEC in Sect.~\ref{Sec.:description_condensate}, we can use the same techniques of kinetic theory in order to study the relaxation of the FDM and the macroscopic objects. We describe bellow briefly this analysis. This was introduced in~\citep{BarOr} and the reader should consult this reference for more details. There are also different ways to model dynamical friction, which we will comment bellow.

\vspace{0.3cm}
We want to describe the relaxation between the FDM and macroscopic objects inside the condensate core.
Therefore, we can think about this as a test particle, or a contaminant, that is present in the condensate\footnote{We are going to see this modelled from a microscopic theory in Section 4.2.4 in the case of the superfluid.}. We want to see how stochastic density fluctuations in the FDM core lead to diffusion of the velocity of the macroscopic object\footnote{In this section I restore the $\hbar$ factors, sine the relation with the classical limit is more subtle. See~\citep{BarOr} for more details.}.

Using kinetic theory one can treat the problem as a diffusion of a test particle in a fluctuating density field. In the case of the FDM, the FDM is represented by a wavefunction which is a collection of plane waves, $\psi(\mathbf{r},t) = \int d\mathbf{k}\, \varphi(\mathbf{k}) exp(i\mathbf{k.r} - i \omega (k) t)$, with $\omega=\hbar^2 k^2/2m$, interacting with a zero-mass particle. In this case we can ignore gravitational interactions between the FDM particles and only focus in the interaction of the FDM and the classical test particle.
The FDM can be treat in this way inside the core where it is condensed. 

Computing the diffusion coefficients for this system of a test particle in the FDM condensate, it was found that it yields the same diffusion coefficients as the ones for a classical two-body relaxation where the test particle interacts with a homogeneous classical density field composed by particles with mass $m_p$, but with a different effective mass for the  FDM "particles" $m_{\mathrm{eff}}$, a different distribution function $F_{\mathrm{eff}}$ and a different Coulomb logarithm $\log \Lambda_{\mathrm{FDM}}$:
\begin{equation}
m_{\mathrm{eff}} = \frac{(2\pi \hbar)^3}{m^3} \frac{\int d\mathbf{v}\, f^2(\mathbf{v})}{\int d\mathbf{v}\, f(\mathbf{v})} \,, \qquad f_{\mathrm{eff}} = f^2(\mathbf{v})\frac{\int d\mathbf{v}\, f(\mathbf{v})}{\int d\mathbf{v}\, f^2(\mathbf{v})}\,,
\end{equation} 
where $f(\mathrm{v})$ is the FDM distribution function and the effective distribution function is normalized like $\rho_{\mathrm{FDM}}=\int d\mathbf{v}\, f_{\mathrm{eff}}(\mathbf{v})$. The Coulomb logarithm is $\log \Lambda_{\mathrm{FDM}} \approx \log (2b_{\mathrm{max}}/\lambda_{\mathrm{dB}}(\sigma))$, where $b_{\mathrm{max}}$ is the maximum scale of encounter and $\sigma$ is the dispersion velocity.

So the picture is that for understanding the effects of a macroscopic object in the  FDM core one can think as the FDM was composed by effective particles with mass $m_{\mathrm{eff}}$, called FDM "quasiparticles" in~\citep{BarOr}\footnote{Careful here that the term quasiparticle was used for the phonon and they do not have the same meaning as here.}. If this quasiparticle has a mass that  is larger than the mass of these macroscopic objects $m_t$, this leads to the object to either lose energy or otherwise to gain energy from the encounters with these quasiparticles. These two processes are the cooling (or dynamical friction) and heating. 

Thus, if $m_{\mathrm{eff}} \gg m_t$, the fluctuations in the FDM field inject energy into the orbit of the macroscopic object. This injection increases the velocity dispersion of the macroscopic object, which in turn increases its size and we have heating. The heating time scale is defined as:
\begin{equation}
\tau_{\mathrm{heat}} = \frac{3 \sigma^3}{16 \sqrt{\pi} G^2 \rho_{\mathrm{b}} m_{\mathrm{eff}}\, \log \Lambda_{\mathrm{FDM}}} =  \frac{3 m^3 \sigma^6}{16 \pi^2 G^2 \rho_{\mathrm{b}}^2 \hbar^3 \, \log \Lambda_{\mathrm{FDM}}}
\end{equation}
For times smaller then  $\tau_{\mathrm{heat}}$,  the system can gain energy from these FDM fluctuations and heating takes place.
For values of the effective mass in halos, we can see that the quasiparticle is more massive than stars. This heating of stars either in a cluster or in the disk of galaxies can lead to observational effects. 

If $m_{\mathrm{eff}} \ll m_t$, then the mass of the macroscopic object is bigger than the FDM quasiparticle, and the macroscopic object loses  energy from its orbit to the FDM field.  This process can be interpreted as dynamical friction. Since the macroscopic object is losing energy this process is also known as cooling, with a cooling time given by
\begin{equation}
\tau_{\mathrm{cool}} =  \frac{3 \sigma^3}{8 \sqrt{2\pi} G^2 \rho_{\mathrm{b}} m_{\mathrm{t}}\, \log \Lambda_{\mathrm{FDM}}}
\end{equation}
This process usually happen for more massive objects like globular clusters merging with the center of the galaxy or black holes.

A similar analysis is done in~\citep{El-Zant:2019ios,El-Zant:2016byp}, where they confirm the results from~\citep{BarOr} and extend the model to include the effect density perturbations coming from  stellar winds, supernova explosions or active galactic nuclei.
The consequence for super massive black holes in a FDM halo was studied in~\citep{El-Zant:2020god} which can lead to its ejection from the core. This effect can be used to put bounds in the FDM mass.

We are going to discuss bellow some specific examples where these dynamical effects take place.

%%%
\vspace{0.3cm}
\subparagraph{Gravitational heating} \mbox{}\\
%%%

We can now see here how the heating affects a stellar population.
We show this for a prototype case of a spherical stellar system of radius $r_{\star}$ in a FDM gravitational potential. The initial distribution function of the stellar system is Maxwellian with a velocity dispersion $\sigma_t$. Since star have $m_t = m_{\mathrm{star}} \ll m_{\mathrm{eff}}$, only heating  takes place.

The heating is important when $\tau_{\mathrm{heat},\star}$ is of order of $1/3$ of the age of the galaxy $\tau_{\mathrm{age}}$, where 
\begin{equation}
\tau_{\mathrm{heat},\star} \approx \frac{2.08 }{\log \Lambda_{\mathrm{FDM}}} \, \left( \frac{r_{\star}}{1\, \mathrm{kpc}} \right)^4  \left( \frac{m}{10^{-22}\, \mathrm{eV}} \right)^3  \left( \frac{m}{200\, \mathrm{km/s}} \right)^2 \, \mathrm{Gyr}\,. 
\end{equation}
In this case we have a increase in the velocity dispersion and the system is going to expand\footnote{One thing to be attentive here is that heating or diffusion is the term used for the injection of energy in the orbit of the macroscopic system by the FDM. This might cause an increase in the velocity dispersion, which leads to an expansion of the system, or not depending on the radial profile of the galaxy (see~\citep{BarOr} for more details).}. This happens in scales
when  $r_{\star} < r_{\mathrm{heat},\star}$ with
\begin{equation}
r_{\mathrm{heat},\star} = 1.13\, \log \Lambda_{\mathrm{FDM}} \left( \frac{\tau_{\mathrm{age}}}{10\, \mathrm{Gyr}} \right)^{1/4} \left( \frac{m}{10^{-22}\, \mathrm{eV}} \right)^{-3/4}  \left( \frac{m}{200\, \mathrm{km/s}} \right)^{-1/2} \, \mathrm{kpc}\,.
\end{equation}

%FDM fluctuations adds energy to the stellar population, thereby causing it to expand, but the velocity dispersion of the stars may either grow or decay as a result of this expansion depending on the radial profile of the gravitational potential of the galaxy.

For the comparison with observation that we are going to show in Section 5, we are interested in the cases where this effect leads to a increase in the dispersion relation of the star, expanding the stellar system. This is going to be studied in the disks of galaxies and also in a star cluster, like Eridanus II.

%Jens: Relaxation

\vspace{0.3cm}
\subparagraph{Dynamical friction} \mbox{}\\
%%%

The change in dynamical friction is one of the most interesting consequence in BECs and superfuids. This emergent phenomenology can lead to consequent observations that might reveal the characteristics of those systems. It is interesting to see how dynamical friction behaves in the presence of a BEC core, as in the case of the FDM. The dynamical friction in a BEC is not expected to change as dramatically as in a superfluid, which has no friction, but some change is expected nevertheless. 

It is expected that the FDM changes this prediction because of three phenomena: (i) change in the rate of orbital decay because of the presence of the condensed core; (ii) since the FDM produces a homogeneous core, a mechanism similar to the "core stalling"  observed in N-body simulations can take place and  reduce or eliminate drag from dynamical friction; and (iii) the way  dynamical friction is calculated must be modified by the presence of an object with large de Broglie wavelength, a quantum mechanical extension to the calculation of dynamical friction must be done. 

An interesting puzzle that can potentially be explained by a modified dynamical friction is the puzzle of the Fornax globular clusters. From the standard dynamical friction expected for CDM and baryons it is expected that globular clusters orbiting Fornax should have rapidly fallen towards its center to form a stellar nucleus. However, there is no signal of mergers and we detect $5$ globular clusters orbiting Fornax. 

In \cite{Hui:2016ltb} only the last effect is described and simulated for different parameters the orbital decay times for Fornax in CDM and in the FDM cases. They found that in FDM the orbital decay time is longer, and four of the five decay times simulated are bigger than $10\, \mathrm{Gyr}$ or more, thus explaining the puzzle for why the globular cluster in Fornax survived. More simulations and observations are needed in order to confirm this claim,  but the FDM model seems to address the dynamical friction puzzle. The ideal is to have the microscopic theory describing dissipation in the FDM model.

These qualitative results are confirmed in~\citep{BarOr} using the classical two-body relaxation, showed above. 
See also results for how dynamical friction alters inspiral systems~\citep{BarOr}.

Finally in \cite{Lancaster:2019mde}, a detailed analytical and numerical study of dynamical friction in the FDM model was performed. To describe the dynamical friction in this model, they describe the dissipation that a perturber moving in a condensate causes. They work this dissipation theory for point-sources (satellites), extended satellites and point-like satellite in a FDM background with finite velocity dispersion. This analytical theory is then verified by their numerical-simulation that solves the Schr{\"o}dinger--Poisson system in the presence of such perturber satellite, showing good agreement with the analytic methods. This framework is applied to the cases of the Fornax globular cluster, but also to the Sagitarius (Sgr) stream and the Large Magellanic Clouds (LMC) (we will see about those systems in more detail in Sect.~\ref{Sec.:Observations}). For the Fornax, they find that if the mass of the FDM model is $m \gtrsim 10^{-21}\,\mathrm{eV}$ this model stops explaining the Fornax globular cluster merging times, which is in agreement with the mass bounds necessary for the FDM to solve the small scale problems. For Sgr and LMC it is found that the dynamical friction on those are described by the classical limit, described by the Chandrasekhar formula. More simulations need to be done to confirm this, and it is very important to understand the dynamical friction in this regime since these bodies have a strong influence in the MW, and are the target of many studies and observations. These  results already show that this modified  dynamical friction in the FDM model  can maybe explain some interesting astrophysical observations which might be a good opportunity to measure the ULDM on galactic scales.
%We review the classical treatment of dynamical friction before presenting analytic results in the case of FDM for point masses, extended mass distributions, and FDM backgrounds with finite velocity dispersion. We then test these results against a large suite of fully non-linear simulations that allow us to assess the regime of applicability of the analytic results

%%%
\paragraph{Clusters: most massive halos}\mbox{}\\
%%%

The FDM can also affect clusters. For distances larger than the de Broglie wavelength of the FDM, it is expected that DM behaves as standard CDM and that the halo enveloping the soliton has a NFW profile.  This can be seen numerically for the mass range $10^9 \,M_\odot \lesssim M_{\mathrm{vir}} \lesssim 10^{11} \, M_\odot$ \citep{numerical_fuzzy_1,numerical_fuzzy_2,numerical_fuzzy_3,numerical_fuzzy_4}, which gives an estimate for the mass of the central soliton. Extending this relation to larger halos, the FDM predicts that in the inner regions of clusters there will be a condensed core, a soliton, with mass:
\begin{equation}
M \simeq 1.3 \times 10^{10} M_\odot   \left( \frac{10^{-22}}{m} \right)     \left( \frac{M_{\mathrm{vir}}}{10^{15} M_\odot} \right)^{1/3} \,.
\end{equation}
which is still below the maximum mass  for the soliton calculated above for a galaxy. The corresponding half-mass radius is:
\begin{equation}
R_{1/2} \simeq 25 \, \mathrm{pc}  \left( \frac{10^{-22}}{m} \right)     \left( \frac{M_{\mathrm{vir}}}{10^{15} M_\odot} \right)^{-1/3}\,.
\end{equation}
So, the presence of this soliton with this mass and size would be a prediction of the FDM model. But a question that remains to be answered is the following:  is the presence of such solitons in the interior of clusters halos compatible with merging cluster like the Bullet cluster or the anit-Bullet cluster?

In \citep{Hui:2016ltb} they ask the question if  solitons in the center of the galaxies have not been misinterpreted as super massive black holes. They compare the mass of the central dark region measured from Virgo and show that this is similar to the mass of the soliton core in a galaxy like Virgo for a mass of the FDM particle of $ m \sim 10^{-22} \, \mathrm{eV}$. However, since the observations from the Event Horizon Telescope of the black hole in the center of M87 were released, this hypothesis seem to be almost excluded, and it is indeed a super massive black hole that inhabits the center of this galaxy. We have to wait for more data to confirm this, but this is an exciting measurement that can also be used to test the FDM hypothesis.

Another interesting fact is that we know that the galaxies host a super massive black hole in their center. For this reason in \citep{Hui:2016ltb}  they investigate the possibility of a super massive black hole to be created in the center of a soliton. Apparently, the black holes do not grow for the fiducial mass $m_{22}$, in a condensate core. Their creation only starts being significant for $m \gtrsim 5 \times 10^{-22} \, \mathrm{eV}$, which is in tension with other bounds in the mass, like the one to solve the missing satellites problems.

\subsubsection{Addressing the small scale challenges}

Perhaps even more interesting than solving the small scale problems of the CDM paradigm, it is the new and rich phenomenology that the ULDM present. But the ULDM models can also address these small scale challenges and reconcile the small scales observations with the CDM paradigm. Usually the scales where the non-CDM and interesting phenomenology happens coincides with the scales where these models present modifications in the small scales that are necessary to explain the small scale observations.

We are going to discuss now how this new phenomenology  on small scales presented by the FDM can address the small scale challenges and what are the conditions in the parameter of this class of models for that to happen. 
We are only going too discuss the FDM case now, and the conditions in the mass, but one can think that a similar analysis can be done for the case of SIFDM. 

\paragraph{FDM} \mbox{}\\

We want to determine the mass of the FDM candidate.  From the discussion above, we saw the there is a bound for the mass in order to condensate in galaxies $ 10^{-25} \lesssim m \lesssim  \mathrm{eV}$ for a typical MW-like galaxy. Later we saw that, for masses of order of the usual QCD axion mass, around $m \sim 10^{-5} \mathrm{eV}$, the stable configurations are very localized and small, far from galactic scales. With that we can already see that $m \ll 10^{-5} \mathrm{eV}$ for the FDM model. Now, we are going to see other conditions that can bound the mass and show the mass range for the FDM particle that can address the small scale challenges.

\begin{itemize}
\item \textbf{Halos: minimum size, maximum density and the cusp-core problems:}

The general idea why FDM (and all ULDM models) can solve the cusp-core problem is because these systems naturally predict a core in the center of galaxies. In this core there is no structure formation (Jeans instability), and for that reason they might prevent the formation of a cusp in the center of the galaxy. We are going to investigate here how and for which mass the FDM core size and density can address the cusp-core problem.

Since we want the non-CDM behaviour to happen inside galaxies, the de Broglie wavelength of the FDM inside galaxies cannot exceed the size of the galaxy, given by the virial radius: $\lambda_{\mathrm{dB}} < R = GM/v^2$. Therefore $R \gtrsim \frac{1}{G M m^2}$, where $M$ is the mass of the galaxy. We can write that in terms of the half-radius, radius where half of the mass of spherically symmetric system is contained~\citep{Hui:2016ltb} 
\begin{equation}
R_{1/2} \gtrsim 0.335 \left( \frac{M}{10^9 \, M_{\odot}} \right)^{-1}  \left( \frac{m}{10^{-22} \, \mathrm{eV}} \right)^{-2} \, \mathrm{kpc}\,.
\label{Eq.:half_radius}
\end{equation}
This bound on the radius is compatible with the half-light radii inferred from the densities of 36 Local group dwarf spheroidals \citep{McConnachie:2012vd} if the mass of the FDM particle is  $m \sim 10^{-22} \mathrm{eV}$.

With the above condition, we can also compute the upper bound in the central density:
\begin{equation}
\rho_{\mathrm{c}} \leq 7.05 \,   \left( \frac{M}{10^9 \, M_{\odot}} \right)^{4} \left( \frac{m}{10^{-22} \, \mathrm{eV}} \right)^{6} \, M_\odot \,  \mathrm{pc}^{-3}\,.
\label{Eq.:central density}
\end{equation}
If we compare this bound to the observations from $8$ dwarf spheroidals, and we can see for the density to be comparable to the one measured for these dSph,  the FDM mass needs to be $m= 8^{+5}_{-3} \times 10^{-23} \, \mathrm{eV}$ for Draco and $m= 6^{+7}_{-2} \times 10^{-22} \, \mathrm{eV}$ for Sextans~\citep{Chen}. For those masses, the distribution at the center of the galaxies seem to be cored, alleviating the cusp-core problem.

Therefore, it is necessary a mass of order $m \sim 10^{-22} \, \mathrm{eV}$ in order for FDM to solve the cusp core problem. And we showed that this is compatible with the measurements from dwarf galaxies. However, we are going to discuss in Section 5.2 that some studies dispute this conclusion.
 
\item \textbf{Lower bound on the FDM halo masses, and the missing satellites and too big to fail problem:}

As we saw before, for the self-gravitating FDM systems, since gravity is attractive, we have coherence on a finite scale. The size of this core depends on the mass, being larger as the mass gets smaller. So the smallest radius to be produced in the FDM model are determined by the smallest mass allowed for the particle. 

Having a limit for the smallest cores that can be created has important consequences in the abundance of low mass halos, and it is going to be different in this model than what is given by $\Lambda$CDM. We can see that by calculating the smallest structures formed in the FDM model. This is given when $\lambda = \lambda_{\mathrm{J}}$, where $\lambda_{\mathrm{J}}$ represents the last scales that can suffer gravitational instability. With the Jeans length, we can calculate the Jeans mass:
\begin{equation}
M_{\mathrm{J}} = \frac{4\pi}{3} \rho \left(  \frac{1}{2} \lambda_{\mathrm{J}} \right)^3 = 1.5 \times 10^7 M_\odot \, (1+z)^{3/4} \,  \left(  \frac{\Omega_{\mathrm{FDM}}}{0.27} \right)  \left( \frac{H}{70 \mathrm{km/s/Mpc}}\right)^{1/2}    \left( \frac{10^{-22} \, \mathrm{eV}}{m} \right)^{3/2}\,.
\label{Eq.:minimum_mass}
\end{equation}
This is the minimum mass of substructure created in the FDM model. This is in contrast with CDM, where halos with mass below $\sim 10^{8} M_\odot$ are  highly created, with abundance $dn(M_{\mathrm{h}}) \propto M_{\mathrm{h}}^{-2} dM_{\mathrm{h}}$.  In that sense, the missing satellites problem is addressed in the FDM model, since halos of smaller masses than $M_{\mathrm{J}}$ are not created, if  $m = 10^{-22} \, \mathrm{eV}$. If the mass of the FDM is smaller than that, sub-halos with smaller mass will form, and the FDM model is not going to address the missing satellite problem anymore. Therefore this shows that in order to solve this problem, the FDM has a preferred mass of around $10^{-22}\, \mathrm{eV}$.
We are going to see below that tidal disruption can also act suppressing small mass halos, aiding FDM in solving the missing satellites problem. The too big to fail problem is also addressed by the FDM, since we have a mechanism to explain the fact that low-mass sub-halos are not formed, making it not necessary to invoke mechanisms that creates the too-big-to-fail problem as a by product.
\end{itemize}

\vspace{0.2cm}
Summarizing, in order to address the small scale challenges, which means presenting a cored density profile inside galaxies and suppressing the formation of small mass sub-halos which would explain the smaller amount of satellites, the mass of the FDM needs to be of the order of $\sim 10^{-22} \, \mathrm{eV}$. A component with that mass and presenting the dynamics of the FDM class of models would behave like CDM on large scales and present the observed structures on small scales. 

However, we are going to see in Section 5 that, although for many years this mass range was available as a possibility for the mass of the FDM, the latest observations have been challenging this mass, and showing that the FDM has to probably have a heavier mass than the one necessary to address the small scale problems.

%%%%%%%%%%%%%%%%%%%%%%%%%%%%%%%%%%%%%%%%%%

%%%%%%%%%%%%%%%%%%%%%%%%%%%%%%%%%%%%%
\subsection{DM Superfluid}
\label{Sec.:DM_superfluid}
%%%%%%%%%%%%%%%%%%%%%%%%%%%%%%%%%%%%%

In this section we are going to describe the third category of models of ULDM,  the DM superfluid. In previous sections we saw the small scale problems of $\Lambda$CDM and how MOND empirical law offered a very good fit to the rotation curves of galaxies and the scaling relations that emerge from the dynamics of galaxies, which might be challenging in the context of $\Lambda$CDM. However, as we saw there is no present framework that can explain MOND, given that the initial proposed theory, the full MOND, and its extensions present serious problems. We present here an alternative model to  DM that has the goal of reconciling CDM and MOND: the DM superfluid. This model intends to not only solve the problems that the previous models attempt to address but also offer a mechanism to describe MOND on small scales.  In this framework, DM behaves as standard CDM on large scales, while the MOND dynamics emerges on galactic scales. And this is possible through the physics of superfluidity.

The idea of the DM superfluid model, is that DM forms a superfluid on galactic scales, where superfluidity arises upon condensation. This superfluid core present in the inner regions of galaxies not only addresses the small scale challenges of $\Lambda$CDM in a similar way as the previous models, but the superfluid described in these regions behaves following a different dynamics which reproduces the MOND behaviour. This is possible given that this superfluid is described by a Lagrangian similar to thee one from the MOND theory which is allowed given the EFT description of the superfluid. In this way, the long-range correlation present in MOND is going to be given by the behaviour of the phonons which mediate long range forces. Outside the superfluid core, DM behaves as normal matter as in the previous models. This is the general vision how the DM superfluid model attempts to describe the behaviour of DM in galaxies.

In the following subsection we will construct the DM superfluid theory showing first in which conditions DM condensates on galactic scales, following that we will present the theory that describes this superfluid phase. With that in hand we can calculate the halo profile and rotation curves in order to compare with data and check the fit of the theory.  We present how this model explain many astrophysical systems and possible predictions. After that we show the limits of validity of this description and its relativistic completion. We briefly describe how the cosmology works in this model.

%%%%%%%%%
%%%%%%%%%
%%%%%%%%%
This model is constructed by using the fact that in a generic superfluid we can reproduce the exact action as in MOND. This is a very specific model and serves as a toy model for the understanding of theories of DM that present an emergent dynamics on small scales. It is important to point out that what is important is for the model to be able to explain the rotation curves and scaling relations described by galaxies. This translates into a theory that exhibits long-range correlation on Galactic scales. We present here a very specific example of a model when this occurs, where the way the long-range correlation is obtained is by restricting the Lagrangian to behave like MOND on small scales, but this does not have to be the case.
The search for a more general theory where this emerges, with a known microphysics is the final goal. Such a construction is currently being searched.

\subsubsection{Conditions for DM condensation}

Before describing how the DM superfluid behaves inside galaxies, we need to determine in which conditions DM condensates into a Bose--Einstein condensate in galaxies. As we saw in the previous section, two conditions need to be met for condensation: first, we need that all the particles are in a single coherent quantum state, described by a single wavefunction of the condensate; a second condition is that the DM particles are in thermal equilibrium, in order to be described by a Bose distribution.

In this section we want to obtain a rough estimate of the bounds in the  parameters of the model in order to obtain this condensed core in the inner parts of the galaxy. For that, for simplicity, we use the criteria for weakly interacting gases.

\paragraph*{Condensate wavefunction and thermalization}  \label{Sec.:conditions} \mbox{} \\

We showed in the beginning of this section the condition on the mass for the ULDM particles to condense in galaxies, showing that they should be in the range $10^{-25}\, \mathrm{eV} \lesssim m \lesssim 2 \, \mathrm{eV}$. This is an approximate condition for the case of DM in galaxies, but it gives us an order of magnitude estimation for the mass of the DM superfluid particles. However, if we remember from Sect.~\ref{Sec.:BEC_superfluids}, there is a second condition  for condensation.

The second condition to form a condensate is that the particles are in thermal equilibrium.  The condition to achieve thermal equilibrium is that the time scale of thermalization must be smaller than the time scale where dynamical processes happen in the halo, the dynamical time.  If this condition is satisfied and thermal equilibrium is achieved, the condensate is coherent in the entire halo. 
The time scale of thermalization if given by the inverse of the self-interaction rate, and the condition for thermalization is given by:
\begin{equation}
\Gamma \sim \Upsilon v_{\mathrm{vir}} \rho_{\mathrm{vir}} \frac{\sigma}{m} \, \lesssim \, t^{-1}_{\mathrm{dyn}} = \left( 3\pi / 32 G \rho \right)^{-1/2} \,,
\end{equation}
where $\Upsilon \sim \frac{\rho_{\mathrm{vir}}}{m} \frac{(2\pi)^3}{(4\pi /3)(mv)^3}$ is the Bose enhancement factor, which tells you if a boson is already in the state, the probability to another boson to be in that state will be enhanced by a factor of $\Upsilon$. The dynamical time is taken here as the time it takes to a sphere of density $\rho$ to collapse due to gravity. This condition gives a bound in the self-interaction cross section:
\begin{equation}
\frac{\sigma}{m	} \gtrsim 0.3 \left( \frac{m}{\mathrm{eV}} \right)^4 \frac{\mathrm{cm}^2}{g}\,,
\end{equation}
where we assumed $z_{\mathrm{vir}}=2$ and $M = 10^{12} M_\odot$.  If we want that our self-interaction satisfies the merging-cluster bound \citep{SIDM1,SIDM2,SIDM3}, which is $\sim 1$cm$^2$/g, this gives another bound in mass of the superfluid: $m \lesssim  \mathrm{eV}$.

\begin{figure}
\centering
\includegraphics[scale=0.5]{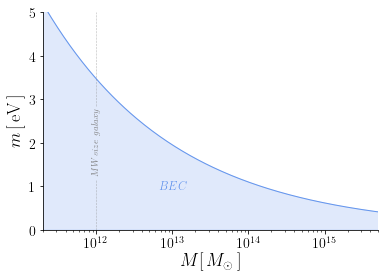}
\includegraphics[scale=0.5]{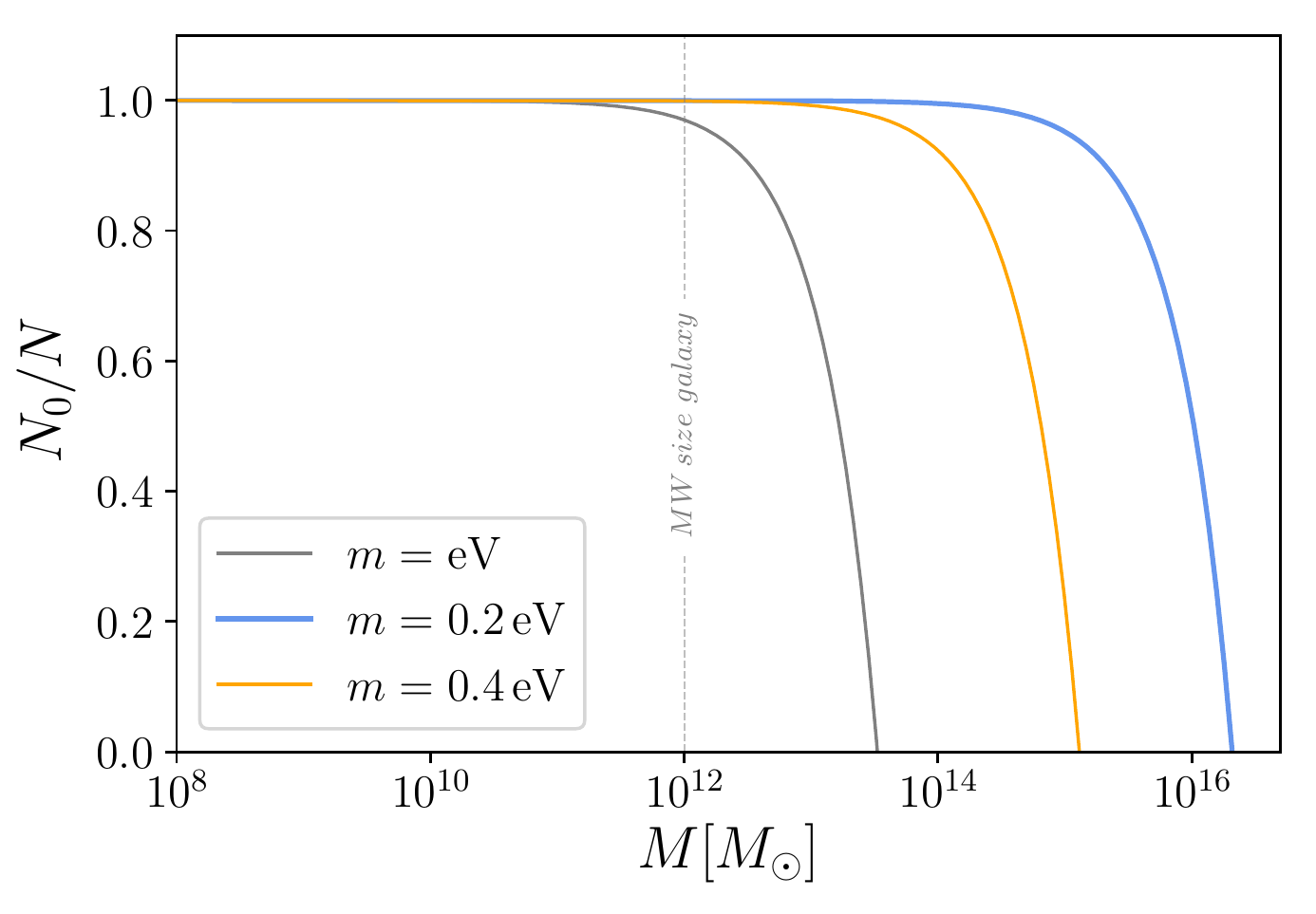}
 \caption{\textit{Left panel:} Mass of the superfluid particle with respect to the mass of the halo. The filled are represents the regime where DM is condensed and behaves as a superfluid, for $z_{\mathrm{vir}}=2$. \textit{Right panel:} Approximate calculation form \citep{Berezhiani:2015bqa} for the fraction of the particles in the condensate versus in the normal state, for a series of sub eV masses in accordance with our bounds. We assumed $z_{\mathrm{vir}}=0 $.}
\label{Fig.:fraction_cond}
\end{figure}

From these conditions, we can obtain a few properties of the our DM superfluid condensate:
\begin{itemize}
\item \textbf{Critical temperature:}
With DM  in thermal equilibrium, the temperature can be obtained by the equipartition theorem: $k_{\mathrm{B}} T = m \langle v \rangle^2 /2$, which is valid for temperatures smaller than the critical temperature. Above that temperature, the condensate is broken. The critical temperature $T_{\mathrm{c}}$ is associated with the ``critical'' velocity $v_{\mathrm{c}}$, than can be read when we saturate the bound  (\ref{cond1}):
\begin{equation}
T_{\mathrm{c}} \sim 6.5 \left( \frac{ \mathrm{eV}}{m}\right)^{5/3} (1+z_{\mathrm{vir}})^2 \, \mathrm{mK}\,.
\label{T_crit}
\end{equation}
With that, the temperature in a halo of mass $M$ is given by:
\begin{equation}
\frac{T}{T_{\mathrm{c}}} \sim 0.1 (1+z_{\mathrm{vir}})^{-1} \left( \frac{m}{\mathrm{eV}} \right)^{8/3}  \left( \frac{M}{10^{12} M_\odot}\right)^{2/3}
\end{equation}
\item \textbf{Condensate fraction:} 
As we saw on Sect.~\ref{Sec.:BEC_superfluids} at $T=0$ it is expected that almost all the particles are in the condensate. However, at finite  but subcritical temperature, as seen in Landau's theory \citep{Landau}, it is expected that the fluid is going to be a mixture of superfluid and normal fluid, with the majority in the superfluid. Borrowing from the non-interacting BEC description, this can be estimated as:
\begin{equation}
\frac{N_{\mathrm{cond}}}{N}= 1-\left( \frac{T}{T_{\mathrm{c}}} \right)^{3/2} \sim 1- 0.03 (1+z_{\mathrm{vir}})^{-3/2} \left( \frac{m}{\mathrm{eV}} \right)^4  \left( \frac{M}{10^{12} M_\odot}\right)\,.
\end{equation}
This formula is only valid for free-particles, and a particle with interaction and trapped in the gravitational potential has a different power than $3/2$, but it serves as an estimate. We can see in the Fig.~\ref{Fig.:fraction_cond} the condensate fraction with respect to the mass of the halo for different masses of the DM particle. From this we an see that for a MW-like galaxy for any mass smaller than $\mathrm{eV}$, the particles are in the condensate, while for higher masses of clusters, for example, this is not true. Therefore, the mass range from the bound (\ref{Eq.:Up_low_bound_m}) describes a condensate that condenses on galactic scales.
\end{itemize}

\vspace{0.3cm}

The above conditions were obtained assuming that the condensate will take the entire halo. However, as mentioned in Sect.~\ref{Sec.:halos_theory}, virialization occurs through violent relaxation, which is an out-of-equilibrium process. In this way, the DM superfluid cannot thermalize. What should happen is that first, the halo virialized and the profile is the expected NFW. After this process, DM particles start to enter thermal equilibrium in the inner, most central regions of the condensate, where the interaction is more pronounced. In this way, the halo would have an inner region ($r<R_T$) where DM is in a condensed state surrounded by the outer part of the halo ($r>R_T$) that follows the NFW profile \citep{Berezhiani:2017tth}.  Since in this model the goal is to be able to describe the rotation curves of galaxies, $R_T$ needs to be larger than the radius where the circular motion of stars and gas is observed. For $r > R_{\mathrm{T}}$, the density profile of the halo follows the NFW profile, $\rho \propto r^{-3}$. So we can rewrite the density and velocity with respect to the virial quantities used above: $\rho(r)=\rho(R_{200}) \left( R_{200}/r \right)^3$, where for a NFW we can estimate $\rho_{200}/\rho(R_{200}) \sim 5$. With that, the thermalization bound becomes:
\begin{equation}
\frac{\sigma}{m	} \gtrsim 0.2 \left( \frac{m}{\mathrm{eV}} \right)^4  \left( \frac{M}{10^{12} M_\odot}\right)^{2/3} \left( \frac{r}{R_{200}} \right)^{7/2} \frac{\mathrm{cm}^2}{g}\,,
\end{equation}
which tells us that it is easier to reach thermal equilibrium in the center of the galaxies where the density is higher. This translates into a bound to the thermalization radius:
\begin{equation}
R_T \lesssim 310 \left( \frac{m}{\mathrm{eV}} \right)^{-8/7}  \left( \frac{M}{10^{12} M_\odot}\right)^{1/7} \left( \frac{\sigma/m}{\mathrm{cm}^2/g} \right)^{2/7} \mathrm{kpc}\,.
\end{equation}
For a MW-like galaxy with $M=10^{12} M_\odot$ if we can measure the the circular velocity up to approximately $60\,\mathrm{kpc}$, this will translate into a bound for the mass: $m \lesssim 4.2 \left( \frac{\sigma/m}{\mathrm{cm}^2/g} \right)^{1/4}\, \mathrm{eV}$.

\subsubsection{Superfluid dynamics}

Since we have determined that DM condenses and forms a supefluidity in the central regions of the halo, we now need to describe the evolution of this superfluid inside this region. We need to determine the dynamics of the superfluid in order to be able to calculate the profile of the region of the halo comprising the superfluid and with the calculate the rotation curves of galaxies. In this section we will describe the effective field theory of superfluids and show how this is theory reproduces MOND at small scales.

As we saw in the previous section, a superfluid at low-energies is described by the effective Lagrangian that is invariant under shift and Galilean symmetries:
\begin{equation}
\mathcal{L}_{T=0}=P(X)\,, \qquad X=\dot{\theta} +\mu -m \Phi - \frac{(\nabla \theta )^2}{2m}\,,
\label{P(X)}
\end{equation}
where $\Phi = -G M(r)/r$  is the external gravitational potential for a spherical symmetric static source. The thermodynamic pressure is given by $P$.

We want our theory to describe the MOND dynamics at the regions where it is superfluid. Given this general Lagrangian for the phonons (\ref{P(X)}), we want it to describe the MOND Lagrangian (\ref{L_MOND}). For this, we conjecture that our phonon action is given by:
\begin{equation}
\mathcal{L}_{\mathrm{DM},T=0}=\frac{2 \Lambda (2m)^{3/2}}{3} X \sqrt{|X|}\,.
\label{L_SF}
\end{equation}
This fractional power might seem strange from the point of view of a quantum field theory of fundamental fields, leading to superluminal behaviour and caustics. However, as a theory for the phonons this is not problematic and it determines uniquely the equation of state of the superfluid. As we can see for the condensate, the background, where $\theta=\mu t$,  the pressure is given by the Lagrangian density:
\begin{equation}
P(\mu) = \frac{2\Lambda}{3} (2m\mu)^{3/2}\,, \qquad \Rightarrow \qquad P=\frac{\rho^3}{12 \Lambda^2 m^6}\,,
\label{EoS}
\end{equation}
where, in the non-relativistic regime, $\rho=mn$ and  $n=\partial P / \partial \mu$ is the number density of condensed particles. As expected from the result from MOND, this Lagrangian gives us an EoS for the superfluid $P \propto \rho^3$, which is what we wanted to reproduce MOND.

To evaluate the excitation spectrum, we write the action for the phonon excitations $\phi$, that can be obtained by expanding (\ref{P(X)}) to quadratic order. Neglecting the gravitational potential:
\begin{equation}
\mathcal{L}^{(2)}= \frac{(2m)^{3/2}}{4\mu^{1/2}} \left( \dot{\phi}^2 - \frac{2\mu}{m} (\nabla \phi)^2 \right)\,,
\label{quad_phonon}
\end{equation}
from where we can infer the sound speed of the phonon excitations:
\begin{equation}
c_{\mathrm{s}} = \sqrt{\frac{2 \mu}{m}}\,.
\end{equation}

However, only those ingredients are not enough to reproduce a MOND-like force and a coupling between the phonons and the baryon density needs to be introduced:
\begin{equation}
\mathcal{L}_{\mathrm{int}} = -\alpha \frac{\Lambda}{M_{\mathrm{pl}}} \theta \rho_{\mathrm{b}}\,,
\end{equation}
where $\alpha$ is a dimensionless coupling constant. Although necessary in order to obtain the MOND regime, this interaction Lagrangian breaks shift symmetry softly, only at the $1/M_{\mathrm{pl}}$ level. This term is here considered as a phenomenological term in order to reproduce MOND.
In this way, this superfluid theory has 3 parameters: the mass $m$, the scale $\Lambda$ and the coupling $\alpha$.  

The present form of the Lagrangian to obtain MOND is not the only way of obtaining the MOND behavior in the context of the DM superfluid model. In \citep{Khoury:2016ehj} it was used higher order corrections to generate the non-relativistic MOND action, which is inspired in the symmetron mechanism. Using the same Lagrangian (\ref{L_SF}) as the leading order Lagrangian, higher order corrections involving gradients of the gravitational potential are added to effectively modify the gravitational force.  This results in the spontaneous breaking of a discrete symmetry. The symmetry is broken for small accelerations leading to MONDian gravity, and is restored in the limit of large acceleration leading to Newtonian gravity. In this theory the shift symmetry of the entire system is maintained. A difference from the present mechanism, as we are going to see later, is that cosmologically all the DM is in the normal phase, behaving like CDM, and reproducing all the results from $\Lambda$CDM. Here we will describe the method of adding a photon-baryon coupling since this was studied in more detail in the literature.

\paragraph{Finite temperature} \mbox{}\\

The theory developed above is valid for a $T=0$ superfluid. However, in reality, the DM in galaxies has a non-zero temperature. As we mentioned in Sect.~\ref{Sec.:BEC_superfluids}, for finite temperatures, this Lagrangian needs to receive finite temperature corrections. In Landau's model the finite temperature superfluid consists is described as a two-fluid theory where a  superfluid component and a normal component are present. Those components must interact with each other. Following this idea from Landau's theory, at lowest order in derivatives, we can write the general form of the EFT at finite temperatures and finite chemical potential as a function of three scalars \citep{Nicolis}:
\begin{equation}
\mathcal{L}_{T\neq 0} = F\left( X, B, Y \right)\,,
\end{equation}
where $X=X(\theta)$ was defined before with respect to the superfluid variables. The other new components are: $B$ is defined with respect to the normal fluid three Lagrangian coordinates $\varphi^I (\vec{x},t)$; and $Y$ represents the scalar product of the  normal and superfluid velocities:
\begin{equation}
B \equiv \sqrt{\det \partial_\mu \varphi^I \partial^\mu \varphi^I} \,, \qquad Y \equiv u^\mu \left( \partial_\mu \theta +m \delta_\mu^0 \right) \simeq \mu-m\Phi+\dot{\phi}+ \vec{v} \cdot \nabla \phi\,,
\end{equation}
where $u^\mu$ is the unit 4-vector from $\varphi^I (\vec{x},t)$, and in the last equality of $Y$ we have taken the non-relativistic limit, so $\vec{v}$ is the velocity vector of the normal fluid component.

% Say somewhere that when we try with the T=0 , the perturbations around this zero-temp static background are unstable, ghost-like.

There are many ways to construct the finite temperature operators. Our restriction is that we want our finite-temperature theory to generate the expected MOND profile. To construct such a Lagrangian requires first-principle knowledge of the microphysics of the superlfuid. Since we still do not have a fundamental description of the DM superfluid model, we proceed empirically. We suggest the following finite-temperature Lagrangian for the model:
\begin{equation}
\mathcal{L} = \frac{2 \Lambda (2m)^{3/2}}{3} \, X \sqrt{|X-\beta Y|} - \alpha \frac{\Lambda}{M_{\mathrm{pl}}} \phi \, \rho_{\mathrm{b}}\,,
\label{L_temp}
\end{equation}
where the finite temperature effects are parametrized by a dimensionless constant $\beta$. When $\beta \rightarrow 0$, we recover the $T=0$ result; we are using the fiducial value $\beta=2$. We included the interaction term so we could represent the entire action of the model that we are going to use next.

\subsubsection{Halo profile}

With the Lagrangian of the theory, we can evaluate the halo profile in the superfluid region and, after matching with an outer NFW profile, calculate the rotation curves of galaxies. And this is what we are going to do in this section: estimate the halo profile. This will be done in steps. First, we estimate the DM halo profile taking into account only the density coming from (\ref{L_SF}). Next, we include the baryons, by calculating the profile for the full action including interaction. We are going to use here the finite-temperature effective action (\ref{L_temp}), since in the case of the $T=0$ the perturbations around a static background configuration suffer from a ghost-like instability. Although phenomenological, it retains the features of the initial superfluid Lagrangian and can give a more realistic description of the system.

\paragraph{DM halo profile} \mbox{}\\

We can now calculate the density profile of the \emph{condenstate}, in the superfluid region, assuming that we have only dark matter and no baryons for simplicity. This is the halo profile given by the different equation of state that the superfluid has: $P \propto n^3$, given by equation (\ref{EoS}). This analysis is almost the same for the zero-temperature and finite temperature cases, with accounts for the replacement:$\Lambda \rightarrow \tilde{\Lambda}=\Lambda \sqrt{\beta - 1}$. Assuming hydrostatic equilibrium, for a static and spherically symmetric halo, the pressure and acceleration are related by:
\begin{equation}
\frac{1}{\rho(r)} \frac{dP(r)}{dr}=-\frac{d\Phi (r)}{dr} = -\frac{4\pi G}{r^2} \int_0^r dr^{'} \, r'^2 \rho(r^{'})\,.
\end{equation}
By making a change of variables $\rho(r)=\rho_0 \Xi$ and $r=\xi \left[ \rho_0 /  (32 \pi G \tilde{\Lambda}^2 m^6) \right]^{1/2}$, where $\rho(r=0)=\rho_0$, this equation reduces to the Lane-Emden equation (with $n=1/2$),
\begin{equation}
\frac{1}{\xi^2} \frac{d}{d\xi} \left( \xi^2 \frac{d \Xi}{d\xi}  \right) = - \Xi^{1/2}\,.
\end{equation}
Choosing boundary conditions $\Xi (0) = 1$ and $\Xi^{'}(0)=0$, we can numerically solve this equation. We can see from the change of variables that  the size of the condensate and the central density are given by \citep{Lane-Emden},
\begin{equation}
R= \xi_1 \sqrt{\frac{\rho_0}{32\pi G \tilde{\Lambda}^2 m^6}}\,, \qquad \rho_0 = \frac{M}{4\pi R^3} \frac{\xi_1}{|\Xi^{'} (\xi_1)|}\,,
\end{equation}
where at $\xi_1$ the numerical solution vanishes. From the numerics $\xi_1 \sim 2.75$ and $\Xi^{'} (\xi_1) \sim -0.5$ gives the following halo radius and central density:
\begin{align}
\rho_0 &\sim \left( \frac{M_{\mathrm{DM}}}{10^{12} M_\odot} \right)^{2/5}    \left(\frac{m}{\mathrm{eV}} \right)^{18/5}    \left(\frac{\Lambda}{\mathrm{meV}} \right)^{6/5} \left( \beta -1 \right)^{3/5}  \, 10^{-24}  \, \mathrm{g/cm}^3\,, \label{rho_0} \\
R &\sim \left( \frac{M_{\mathrm{DM}}}{10^{12} M_\odot} \right)^{1/5}    \left(\frac{m}{\mathrm{eV}} \right)^{-6/5}    \left(\frac{\Lambda}{\mathrm{meV}} \right)^{-2/5}  \left( \beta -1 \right)^{-1/5} \, 45 \, \mathrm{kpc}\,,
\end{align}
With that, we can determine the chemical potential $\mu = \rho^2 / \left( 8 \Lambda^2 m^5 \right)$. For $m \sim \mathrm{eV}$ and $\Lambda \sim \mathrm{meV}$, we obtain realistic core sizes, which are of sizes that cover a big part of the halo, as we wanted. For this reason, we choose the fiducial values:
\begin{equation}
m=0.6 \,  \mathrm{eV}\,, \qquad \Lambda=0.2 \, \mathrm{meV}\,.
\label{fiducial}
\end{equation}
For these values, we have a \emph{cored} density profile with a condensate core of radius $158 \, \mathrm{kpc}$ for $M_{\mathrm{DM}}=10^{12} M_\odot$. The condensate does not make the entire halo, but we expect that this condensed core is surrounded by a NFW profile. The central density obtained is smaller than the expected from CDM simulations, which is preferred by observations. In this way the DM superfluid offers a simple resolution to the cusp-core  and the "too big to fail" problems. We will see these results in more details in Sect. \ref{Sec: Observational Signatures}.

%% This seems to be too much following their article.

\paragraph{Including baryons} \mbox{}\\

Now, we derive the condensate profile in the presence of baryons. We expect that there is this extra acceleration due to the interaction to baryons. This comes from the dynamics of the phonon excitation $\phi$ given the the Lagrangian (\ref{L_temp}). We are going to assume a static, spherically symmetric approximation: $\theta=\mu t + \phi(r)$. The equation of motion for the phonon is given by,
\begin{equation}
\nabla \cdot \left( \frac{(\nabla\phi)^2 -2m\hat{\mu}}{\sqrt{(\nabla \phi)^2 -2m\hat{\mu}}} \, \nabla \phi \right)= \alpha \frac{\rho_{\mathrm{b}}}{2M_{\mathrm{pl}}}\,,
\label{EOM_phi}
\end{equation}
where $\hat{\mu}\equiv \mu-m\Phi$.  If we ignore the homogeneous curls term, in the limit where $(\nabla \phi)^2  \gg 2m\hat{\mu}$ the solution is
\begin{equation}
|(\nabla \phi)|\, (\nabla \phi)  \simeq \alpha M_{\mathrm{pl}} \vec{a}_{\mathrm{b}}\,,
\end{equation}
where $\vec{a}_{\mathrm{b}}$ is the Newtonian acceleration due to baryons only. Then acceleration mediated by $\phi$ is,
\begin{equation}
\vec{a}_\phi = \alpha \frac{\Lambda}{M_{\mathrm{pl}}} \qquad \Longrightarrow \qquad  a_\phi = \sqrt{\frac{\alpha^3 \Lambda^2}{M_{\mathrm{pl}}} \, a_{\mathrm{b}}} = \sqrt{a_0 a_{\mathrm{b}}}\,,
\label{accel_phonon}
\end{equation}
for $a_0=\alpha^3 \Lambda^2/M_{\mathrm{pl}}$, which is exactly the acceleration expected in the deep MOND regime, as showed in Sect. \ref{Sec.:MOND}. In the regime $(\nabla \phi)^2  \ll 2m\hat{\mu}$, we recover the Newtonian acceleration given by the baryons. So, in this model, the total acceleration is given by $\vec{a}_{\mathrm{b}}$, $\vec{a}_\phi$, and also $\vec{a}_{\mathrm{DM}}$ the Newtonian acceleration from the DM halo itself (obtained in the previous section), since we have DM in this model (different than MOND).

\paragraph{Halo profile algorithm} \mbox{}\\

Having developed the theory of the superfluid DM above, now we want to evaluate the density profile of the DM halo and the rotation curves, and compare it with the data to make a first proof of concept of the model. To evaluate the rotation curve, we need to determine the circular velocity with respect to the radius.

As discussed in our model the galaxy contain a superfluid core in the central region of the galaxy surrounded by a NFW profile envelope. So in order to calculate these quantities for the galaxy we first need to evaluate them inside the superfluid core, and then at $R=R_{\mathrm{NFW}}$ match the density and the pressure obtained for the superfluid $\rho_{\mathrm{SF}}$ and $P_{\mathrm{SF}}$, to the ones given by the full NFW profile.

For that, we need to evaluate these quantities in the superfluid phase. In order to obtain the halo density profile, we need to determine the total mass of the halo $M (r)$. The rotation curve is the circular velocity with respect to the radius, given by $a=v^2_{\mathrm{circ}} (r) /r$ where $a=\partial \Phi / \partial r$. So we need to determine the gravitational potential $\Phi$ in order to calculate the rotation curve and, also to determine $M(r)$. The Poisson equation in the superfluid region is given by:
\begin{equation}
\nabla^{2} \Phi = 4\pi G \left(  \rho_{\mathrm{SF}} + \rho_{\mathrm{b}} \right)\,.
\label{poisson_total}
\end{equation}
The baryon density is given by the observations, while the superfluid density we can obtain from our theory by differentiating our Lagrangian (\ref{L_temp}) with respect to $\Phi$,:
\begin{equation}
\rho_{\mathrm{SF}} = \frac{2\sqrt{2} m^{5/2} \Lambda \left[ 3 (\beta -1) \hat{\mu}+(3-\beta) \frac{(\vec{\nabla} \phi)^2}{2m} \right]}{3\sqrt{(\beta -1) \hat{\mu}+\frac{(\vec{\nabla} \phi)^2}{2m}}}\,,
\end{equation}
where we can see that $\rho_{\mathrm{SF}} = \rho_{\mathrm{SF}} (\Phi, \phi)$. So, in order to solve the Poisson equation, we need the equation for $\phi$, which is given by its  equation of motion (\ref{EOM_phi}). The system of equations we need to solve is given by (\ref{poisson_total}) and (\ref{EOM_phi}), which can be very intricate to solve. One approximation that can be done to simplify this is to assume that baryon distribution is spherically symmetric (which we know it is not true, but used as a simplification). With that, the system can be solved numerically. This is done in \citep{Berezhiani:2017tth}. After having this, this solution needs to be matched to the NFW profile that describes the outskirts of the halo. With that, it is possible to evaluate the density profile and the rotation curves of galaxies.

\subsubsection{Observational consequences} \label{Sec: Observational Signatures}

\begin{table*}
\caption{\label{observationssummary}Summary of observational consequences of superfluid DM from \citep{Berezhiani:2017tth}, showing the behaviour of this model in each of the systems listed.}
\centering
\small
\begin{tabular}{lll}
 \toprule
System & Behaviour\\
\hline
\textbf{Rotating Systems} & \\
    Solar system & Newtonian \\
    Galaxy rotation curve shapes & MOND (+ small DM component making HSB curves rise)\\
    Baryonic Tully--Fisher Relation & MOND for rotation curves (but particle DM for lensing) \\
    Bars and spiral structure in galaxies & MOND  \\
  \textbf{Interacting Galaxies} & \\
    Dynamical friction & Absent in superfluid core \\
    Tidal dwarf galaxies & Newtonian when outside of superfluid core \\
  \textbf{Spheroidal Systems} & \\
    Star clusters & MOND with EFE inside galaxy host core --- Newton outside of core \\
    Dwarf Spheroidals & MOND with EFE inside galaxy host core --- MOND+DM outside of core \\
    Clusters of Galaxies & Mostly particle DM (for both dynamics and lensing) \\
    Ultra-diffuse galaxies & MOND without EFE outside of cluster core \\
  \textbf{Galaxy-galaxy lensing} & Driven by DM enveloppe $\implies$ not MOND \\
    \textbf{Gravitational wave observations} & As in General Relativity\\
\end{tabular}
\end{table*}
%\footnotetext{EFE stands for  external field effect, and it reveals an interesting characteristic of the superfluid DM model. This effect is an example from kinetic screening in scalar field theories, where in the presence of gradient interactions the scalar acceleration, given by the non-linearities in the scalar field gradient, can suppress scalar field effects in gravity, making the system behave as standard gravity theory. This effect was an essencial aspect of MOND, but in the DM superfluid model this effect is present in the superfluid cores, coming from the phonon non-linearities. With this, we can see that for satellite galaxies that are inside the superfluid core should follow a MOND-like dynamics, while more distant satellites are outside the core, and therefore should follow Newtonian dynamics. We discuss some o }

In this section we will describe the main observational consequences of the superfluid DM. A summary of all the effects already worked out can be seen in Table \ref{observationssummary}. This table compiles a list of the behaviour that this model has in different systems. We describe in this section some of those results, but point to \citep{Berezhiani:2017tth} for a detailed explanation of each of those cases. 

In Table 1, EFE stands for  external field effect, and it reveals an interesting characteristic of the superfluid DM model. This effect is an example from kinetic screening in scalar field theories, where in the presence of gradient interactions the scalar acceleration, given by the non-linearities in the scalar field gradient, can suppress scalar field effects in gravity, making the system behave as standard gravity theory. This effect was an essencial aspect of MOND, but in the DM superfluid model this effect is present in the superfluid cores, coming from the phonon non-linearities. With this, we can see that for satellite galaxies that are inside the superfluid core should follow a MOND-like dynamics, while more distant satellites are outside the core, and therefore should follow Newtonian dynamics. For more details, see~\citep{Berezhiani:2017tth}.

Since it is going to be used a lot in this section, we remind Landau's conditions for superfluidity is that the fluid velocity ($v_{\mathrm{s}}$) is smaller than the superfluid sound speed $c_{\mathrm{s}}$, $v_{\mathrm{s}} < c_{\mathrm{s}}$.

\begin{itemize}
\item \textbf{Galaxy rotation curves:}

In \citep{Berezhiani:2017tth} the rotation curves of IC 2574, a low surface brightness galaxy, and UGC 2953, a high surface brightness galaxy, were numerically calculated using the method developed above, as a proof of concept of the galactic dynamics that the DM superfluid is able to reproduce. Since for the theoretical predictions a spherical baryonic distribution was assumed to simplify the calculations, and since this is far from the actual distribution of baryons, in these calculations a hybrid method, mixing the results calculated with the spherical distribution, was implemented. In this, the acceleration is corrected for the actual distribution leading to:
\begin{equation}
\vec{a}_{\mathrm{hybrid}} = \vec{a}_{\mathrm{b,real}}+\vec{a}_{\mathrm{DM}}+\vec{a}_{\mathrm{phonon}}\,,
\end{equation}
where $\vec{a}_{\mathrm{b, real}}$ is the acceleration computed from Poisson's equation for a non-spherical baryon distributions; $\vec{a}_{\mathrm{DM}}$ the Newtonian acceleration from the DM halo using spherical baryon distribution; and $\vec{a}_{\mathrm{phonon}}$ from (\ref{accel_phonon}) sourced by $\vec{a}_{\mathrm{b, real}}$, but with Newtonian potential from the spherical case. Although calculated in the hybrid method, $\boxed{\vec{a}_{\mathrm{phonon}} \sim \sqrt{a_0 a_{\mathrm{N,b}}}}$ as expected in MOND regime.

The fiducial parameters used for this numerical calculation were $m = 1\mathrm{eV}$, $\left( \sigma/m \right) = 0.01 \, \mathrm{cm}^2/\mathrm{g}$, which are optimal for having a superfluid core that encompasses the baryonic disk of the galaxy, while still within the bounds to agree with cluster observations; $\Lambda m^3 = 0.05 \, \mathrm{meV} \times \mathrm{eV}^3$; and $\alpha = 5.7$. The rotation curves can be seen in Fig.~\ref{Fig.:rotation_curves}.

\begin{enumerate}
\item \textit{LSB galaxy:} As pointed out before since these type of galaxies are DM dominated, the rotation curves from LSB are expected to have a slow raise before reaching the plateau region. As we can see in the \textit{left panel} of Fig.~\ref{Fig.:rotation_curves}, our model reproduces the observed rotation curve for IC 2574, represented by the orange points, very precisely for the parameters chosen. 

The size of the superfluid core obtained for this galaxy is $R_{\mathrm{SF}} \sim 40 \, \mathrm{kpc}$, which here is represented by the NFW radius where the profile is matched with a NFW profile and has a close value to $R_T$. Relative to $R_{200} \sim 57 \, \mathrm{kpc}$ for this galaxy, the superfluid core is relatively large encompassing $58\%$ of the total DM mass of the halo.

\item \textit{HSB galaxy:} The rotation curve features of HSB galaxies are known to be hard to be reproduced. We saw that MOND empirical theory is successful in reproducing those features. It is interesting to see if the superfluid DM model is also able to reproduce it. The rotation curve for UGC 2953 is shown in the \textit{right panel} of Fig.~\ref{Fig.:rotation_curves}, using the same conventions as for the LSB. The radius obtained for the superfluid core in this case is $R_{\mathrm{SF}} \sim 79\,  \mathrm{kpc}$, which is small in comparison to $R_{200} \sim 245 \, \mathrm{kpc}$. Only $24\%$ of the total mass of DM is in the superfluid core. The difference from the LSB results is the red curve, where the total DM mass is set by the $\Lambda$CDM abundance matching value of $M=65 \, M_\odot$. For the red curve, we get a bigger superfluid radius, $R_{\mathrm{SF}}=93 \, \mathrm{kpc}$, which is still significantly smaller than $R_{200} = 446 \, \mathrm{kpc}$.  The rotation curves seem to fit the data well, showing a smaller value but still compatible with observations for the velocity in the point where the curve turns to flat. Also, the superfluid DM show a slight rise in the end of the rotation curve, which is compatible to the data but not existent in MOND.
\end{enumerate}

\begin{figure}[htb]
\centering
\includegraphics[width=0.48\textwidth]{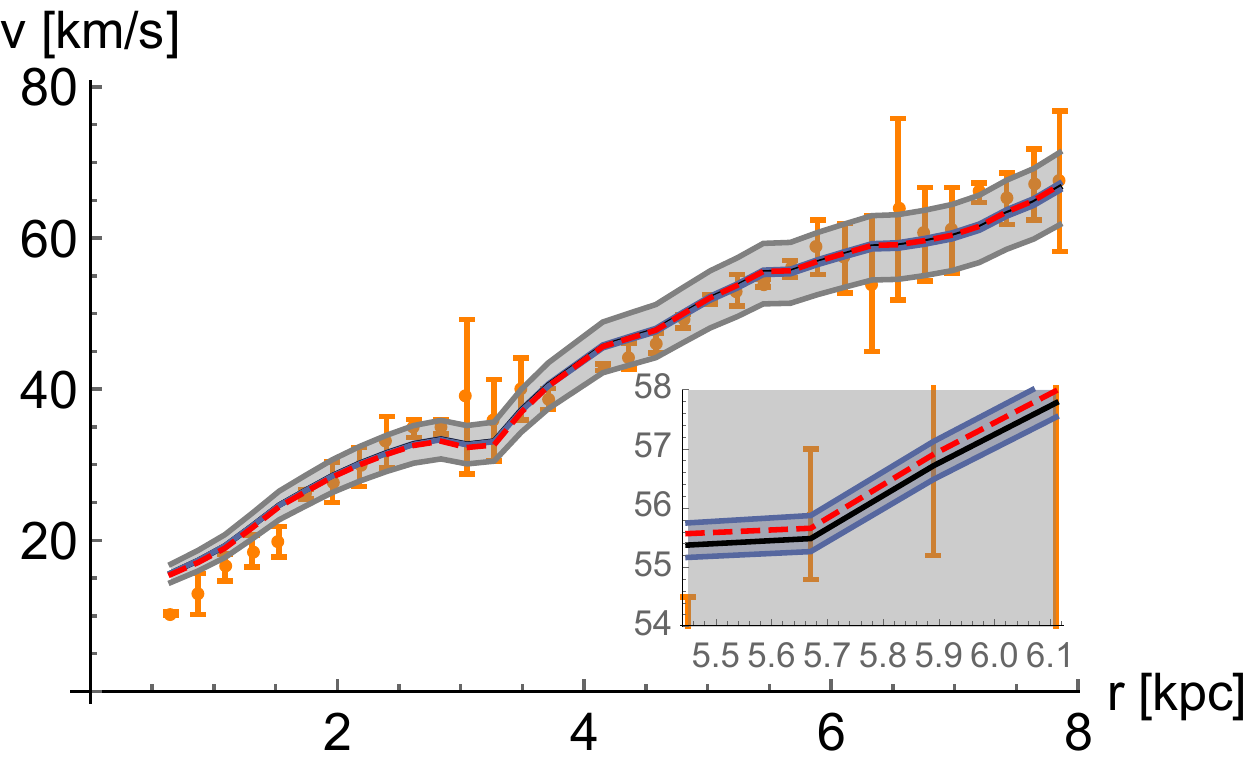}
\includegraphics[width=0.48\textwidth]{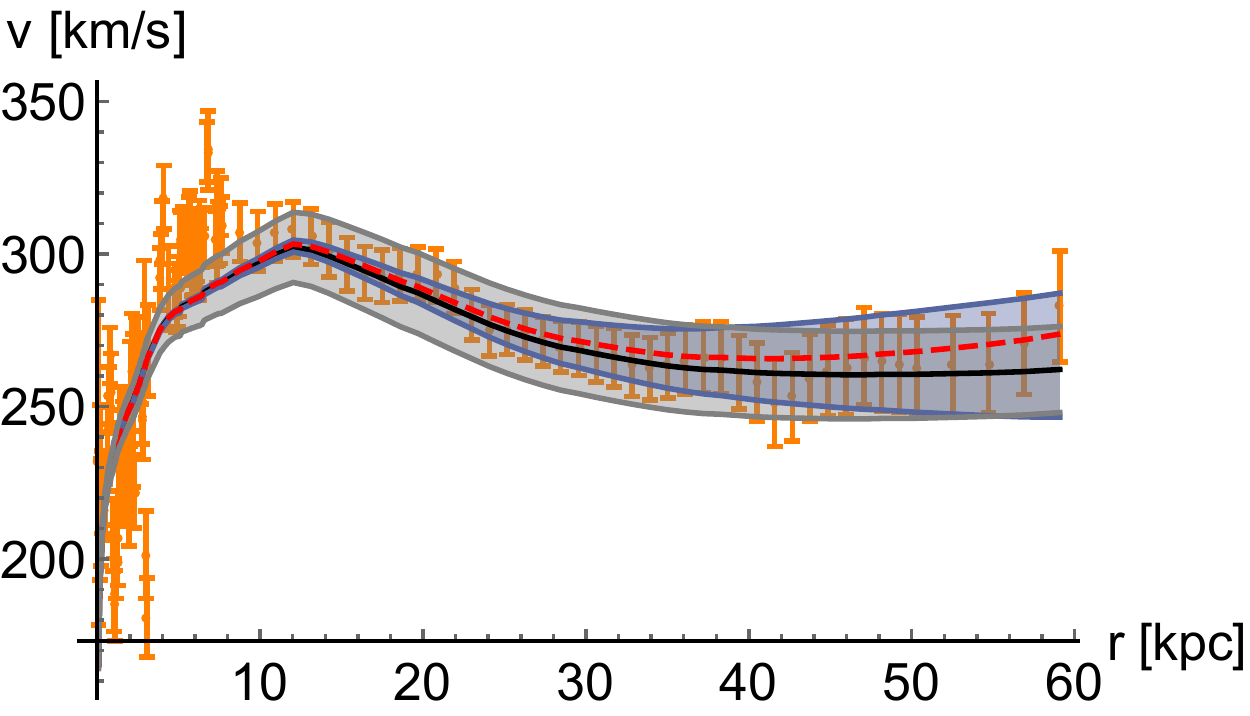}
 \caption{Predicted rotation curves evaluated in \citep{Berezhiani:2017tth}. \textit{Left panel:} Rotation curve for the LSB galaxy IC2574. The orange points are data from \citep{Lelli:2016zqa} assuming a distance of  $\sim 3\, \mathrm{Mpc}$ \citep{Tully:2007ue}, the black and red curves are the predicted curves for $M_{\mathrm{DM}}= 20 M_{\mathrm{b}}$ and $50M_{\mathrm{b}}$. The gray band corresponds to two values of $a_0 \in (0.6, 1.2) \times 10^{-8}$ and the blue band two values of $\Lambda \sim (0.02,\, 0.1)$meV. \textit{Right panel:} Rotation curve of UGC~2953. The orange points are data from \citep{Noordermeer:2007ux} with all the parameters like in the left panel figure, but the red curve where $M_{\mathrm{DM}}=65 M_{\mathrm{b}}$.}
\label{Fig.:rotation_curves}
\end{figure}

In general, it seems that the superfluid model reproduces the rotation curves of LSB and HSB galaxies. 
Also, the BRTF relation is satisfied, as expected. Of course, this calculation shows a proof of concept and the rotation curves of many more galaxies with different characteristics need to be fitted, also to help determine the parameters of the theory, which were chosen here. However, it is expected that the behaviour of the rotation curve is similar to the fits shown above for different types of galaxies inside the superfluid core, only changing the size of the core depending on the galaxy.
%The photon-baryon coupling inside the core guarantees the behavior following MOND dynamics which will give a similar behavior for the rotation curve for different galaxies.
%However, we expect that this model can fit the rotation curve of galaxies of different masses given that the photon-baryon coupling inside the core, simply changing the size of the core with respect to the galaxy.%
%an already anticipate that we expect this to work for galaxies with very different masses, since the phonon-baryon coupling guarantees this behavior for the rotation curve within the core; outside the core its is going to be very different.

\item[]

\item \textbf{Dynamical friction:} 
From the very definition of superfluidity, flow without friction, we can expect that in these models in the inner regions of galaxies, where superfluidity emerges upon condensation,  dynamical friction to be absent. This might lead to interesting astrophysical consequences and help understand some puzzles with CDM \citep{Hui:2016ltb,Ostriker:1998fa}, while testing the DM superfluid model.
%Many effects in this DM scenario have different predictions given the absence of friction when DM behaves as a superfluid. 

One example of an observation that can be explained by this characteristic of superfluids is the velocity of galactic bars in spiral galaxies, which are expected to have been slowed down by dynamical friction, but are measured to be nearly constant which is consistent with no dynamical friction.

Another interesting puzzle directly linked with dynamical friction is the Fornax globular clusters, as we already mentioned above.   In the presence of a superfluid in the halo, given the absence of dynamical friction, these globular cluster should not necessary have merged with Fornax. The effect expected for the case of a superfluid is more pronounced than in the FDM model, for example. This shows that these type of system can offer an opportunity to test these ULDM model.

%We are going to see, in the following examples, other astrophysical consequences of this absence of friction in a superfluid.
%There is also speculation that this absence of dynamical friction can lack of feature in the two-point correlation function of luminous red galaxies, consequences for the history of the Local Group, among others. The quantitative analyses of dynamical friction is currently under investigation.atinos to te

However to use these observations to test these models we need to really understand what is the behaviour of dynamical friction in the DM halo. For that a microscopical description of a superfluid theory with dissipation is necessary, and that is what is shown next.

\vspace{0.3cm}

The simple picture that in a superfluid there is no friction is a simplification. A superfluid can suffer dissipation, when its internal degrees of freedom, are excited out of the condensate, resulting in a mixture of superfluid and normal particles, or even the complete depletion of the condensate. In those cases, then the system exhibits friction. To fully understand how friction and dissipation in the superfluid takes place, one needs to work out the superfluid theories presented in Sect.~\ref{Sec.:BEC_superfluids} in the presence of dissipation. To describe dissipation, one needs to study the motion of an impurity, a particle moving in the superfluid represented by a real scalar field $\chi$. In general lines, dissipation is described in the following way: if an object passes through this superfluid, this is called an impurity. When this happens, two things can happen: (i) if the object is moving through the condensate with a velocity bigger than the superfluids sound speed, $v>c_s$, dissipation of the superfluid takes place, given that the moving object transfers energy to the internal degrees of freedom of the superfluid. At low energies, the only accessible degree of freedom is the phonon, so the passage of such an impurity excites phonons out of the superfluid and the radiation of phonons occurs. The rate of phonon emission,describes the dissipation of the superfluid. The If the impurity passes through the condensate with subsonic speed, then there is no dissipation and the particle travels without friction.

This was discussed in detail in \citep{Berezhiani:2019pzd}. The regime of validity of such a theory in the presence of dissipation is discussed in \citep{Berezhiani:2020umi}, and it is an extension to the discussion presented in Sect.~\ref{Sec.:BEC_superfluids}, where we assumed the limit without dissipation.

The theory with dissipation was worked out for the simplest superfluid example, the interacting BEC described by  the microscopic Lagrangian of a self-interacting complex field (\ref{L_2body}), with the presence of an impurity, 
\begin{equation}
\mathcal{L}= - |\partial_\mu \Psi|^2 - m^2 |\Psi|^2 - \frac{g}{2} |\Psi|^4 -\frac{1}{2} (\partial \chi)^2 - \frac{1}{2} M^2 \chi^2 - \frac{1}{2} g_{\mathrm{int}} \, \chi^2 |\Psi|^2 \,.
\end{equation}
We are going to work with the Lagrangian which actually describes the SIFDM model, but that is the simplest model to understand superfluidity.
We are interested, as before, in the non-relativistic case and low energy regimes. To study the dissipation of phonons, we perturb this Lagragian and work with the linear theory. The process that we want to study is the dissipation of the superfluid radiating phonons caused by the motion of the impurity, with $\Phi$, the Newtonian potential behaving as the mediator of this process. This can be described by the process (at first approximation) $\chi \rightarrow \chi + \pi$, and the rate of this process can be computed.

As discussed in \citep{Berezhiani:2020umi}, it is important to reach  the correct result to consider the higher order derivatives of the phonon effective action, like we did in (\ref{L_2body}) that gives rise to the higher order $k^4$ term in the dispersion relation, together with  the higher order terms involving the impurity field. 
With that it is possible to calculate the energy dissipation,
\begin{equation}
|\dot{E}| = \int \omega_k \, d\Gamma \propto \frac{n\, g_{\mathrm{int}}^2 \,  k^4_{*} }{m^3M^2 V}\,,
\end{equation}
where $d\Gamma= (q/E_{\chi}^{\mathrm{in}}) |\mathcal{A}|^2\, \delta^{(4)}(p_{\chi}^{\mathrm{in}}-p_{\chi}^{\mathrm{f}}-p_{\pi}^{\mathrm{f}}) \, (d^3 p_{\chi}^{\mathrm{f}}/E_{\chi}^{\mathrm{f}}) (d^3 p_{\pi}^{\mathrm{f}}/E_{\pi}^{\mathrm{f}})$ is the rate of the process described above, with `in' indicating the initial values before the collision and `f' indicating final values. The initial momentum of the impurity is $p_{\chi}^{\mathrm{in}}=(M+MV^2/2, \, M \mathbf{V})$ and with $k_{*}$ given by $k^2_{*}/2M + \omega_{k_{*}}=k_{*}V$. With that, we can determine the friction force in the system:
\begin{equation}
|F|=\frac{|\dot{E}|}{V}=\frac{mng^2_{\mathrm{int}}}{M^2} \frac{(V^2-c^2_{\mathrm{s}})^2}{V^2}\,.
\end{equation}
This shows that the friction force is not discontinuous, given having friction or no friction in the case of having a superfluid or not, as suggested by Landau, but actually it varies monotonically with the velocity. In the limit where $V$ is equal to the sound speed, then the friction force vanishes, as expected for a superfluid. 

If we include gravity in this system, we have an extra term coming from the coupling to gravity which modifies the dispersion relation for the superfluid as shown in Sect.s~\ref{Sec.:Interacting BEC} and \ref{Sec.:SIBEC}
\begin{equation}
\omega_k^2=-m^2_{\mathrm{g}} + c^2_k k^2 +\frac{1}{4m^2}k^4\,,
\end{equation}
where there is an additional tachyonic mass term from the gravitational contribution, given by $m^2_{\mathrm{g}} \equiv 4 \pi G \rho_0$. The presence of this term modifies the Jeans scale and Jeans instability occurs when  $k > k_J=2m^2 c^2_{\mathrm{s}} (-1+\sqrt{1+(m^2_{\mathrm{g}}/m^2c^4_{\mathrm{s}})})$.  In this case, as shown in \citep{Berezhiani:2019pzd}, the force evolves monotonically with the velocity. However, it never reaches zero friction for subsonic velocities, when there is the superfluid, because of the Jeans instability. This shows that the dynamical friction in a superfluid is more complex than the simple dichotomy of absence or presence of friction if there is or not a superfluid. This is an active field of research and might lead to interesting observational consequences for the DM superfluid and SIFDM, which is the model worked out here.

\item[]

\item \textbf{Galaxy clusters:} In a simple way, following the analyzes in Sect.~\ref{Sec.:conditions}, clusters have large dispersion velocities, and at large distances, of order of $R_{200}$, $\Upsilon$ is going to be small and thermal equilibrium cannot be achieved. The DM in clusters is in the normal phase.
However, as we saw for galaxies, in the central regions the density increases and thermal equilibrium may be achieved. In clusters only a very small amount can be in the superfluid state, since observations exclude that clusters are largely in the superfluid regime. We can then see the bounds in our mass in order to have a small amount of superfluid component in clusters that is not in tension with data. We assume that $R_T/R_{200} \lesssim 0.1$, which gives, using the relations from Sect.s~\ref{Sec.:conditions} and~\ref{Sec.:halos_theory}: 
\begin{equation}
R_T \lesssim 200 \left( \frac{M}{10^{15} M_\odot} \right) ^{1/3} \, \mathrm{kpc}\,.
\end{equation}
 
 We can now repeat the analysis of thermal equilibrium done in Sect.~\ref{Sec.:conditions}. However, for such a small $R_T$ in comparison to the cluster size, we use the full NFW profile for the halo. This yields a constraint  in the mass of the DM particles:
 \begin{equation}
 m \gtrsim 2.7 \left( \frac{\sigma/m}{\mathrm{cm}^2/g} \right)^{1/4} \mathrm{eV}\,.
 \end{equation}
This combined with the condition from thermalization in galaxies gives the allowed range for the DM mass:
\begin{equation}
2.7 \, \mathrm{eV} \lesssim m  \left( \frac{\sigma/m}{\mathrm{cm}^2/g} \right)^{-1/4} \lesssim 4.2 \, \mathrm{eV} \,
\label{cluster}
\end{equation}
From the tightest constraints from approximately 30 merging systems \citep{SIDM3},  $\sigma /m \lesssim 0.5 \, \mathrm{cm}^2/\mathrm{g}$. This value is in accordance with the one from Sect.~\ref{Sec.:conditions}, and from the constrain above it gives a DM mass between $1.5 \, \mathrm{eV} \lesssim m  \lesssim 2.4 \, \mathrm{eV}$.
For DM superfluid in this mass range, we have condensation inside galaxies and the condensation in the interior of cluster happens for very small radius, appearing not to be in conflict with what is expected from observations. This constraint can be made broader by assuming a more realistic and not constant cross section.  A quantitative analysis via numerical simulations would be ideal to check this result.

\item[]

\item \textbf{Galaxy mergers:} The behaviour of merging galaxies is an interesting question, given the superfluid nature, the absence of friction, proposed for the inner core of galaxies. In the absence of friction, it is expected that the merger would make the galaxies pass through each other without interacting. But the existence of these superfluid phases in these merging systems is going to depend on the comparison between the infall velocity for the merging galaxy and the sound speed of the phonon, given by the Landau criteria.
\begin{itemize}
\item $v_{\mathrm{infall}} \gtrsim c_{\mathrm{s}}$ - In this regime, the halos are driven out of equilibrium, so coherence of the condensate is broken and the halo will be in the normal phase. The merging process will proceed as in $\Lambda$CDM, where mergers are fast due to  dynamical friction. Thermal equilibrium and condensation will be achieved in the merged halo after some time.
\item $v_{\mathrm{infall}} \lesssim c_{\mathrm{s}}$ - In the case of subsonic velocities, the DM halo is in the superfluid phase, and the superfluid cores will pass through each other with almost no dissipation. In this case, dynamical friction is reduced taking a much longer time to the system to merge, and possible multiple encounters. 
\end{itemize}
In our case, the phonons have sound speed $c_{\mathrm{s}}= \sqrt{2\mu/m}$, and for the fiducial values adopted (\ref{fiducial}), $c_{\mathrm{s}} \sim 220 \, \mathrm{km/s}$ for a $10^{12} M_\odot$ halo. This needs to be compared with the infall velocities of galaxies of a merging system to see how the merger dynamics proceeds.

\item[]

\item \textbf{Merging Clusters:} 

The Bullet cluster is a system of two merging cluster that was very well investigate observationally.  It represents one of the best evidences of the existence of DM (and against alternatives like MOND). This is seen by a segregation in the position of the  mass peak (highest concentration of total matter) given by lensing that probes all the matter content, and the one from X-ray measurements, which measures the baryonic matter. This is consistent with the CDM picture, where the DM in the merging processes due to its negligible interaction passes through almost without interaction, while the baryons are slowed down. This poses a problem for theories that do not have DM, like full MOND.

By construction, and as it was seen above, in this model clusters do not develop a condensed core of cluster size and have most of the DM in the normal phase. However, the galaxies inside the clusters have a condensed core, and the cluster can develop smaller sized cores in its inner regions. Therefore, when clusters merge, the presence of a core or not also depends if the merger is  subsonic or supersonic, obeying or not Landau's criteria like we saw for the merging galaxies.

The outcome of the merging depends on the infall velocities, which determines if most of the DM is in the superfluid phase or in the normal phase  in each of the merging clusters.  If the infall velocities are subsonic, the superfluid cores are present, and most of the DM will be in the superfluid phase. Any collision between a cluster where DM behaves like a superfluid will follow without dissipation, with the clusters pass through each other without friction. Now, the DM in the normal phase presents self-interactions. Therefore the collision of two clusters in the normal phase would be slowed down due to these interaction.  

In the case of the Bullet cluster \citep{Berezhiani:2015pia}, in order to be consistent with observations, at least the sub-cluster must be in the superfluid phase. As we can see, the sound speed of the phonon for the sub-cluster ($M_{\mathrm{sub}} \sim 10^{14} M_\odot$) is, for our fiducial values, $c_{\mathrm{s,sub}} \sim 1400 \, \mathrm{km/s}$, while for the main cluster ($M_{\mathrm{main}} \sim 10^{15} M_\odot$) is $c_{\mathrm{s,main}} \sim 3500 \, \mathrm{km/s}$. The relative velocity between the clusters is $\sim 2700\, \mathrm{km/s}$ \citep{Bullet1,Bullet2}. If we take this to be the infall velocity, we can see that the sub-cluster is in the superfluid phase, while the main cluster is in the normal phase. With that, the clusters will merge without dissipation and pass through each without friction, as it is expected from observations.

For the Abell 520 ``train wreck'' \citep{Abel1,Abel2,Abel3,Abel4}, another merging cluster system, the DM superfluid model predicts a subsonic merger, with two peaks representing the superfluid component, compatible with the lensing map, and a peak during the normal component, coming from the X-ray luminosity peak. This shows that the DM superfluid framework can accommodate not only the dynamics on galactic scales, but also explain clusters and its merger events.

%An interesting puzzle is that of Fornax globular clusters. From dynamical friction it is expected that globular clusters orbiting Fornax should have rapidly fall towards its center to form a stellar nucleus . However, there is no signal of mergers and 
 
\item[] 
 
\item \textbf{Gravitational lensing:}

In the case of the full MOND theory, or its relativistic completion TeVeS, because of the absence of DM to be able to explain the relativistic regime makes necessary the introduction of a complicated non-linear term between the scalar field of the theory and baryons, which should also couple to a time-like vector field in order to give the correct gravitational potential to be able to explain gravitational lensing.

In the case of the DM superfluid, since the theory has DM, we have the superfluid component described by the phonon scalar field, and we have the normal component which provides the time-like vector field $u^{\mu}$.  The gravitational potential is then sourced by both dark matter and baryons, as expected.

As we have that the superfluid core resides in the inner part of the galaxy, surrounded by an NFW envelope, gravitational lensing will come primarily from this NFW outter part. 

Recently, the DM superfluid model was studied in the context of strong lensing \citep{Hossenfelder:2018iym}.

\item[]
 
\item \textbf{Gravitational waves:}

In the superfluid DM, different than in MOND, the superfluid core is locates in the inner regions of the halo and the outskirts of the halo have a NFW profile. So the gravitational lensing signal comes from this outer part of the halo and it behaves like in the case of GR+CDM. This means that photons and gravitons propagate at the speed of light travelling along the same geodesics. This is in agreement with the recent constraints from the gravitational waves from neutron stars merger GW170817 \citep{GW1}, which rule out relativistic completions of full MOND \citep{GW2}.

The implications for the gravitational waves in the case where the phonon has a non-vanishing sound speed was considered in \citep{GW3}, together with its observational effects in future GW experiments.

Specifying the microphysics of the DM superfluid particle can also yield other signatures in the produced gravitational waves, like chirality, as done in \citep{GW4}.
\end{itemize}

\subsubsection{Validity of the EFT} \label{Sec.:4_validity}

In this section we are going to scrutinize the validity of the EFT construction used, verifying the regimes where this leading order EFT is valid, and the regimes where the theory obeys the Landau criteria for superfluidity.

\paragraph{Higher-order derivatives} \mbox{} \\

First, we need to check if in our regime and for the  parameters of the model, it is valid to ignore higher order terms in the EFT. As we saw above, the EFT is constructed by including all the terms which are invariant under shift symmetry. We retained only the first order contributions, given that we are working in the low-energy limit. 
Higher order terms involve more than one derivative per field. Higher order contributions to the quadratic Lagrangian for the phonon (\ref{quad_phonon}), can contain terms of the form:
\begin{equation}
\mathcal{L}_{\mathrm{higher-order}} \supset \Lambda m^{3/2} \mu^{\frac{3}{2}-n} \partial^n \phi^n \sim \left(\Lambda m^{3/2} \mu^{3/2} \right)^{1-\frac{n}{2}} \partial^n \phi^n_{\mathrm{c}}\,,
\end{equation}
where $\partial \rightarrow \partial_t \, \mathrm{ or  }\, c_{\mathrm{s}} \vec{\nabla}$, and $\phi_{\mathrm{c}} = \Lambda^{1/2}m^{3.4} \mu^{-1/4} \phi$ is the canonical variable. The scale that controls these higher order terms is given by $\Lambda_{\mathrm{s}} = \left(\Lambda m^{3/2} \mu^{3/2} \right)^{1/4}$, which we call the strong coupling scale. 
So, higher order corrections can be neglected when:
\begin{equation}
\frac{1}{\Lambda_{\mathrm{s}}} \frac{\partial^2_r \phi}{\partial \phi} \sim \frac{1}{\Lambda r} \ll 1\,. 
\end{equation}
This is the general condition for ignoring the higher order corrections in a EFT given by the Lagragian we described here. Given this, we an easily see that the approximation of ignoring these terms breaks for small sound speeds.

However, specializing to the parameters of the DM superfluid model described here, and using  the profile obtained in (\ref{rho_0}), which determines $\mu$, the strong coupling scale is given by the DM superfluid model:
\begin{equation}
\Lambda_{\mathrm{s}} \sim \mathrm{meV}  \left(\frac{M_{\mathrm{DM}}}{10^{12} M_\odot} \right)^{3/10}  \left(\frac{m}{\mathrm{eV}} \right)^{6/5}    \left(\frac{\Lambda}{\mathrm{meV}} \right)^{2/5} \,.
\end{equation}
For the fiducial parameters, $\Lambda_{\mathrm{s}} \sim \mathrm{meV}$. So, higher derivatives are suppressed if $r \gg 0.2 \mathrm{mm}$, which is clearly satisfied on astrophysical scales.

\paragraph{Criteria for condensate coherence} \mbox{} \\

An important criteria to verify the validity of the superfluid description we are using is to check if our superfluid obeys the Landau criteria. As we saw in Sect.~\ref{Landau}  the criteria for the system to transports charge without dissipation, leading to the coherence of the BEC to be maintained, is that the velocity of the superfluid is smaller than the critical velocity:
\begin{equation}
v_{\mathrm{s}} \ll v_{\mathrm{c}} \sim \left( \frac{\rho}{m^4} \right)^{1/3}\,.
\label{coherence_SF}
\end{equation}
 where the critical velocity must be non-vanishing. This condition is equivalent to the condensation conditions showed in the panel in Sect.~\ref{Sec.:ideal_BEC}.  In the case of the DM superfluid, in Sect.~\ref{Sec.:conditions}, we already evaluated the conditions for DM to be condensed in the center of galaxies.
 
We need to estimate $v_{\mathrm{c}}$ locally.  For that, we can use the halo mass density $\rho = (2m)^{3/2} m \Lambda \sqrt{|X|} \sim 2m^2 \Lambda \sqrt{\kappa}$, where we assumed MOND regime in the last equality and $\kappa = m \hat{\mu}$, which gives us:
 \begin{equation}
 v_{\mathrm{c}} \sim 0.025 \left(\frac{M_{\mathrm{b}}}{10^{11} M_\odot} \right)^{1/6}  \left(\frac{m}{\mathrm{eV}} \right)^{-2/3}    \left(\frac{\Lambda}{\mathrm{meV}} \right)^{2/9}     \left(\frac{\mathrm{kpc}}{r} \right)^{1/3}\,,
 \end{equation}
 where $M_b$ is the baryon's mass.
The superfluid velocity is given by $v_{\mathrm{s}} = \partial_r \phi / m \sim \sqrt{\kappa}/m$, which yields:
\begin{equation}
v_{\mathrm{s}} \sim  \left(\frac{M_{\mathrm{b}}}{10^{11} M_\odot} \right)^{1/2}  \left(\frac{m}{\mathrm{eV}} \right)^{-1}    \left(\frac{\Lambda}{\mathrm{meV}} \right)^{-1/3}     \left(\frac{\mathrm{kpc}}{r} \right)\,.
\end{equation}
With that, using(\ref{coherence_SF}) and assuming spherical symmetry, we an determine thee radius where superfluidity can occur:
\begin{equation}
r \gg  \left(\frac{M_{\mathrm{b}}}{10^{11} M_\odot} \right)^{1/2}  \left(\frac{m}{\mathrm{eV}} \right)^{-1/2}    \left(\frac{\Lambda}{\mathrm{meV}} \right)^{-5/6}     \mathrm{kpc}\,.
\end{equation}
We can see that this condition is satisfied in the central regions of galaxies, and we have coherence of the condensate and superfluidity in those scales.

\paragraph{Solar system}\mbox{} \\

We can check this criteria for condensate coherence locally to verify in which regions and regimes we have DM in the superfluid phase. Therefore, we can apply the criteria for condensation developed above and the radius where locally the superfluid can occur to our solar system.

At solar system scales the bounds on deviation from standard Newtonian gravity are very tight, and these measurements do not allow deviations from the Newtonian dynamics. Full MOND is in tension with these bounds. However, the DM superfluid scenario fits well into the Solar system bounds. We can see that using the coherence bound for the condensate (\ref{coherence_SF}). 
For that we need to evaluate $v_{\mathrm{s}}$ and $v_{\mathrm{c}}$ for solar system quantities.

Given the local gradients of the phonon in the vicinity of the Sun ($M_{\mathrm{b}} = 1 \,M\odot$), the superfluid velocity is given by:
\begin{equation}
v_{\mathrm{s}}^\odot = 5  \left(\frac{m}{\mathrm{eV}} \right)^{-1}    \left(\frac{\Lambda}{\mathrm{meV}} \right)^{-1/3}  \frac{\mathrm{AU}}{r} \,,
\end{equation}
where $r$ is the distance to the Sun and AU is the astronomical unit, the average distance between the Earth and the Sun. The critical velocity of the Milky Way galaxy (for $M_{\mathrm{b}} = 3 \times 10^{11} M_\odot$) evaluated at our solar system ($r \sim 8 \, \mathrm{kpc}$) is:
\begin{equation}
v_{\mathrm{c}}^{MW} \sim 0.02  \left(\frac{m}{\mathrm{eV}} \right)^{-2/3}    \left(\frac{\Lambda}{\mathrm{meV}} \right)^{-2/9}\,. 
\end{equation}
We can see that the coherence bound $v_{\mathrm{s}}^\odot \ll v_{\mathrm{c}}^{\mathrm{MW}}$ is obeyed for distance much larger than the solar system scales:
\begin{equation}
r \gg r^{\odot}_{\mathbf{s}} = 250 \left(\frac{m}{\mathrm{eV}} \right)^{-1/3}    \left(\frac{\Lambda}{\mathrm{meV}} \right)^{-5/9} \mathrm{AU}\,.
\end{equation}
This shows that on solar system scales, the presence of the Sun causes the local phonon gradients to be large, making the local superlfuid velocity $ v_s \sim \phi'/m$, to be larger than the local critical velocity for a region around the solar system with a radius $r^{\odot}_{\mathbf{s}}$ from the Sun. 
Therefore, on distances like the solar system, $r < r^{\odot}_{\mathbf{s}}$, DM is in the normal phase since the condensate loses its coherence, and obeys standard Newtonian gravity.

\subsubsection{Relativistic completion}

As we saw in Sect.~\ref{Sec.:Interacting BEC}, the description of a superfluid is given by a weakly self-interacting field theory with global U(1) symmetry. The symmetry is spontaneously broken by the superfluid ground state of a system at chemical potential $\mu$. In the previous section, where we defined this field theory for superfluids, we added a 2-body self-interaction, $  g-3 |\Psi|^4$. This gives an equation of state $P \propto n^2$. As we saw in the previous section, the pressure that describes the interaction in the Madelung equations has the form of the pressure of a barotropic fluid. For a three-body interaction, the equation of state is given by $P \propto n^3$. For the DM superfluid, in order to reproduce MOND, we wanted to have a theory that gave $P \propto n^3$. So one might think that the DM superfluid could be described by the microscopic theory of an interacting BEC with three-body interaction. However, we are going to show now that this is in fact not the case, since those theories give a Lagrangian with different signs.

\paragraph{3-body interaction} \mbox{} \\

Lets consider now like before that the self interacting theory with U(1) symmetry that gives us the superfluid has a 3-body interaction, instead of a 2-body one. The relativistic action of this theory is given by:
\begin{equation}
\mathcal{L} = -|\partial \Psi|^2 - m^2 |\Psi|^2-\frac{g_3}{3} |\Psi|^6\,,
\end{equation}
where $g_3>0$ for stability. Like before, this theory conserves particle number. Since we are interested in the non-relativistic (NR) theory, replacing $\Psi=\psi e^{imt}$ and taking the NR limit gives us:
\begin{equation}
\mathcal{L}=\frac{i}{2} \left( \psi \partial_{t} \psi^{*} - \psi^{*} \partial_t \psi \right) - \frac{|\nabla \psi|^2}{2m}-\frac{g_3}{24m^3} |\psi|^6\,.
\label{NRL}
\end{equation} 
With that, we can calculate the equation of motion, which gives us the Schr\"{o}dinger's equation:
\begin{equation}
-i\partial_t \psi + \frac{\nabla^2 \psi}{m}	-\frac{\lambda}{8m^3}|\psi|^4 \psi = 0\,.
\end{equation}
The condensate is described by the background solution, at zero temperature: $\psi_0 = \sqrt{2m}\, v e^{i\mu t}$, where $\mu = \lambda v^4/2m$. The excitations are given by:
\begin{equation}
\psi = \sqrt{2m} \, (v+\rho) e^{i(\mu t + \phi)}\,,
\end{equation}
where $\rho$ is the gapless mode and $\phi$ is the Goldstone boson associated with the broken U(1). At low energies, we substitute this into (\ref{NRL}) and  integrate out the gapless mode:
\begin{equation}
\mathcal{L} = \frac{4}{3} m \left( \mu+\dot{\phi}-  \frac{(\nabla \phi)^2}{2m} \right) \left[ \frac{2m}{\lambda} \left( \mu+\dot{\phi}-  \frac{(\nabla \phi)^2}{2m} \right)   \right]^{1/2}\ = \frac{4}{3} \left( \frac{2}{\lambda} \right)^{1/2} m^{3/2} \, X\sqrt{X}\,,
\end{equation}
which is the action to leading order in the derivative expansion, with $X=\mu+\dot{\phi}- (\vec{\nabla}\phi)^2/2m$. This is very promising since the theory with a 3-body interaction gives a low-energy Lagrangian with the same exponent as the one we need for MOND and for the effective Lagrangian $P(X)$ for the EFT of superfluids. However, it has the opposite sign, given that $g_3 > 0$!  As we saw before, the limit where $g_3 < 0$ is unstable, and we cannot have condensation on all scales, superfluidity and MOND.

\paragraph{Phenomenological relativistic Lagrangian} \mbox{} \\

So, the expected description as a theory with 3-body processes is does not work for the DM superfluid model, where we want to recover MOND behavior in galaxies. It was phenomenologically proposed in \citep{Berezhiani:2015bqa} a relativistic Lagrangian that is able to reproduce our expected Lagrangian (\ref{L_SF}) in the non-relativistic regime, which is given by:
\begin{equation}
\mathcal{L}=-\frac{1}{2} \left( |\partial_\mu \Psi| + m^2 |\Psi|^2 \right) - \frac{\Lambda ^4}{6 \left( \Lambda_{\mathrm{c}}^2 + |\Psi|^2 \right)^6} \left( |\partial_\mu \Psi| + m^2 |\Psi|^2 \right)^3\,.
\end{equation}	
The scale $\Lambda_{\mathrm{c}}$ was introduced in order for the theory to admit $\Psi=0$ vacuum.  It is easy to see that this action reduces, in the non-relativistic limit and when $\Lambda_{\mathrm{c}} \ll |\Psi|^2$, this action gives (\ref{L_SF}). The condition for MOND, given by $\Lambda_{\mathrm{c}}$ can be rewritten as $|X| \gtrsim \Lambda_{\mathrm{c}}^4/(2m\Lambda^2)$, which corresponds to:
\begin{equation}
a_{\phi} \gtrsim \frac{\Lambda_{\mathrm{c}}}{\alpha^2 \Lambda} a_0\,,
\end{equation}
where $a_\phi$ is the acceleration from the phonon that can be obtained from the action and given by $a_\phi=\alpha (\Lambda/M_{\mathrm{pl}}) \phi^{'}$. According to observations, the deep MOND regime is very accurate for $\sim a_0/10$, which poses a bound for $\Lambda_{\mathrm{c}}$.

This theory presented here is a phenomenological  relativistic version of the DM superfluid. However, it would be interesting to have a relativistic complete microscopic theory for the superfluids.

\subsubsection{Cosmology}

After working out the galactic behavior of the DM superfluid model, we need to understand what happens on cosmological scales in this model.  In this section we show how DM superfluid behaves cosmologically.  Since we do not have a proper relativistic theory that we can use to describe cosmology, we make some estimates in order to understand the behaviour of DM on cosmological scales in this theory.

The first question we would like to answer is if DM is in the superfluid or normal phase cosmologically. We saw in Sect.~\ref{Sec.:conditions} that the critical temperature of the DM superfluid is given by (\ref{T_crit}), and $T/T_{\mathrm{c}}$ today is around $10^{-2}$ for massive galaxies ($M \sim 10^{12} M_\odot$). Cosmologically, the temperature is much colder. We can estimate given that ultra-light candidates for DM, like the DM superfluid, are non-thermal relics.  They can be generated, for example, through a vacuum displacement mechanism (see below for a definition of this mechanism) like the axion. So, the particles are created when $H_i \sim m$, which corresponds to a temperature for the photon-baryon plasma:
\begin{equation}
T_i^{\mathrm{b}} \sim \sqrt{m M_{\mathrm{pl}}} \xrightarrow[m \sim \mathrm{eV}]{}  \, 50 \mathrm{TeV}\,,
\end{equation}
which is around the weak scale! 

With that, we can rewrite the condition for thermalization (\ref{cond1}), given that the velocity and density redshifts as $v\propto a^{-1}$ and $\rho \propto a^{-3}$. At matter-radiation equality, we can write this condition as:
\begin{equation}
m \sim \rho_{\mathrm{eq}}^{1/4} \ll \left( \frac{\rho_{\mathrm{eq}}}{v^3_{eq}} \right)^{1/4}\,,
\end{equation}
where $\rho_{\mathrm{eq}} \sim 0.4 \, \mathrm{eV}^4$ and by using that $v_{\mathrm{eq}}  = v_i a_i / a_{\mathrm{eq}}  \sim \mathrm{eV}/ \sqrt{m M_{\mathrm{pl}}}$ is much smaller than one, while $v_i \sim 1$ since it was created deep into the radiation era. Since $T/T_{\mathrm{c}}  = (v/v_{\mathrm{c}})^2$, we have that:
\begin{equation}
\left( \frac{T}{T_{\mathrm{c}}} \right) _{\mathrm{cosmo}} \sim v_{\mathrm{eq}} \left(  \frac{m}{\mathrm{eV}}\right)^{8/3} \sim 10^{-28} \left(  \frac{m}{\mathrm{eV}} \right)^{5/3} \,.
\end{equation}
So, cosmologically, all the DM is in the superfluid state. Once this DM is formed, if condenses and behaves as a superfluid. However, one question than comes to mind: as we saw previously, in the DM superfluid model in the superfluid phase the dynamics is give by a MONDian dynamics, instead of Newtonian. Then the question is: does DM behaves differently than Newtonian on cosmological scales, which can be a problem to reproduce some known results in cosmology? We can see that this is not the case.

The cosmological temperatures are many orders of magnitude different than the temperatures on galaxies. For the EFT built for the superfluid to be valid on such different scales, the parameters of the EFT $\Lambda$ and $\alpha$ need to evolve with the temperature. This dependence is estimated in \citep{Berezhiani:2015bqa} by making some phenomenological statements  for the theory to match both regimes. A consequence of the introduction of this variation is that the critical acceleration, given by $a_0$ in galaxies, is now temperature dependent. Therefore, on cosmological scales the critical acceleration of the theory has a much smaller value than the one from galaxies: $a_0^{cosmo} \ll 10^{-4} \, a_0$. This has an important consequence: although the DM superfluid is condensed on cosmological scales, the gravity is highly Newtonian on those scales.

This shows to us that a very compelling feature of this model: at the same time it describes the small scale behavior, given by a MOND-like dynamics, it also recovers the large scale successes of CDM.  It also recover the expected CDM behaviour in clusters, and also in the vicinity of star like in the solar system, where the DM is in the normal phase behaving like particle DM. 

\vspace{0.5cm}

To close the DM superfluid section, we can summarize the global behaviour that this model has on all the scales. This together with Table~\ref{observationssummary} describes the phenomenology of the DM Superfluid class.
\begin{table}[htb]
\centering
  \begin{tabular}{l@{\,$\longrightarrow$\,}l@{\,$\longrightarrow$\,}l@{\,$\longrightarrow$\,}l}
%    \toprule
       \begin{tabular}{c}
       \textbf{Cosmological Scales} \\ \textcolor{DarkOrchid}{Condensate} \\ \textit{No} MOND
	   \end{tabular}          & 
       \qquad        
       \begin{tabular}{c}
       \textbf{Clusters} \\ \textit{Mostly no} condensate \\ No MOND
	   \end{tabular}   & 
         \begin{tabular}{c}
       \textbf{Galaxies} \qquad \\ \textcolor{DarkOrchid}{Condensate} \\ \textcolor{RoyalBlue}{MOND}
	   \end{tabular}   & 
        \begin{tabular}{c}
       \textbf{Solar System} \\ \textit{No} condensate \\ \textit{No} MOND
	   \end{tabular}     
%     \bottomrule
  \end{tabular}
\end{table}

%%%%%%%%%%%%%%%%%%%%%%%%%%%%%%%%%%
\subsection{Simulating ULDM models}
\label{Sec.:simulations}
%%%%%%%%%%%%%%%%%%%%%%%%%%%%%%%%%%

We described above the main characteristics of the ULDM models. We showed how we expect the small scale structures to be suppressed in this model by computing quantities in the linear limit, and showed how this model presents a core solution for a simplified model of the halo. 
However, to study the formation of structures  at different scales, and the formation of galaxies which are highly non-linear processes, one needs to resort to simulations. 
%However,  the formation of structures  at different scales, we need to describe the full non-linear regime. 
%However,  to study the formation of structures  at different scales, specially the small scales of halos and galaxies which are highly non-linear, one needs to resort to simulations. 
Cosmological simulations have been one of the biggest tools for the understanding of the non-linear formation and evolution of structures and  galaxies in the past few years, modelling diverse scales and physical process that are present in those processes.
Therefore, to better understand how different the structures are going to be in the ULDM models, together with modelling some structures that are exclusively present in these constructions like the presence of cores, interference and vortices, we need to resort to cosmological simulations (for a summary of the current FDM simulation, see \citep{Zhang:2018ghp,Li:2018kyk}).

The traditional simulation methods present in the literature to study the formation of structures, like N-body simulations or hydrodynamical simulations, cannot be readily applied to the case of ULDM since they do not take into account the wave nature of these models. As we saw above it is this wave nature that can lead to important observational consequences present in these models of DM.

There are two approaches to simulate the ULDM models. One solves the Schr{\"o}dinger--Poisson system, composed by the Gross--Pitaevskii equation for a given ULDM model coupled to the Poisson equation; and the other is given by solving the hydrodynamical equivalent of the GP equation, the Madelung equations. Each of those approaches have advantages and disadvantages, so they can be considered complimentary.

\vspace{0.2cm}
\begin{center}
\boxed{
\begin{tabular}{l|l}
Schr{\"o}dinger--Poisson & \hphantom{0000} Hydrodynamical-Madelung  equations \\
$\displaystyle i \dot{\psi} = -\frac{1}{2m} \nabla^2 \psi +  m\Phi \psi + {\color{BlueViolet} \frac{g}{8m^2} \, |\psi|^2 \psi + \frac{g_3}{12m^3} \, |\psi|^4 \psi + \cdots}$  &  \hphantom{0000} $\displaystyle \frac{\partial \rho}{\partial t} + \nabla \cdot (\rho \mathbf{v}) = 0$ \\
$\displaystyle \nabla^2 \Phi = 4 \pi G\, m \left(  |\psi|^2 - |\bar{\psi}|^2  \right)$ & \hphantom{0000} $\displaystyle \frac{\partial \mathbf{v}}{\partial t} + \rho \, (\mathbf{v} \cdot \nabla)\mathbf{v} = -\frac{1}{m} \nabla \left( P_{\mathrm{QP}} + \Phi + {\color{BlueViolet}  P_{\mathrm{int}}} \right)$\,.
\end{tabular} 
}
\end{center}

\vspace{0.5cm}

In the absence of interaction, we have the Fuzzy DM model, and in the presence of interaction, the terms marked in blue, we have the SIFDM model. The interactions can be a two-body or three-body interactions, like showed respectively in the last two terms of the GP equation above, or even higher order body collisions represented in the ellipsis, if this is allowed in the system. These describe fluids with different equation of state. In the Madelung equations, the interactions are represented by the interaction pressure term $P_{\mathrm{int}}$, which has the form of a polytropic fluid. This polytropic fluid can describe different fluids with different equations of state, that arise depending on the type of interaction.
So one can simulate either one of those models using these system of equations\footnote{Theoretically, one could also simulate a gravitationally bounded DM superfluid model by evaluating its the equation of motion in the NR regime, obtaining a equation analogous to the GP equation, but with a more complicated form. There is not, to our knowledge any group that is performing simulations of those models at the moment. We might expect that the more complicated form of the equations might render a more computationally expensive simulation. For this reason we stick for now to the simulations of the other classes of models.}. 

\begin{figure}[htb]
\centering
\includegraphics[width=0.4\textwidth]{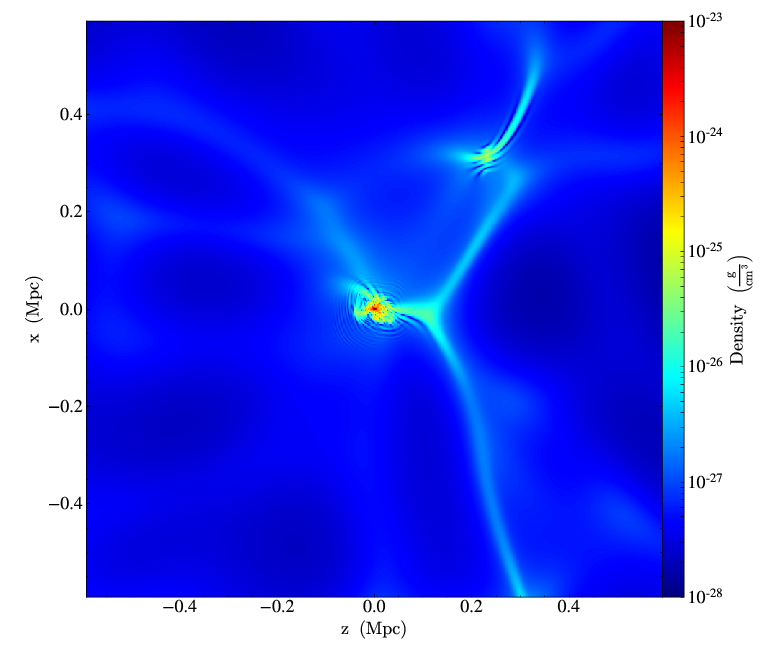}
\includegraphics[width=0.4\textwidth]{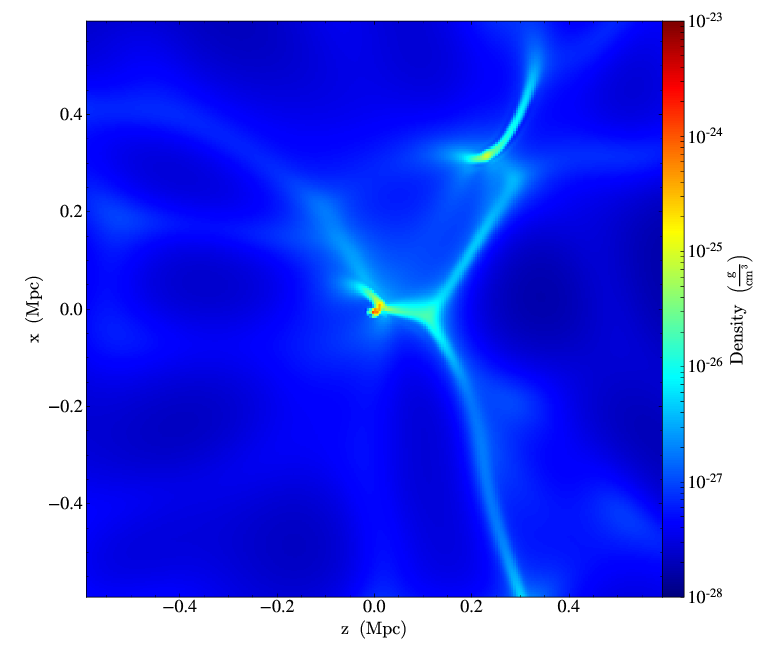}
 \caption{Figure from \citep{Li:2018kyk} shows wave simulation (\textit{left panel}), where the SP system is solved, and a fluid simulation (\textit{right panel}), where the Madelung equations are solved, of  the FDM model with mass $m = 10^{-23}\,\mathrm{eV}$. This represents a slice at redshift $z=5$ of the density, given by the color code, and it shows how the wave simulation can resolve the small scales by showing the interference patterns, that are coarse grained by the fluid simulations. }
\label{Fig.:sims_comparison}
\end{figure}

For the simulation that solve the Schr{\"o}dinger--Poisson equations, also called wave simulations, there are many groups that are attempting to solve this system using different methods \citep{numerical_fuzzy_1,numerical_fuzzy_2,SP_solver3,SP_solver4,SP_solver5,Garny:2019noq}. In general this approach is very good to describe the small scales, being able to resolve the small structures and taking into account the wave nature of the condensate, as we can see in the left panel of Fig.~\ref{Fig.:sims_comparison}. With that, this approach predicts the correct and expected structures on small scales. Being able to resolve the smaller scales, this simulation can resolve and show  the presence of cores in the halos, the granular interference structure in the condensate or the presence of vortices. However, this approach is very demanding numerically, since it requires a more finely resolution in order to resolve scales of the order of the de Broglie wavelength. This makes these simulations to be much smaller in size than the fluid ones, not describing cosmological scales or being able to span many decades in redshift.

%%%%%%%%%%%%%%%%%%%%%%%%%%%%%%%%%%%%%%%%%%
%%%%%%%%%%%%%%%%%%%%%%%%%%%%%%%%%%%%%%%%%

The simulations that solve the Madelung equations, the fluid simulations, have the advantage of being able to use the already written and well explored hydrodynamical codes available in the literature. They can be implemented by adapting those known codes to the case of FDM or SIFDM. The difference from a normal fluid simulation is the presence of the quantum pressure term, $\nabla P_{\mathrm{QP}}= -n \nabla Q$, where $Q=(\hbar^2/2m) (\nabla^2 \sqrt{\rho})/\sqrt{\rho}$. This term is singular when the density is zero, and the quantum pressure is not well defined in this regime. This restriction translate into those simulation not being able to resolve the smallest scales, coarse graining through the granular structure or any other substructure expected in these models. This leads to fluid simulations predicting a more pronounced gravitational collapse leading to an enhancement bigger than expected in the power spectrum at small scales~\citep{Li:2018kyk}.

The advantage of the fluid approach, though, is not only being implemented using the already mature hydrodynamical codes, but also being able to run much larger simulations than in the wave case, since it is less computationally expensive. With this method cosmological size simulations are possible. Many groups have been working on simulating the FDM using the fluid approach, with some variations in the implementations of the codes and the solvers~\citep{numerical_fuzzy_4,fluid_sim1,fluid_sim3,fluid_sim4}.

Many research groups are attempting to perform those simulations so we can better understand  the behaviour of the FDM model and reveal possible smoking gun signatures of this model.  
Those simulations are crucial so we can understand and better search for these signatures on observations. For this it would be interesting to have the small and large scales of the simulations resolved. Since the fluid simulations are good to describe the large scales and the wave simulations the small scales, some groups are exploring the possibility of having  hybrid simulations where both methods are considered for the scales they work better~\citep{Li:2018kyk}. In~\citep{Veltmaat:2019hou} another hybrid method is considered where N-body simulations are used to simulate the cosmic web, while the wave simulation is implemented in the inner halo. 

Another simulation that is also hybrid is the {\tt{AxioNyx}} simulation~\citep{Schwabe:2020eac}. This simulation actually mixes different dark matter models, having a Schr{\"o}dinger-Poisson solver built on a cosmological N-body simulation of CDM and baryons to simulate self-gravitating mixed fuzzy and CDM. This allows to include in this simulation baryonic effects and some astrophysical processes. With this simulation one can study spherical collapse and core formation in this mixed DM context. This mixed DM nature of the simulation can be seen in a suppression in the CDM collapse due to the FDM fraction, while CDM delays the the FDM collapse shrinking the Jeans scale. It was also found that both FDM and CDM evolution respond to the same gravitational potential, although in the center of the overdensities, where solitons form depending on the granule mass which is determined by the FDM fraction, a large fraction of FDM particles of about $10\%$ of the total DM is present.

More work in the presence of interactions, describing the SIFDM, would also be welcomed, since this would reveal more about the superfluid nature of this DM scenario and the possible consequences of having a superfluid core in the inner regions of galaxies.

Simulations of ULDM models are a fast moving and essential field to study these models, and the current advances are very exciting, which makes us look forward to new results in the near future. 

%NR, but there is also relativistic...long literatura
We are focusing in this review on dark matter, in the late time non-relativistic limit 

\paragraph*{SIFDM/Superfluid - \qquad}

There are a smaller number of simulations for the case of the self-interacting ULDM.  It is important to also consider systems where self-interaction is present, on top of gravitational interaction, since these systems contain a very rich phenomenology.   Depending on their sign these models present different clustering scales, soliton formation, which in this case can be driven by the self-interaction instead of gravity alone, and its rates, times and sizes; and effects like dynamical friction work differently. 
% When self interaction included, 3 scales

\begin{figure}[htb]
\centering
\includegraphics[width=0.45\textwidth]{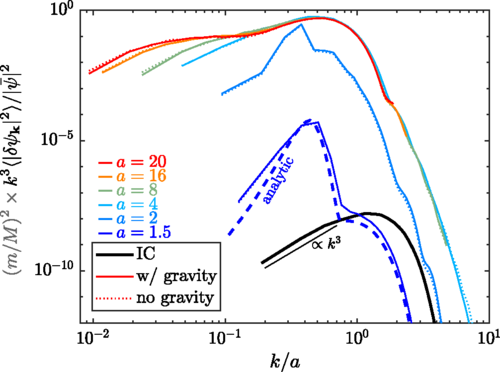}
\includegraphics[width=0.44\textwidth]{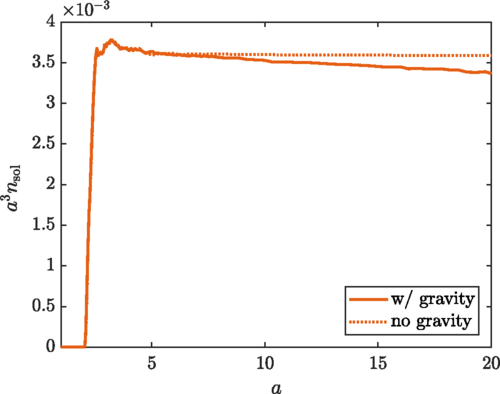}
 \caption{Figures from~\citep{Amin:2019ums}. \textit{Left panel:} Power spectrum of $\Psi$, with adiabatic vacuum fluctuations initial condition, for the case with and without gravity.Growth of perturbations occur first through self-interaction instability, backreacting in the condensate, followed by soliton formation. \textit{Right panel:} Comoving number density of solitons in simulations with and without gravity. The fact that when gravity is included effects like mergers occur makes the number of solitons to be smaller in the case gravity is included.}
\label{Fig.:soliton_interaction}
\end{figure}

% Mustafa's paper
In~\citep{Amin:2019ums}, the system of a non-relativistic scalar field  in the presence of two-body self-interaction in an expanding universe was investigated. The main goal of the authors is to study the formation, clustering and collision of solitons when their formation is controlled by attractive self-interactions, on top of the gravitational part. They work in the regime where the scales of the problem present the following hierarchy: $m \ll M \ll m_{pl}$, where $m$ is the mass of the ULDM particle, $M$ controls the interaction scale, and the reduced Planck mass determines the strength of gravity.
In this work they solve numerically the Schr{\"o}dinger-Poisson system in the presence of interaction in $3+1$ dimensions in the presence of expansion, with cosmological initial conditions. They compare the result of this simulations with analytical results calculated for the soliton formation time and length scales, the soliton distribution and two-point function of the clustering of the solitons, showing good agreement between both.  The problem then has two instability scales from the self-interaction and from gravity, as shown separately in Section 4.1. We show in left panel of Figure~\ref{Fig.:soliton_interaction} the power spectrum of the scalar field. The formation of solitons controlled by the interactions is faster than under gravity alone, as seen in the right panel of Figure~\ref{Fig.:soliton_interaction}.  In the presence of those two components, the solitons scatter, merge and form binary systems, the last only present in the presence of gravity. This shows that the system with interactions presents significant different phenomenology than in the case of FDM.  Therefore, more efforts to simulate these interacting BECs is necessary, specially in the case of repulsive interaction.
%
%%%%%%%%%%%%%%%%%%%%%%%%%%%%%%%%%%%%%%%%%%
%%%%%%%%%%%%%%%%%%%%%%%%%%%%%%%%%%%%%%%%%

Another interesting discussion present  in this paper is the connections to the equivalent relativistic system: a relativistic non-linear Klein--Gordon equation. In the relativistic system the equivalent to the soliton is the oscillon.  In this review we are interested in DM, so discussing simulations of relativistic system is not in the scope of the paper. But it is worth mentioning that there is a large body of literature in this topic (see for example~\citep{Kolb:1993hw,Amin:2011hj,Lozanov:2019ylm}). They describe the mainly the early universe evolution, but are directly applicable in the context of ULDM.

\paragraph*{Two-fluid approach: \qquad} As we mentioned in Sect.~\ref{Sec.:BEC_superfluids}, based on the ideas by Landau, the correct description of a finite temperature superfluid is using the two fluid model, where there is a normal and a superfluid components. The description used above for the SIFDM and DM superfluid extend the limit of usage of the zero temperature description of the superfluids when describing DM in a halo, which should be treated with a finite temperature approach. An effective extension to include  finite-temperatures was attempted for the DM superfluid case.
When simulating the superfluid model, one should use the equations for a superfluid in the two-fluid formalism, where the system is described as a mixture of a superfluid component (represented with subscript '$s$') and a normal fluid (represented with subscript '$n$'). The superfuid is the component that flows without friction while the normal fluid is the only of the two that transports entropy and thermal energy.

This is what it was done in \citep{Hartman:2019hvl}. They want to study structure formation in the a model where DM forms a superfluid. For this reason, they solve the hydrodynamical equations that describe this superfluid. The finite-temperature hydrodynamical equations for the simplest superfluid, the weakly interacting Bose gas in a trapped potential, are given by \citep{hydro_twofluids1,Chapman:2014hja}:
\begin{align}
\frac{\partial \rho}{\partial t} + \nabla \cdot \vec{j} = 0\,,  \qquad \frac{\partial S}{\partial t} + \nabla \cdot (S\vec{u}_n) = 0\,,  \qquad \frac{\partial \vec{u}_s}{\partial t} + \nabla \left( \mu + \vec{u}_s^2/2 \right) = - \nabla \Phi\,,\\
\frac{\partial \vec{j}}{\partial t} + \nabla P + \rho_s (\vec{u}_s \cdot \nabla)\vec{u}_s+ \rho_n (\vec{u}_n \cdot \nabla)\vec{u}_n + \vec{u}_s \left[ \nabla \cdot (\rho_s \vec{u}_s) \right] + \vec{u}_n \left[ \nabla \cdot (\rho_n \vec{u}_n) \right] = - \rho \nabla \Phi\,, \\ 
\frac{\partial E}{\partial t} + \nabla \cdot \left[ \left( U + \frac{1}{2} \rho_n u_n^2 + P \right) \vec{u}_n + \frac{1}{2} \rho_s u_s^2 \vec{u}_s + \mu (\vec{u}_s - \vec{u}_n) \right] = - \vec{j} \cdot \nabla \Phi \,,
\end{align}
where $\rho=\rho_s + \rho_n$ is the fluid mass density, $S$ the entropy density, $\vec{u}$ is the velocity,  $\vec{j} = \rho_n \vec{u}_n+ \rho_s \vec{u}_s$ is the momentum density, $E$ is the energy density E, $\Phi$ the gravitational potential, following the Poisson equation and $\mu = [P + U -ST -(\vec{u}_s - \vec{u}_n)^2/2]/\rho$  is the chemical potential ($P$ is the pressure, $U$ the internal energy density and $T$ the temperature).

In this set of equations, the authors set the trapping potential to be the gravitational potential and these equation can be re-written in an expanding background by transforming to super-comoving coordinates and using $\vec{v} = \vec{u}-H\vec{r}$. In this way, the model describes the finite-temperature hydrodynamical equations for the SIFDM model. The authors test the two-body and three-body interactions for comparison.

One challenge of this approach is that the hydrodynamical equations do not include the Landau criteria in them. To include this explicitly one needs to add dissipative terms. Since this is not completely worked out analytically, to do that one has to make some assumptions and assume a form for these terms.
A very big step in including dissipation in this theory was worked by \citep{Berezhiani:2019pzd,Berezhiani:2020umi} and shown above in the ``dynamical friction'' subsection of the DM superfluid. This is a work in progress. Until this is worked out in detail, and avoiding adding unknown dissipative terms, the authors chose to do this numerically imposing Landau's condition at every position in the simulation.  

They numerically integrate the hydrodynamical equations from redshift $z=100$ until today. The authors found from the simulation that, in  SIFDM model, the growth of structure proceeds less efficiently than in CDM, as expected for the ULDM models, although more efficiently than expected, with the suppression more pronounced on small scales and at high temperatures. They also study the role of the interaction strength and of the equation of state. 

This numerical simulation presents some limitations giving some of the assumptions and limitations inherited from the hydrodynamical formalism. For example, in this approach we cannot see the complete dissipation of the superfluid.  Also, correctly adding the dissipation physics would be a big improvement in this description. However, this simulation represents a very important step towards simulating more realistic superfluids, which already showed to lead to interesting observations consequences.

%%%%%%%%%%%%%%%%%%%%%%%%%%%%%%%%%%
\subsection{ULDM as dark energy} 
%%%%%%%%%%%%%%%%%%%%%%%%%%%%%%%%%%

Ultra-light fields can also behave as dark energy (DE), depending on their mass and on the different theory they are applied. We can see this in two different cases where the ultra-light field can be used to explain the acceleration of the universe.

\subsubsection{Fuzzy DM} 

In the case of fuzzy DM, where we have a  (non-interacting) ultra-light scalar field in FRW universe, the  field can behave as dark matter, early dark energy or dark energy depending on the mass of this field. The behaviour of the field depends on how its mass is related to the Hubble parameter.
At early times, the ultra-light field has $ m \ll H$. In this regime, the field is almost frozen and behaves as DE with $w \approx -1$. As $m \sim 3 H$, the field starts to coherently oscillate around the potential minimum and to behave as DM, where the equations of state averaged over the oscillations approaches zero. Depending on the mass, the oscillating DM phase can happen at different times of the evolution of the universe. 

If the oscillating phase happens after radiation-matter domination, the ultra-light field behaves like DM. From \citep{Hlozek:2014lca}, we see this happens for $m \gtrsim 10^{-27} \, \mathrm{eV}$.  This bound comes from observations from the CMB and LSS galaxy clustering, and in this regime the ultra-light field can oscillates before the present day and the field redshifts as DM. More generally, for $m \lesssim 10^{-32} \, \mathrm{eV}$ we can still have the ultra-light field behaving as DM, since the field starts oscillating before the present time, but in this case the ultra-light field can only make a fraction of the DM.
%In this case only a fraction of DM cold be given by this field, and only if $m \gtrsim 10^{-27} \, \mathrm{eV}$, the ultra-light field behaves as DM since radiation-matter equality.

For $m \sim  10^{-33} \, \mathrm{eV} \approx H_0$, the field behaves like a  \emph{quintessence} field, if in the presence of a potential, and it can be the responsible for the late time acceleration. Since the field is virtually frozen until the present time, or slowly-rolling the potential, with almost constant density, and it behaves very closely to a cosmological constant. With observations from CMB, we can see that in this case the ultra-light fields have a maximum bound on the energy density compatible with the expected amount of DE in the universe.
For masses around $10^{-32} \, \mathrm{eV} \lesssim m \lesssim 10^{-27} \, \mathrm{eV}$, the ultra-light field behaves like DE earlier than what it is expected, and can be thought as an early DE component.
%They can still behave as DE for masses around $10^{-32} \, \mathrm{eV} \lesssim m \lesssim 10^{-30} \, \mathrm{eV}$, but in this case the ultra-light field cannot make a significant fraction of the dark energy.

\subsubsection{Superfluid DM - Unified superfluid dark sector}
\label{Sec.:unified}

There is another way to explain the  late time acceleration using these ultra-light fields, where the acceleration is not given by this behaving like a quintessence field. This can be done in the context of the DM superfluid, where the dark energy behaviour is yet another manifestation of the same superfluid that emerges at cosmological scales at late times, as presented in \citep{Ferreira:2018wup,Ferreira:2019wqg}.

In the previous case and in the case of quintessence,  a ultra-light field is a component with a very small mass $m \approx 3 H_0$, that dominates around the present times and drives the acceleration. Differently, in the framework we present here, the dark sector is composed only by DM described by a superfluid, without DE. The late acceleration emerges from the dynamics of this superfluid, and we have a \emph{unified} model for the behaviour of DM and DE. We can see how this arises from the model described below.

BECs and superfluids in the laboratory are usually made of atomic species. After the discovery of BEC and superfluidity, one big evolution in the study of these systems was to study mixtures of condensates where atoms that compose the fluid that condenses might be at different atomic configurations. This allowed researchers to explore the richness of the internal structure of the atoms that compose the superfluid, describes a more realistic system, where atomic transitions are allowed to happen, and also the different dynamics that appears in systems where more than one species of condensate and superfluid is present.

In this entire section we are using the same formalism as the one defined for the DM superfluid (and the EFT of superfluids from Section 3.4. In the model we present here, we assume that the superfluid  is composed by two different species, which can be represented by the ground ($\Psi_1$) and excited ($\Psi_2$) state of the dark atom that composes the superfluid. These species interact through a Josephson interaction \citep{Josephson,Tommasini}\footnotemark, which is a contact interaction between the components of the superfluid that has the simple form $\mathcal{L}_{\mathrm{int}}\propto -(\Psi^{*}_1 \Psi_2+\Psi_1 \Psi^{*}_2)/|\Psi_1||\Psi_2|$. So the Lagrangian of this theory is the Lagrangian for the EFT of superfluids, given by a non-canonical kinetic term, plus the term coming from the interaction. This interaction leads to an oscillatory potential for the low-energy Lagrangian of the phonons,
\begin{align}
  &\mathcal{L} = P_1(X_1)+ P_2(X_2) - (1-2\Phi) \, V(\theta_1,\theta_2 )\,, \quad \mathrm{with} \nonumber\\
  &V(\theta_1,\theta_2 )=M^4 \, [1+ \cos (\theta_2 - \theta_1 + \Delta E \, t)]\,,
\end{align}
From the form of the interaction term, the oscillatory potential for the phonons is given by a cosine potential,
%5\begin{equation}
%V(\theta_1,\theta_2 )=M^4 \, [1+ \cos (\theta_2 - \theta_1 + \Delta E \, t)]\,,
%\end{equation}
where $M$ is the explicit symmetry breaking scale coming from the interaction Lagrangian that breaks softly the shift symmetry of the phonon action, and that has to be of the order of $M^4 = 2 M^2_{\mathrm{pl}} H_0^2 \approx \mathrm{meV}$, in order to drive the late time acceleration.
The parameter $\Delta E$ is the energy gap between the two species, between the ground and first excited state of the component of the superfluid.
 
%%% SOLVE PROBLEM WITH FIGURE
\begin{figure}[htb]
\centering
\includegraphics[scale=0.5]{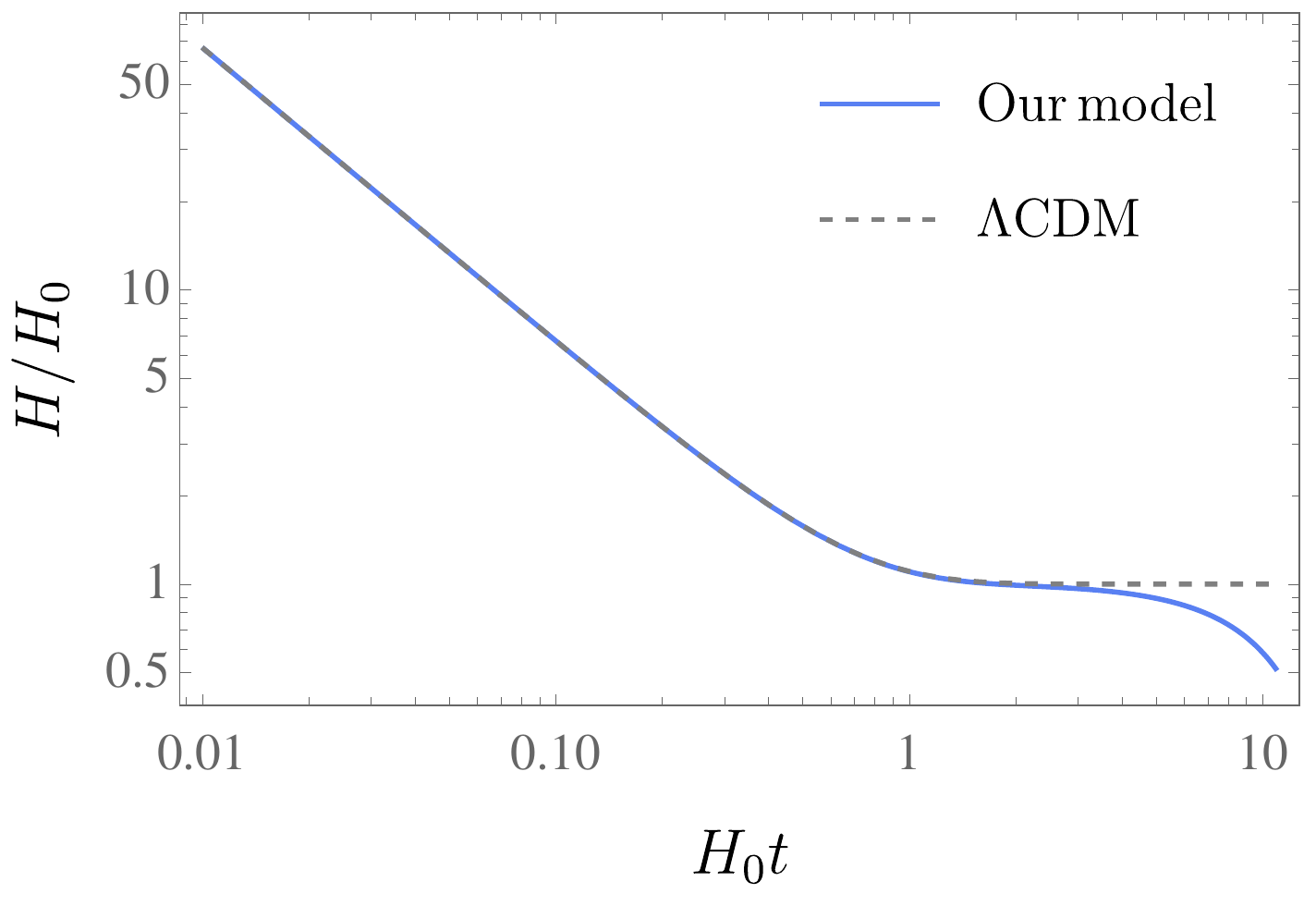}
\includegraphics[scale=0.38]{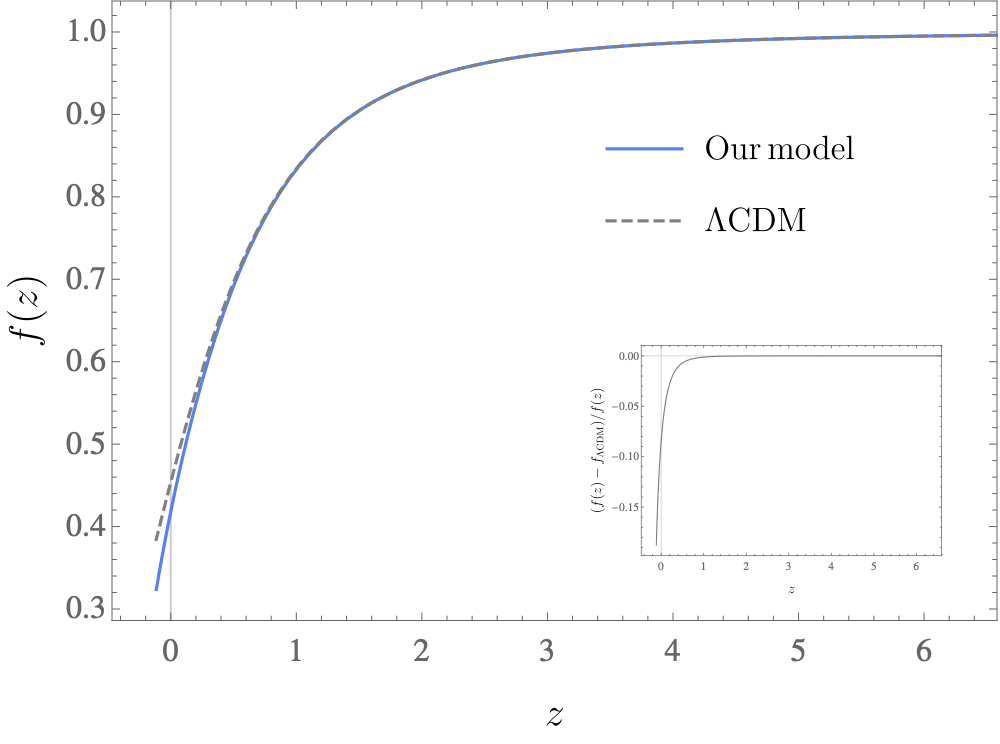}
 \caption{\textit{Left panel:}Evolution of the Hubble parameter for the unified DM-DE model with superfluids in comparison to $\Lambda$CDM. We can see that both model describe the same cosmology given by a period of deceleration where the universe is matter dominated, followed by an acceleration period, around present times. The unified model deviates from the $\Lambda$CDM evolution close to present and for future times, where the action of the oscillatory potential is more pronounced. \textit{Right panel:} Evolution of the growth factor in the unified model in comparison to $\Lambda$CDM, and percent difference, showing that those differ for present times given the potential that describes the accelerated expansion period.}
\label{Fig.:Hubble_unified}
\end{figure}
%\vspace{0.35cm}

Considering the approximation $\Delta E \ll m_i$, we can see that the superfluid has two distinguished behaviours: one degree of freedom that behaves like dust, described only by the non-canonical kinetic term that behaves like DM in the non-relativistic regime as the DM superfluid model, and one that evolves under the influence of the potential, like what is expected from a field that behaves like DE. The cosine potential is similar to the pNGB models of DE \citep{Frieman:1995pm,Kaloper:2005aj}, and it is a special potential for explaining DE since it only softly breaks the shift-symmetry, and the flatness of this potential is still approximately protected against radiative corrections, which is one big problem in quintessence models of DE. The late time acceleration behaviour from this DM superfluid can be seen in the evolution of the Hubble parameter in the NR limit,
\begin{equation}
2\dot{H}+3H^2=V(\theta_1,\theta_2)/M^2_{\mathrm{pl}}\,.
\end{equation}
From the left panel of Figure~\ref{Fig.:Hubble_unified} we can see that we have a decelerated evolution, following the behaviour of DM,  followed by a period of accelerated expansion at present time. Therefore, this model behaves like what is expected by DE, even without the presence of a specific component responsible for the acceleration, and being a model of DM alone. At future times, this models deviates a lot from the predictions of $\Lambda$CDM, as the cosine potential becomes important. 

Although the evolution in this model is very close to $\Lambda$CDM, this model presents distinct predictions. This can be seen by computing the perturbations in this model. One of those predictions is the growth factor, that in this model deviates from the $\Lambda$CDM one by around $10\%$ at present times, as presented in right panel of Figure~\ref{Fig.:Hubble_unified}. Future galaxy surveys might be able to test this deviation.

\footnotetext{The Josephson interaction or Rabi coupling \citep{Josephson,Mahan} is very common in many systems in condensed matter systems. It is a contact interaction that  represents a long-range phase coherence between components, leading to conversion between the different species. This is used in many systems leading to the very well known Josephson effect, but it is also present in other effects studied theoretically and experimentally \citep{J1,J2,J3,J4,J5,J6,J7}.}

%%%%%%%%%%%%%%%%%%%%%%%%%%%%%%%%%
%  Observations
%\section{Cosmological and astrophysical constraints, and addressing the small scale problems}

%\section{Cosmological and astrophysical constraints, and new windows of observation: \\ Solving the small scale problems, constraints and new phenomenology on small scales}

%\section{Cosmological and astrophysical observations, and new windows: solving the small scale problems, constraints and forecasts}
\section{Cosmological and astrophysical constraints, and new windows of observation}
\label{Sec.:Observations}

Now that we have described our ULDM classes and showed the consequences that these models might have in cosmology and astrophysics, in this section we are going to show some of the constraints obtained for the parameters of these theories when the different phenomenology of these models is tested with data.

For most part of this section the constraints are going to be for the FDM model. This model has been much more explored in the literature than the others, not only because it has been introduced first, but also since it has only one parameters $m$ (we are assuming in general that the ULDM is the total mass of the universe, unless stated otherwise).  
However, we are also going to show some constraints obtained for the SIFDM and the DM superfluid models. In each part it is stated for which model the constraints are obtained.

\begin{figure}
\centering
\includegraphics[scale=0.8]{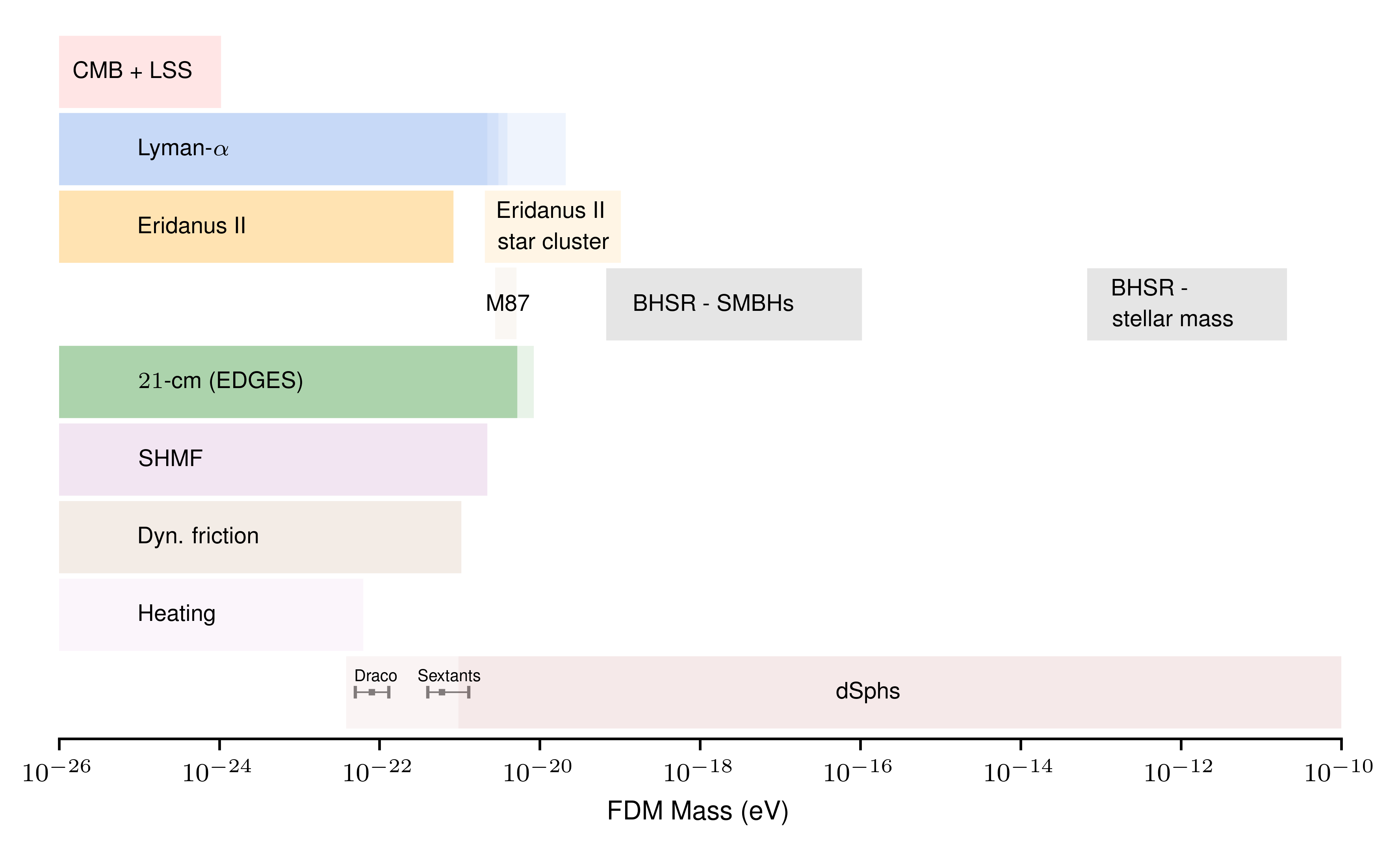}
 \caption{Summary of most of the constraints on the mass of the FDM particle discussed in the section\protect\footnotemark. These bounds assume that FDM makes most of the DM in the universe. In this figure, the shaded regions represent the excluded regions. The CMB and LSS bounds come from~\citep{Hlozek:2014lca,Hlozek:2017zzf} using \textit{Planck} (2015) TT CMB auto-power and the WiggleZ galaxy-galaxy auto-power spectrum. The Lyman-$\alpha$  constrains correspond different analysis made in the literature coming, from the darker to lighter, from~\citep{Nori:2018pka,Lyman2,Lyman1,Rogers:2020ltq}, respectively.  The Eridanus II constraint are both for its existence and for the survival of its star cluster from \citep{Marsh:2018zyw}. The next line presents the constraints from black hole superradiance (BHSR). The first constraint comes from bounds on the spin of the supermassive BH (SMBH) in M87, from the measurments obtained by the Event Horizon Telescope~\citep{Davoudiasl:2019nlo}. The second set of bounds  comes from~\citep{Stott:2018opm}, which presents the stringiest bounds from BHSR of ultra-light particles from stellar BHs and from SMBHs. The global 21-cm signal detected by the EDGES team can also be used to put bounds on the mass of FDM as shown in \citep{21cm_0,Schneider:2018xba}. The next row refers to bounds on the FDM imposed by testing the suppression of the sub-halo mass function in comparison with the SHMF from WDM models constrained using strong gravitational lensing of quasars and from fluctuations in stellar streams~\citep{Schutz:2020jox}. In~\citep{Lancaster:2019mde} they compute the different description that dynamical friction has for the FDM and apply this to the Fornax globular cluster. The next bound comes from another dynamical effect, which is heating of the MW disk, that can be constrained measuring the velocity dispersion of stars in the solar neighbourhood~\citep{FDM_vertical}. We also include two constraints in the mass assuming that the measured central density of dSphs, Draco and Sextants should match maximum FDM core size, which should be smaller then the virial radius of these galaxies~\citep{Chen}. This row also contains the results from the reanalysis of the bounds from dSphs from~\citep{Gonzales-Morales:2016mkl} starting at the lighter region, and~\citep{Safarzadeh:2019sre} the darker shaded region. }
\label{Fig.:constraint_mass_FDM} 
\end{figure}

We summarize most of these constraints on the mass FDM in Figure~\ref{Fig.:constraint_mass_FDM}.   As we can see in this figure, this set of (current) constraint, if they hold, strongly suggests that an FDM with mass of order of $10^{-22}\, \mathrm{eV}$, which was proposed as the ideal mass that introduces interesting new phenomena on small scales and that addresses the controversies that appear in those scales, is \textit{strongly challenged}. For the heavier masses that seem to be allowed now, the phenomenology of the FDM is closer to the one from CDM. 

In the plot we presented only some of the constraints present in the literature. Other bounds obtained from other observables testing different astrophysical consequences of the ULDM models is  presented bellow, together with a description of the bounds of the figure\footnote{The bounds presented here assume that almost all the DM is composed by FDM. If one wants to see an equivalent figure that takes into account the the fraction of the FDM, see Figure 1 from~\citep{Grin:2019mub}. This reference also presents a very good review of the gravitational probes of FDM.}.

We just want to emphasize that this bounds are for the FDM model only, and the SIFDM and the DM superfluid have other sweet spots for the mass of their ultra-light particle. These two models currently are weakly constrained with not a lot of work done in the literature to constraint the parameter of these models. For the DM superfluid, as we saw in Section 4.2, the mass is constrained to be $1.5 \, \mathrm{eV < m < 2.4 \, \mathrm{eV}}$ coming from the thermalization condition in the halo with cross-sections that are allowed by measurements from galaxy cluster mergers.

\footnotetext{For other versions of this figure and a notebook to generate it check~\url{https://github.com/elisaferreira/figure_mass_FDM}.}

\subsection{Cosmological constraints: CMB and LSS}

We are first going to talk about the constraints and forecasts that can be obtained on the ULDM parameters using cosmological observations.

\subsubsection{CMB and LSS}

\paragraph{FDM} \mbox{}\\

We saw in Section 4.1.4 how the angular temperature power spectrum and the matter power spectrum can be affected by the FDM. We show now constraints obtained in this class of models using measurements of the CMB power spectrum and of the matter power spectrum. These constraints are obtained in mainly in two articles. In~\cite{Hlozek:2014lca}, the authors investigated that using a combination of CMB data from the Wilkinson Microwave Anisotropy Probe (WMAP), \textit{Planck} satellite, and also from ground CMB experiments like the Atacama Cosmology Telescope, and South Pole Telescope, and galaxy clustering data from the WiggleZ. And in~\citep{Hlozek:2017zzf} this analysis was updated, and some additional effects were tested, using a combination of data from \textit{Planck} (2013) temperature power spectrum and the WiggleZ galaxy-galaxy auto-power. In these references they investigate the FDM model, where only one particle is responsible for the FDM.

In those two papers they investigate the effects in the CMB and in the matter power spectrum of a large range of FDM masses, encompassing masses where the ultra-light particles behave as dark energy. These two papers also investigate effects that could come from specific models inside the FDM, like the constraining the axion isocurvature modes,  and the spontaneous symmetry breaking scale, which we are not going to discuss in this review.

In these references, the data from CMB and from LSS were combined in order to obtain the constraints. The combination of this data is important to make the constraints on the higher mass end of the FDM tighter, and it is driven mainly by the LSS tight constraints for $k \sim 0.1\, h\, \mathrm{Mpc}^{-1}$. 

\begin{figure}
\centering
\includegraphics[scale=0.4]{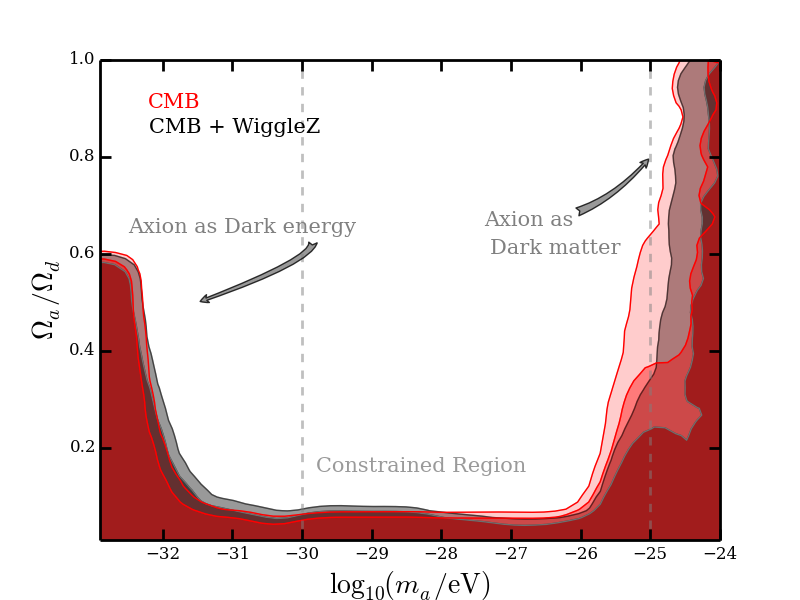}
 \caption{Figure from \cite{Hlozek:2014lca} that shows the $2\sigma$ and $3\sigma$ of the mass fraction $\Omega_{\mathrm{ultra-light}}/\Omega_{\mathrm{d}}$ in function of mass. The regions show the constrained region $\Omega_{\mathrm{ultra-light}}/\Omega_{\mathrm{d}} \lesssim 0.05$ at $95\%$, where $\Omega_{\mathrm{d}}$ is the total dark-matter density fraction.  CMB-only constraints are the red regions, while grey regions include large-scale structure data. }
\label{Fig.:Rene}
\end{figure}

For the low mass end of the FDM, the combination makes the constraints weaker. This results show that, if one wants a DM component that can be responsible for all the DM in the universe, then
\begin{equation}
m \gtrsim 10^{-24} \, \mathrm{eV}\,. 
\end{equation}
If $m \lesssim 10^{-32} \mathrm{eV}$ the ultra-light field can behave as dark energy. This constraints are driven mostly by the expansion of the universe given that a component with this mass modifies the evolution of the universe  after matter-radiation equality and can be strongly constrained by CMB (we will discuss FDM as DE  in  more details in the end of this section). The most up to date constraints are present in reference~\citep{Hlozek:2017zzf}, but we show in Figure~\ref{Fig.:Rene} a very explanatory plot from~\cite{Hlozek:2014lca} where we can see these constrains. The allowed region is the red shaded region.

The region where $10^{-32}\,\mathrm{eV} < m < 10^{-23} \, \mathrm{eV}$ is interesting since is allows to constrain the model a lot. This might not be the region where the ultra-light field can have a sizable DM component, but it allows to put percent level constraints the fraction of the FDM, with CMB being the best gravitational probe in this regime.  In this region FDM cannot be more than $5\%$ of the total DM: $\Omega_{\mathrm{FDM}}/(\Omega_{\mathrm{DM}} +\Omega_{\mathrm{FDM}})  \leq 0.05$ and $\Omega_{\mathrm{FDM}} h^2 \leq 0.006$ with $95\%$ confidence level.

Future CMB experiments, CMB S4, will be much more sensitive to the energy density of the ultra-light particles~\citep{Future_CMB_ULA}. They will be capable of probe different imprints  that  ultra-light particles can leave in the CMB  in the range of masses $10^{-32}\,\mathrm{eV} < m < 10^{-23} \, \mathrm{eV}$. This range is particularly interesting to probe many different aspects of the microphysics of the models that belong to the FDM class, like the axion and ALPs. In the higher mass range of the above interval, this next generation of CMB experiments can provide constraints in the mass that are competitive to small scale observables like dwarf galaxies abundances and mass-halo profiles.

\textit{Optical depth:} The suppression of the structure formation present in the FDM model leads to a possible smaller amount of galaxies at high-z, changing the reionization history. We saw that the optical depth can be changed by the FDM (\ref{optical_depth}). We can then use the optical depth measured by the CMB, $\tau(r_{\mathrm{rec}})$, to constrain the mass of the FDM. This was done in~\citep{HMF_2} where they use the value of the optical depth from \textit{Planck} + WMAP~\citep{Spergel:2013rxa}. They found that a mass $m \lesssim 10^{-23}\, ]\mathrm{eV}$ for all the DM to be FDM is excluded, depending on the model chosen which entails details of the reionization and the luminosity function, at more then $3\sigma$. The standard mass of $10^{-22}\,\mathrm{eV}$ is challenged by these high-z measurements, with results from this mass being on the edge of the allowed parameter space. This shows that non-linear high-z measurements can also be used to constrain the FDM mass, and other ULDM models. However, these bounds carry a lot of astrophysical uncertainties and so need to be considered carefully.

This study opens up an interesting avenue to constrain the FDM model through the modified reionization using CMB. The epoch of reionization can be better constrained by measuring the kinematic Sunyaev–Zel’dovich (kSZ) effect~\citep{kSZ}. The amplitude of the kSZ is sensitive to the duration of the reionization, and could be used to put bounds in the FDM mass. Experiments like Advanced ACTPol (AdvACT) have an improved measurement of the kSZ and help constrain the FDM.

\subsubsection{Lyman-\texorpdfstring{$\alpha$}{a}}

\paragraph{FDM} \mbox{}\\

Recent investigation of FDM models in light of Lyman-$\alpha$ forest finds new constraints on FDM mass~\citep{Lyman1,Lyman2,Lyman_alpha}. It puts a bound in the mass of FDM in the case where more than $30\%$ of the DM being composed by this scalar field of $m \gtrsim 10^{-21} \, \mathrm{eV}$. This value is larger than the value necessary for the FDM model to solve the small scale problems of $\Lambda$CDM, and puts some tension in the FDM model.

%Lyman-$\alpha$ forest is an important probe of the matter spectrum on small scales, on scales of order $0.5 \mathrm{Mpc}/h \lesssim \lambda \lesssim 100 \mathrm{Mpc}/h$. It is produced by the absorption of the light from quasar by clouds of neutral hydrogen localized in the line of sight between the quasar and us.

The Lyman-$\alpha$ forest is produced by the absorption of the light from quasar by clouds of neutral hydrogen localized at low-redshifts, in the line of sight between the quasars and us. For this reason, this  is an important probe of the matter spectrum on small scales, on scales of order $0.5 \, \mathrm{Mpc}/h \lesssim \lambda \lesssim 100 \, \mathrm{Mpc}/h$.

\begin{figure}
\centering
\includegraphics[scale=0.5]{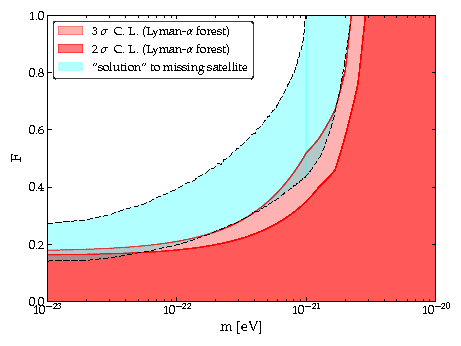}
\includegraphics[scale=0.5]{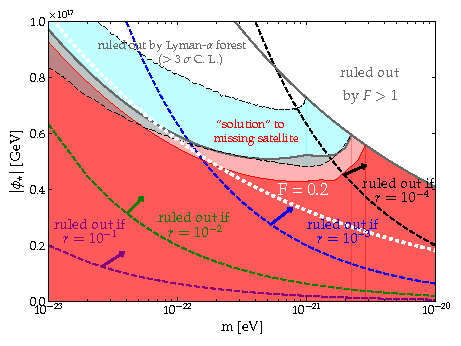}
 \caption{\textit{Left panel:} Shows the constraint on the mass of the FDM and the fraction of the total DM mass from Ly-$\alpha$ forest measurements. \textit{Right panel:} Constraints in the mass of the FDM and the value of the displaced field. This is combined with cosmological constraints, shown by the dashed lines for different tensor to scalar ratios. The region in parameter space where the missing satellite problem is solved for the FDM model, is shown by the cyan stripe. The white dotted contour represents the line where FDM constitutes only 20\% of the total DM. Image reproduced with permission from \cite{Lyman_alpha}, copyright by xxx. }
\label{Fig.:Lyman_alpha}
\end{figure}

In this work, data from the XQ-100 survey was used which refers to $100$ medium resolution spectra with emission redshift $3.5 < z < 4.5$. This data is compared against a simulation of the FDM with different masses and abundance today. On non-linear scales, quantum pressure is added. The result is shown in the left panel of Fig.~\ref{Fig.:Lyman_alpha}. The right panel of this figure we also see the impact of the constraints obtained in cosmology and in the astrophysical implications. In cosmology, the constraints obtained give a bound in the value of the displaced field, assuming that the genesis mechanism for this light field is vacuum displacement. Combining this data with CMB data, they also derive bounds on inflation, more specifically on $r$  the tensor to scalar ratio for an inflationary epoch in the presence of FDM.
They also show how this bound impacts the resolution of the small scale problems presented by FDM. The cyan line indicates the bound where the missing satellites problem is solved by FDM. The constraint is very tight and it shows a tension with the Lyman-$\alpha$ measurements.

A possible caveat from this analysis, and of the other in \cite{Lyman1,Lyman2}, is that they use hydrodynamical simulations, and they mostly neglect quantum pressure. However, quantum pressure can be very important and play a vital role in structure formation, which is what the analytical behaviour seems to show us \citep{Zhang:2017chj}.  In a follow up paper \citep{Nori:2018pka}, quantum pressure was included in the simulations and its effect on the LSS evolution is studied. It is found a constraint in the mass of the FDM model of $ m \sim 2.1 \times 10^{-21} \, \mathrm{eV}$, which is compatible with the ones obtained in the analyses that do not include quantum pressure \citep{Lyman1,Lyman2,Lyman_alpha}. It is found that quantum pressure does not affect the LSS in these redshifts and scales relevant for Lyman-$\alpha$, not affecting the bound on the mass largely. These simulations also allow the authors to study the properties of the halos formed in this model, showing the differences in their distribution and shape in comparison to CDM.

Recently in~\citep{Rogers:2020ltq}, a reanalysis of the Lyman-$\alpha$ data is presented. This is based on a method that emulates the power spectrum from a "training" simulation constructed using Bayesian optimization, which  then is fed in the MCMC sampling of the parameter space. The authors claim that this emulator makes less assumptions than the usual interpolation techniques and for this reason presents a better statistical modelling of the power spectra. With this technique they obtain a even higher bound for the FDM particle: $m > 2 \times 10^{-20} \, \mathrm{eV}$, which disfavours the strongly  $10^{-22}\, \mathrm{eV}$ canonical FDM mass.
This reference presents the newest and most complete study of the bound on the FDM model using Lyman-$\alpha$, and it confirms the tension with the value of the mass that is necessary in order for the FDM model to address the small scale challenges.
 
However, as it is pointed out in these references, the bounds obtained above depend on how the intergalactic medium (IGM) is modelled.  It is expected, for example, that as reionization proceeds in a spatially inhomogeneous manner, fluctuations in the temperature and ionization must be present, and, therefore, the IGM model should consider this effect.
As pointed out in~\citep{Hui:2016ltb} and some of the references above, differences in this modelling like missing the addition of these fluctuations could drastically change these bounds. In the recent analysis from~\citep{Rogers:2020ltq}, however, where they marginalize over physically-consistent IIGM models with different temperature and ionization histories it is argued that this can actually tighten the constraint. They also claim that the current data might be only marginally sensitive to these different modellings of the IGM. How to model an the impact of the IGM modelling in the FDM bounds, remains, then, a question that needs further investigation.
Therefore, new and independent analysis needs to be done in order to confirm if the intermediary to small scales hold more information about these models. Another probe that can help with that is the 21-cm from neutral hydrogen, since it probes even smaller scales. 

Another thing that could change these bounds is the properties of the FDM model.  In~\citep{Leong:2018opi} it was pointed out that for FDM models with a axion-like cosine potential, different initial conditions can yield different bounds on the FDM mass when using the Lyman-$\alpha$ data. Different than in the standard case considered until now where we considered small angles (field values - see discussion bellow (\ref{axion_action}), in the extreme axion misaligned angle is considered,  the transfer function presents a bump for small scales. This affects the Lyman-$\alpha$ flux power spectrum. With this initial conditions, the mass of this  FDM scenario necessary to explain the Ly-$\alpha$ data is of order $10^{-22} \, \mathrm{eV}$, $>10$ times bigger than for the case of standard FDM initial conditions considered above (see~\citep{Arvanitaki:2019rax} for other phenomenology of this extreme FDM model). This shows that the bounds also depend on the properties of the FDM model.

%%%%%%
%\citep{Leong:2018opi}
%%%%%%

%%%% CHANGE PLACE!!
%%%% This bound, together with the other bound from strong lensing and stellar streams, come from very different measurements that are in principle subjected to different systematics. This points that this tension should be considered seriously and studied in more detail.

\subsubsection{21-cm cosmology}

\paragraph{FDM}\mbox{}\\

As discussed above, the ULDM models give a suppression of the matter power spectrum on small scales. Those scales can only be marginally probed by the cosmological probes like CMB, LSS, cluster abundance, Ly-$\alpha$ forest. Those measurements can only constraint the structures on scales $k \leq 10\, \mathrm{Mpc}^{-1}$, not being able to probe the smaller scales. One new window of observation, the 21-cm line from neutral hydrogen (HI), promises to allow us to probe the matter power spectrum in much smaller scales, $k \geq 10\, \mathrm{Mpc}^{-1}$, on scales even smaller than the  Ly-$\alpha$ forest. This is possible since neutral hydrogen is only present in the universe, after reionization,  inside dense clouds in damped Ly-$\alpha$ systems, which are small objects ($k \sim 10^2\, \mathrm{Mpc}^{-1}$). In this way the 21-cm HI signal is a biased tracer of the galaxies, and consequently of the underlying matter distribution in such small scales. 
Therefore,  measuring the global 21-cm HI signal together with its fluctuations can gives information about the still largely unconstrained small scale matter power spectrum.

However, at those scales these system are dominated by astrophysical process, making it difficult to disentangle the behaviour of DM from those processes. To obtain cosmological information from these measurements is a challenge. Some recent studies \citep{Munoz:2019hjh} forecast that the matter power spectrum can be measured from the global 21-cm HI signal for an experiment with parameters close to EDGES \citep{EDGES}, with a precision of $\mathcal{O}(10\%)$ integrated over the scales $k = (40-80)\, \mathrm{Mpc}^{-1}$, after imposing priors on the astrophysical effects like star formation rate and feedback amplitude. They also parametrize the effect of foregrounds, like the Galactic foreground, that plagues all 21-cm experiments and might represent a huge limitations for them if not well mitigated. Detecting the 21-cm HI fluctuations is a much harder task. Large interferometer experiments like HERA (Hydrogen Epoch of Reionization Array) \citep{HERA}, LOFAR (LOw-Frequency ARray) \citep{LOFAR}, LWA (Long Wavelength Array) \citep{LWA}, and SKA (Square-Kilometer Array) \citep{SKA} have the goal of  measuring this signal from the epoch of reionization (EoR) and also late times. For a HERA-like experiment it was found that the matter power spectrum can be probed with an accuracy of $\mathcal{O}(10\%)$ integrated over the scales $k = (40-60)\, \mathrm{Mpc}^{-1}$ and $k = (60-80)\, \mathrm{Mpc}^{-1}$. The measurements of the fluctuations probe the evolution of the matter density tomographically, carrying more information about the scale and redshift dependency of the HI signal,  bringing more information on the astrophysical processes. This makes the parameters of these astrophysical processes to be better disentangled from the HI signal, allowing this constraint in the power spectrum be less dependent on the astrophysics in these regions. 

Specifically for the case of ULDM models, forecasts using 21-cm HI signal were made in \cite{21cm_0,21cm_1,21cm_2}, and they specialize in the FDM model.
Reference \cite{21cm_0} studies how the recent EDGES measurement of the global 21-cm HI  signal \citep{EDGES} can constraint the FDM model. The global signal is the average radio signal from 21-cm redshifted emission from  $z \sim 15-20$ in the case of EDGES. This measurement showed an absorption profile that had an amplitude two times bigger than the expected. This higher amplitude indicates that already at redshift $z=20$, there was significant star formation, which leads to a also significant Ly$\alpha$ background. This fact shows that the smallest structures, and consequently the power spectrum on small scales, cannot be largely suppressed. This puts constraints in models of DM that have the feature of suppressing the small scale structures, like the FDM (or any ULDM model). This measurement alone is capable of putting a challenging constraint in the mass of the FDM particle: $m \geq 5 \times 10^{-21}$. A similar analysis is performed in~\citep{Schneider:2018xba}, where using conservative limits of the stellar to baryon fraction and minimum cooling temperature motivated by hydrodynamical simulation puts a comparable bound in the FDM particle mass: $m \geq 8 \times 10^{-21}$.
To obtain this constraint some assumptions on the star formation, and on the halo mass profile had to be made. Given the importance of this result for the FDM models, the bounds obtained from this data should be explored further, as well as the future data from 21-cm signal.

To better understand how the HI signal is affected by the FDM model, in \cite{21cm_1} they study of the impact of FDM models in the 21-cm HI signal from the cosmic down and EoR analytically, together with some forecasts for future experiments. They use an analytic model which take into account the Ly-$\alpha$ coupling, X-ray heating and ionization to study the 21-cm in a $\Lambda$FDM cosmology. They find that suppression of structure from FDM models, which makes small sub-halos absent in this model, has the effect of postponing the formation of sources and the reionization of neutral hydrogen. This delay changes the global 21-cm  signal showing a smaller absorption feature than expected from $\Lambda$CDM. The amount of suppression allowed considering the results from the EDGES experiment puts a lower bound on the mass of the FDM model, with $m \geq 6 \times 10^{-22}\, \mathrm{eV}$, which might already be considered challenging for the FDM to solve the small-scale problems, but marginally. They also show the potential of a SKA-like and LOFAR experiments to test these models in the future. This shows us that future experiments will be able to confirm the important bound of the FDM model imposed by the EDGES.

In \cite{21cm_2}, they analyze the impact of measurements of the 21-cm forest, which alternatively from the tomographic and power spectrum techniques to use 21-cm HI signal, proposes to use the 21-cm narrow absorption features from the IGM cause by high-z loud radio sources or collapsed objects, like minihalos. The 21-cm forest is expected to be measured by SKA. In this reference they show that the impact of this measurement can also constraint the mass of the FDM model, and that this is degenerate with the fraction of FDM that composes the DM. 

For post-reonization HI signal from $0 < z < 3$, that can be measured using the intensity mapping technique, a study of the forecast of the possible constraints in the FDM model was presented in~\citep{Bauer:2020zsj}. This analysis shows how this signal can be used as a powerful probe of the halo formation, since the halo abundance is changed if a fraction of the DM is given by the FDM. They forecast the constraints in the mass of the FDM for a SKA1MID-like IM experiment. They find that the fiducial value adopted for the FDM ($m = 10^{-22}\, \mathrm{eV}$) can be constrained at the $10\%$ level when the 21-cm data is combined with CMB data from the Simons Observatory.

The possibility of  21-cm HI signal to constrain alternative models to CDM, like WDM for example, was studied in many references \citep{DM_21_1,DM_21_2,DM_21_3,DM_21_4,DM_21_5}. They show how the signal changes for different DM models and also show how measurements like the one from EDGES can put constraints in the mass of WDM.

The study of the capabilities of 21-cm experiments  to give us cosmological information is an active field of study and the references above are just some examples of those efforts. These studies give us hope that, in the near future, this new window of observation will allow us to probe the still unconstrained small scales, helping elucidate the nature of the DM component.

\subsection{Astrophysical constraints and new windows of observation}

We presented above some cosmological  observations that help to constraint the ULDM models. In the past few years, there has been a huge advance in the observations of the small scales, with new windows of observations being opened that can help determine  the nature of DM. We present in this section the constraints on the ULDM parameters coming from astrophysical observations and present some of these new windows of observations that are still being tested and being developed, but that promise to help testing the ULDM models.

\subsubsection{Local Milky Way observables and stellar streams}

Our galaxy, the Milky Way (MW), is our closest source of information about DM and it is a very good laboratory for studying its behaviour on small scales. Here we present some observations from the MW that promise to help us  test different models of DM. We are in a very special era for observations of the MW and Local group with data coming from many current and future experiments like \textit{Gaia}~\citep{Gaia}, Large Synoptic Survey Telescope (LSST)~\citep{LSST}, Prime Focus Spectrograph (PFS)~\citep{PFS}, WFIRST~\citep{WFIRST}, among others. Using the incredible new data from these observations promises to be revolutionary in the studies of the MW, and hopefully for helping discover the nature of DM.

\paragraph{MW's gravitational potential: \,}

Very quickly we would like to point out an observable that can help us test DM. Knowing the shape, mass and distribution of the halo in the MW can gives us clues on the DM model since different DM models predict distinct shapes for the halos. 
To understand the halo we need information not only in the inner regions of the halo, but on many different scales up to the virial radius. The measurement of the position and velocity of satellite galaxies and globular cluster can give information for the dynamics in a good range of distance from the center of the MW. The distribution of satellite galaxies is already used to put constraints on the mass of the ULDM models (and other models of DM), as we saw above.

Experiments like \textit{Gaia}, can give us very accurate data for scales much smaller than the virial radius. PFS galaxy archaeology survey will also measure stars in the galactic disk, complementing complementing \textit{Gaia}'s survey. LSST is expected to provide information from stellar tracers on scales close to $R_{\mathrm{vir}}$, being able to measure many new satellites that are fainter and more distant than the known today, extending the determined halo mass function by three orders of magnitude. 

This is linked to the study of streams discussed above, since streams given their long range in the halo, can give us information on the gravitational potential of the halo for of even larger scales.

\paragraph{Dwarf spheroidals }\mbox{}\\

Dwarf spheroidals (dSphs) are good laboratories to study ULDM models. Those small galaxies are DM dominated and allow us to study the behaviour of DM in an environment with small influence from baryonic effects. They can be used to probe the three classes of effects we saw in Section 4.1.1: the suppression of the power spectrum, effects coming from the core structure inside the galaxy, and dynamical effects.

The suppression of the power spectrum present in the FDM model, leads to a suppression in the low-mass halos. So the FDM model predicts halos with a minimum mass. In~\citep{Nadler:2019zrb}, they used the minimum mass of detected halos from the observed population of satellites in the MW, and found that in order for it to be within the bound (\ref{Eq.:minimum_mass}), the mass of the FDM needs to be $m> 2.9 \times 10^{-21} \, \mathrm{eV}$, setting a lower bound from on the mass caused by the linear suppression.

The effects from the presence of the core in the interior of galaxies can also be probed by dSphs. As discussed in Section 4.1.4, classes of models like the FDM present a limit for the size of the cores that they can form which leads to an upper bound in the central density of these cores. We can use measurement the central density and half mass radius of dwarf galaxies to compare with those bounds and constrain the mass of the FDM.

As we already discussed, in~\citep{McConnachie:2012vd}, half-light radii inferred from the densities of 36 Local group dwarf spheroidals was measured and when compared with the bound on the half mass radius predicted for the FDM (\ref{Eq.:half_radius}), obtained a mass around $m \sim 10^{-22} \mathrm{eV}$ so these are compatible.
The density of $8$ dwarf spheroidals has been measured in~\citep{Chen}. Comparing these central densities  measured with the bound (\ref{Eq.:central density}) it was shown that for the central density from FDM to match the measured ones from the dSphs Draco and Sextants, the mass of the FDM needs to be
 $m= 8^{+5}_{-3} \times 10^{-23} \, \mathrm{eV}$ for Draco and $m= 6^{+7}_{-2} \times 10^{-22} \, \mathrm{eV}$ for Sextans. For those masses, the  FDM leads to a cored   distribution at the center of the galaxies, alleviating the cusp-core problem. This shows that dSphs can be used to put bounds on the mass of FDM.
 
 Nevertheless, new studies have been reviewing these bound. They challenge many aspects of this result including the analysis made presenting some reanalysis, and the assumption of sphericity of the halo of these galaxies.
 
 It is suggested in~\citep{Gonzales-Morales:2016mkl} this analysis might be giving biased values for the FDM mass. The reason for that is because there is a degeneracy between the mass density profile and the anisotropy of the velocity dispersion. When using Jeans analysis to obtain the halo parameters, like the FDM mass, from dSph galaxies that we do not know the density profile, leads to a biased determination if this mass. Therefore, in this paper they use mock catalogues of dSphs hosted in a FDM halo and they conclude that the analysis should be fitting the luminosity-averaged velocity dispersion  of the subcomponents. Using this technique for Fornax and Sculptor, they obtain a  bound in the mass of the FDM with $97.5\%$ confidence of $m > 0.4 \times 10^{-22} \, \mathrm{eV}$, which goes in a different direction than the other constraints.

 Another analysis~\citep{Safarzadeh:2019sre} using the half-mass radius and the slope the mass profile of Fornax and Sculptor dwarf spheroidal galaxies, arrives in a different bound for the FDM mass $m \gtrsim 10^{-21} \, \mathrm{eV}$, in order to have the expected density profile and halo mass for those dSphs from observations like dynamical friction. 
 
 It was pointed out in~\citep{Kendall:2019fep} yet another possible limitations of the above analysis. The authors of this study show that the presence of cores with flat density profile in the center of the NFW-halos can actually make the density of large ULDM halos larger than the CDM ones, making the cusp-core problem worse. This happens because solitons obey the inverse mass-radius scaling law, with mass depending on the total mass of the halo. They perform an analysis that takes into account semi-analytically the variability of the core-halo relation showing that this might make this discrepancy less strong for larger halos. However, this shows that many aspects that are crucial for properly describing these systems, like fluctuations and  baryonic effects, are not present in the semi-analytic model.
 
Another study also challenges this result  based on the fact that the DM halo in dSphs might not be spherical~\citep{Hayashi:2019ynr}. This analysis produces less stringent bounds due to uncertainties in the non-sphericity but brings an important characteristic that should be considered about the DM halo of dSphs.

 All of these studies show that we need to have better understanding of the modelling of those halos and their formation, and need broader observations and  numerical simulations, specially including baryons,  in order to understand and test the FDM class using dSphs. However, they also show how powerful these small galaxies can be to constrain the FDM. Measuring the density profile of dSphs is the goal of many future telescopes like PFS and LSST, for example.

Another class of dwarf galaxies that can be used to probe ULDM model are  the ultra-faint dwarf galaxies. These ultra-fainit galaxies present an even larger mass to light ratio than the dwarf galaxies. Thus, they are important for probing small substructures that can be used for testing any DM models that suppress the presence of substructures below a certain  scale (see~\citep{Calabrese:2016hmp} for an example of such analysis using ultra-faint dwarfs).
Experiments like LSST will also make a more complete search for ultra-faint satellites.

\paragraph{Stellar streams}\mbox{}\\

Stellar streams are a stream of stars orbiting a galaxy which are remnants of a tidally disrupted globular clusters or dwarf galaxies, that was torn apart by a more massive system. These streams are usually thin and very long, extending to dozens of kpc across the 3-dimensions of the halo, and wrap around the disrupting galaxy. Streams are good dynamical probes since they are initially cold and very sensitive to the gravitational potential \citep{s1,s2,s3,s4}. This means that the streams when encounter substrutures present in the halo of the galaxies can be influenced by it, causing  dynamical heating, which are changes in the velocities in the stream, but that are very hard to detect. These encounters also cause disturbances in the morphology of the stream, with the formation of gaps which are underdensities caused by the sub-halos encountered \citep{s4}.
Only a part of the sub-halos hosts baryonic matter and can be observed directly, so stellar streams offer the opportunity to detect dark halos invisible by the traditional methods (see also discussion about lensing below). 
%The dynamical heating is hard to detect since the stream is does not have many stars that are bright enough to be measured with precision. 
These gaps contain information about the substructure and can be caused by clumps that are even less massive than a what is expected of a DM sub-halo, showing how sensitive the streams are for detecting substrucutres. 
From these gaps, it is possible to infer the properties of the perturber that caused the gap, determining quantities like its mass, scale radius, relative velocity, and impact parameter. It is estimated that we can observe gaps in streams cause by substructures with mas as low as~\citep{s5,s6,s7} $M \sim 10^{-5} - 10^{-6} \, M_{\odot}$. This is well bellow the limit where those halos are expected to host galaxies, and for this reason cannot be probed by usual methods based on detecting the luminous component.
On top of that, since the stream extend for large distances in the galaxy and outside the galactic plane, streams might contain detailed information about the gravitational potential and its variations of large part of the halo of a galaxy. 
With this, stellar streams are an exciting new probe of substructures that can have important consequences in testing different models of DM.

Since different models of DM predict a different amount of sub-halos, it is argued that the stellar streams can be used to test DM models.  In the case of the ULDM models, which suppress the formation if substructures, having a much bigger size of minimal subhalo allowed to be created in the galaxy halo. Models like WDM and SIDM also have a modified abundance of sub-halos in comparison with CDM. So there is the hope that these different models would imprint very distinct signals in the streams given their different substructure distribution Streams can also have gaps coming from baryonic substructures, so one needs to be very careful in the analyses not to overestimate the presence of sub-halos.

Up to now, 22 MW stellar streams are known, being the Sagitarius stream one the most important.\footnote{Around 4 streams are known in Andromeda and 10 streams are known outside the Local group.} One of the streams that has been used to determine the presence of sub-halos is GD-1.  The MW stellar stream GD-1 was discovered using SDSS maps \citep{s9}, originated from a globular cluster, and it is seen as a $63^{\circ}$ long structure in the North Galactic region. Gaps of scales of approximately $ 10^{\circ}$ were found in this stream using SDSS and these were associated with the encounter with sub-halos by \cite{s10}. 
 This was confirmed by \cite{s13} analyzing the stellar density perturbations from accurate measurements of the morphology \citep{s11}  of this streams using \textit{Gaia} data combined with photometry from Pan-STARRS \citep{s12}. They found that the data indicates that these perturbations should come from sub-halos, and that their abundance and masses are compatible with the expected from CDM from simulations. The error bars of these abundances and expected masses are still large. In order to detect gaps from subhalos from masses $M \sim 10^{-5} - 10^{-7} \, M_{\odot}$ it is necessary a precision in the determination of the radial velocity of $100 - 300 \mathrm{m/s}$.   However, these measurements inaugurate and opened the avenue for searches of substructures using stellar streams.

Novel experiments like PFS and LSST promise to measure the streams from the MW and Andromeda in more detail and with higher precision, with the goal of not only detecting the signals of substructures but to understand in more detail the characteristics of the stream, its formation and the properties of its progenitor. Forecasts for LSST~\citep{LSST_DM_white} show that the lowest mass subhalo that the 10-year LSST data will be able to measure has mass $2 \times 10^{7}\, M_{\odot}$ for a stellar stream with surface brightness of $33 \mathrm{mag\, arcsec}^{-2}$, improving the current bound in three times. LSS can also measure smaller subhalos ($10^{-5}\ M_{\odot}$) since it will allow to access to smaller angular scales, previously dominated by noise, in order to measure the power spectra of the stream.

However,  a recent publication \citep{Ibata_stream}, revises these conclusions and indicate that the features found in GD-1 can be explained by simple epicyclic motion in a smooth Galactic potential as shown by their N-body simulation, without the need of the presence of a subhalo. Therefore, they conclude that the measurement from GD-1 show no evidence for the influence of DM. This shows that obtaining information about DM substructure in the streams might be more complicated than expected, with some degeneracies with other effects, and that more modelling  and understanding of the influences of the MW in the streams are necessary. This is an open topic of research, and all these conclusions need to be studied further and verified by independent groups. But without a doubt, streams are a promising probe for DM, and the next few years are guaranteed to be very exciting in this fast moving field.

\paragraph{Vertical dynamics}\mbox{}\\

Here we are going to show that the vertical dynamics of stars can be used to test predictions coming from different aspects of the ULDM. We are going to show here that it can be use to probe modifications of the dynamics coming from the MOND behaviour of the theory, in the case of the DM superfluid, or as a way to probe the heating of stars cause by the presence of FDM quasiparticles in the halo. The second effect would also be present in the case of the SIFDM or the DM superfluid since both classes would also present a similar relaxation with macroscopic objects, which can be probed by this effect. But this was only studied in the case of the FDM in the literature, and that is what we present here.

\vspace{0.3cm}
\subparagraph{Superfluid DM}\mbox{}\\

Our galaxy holds more information that can be used to probe DM models. There are several other local MW observables that imprint information from the underlying DM that forms the halo.

One example of one observation that can bring information about the different dynamics imprinted by models like MOND and the DM superfluid is presented in \cite{Lisanti1,Lisanti2}.  MOND and the DM superfluid model are constructed to give a very good fit  to the rotation curves of galaxies and to explain the scaling relations, like the MDAR, by having a different dynamics on small scales. On top of that, the DM superfluid model reproduces the expected CDM behaviour on large scales. However, in these papers the success of those models in explaining the dynamics of galaxies is challenged by observations from the MW dynamics.

The radial dynamics is what it is constrained by those models, since it is the information that enters in the rotation curves of galaxies and the scaling relations. Now, if one consider also the vertical velocities in the vicinity of the Sun,  the authors show that then  it remains a challenge for those models to explain the rotation curves and these vertical velocity dispersion data. MOND and the DM superfluid, in slightly different ways, modify the dynamics on small scales, where for low-acceleration regions, a different acceleration than the Newtonian emerges in the system. In the DM superfluid this is caused by the presence of the phonons, as we saw above. This modified acceleration changes the dynamics in the radial and in the vertical directions. And in these articles they constraint both the radial and the vertical velocities dispersion for these models in comparison to CDM.

To constraint the radial dynamics, the very precise data from \textit{Gaia} was used giving data of the circular velocity between $R=[5,\, 18]\, \mathrm{kpc}$. For the vertical dynamics it was used data of K-dwarfs from the SEGUE sub-survey from SDSS (Sloan Digital Sky Survey) were it was inferred the velocity dispersions for three mono-abundance stellar populations. With this data a Bayesian likelihood analysis was conducted. 
%The authors found that the fit to the rotation curve for the DM superfluid model is  overpredicted for the $r > 18\, \mathrm{kpc}$. However, the biggest discrepancy comes from the conclusion that MOND and the DM superfluid models overpredict the vertical acceleration, giving values that are around $15\%$ larger than the ones inferred from the data and larger than the ones predicted by CDM. And for this reason they claim those models are not preferred as the DM model.
The authors found a discrepancy between the vertical acceleration predicted by MOND and DM superfluid model in comparison to the one inferred from the data, giving values that are around $15\%$ larger than the ones inferred from observations and larger than the ones predicted by CDM. And for this reason they claim those models are not preferred as the DM model.

This work establishes an important new observable that should now be taken  into account when constructing DM models, their impact not only in the radial dynamics, but also in the vertical dynamics.

However, as some authors have pointed out this result should be taken carefully. In \cite{McGaugh:2020ppt} it is pointed out that this is a $2 \sigma$ discrepancy and that maybe the data used to infer the vertical velocities implies a local DM density that is the double of the one inferred from the radial dynamics. This would give a halo that is not spherical, like assumed in the analyses of \cite{Lisanti1}.

This brings an important point about the data for the vertical dynamics of the MW that we would like to highlight. The dynamics of the MW is very complex with a  rich accretion  history.  Although the MW, with its thin disk, is considered a very stable galaxy with no major recent dynamical interactions, many observations indicate a major accretion event with Sagittarius (Sgr) dwarf spheroidal. This is seen in the streams that wraps around the MW, and it is expected that this event strongly influenced the dynamics of the MW. 
In \cite{Laporte:2017res} they show a study of the MW's major accretion events with Sgr and the Large Magellanic Clouds (LMC).
They show that this encounter with Sagitarius produces oscillations in the MW disk. These vertical perturbations are an influence of many passages of Sgr, as it falls into the MW.
The influence of  the  LMC also changes the dynamics, but much less than Sgr, with Sgr being the main influence in the dynamics of our galaxy in its recent evolution. This  shows that the dynamics of the MW is complex and has many unknowns.
These vertical oscillations are also found in recent observations both in the Galactic disk and around it, close and far away the solar neighbourhood (see \citealt{Laporte:2017res} for a summary and discussion of those). Therefore, even if the explanation for these oscillations and other features in the Galactic disk is not solely the one above, this rich dynamics of the MW has to be considered when using these observations, specially to constrain models.

In this way there is the possibility that the vertical dynamics data used in \cite{Lisanti1,Lisanti2} has the influence of these oscillations and cannot be used at face value in order to constraint the dynamics of DM. Or, maybe this complexity of features has to be considered as larger error bars in the inferred vertical dispersion. 
Therefore, vertical velocities are a new and important observable to test DM models, but it necessary to understand if this information can be disentangled from the complex dynamics from the accretion history of the MW. When this degeneracy is resolved this will be an important criteria for DM models to satisfy in the MW.

In \cite{Hossenfelder:2020yko} the authors also point out that the model of the DM superfluid used is too simplistic and that the DM superfluid interacts with  baryons which might also change the vertical accelerations. 

\vspace{0.3cm}
\subparagraph{FDM}\mbox{}\\

In the context of the FDM model, the vertical dynamics was also used to constrain the FDM mass in~\citep{FDM_vertical}, where the change in the vertical dynamics comes from a different effect than the one above. The disk of spiral galaxies is modelled approximately by a decaying exponential for the radial and vertical structure. From observations we see that, in general, we have galaxies that have a thin disk or a  thick disk~\citep{Binney_Tremaine}. What determines the thickness of the disk is the vertical velocity dispersion of the stars in the disk, and its linked to how old the star population is in the disk. As we saw in Section 4.1.4, the presence of order one fluctuations in the FDM field, or substructures in the halo, can cause gravitational heating due relaxation between the FDM particles and the stars~\citep{BarOr}. This heating increases the the velocity dispersion of the stars, making the size of the system that contains these stars larger: star clusters larger or making the disk in galaxies thicker. 
%In contrast, the orbits of the stars in the disk are affected by substructures present in the halo, which might lead to heating, increasing the velocity dispersion of the stars, thickening the disk. 
%Different DM models predict a different distribution of substructures (heating of a population of stars in the context of the FDM model was studied in~\citep{BarOr}). 
In this work they investigate how  substructures of the FDM model affect the disk shape. They use simulations that solve the SP system to understand the structure in the FDM model. They find that the subhalos formed in this model, plus the presence of a standing wave in the density profile of FDM, heat the stars in the disk, making the the MW disc thicker, with the second effect affecting more the inner disk. They also find the presence of radius dependent flaring of stars caused by these structures. They compare this with measurements of the velocity dispersion of the MW in its thickest part ($\delta v \sim 32 \, \mathrm{km/s}$), and put a bound in the mass of FDM at $2\sigma$ confidence level of $m > 0.6 \times 10^{-22}\, \mathrm{eV}$. 

%\subsubsection{Dynamical effects }

\paragraph{Dynamical effect: Eridanus II}\mbox{}\\

%We can use other dynamical effects that happen in the halo of the galaxy to constrain the nature of the DM. 
In the previous section we showed how the measurements of the thickness of the disk can probe the velocity dispersion caused by gravitational heating by substructure in the halo in the case DM is the FDM. Here we are going to show that  Eridanus II can be used to probe the same effect of the gravitational heating.

The existence of  old star cluster like Eridanus II can then be used to constrain these FDM fluctuations, and consequently the mass of the FDM particle. But not only the existence of such sub-halo to host Eridanus II is necessary, but also one needs to check the stability of the star cluster contained in this UFD. It is observed in simulations that the central core formed in the inner part of galaxies has oscillations that changes the density inside the core and might affect star formation. Therefore, if the star cluster region is inside the core, which occurs for $m \lesssim 10^{-20} \, \mathrm{eV}$, it is going to be affected by the oscillations and might have a different star formation. Otherwise, the star cluster can extend outside the core and can be subject to the interference patterns present in the halo. The effect of those oscillations in the stars is heating of the star cluster, which can disrupt the cluster in a time scale that should have been observed.  Taking that effect into consideration, if FDM is the total amount of DM, Eridanus II can only form if $m \gtrsim 8 \times 10^{-22} \, \mathrm{eV}$, which is a value of the mass of the FDM particle that is in slight tension with the one necessary for the FDM model. In \cite{Schive:2019rrw} high resolution simulations, where the random walk behaviour of DM inside the core could be seen, it is shown that Eridanus II is bound to the MW, so its halo might suffer tidal disruption. This reduces the oscillations and counteracts the heating. From that, they claim that the bound obtained in \cite{Marsh:2018zyw}, can be evaded and $m \sim 10^{-22} \, \mathrm{eV}$ is allowed and can explain Eridanus II. These studies show how important is to study the effect of oscillations in the core, and this is the goal of future simulations.

\subsubsection{Substructure - strong lensing}

 Different models of DM predict different substructures, from its  abundance to the minimum mass for the possible substructures formed. As we saw in this section, generally all the ULDM models suppress the formation of sub-halos of a certain mass and size. By probing the substructures and its properties, then, we can test different models of DM.  Therefore, the observables that probe substructures represent one of the most important tests for DM on small scales.

We already presented above a probe of substructures present in the halo, the stellar streams. This probe seems to be very sensitive to DM substructures and has the potential in the future to help test DM models. 

However, there is another way of probing directly the presence of substructures, if they are luminous or not, which is gravitational lensing. Gravitational lensing is distortion of light from objects by the presence of a gravitational potential. 
One of  the main observations used to search for substructures is strong gravitational lensing of quasar. Lensed quasar can present multiple images, arcs or even Einstein rings. The presence of substructures modifies the lensed images of quasar changing its morphology and flux ratios, in a way that the substructure can be mapped, as done in \cite{lensing26,lensing27,lensing28,lensing29}, including distortions from sub-halos in the line of sight \citep{Despali:2017ksx}.
There are many efforts to probe DM substructures from strong lensing using different frameworks \citep{lensing30,lensing31,lensing32,lensing33,lensing34}, and machine learning techniques \citep{Alexander:2019puy,Brehmer:2019jyt}  (for a list of other machine learning approach see \citealt{Alexander:2019puy}).

%We showed in the previous subsections, in the context of the FDM model, that substructure probes like strong lensing and stellar streams are already being used to constraint these models, like done in \cite{Schutz:2020jox}.

Different models of DM can also predict different types of substructures, like vortices. They can also predict different shapes for these substructures like dark disks. Dark disks are a unique type of substructure in isolated systems, not expected in CDM. Those dark disks can be formed in the models based on the superfluid DM~\citep{Alexander:2019qsh} and are very distinct signature of those alternative models. Strong gravitational lensing is an important technique to probe these substructures, and one of the techniques that is going to improve the most with the upcoming experiments.

All the other observations presented before probe substructures or other phenomena from systems that contain stars or galaxies. Gravitational lensing and stellar stream can be used even if the sub-halo or dark structure does not have a luminous component. Therefore, these probes offer the possibility of not only probing very light substructures but different dark structures predicted by the ULDM models.

\paragraph{Sub-halo mass function}\mbox{}\\

As we described in Section 4.1.4,  the sub-halo mass function for the FDM model presents a redshift dependent cut for smaller sub-halo masses, which leads to a suppression of the small scale structure.
We can then use probes that are sensitive to the gravitational potential to test the SHMF of the FDM model.

One work that attempts to test the suppression of the sub-halos is  \cite{Schutz:2020jox}.   In this work strong lensing, together with stellar streams, are used to constraint the suppressed sub-halos mass function of the FDM model. A bound on the mass of the FDM particle can be obtained, $m \sim 2.1 \times 10^{-21} \, \mathrm{eV}$, which  is again in tension with the value of the mass where the FDM solves the small scale challenges. A similar analysis was performed in~\citep{Benito:2020avv} using a different fitting form for the sub-halo mass function from~\citep{marsh_20} and comparable constrain for the FDM mass was obtained $m \sim 5.2 \times 10^{-21} \, \mathrm{eV}$

\subsection{UV luminosity function}

%Here we present constraints on the FDM mass that come from measuring the UV luminosity function. 
Here we present constraints on the FDM mass by comparing the predicted cutoff in the luminosity function from the FDM model with observations that probe the UV luminosity function.

The Hubble Ultra Deep Field (HUDF)~\citep{HUDF} is used to search for galaxies at high-z. The high-z galaxies measured by the HUDF can be used to match with the expected UV luminosity function predicted by the FDM. This is done in many studies. In~\citep{marsh_20}, using the HMF fitted from their simulation (\ref{Eq.:HMF_1}), they used the data from HUDF and found that $m \gtrsim 1.2 \times 10^{-22}\, \mathrm{eV}\, (2\sigma )$, using galaxies from $z=6-8$. A very similar analysis was made in~\citep{marsh_22} with similar bounds. In~\citep{HMF_2} this was also analyzed but using the HMF (\ref{Eq.:HMF_2}) and shows that using the observed UV luminosity from HUDF, a mass of $m = 10^{-23}\,\mathrm{eV}$ is excluded with at $>8\sigma$, and therefore $m \gtrsim 10^{-22}\,\mathrm{eV}$ is consistent with HUDF. Obtaining a luminosity function from a full hydrodynamical cosmological simulations of galaxy formation using the initial conditions from the FDM model was done in~\citep{Ni:2019qfa}, they reach a similar conclusion, ruling out $m< 5 \times 10^{-22}\,\mathrm{eV} \, (3\sigma )$.  Combining the HUDF data with deep IRAC
data from Spitzer Space Telescope over the Great Observatories Origins Deep Survey (GOODS) fields, in~\citep{song} they can probe even higher redshifts $z\sim 8$ and show that for $m \sim 10 -5 \times 10^{-22}\,\mathrm{eV}$ the FDM is consistent with their measurements, while  $m< 2 \times 10^{-22} \, \mathrm{eV}$ is inconsistent.

One can also use the Hubble Frontier Field (HFF)~\citep{HFF_1,HFF_2} that observes the gravitationally lensed ultra-faint galaxies. This observation probes the faint end of the luminosity function. This is exactly where it is expected that the FDM changes the luminosity function, so it is a very good observation to put bounds in the FDM mass. This was done in~\citep{FDM_HFF_1}. Since the measurements that they use are based on gravitational lensing and therefore it is possible to measure smaller halos, they construct a new luminosity function extending to smaller values of M in $\Phi (<M_{UV},z) = n(>M,z)$. This new luminosity function does not depend on baryonic physics that takes place at galaxy formation. With that they are able to place a very strong constraint in the mass of the FDM, which must be $m \gtrsim 8 \times 10^{-22}\, \mathrm{eV}$ to be compatible with this data. A similar study~\citep{Leung:2018evj} using HFF finds that $ m\simeq 10^{-22}\,\mathrm{eV}$.

\subsection{Black hole superradiance}

Until now we explored ways of probing DM by the gravitational effect they imprint in the structure and substructure of our universe. We explored in detail the astrophysical tests of DM which happen in environments where we are in the weak field regime, but that are dominated by baryonic effects and complex non-linear physics, but that can still give us hints of the nature of DM.
%Those are indirect ways of probing DM and take into account the different effects they have in the 
We present here a very different way of probing specifically ultra-light fields in strong field environments, far from the linear cosmological scales. 

Ultra-light particles can be largely produced around spinning BHs, a process called BH superradiance \citep{Brito:2015oca}. 
When a BH rotates faster than the angular phase velocity of an incoming wave, it amplifies the energy and angular momentum of the field in its vicinity. This superradiance effect \citep{53,54,55,56} is a natural mechanism to create clouds of ultralight bosons around Kerr BHs (see \citealt{Brito:2015oca} for a review). For ultra-light particles with  Compton wavelengths  of order or larger than the BHs gravitational radius,  they will be efficiently produced by the superradiance, forming a large `cloud' around the BH. This  cloud is a condensate of ultra-light particles created through this instability carrying up to $10\%$ of the BH's mass and angular momentum, diminishing the initial rotation of the BH \citep{117,118,119,120}. 

%%%%%%%%%%%%%%%%%%%%%%%%%%%%%%%%%%%%%%%%%%
%\begin{figure}[htb]
%\centering
%\includegraphics[scale=0.3]{superradiance.png}
 %\caption{Illustration of superradiance, showing the growing mode (2p2p) for the scalar and  (1s1s) for the vector ultra-light fields.  created Credit of the figure: John Stout. Modified from~\citep{Baumann:2019eav}}
%\label{Fig.:superradiance}
%\end{figure}
%Illustration of the dominant growing modes of the scalar (2p2p) and vector (1s1s) gravitational atoms. The small arrows denote the intrinsic spin of the vector field, which allows for superradiant growth even for states without orbital angular momentum.
%%%%%%%%%%%%%%%%%%%%%%%%%%%%%%%%%%%%%%%%%

In the non-relativistic limit, the eigenfunctions of the system are determined by a Schr{\"o}dinger-like equation and the whole set up is sometimes referred to as a \emph{`gravitational atom'}. Superradiance instability depends on the spin of the BH, the mass of the BH and the mass of the ultra-light particle created, where the modes are co-rotating with the BH. Depending on the mass of the BH, from stellar mass BHs to supermassive BHs spanning masses from a few to billions of solar masses,  the ultra-light bosons produced through this mechanism can have masses from $10^{-20}$ to $10^{-10}$ eV \citep{121,122,Stott:2018opm}. The most stringent constraints to date on the mass of ultra-light bosons using superradicance is presented in \cite{Stott:2018opm}, where, for the FDM model, masses from $7\times 10^{-14} \, \mathrm{eV} < m <2 \times 10^{-11} \, \mathrm{eV}$ are excluded with $95\%$ C.L. using stellar mass BHs, and for SMBHs they are excluded in the range $7 \times 10^{-20} \, \mathrm{eV} < m <10^{-16} \, \mathrm{eV}$. These constraints are also valid in the presence of a potential for the axion like mentioned in Sect.~\ref{Sec.:stability} if $f_{\mathrm{a}} \gtrsim 10^{14} \, \mathrm{GeV}$. 
Superradiance can also occur for ultra-light vector fields. In this case particles of mass $10^{-14} \, \mathrm{eV}$ to $10^{-11} \, \mathrm{eV}$ can be created by stellar mass BHs, and $10^{-20} \, \mathrm{eV}$ to $10^{-17} \, \mathrm{eV}$ for supermassive BHs \citep{123}. The case of spin-$2$ particles was also studied in \cite{Brito:2020lup}.
%\footnotetext{In reference~\citep{Davoudiasl:2019nlo}, the data from the  Event Horizon Telescope (EHT) on M87$*$ is used to exclude ranges for the mass of scalar ULDM, $2.9\times 10^{-21} \, \mathrm{eV} < m_s <4.6 \times 10^{-21} \, \mathrm{eV}$, and for vector ULDM, $8.5\times 10^{-22} \, \mathrm{eV} < m_v <4.6 \times 10^{-21} \, \mathrm{eV}$. This reference uses the initial EHT data, and it will be interesting to see how this bound will evolve as more data is obtained.}
%For a BH of mass $M$ and spin $J \equiv aM$, a bosonic field with mass $\mu _a$ and angular frequency $\omega$ experiences superradiance instability if $\omega/m < \Omega_H = a/2Mr_{+}$, where $m$ is the azimutal angular momentum of the field, $\Omega_H$ is the angular velocity if the BH and $r_{+}\equiv M+\sqrt{M^2 - a^2}$ is the  size of the event horizon of the BH.  Superradiant growth requires $m > 0$, i.e., modes that co-rotate with the Kerr black hole.

In reference~\citep{Davoudiasl:2019nlo}, the data from the  Event Horizon Telescope (EHT) on M87$*$ is used to exclude ranges for the mass of scalar ULDM, $2.9\times 10^{-21} \, \mathrm{eV} < m_s <4.6 \times 10^{-21} \, \mathrm{eV}$, and for vector ULDM, $8.5\times 10^{-22} \, \mathrm{eV} < m_v <4.6 \times 10^{-21} \, \mathrm{eV}$. This reference uses the initial EHT data, and it will be interesting to see how this bound will evolve as more data is obtained.

This cloud emits GW, which allow us to probe its presence around BHs. This GW signal could possibly be detected by experiments like LIGO, when coming from stellar BH clouds, and LISA, from supermassive BH clouds \citep{121,122,123}. The signature of this GW will depend if the cloud is made of real particles, which creates a non-axisymmetric cloud, or complex scalar particles, where the cloud is axisymmetric and the emission of GWs  is suppressed. For the cloud made of real bosons given the non-axisymmetric configuration, it can emit GWs when the ultra-light bosons interact with gravitons, or when gravitons change levels, emitting monochromatic GWs \citep{122,Bertone2}. The cloud can also collapse if there is an attractive interaction between those UL particles, emitting GWs in this process.

If the spinning BH and its respective cloud is in a binary BH system, the GW signal is modified due to the presence of the companion presenting a richer GW phenomenology \citep{124,125,126}. The evolution and GW signature of the cloud is modified by the presence of a second BH where the waveform and the amplitude of the signal can be modified and even vanish given the resonant transitions between the growing and decaying modes of the cloud. Sharp features in the GW waveform appear, offering a window to probe the signal from these UL particles using GW experiments.

This is an active field of research with the modelling of these effects, the study of back reaction and the observational signatures still ongoing topics of research. Detecting the GWs coming from these clouds would give us the opportunity to probe ultra-light particles in a range that is very interesting for DM. In this way this observational signature is very relevant to the ULDM models.

\vspace{0.3cm}
%Recently, the possibility of producing part of the DM of the universe by the  evaporation of primordial BHs (PBH) was studied in \cite{Masina:2020xhk}. In the process of their evaporation, PBHs of mass around $0.1\, \mathrm{g}$ to $10^9 \, \mathrm{g}$ which evaporate before BBN, decay into DM particles with masses around $\mathrm{MeV - GeV}$. Although those masses are higher than the ones we are interested in this review, this mechanism serves as another illustration of a different way of creating DM particles from these strong gravity objects that is being studied in the literature.
%\vspace{0.3cm}

As we saw here, those busy and complex environments like galaxies and BHs might still offer, despite their modelling complexity, new ways of probing de fundamental properties of DM. For a review on the same lines for probing fundamental physics in those environments see \cite{Baker:2019gxo}.

\subsection{Probing the wave nature of ULDM}

We have shown above the astrophysical  and cosmological consequences of DM being described by a BEC or a superfluid inside the halos of galaxies. From changing the behaviour of LSS on small scales, to changes related the halos having a minimum mass in those models, and the different profile the wave nature of ULDM leaves in the inner parts of halos, we have been  investigating the gravitational consequences of  the ULDM models.  Some of those predictions might be degenerate with other DM models, like the WDM and SIDM, and even with possible consequences from unaccounted baryonic physics. 
%A series of those measurements together with some unique predictions of the many ULDM might help us test these  models. 

However, given the wave nature of the ULDM, these models present some predictions that are a direct consequence of this wave nature and that are completely distinct from any other DM model. We discuss here  vortices and  interference patterns. These are distinct effects that can appear in galactic scales, and can also be connected in the formation of halos.
%Superfluids when rotate, given their irrotationoal nature, can only rotate by forming an array of vortices, small topological defects that are homogeneously spread in the superfluid. This is seen in laboratory experiments with superfluids and the question is if they arise in the DM halos. Another distinct prediction of those models, and a consequence of the wave nature of the condensates is the formation of quantum mechanical interference fringes coming from the merging processes of those condensate soliton cores, like what is expected in the interference of waves of  quantum mechanical systems. Those two phenomena might also be connected in the formation of halos.

The detection of any of these effects would be a direct evidence of those models of ULDM and a smoking gun for the wave nature of DM. 
For this reason it is interesting to study how these effects arise in each of those classes of ULDM models and understand if they yield observable consequences that allow us to test them.

\subsubsection{Vortices}

%We have shown above many astrophysical  and cosmological consequences of DM being describes by a BEC or superfluid inside the halos of galaxies. From changing the behaviour of LSS on small scales, to changes related the halos having a minimum mass in those models, and the different profile the wave nature of ULDM leaves in the inner parts of halos. 
Until now we neglected the fact that the halo might be rotating. As we saw in Sect.~\ref{Sec.:vortices}, rotating superfluids have an interesting new phenomenology, the appearance of a lattice of quantized vortices that allow for the rotation of this irrotational fluid.  This is a purely quantum mechanical phenomena resulting in  quantized vortices being produced.\footnote{It is also worth noticing that a fluid can also form classical vortices, and the difference between those to the quantum vortices is that the quantum vortices are quantized.} This new phenomenology arising from rotating BEC halos might lead to observable astrophysical consequences that will represent a direct probe of the wave nature of these DM models. The hope is that measuring this unique signature from these models, will make it possible to distinguish this class of models from other alternative DM models which might present some signatures that are degenerate with the ULDM models.

%If DM is made of ultra-light particles, vortices are expected, since DM halos are expect
Disk galaxies are one of the most common galaxies in our universe and those are rotationally supported systems. Therefore, the DM halos from those systems are also expected to be rotating. Galaxy halos acquire angular momentum in their formation via tidal torques coming from the neighbouring  large-scale structures \citep{Peebles:1969jm}. This angular momentum is conserved after those halos virialize, which leads to the rotating supported disks in galaxies. This process is still being fully understood, with N-body and hydro-simulations showing  that the halos of galaxies are expected to have angular momentum. From CDM N-body simulations, the angular momentum obtained, represented by the dimensionless parameter $\lambda = L |E|^{1/2}/GM^{5/2}$ where $L$ is the angular momentum and $E$ the total energy, are on the range  $\lambda \in [0.01,\, 0.1]$ \citep{Barnes:1987hu}. Therefore, when considering realistic halos of DM, one needs to consider rotation, and if DM is made of ultra-light particles, this can lead to the formation of vortices.

Here we show the effects of the halos being rotating in the ULDM, an effect that is not as explored in the literature as expected, but that might present a decisive observational signature for these ULDM models. We saw in Sect.~\ref{Sec.:BEC_superfluids} above there are conditions for the formation of these vortices, depending on the angular momentum of the rotating halo. The presence of vortices in the DM halo can alter some properties of the halo, like the mass distribution or the presence of those substructures in parts of the halo that might lead to observable signatures. Detectability of these vortices or of the effects caused by their presence, and the observational technique used to probe them, will depend on their abundance and size, which should be studied solving the Schr{\"o}dinger--Poisson system. Those present some theoretical solutions and estimates of the properties and formation of these vortices, but mainly vortices have to be studied numerically, with wave simulations like described above.  We present here some of those that investigate this in the SIFDM model, and a rough estimation for the DM superfluid. 

\vspace{0.3cm}
During the final preparation of this review a paper studying the possibility of the formation of vortices in gravitationally bounded BECs appeared \citep{Hui:2020hbq}. In this paper they study the possibility of formation of a vortices in the FDM model and of  other topological defects with different topologies. They also present results of simulations of this system for rotating halos and the possible observational signatures as consequence of the existence of such vortices.

Other topological defects like the creation of strings on the condensate in the DM halo in the SIFDM is investigated in \cite{Harko}.

\paragraph{Self-Interacting BEC} \mbox{}\\

There is a small amount of studies in the literature that investigate the presence of vortices \citep{9,Zinner:2011if,Kain:2010rb} (and recently \citealt{Hui:2020hbq}).  Given that to investigate vortices one needs to study the Schr\"odinger--Poisson equations, the main model where those were studied is the simplest superfluid model given by the SIFDM model. But even in these simplified models it is still hard to find solutions to the GP equation.

The study in \cite{Kain:2010rb} discusses the possibility of the formation of a lattice of vortices in the halo of a galaxy with the same parameters as Andromeda. This aims to show a order of magnitude estimate for a known galaxy, the amount of vortices formed and   put bounds in the mass of the SIFDM particle, and in its interaction. Assuming that a galaxy like M31 has $M \sim 10^{12}\, M_{\odot}$, average radius $R \sim 100\, \mathrm{kpc}$, average density $\rho_{\mathrm{avg}} \sim 10^{-23} \, \mathrm{kg/m}^3$ and angular velocity $\Omega \sim 10^{-16}\, \mathrm{rad/s}$. If vortices are created they cannot be larger than the size of the condensate core, $R_{\mathrm{v}} \leq R_{\mathrm{bec}}$, the SIFDM must have a mass $m \gtrsim 10^{-24}\, \mathrm{eV}$, with radius $R_{\mathrm{v}} \sim 10^{21}\, \mathrm{m} \approx 30\, \mathrm{kpc}$. Therefore, in a halo like the halo of M31, there might be from $1$ to $100$ vortices. For the simplified case of having only one vortex in the halo, the GP equation can be solved in the Thomas--Fermi limit, ignoring the QP term, determining the structure of this cylindrical vortex core, and the critical velocity, showing that Andromeda galaxy could have formed a vortex given its angular velocity.

The reference \cite{Zinner:2011if}, focus in a simplified study of the consequences for virialization and on the rotation curves if a lattice of vortices is present in the FDM model. Depending on the choice of mass and interaction, this vortices can lead to oscillations in the rotation curve of galaxies, which the authors claim resemble what is observed in rotation curve of spiral galaxies.

Now, in \cite{9}, the authors try to study the vortex solutions in more detail,  solving the GP for some assumed halo profiles that are more realistic than in the previous work (see also \citealt{Suarez:2013iw} for a review of vortices in SIFDM). 

First, it is interesting to repeat something we presented in Sect.~\ref{Sec.:vortices}. The presence of a rotation in a condensate in a spherical halo is that without rotation, the BEC wavefunction is real and positive, however the  angular velocity induces a superfluid current, making the wavefunction to be complex, $\psi=|\phi|e^{iS(t,\mathbf{r})}$, with a phase that gives a velocity flow, $v=\hbar \nabla S / m$,  and as we saw in Sect.~\ref{Sec.:vortices}, the fluid velocity is $\mathbf{v}'=\mathbf{v} - \mathbf{\Omega} \times \mathbf{v}$. Therefore, a rotating BEC is a superfluid. As shown in Table 1 from \cite{9}, FDM models, where there is no interactions, do not form vortices. If the interaction is attractive, vortices are also not formed. For this reason we work in the case of SIFDM model with repulsive interaction.

The condition for the formation of vortices is $\Omega > \Omega_{\mathrm{c}}$ or equivalently $L > L_{\mathrm{qm}}$. This imposes a bound in the mass of the SIFDM particles. They work this out and the solutions for the GP equations for two halo models. First, in \textit{halo model A}, they assume a simplified halo model where the density and potential is given by a homogeneous Maclaurin spheroid, which gives a known form for the gravitational potential inside halos. This halo is not irrotational, having $L \gg L_{\mathrm{qm}}$. With that simplified gravitational potential one can calculate the characteristics of a virialized  rotating halo. This allows to put bounds in the mass and the interaction of the SIFDM particle, assuming the critical case where $(L/ L_{\mathrm{qm}})_{\mathrm{c}}$ and $\lambda =0.05$  is the average of  the dimensionless rotation parameter: $m/m_{\mathrm{H}} \geq (m/m_{\mathrm{H}})_{\mathrm{c}} \approx 50$ and $g/g_{\mathrm{H}} \geq (g/g_{\mathrm{H}})_{\mathrm{c}}\approx 2550$.

For a more realistic halo which allows to take into account the compressibility of the fluid, the halo is considered as $(n=1)$-polytropic Riemann-S ellipsoids, \textit{halo model B}, which is irrotational before forming halos $L=L_{\mathrm{qm}}$. 
%The conditions for formation of vortices in this halo follow the ones  from \textit{halo model A}, so there will be formation of vortices in this halo if the conditions showed above are met. 
This shows that if $L < L_{\mathrm{qm}}$, there is no formation of vortices and the rotating BEC can be described by \textit{halo model B}. And for $L \gg L_{\mathrm{qm}}$, there is the formation of, at least one vortex if the quantities are equal to the critical quantities or more in case they exceed these values, with the halo being described by \textit{halo model A}.  The case where $L \gtrsim L_{\mathrm{qm}}$, the halo is described by the \textit{halo model B} again, with a single vortex in the center of the halo. 

This study shows the conditions for the formation of the vortices and shows the complexity that can arise in the presence of rotations and of a vortex lattice.

\vspace{0.3cm}
These studies show important characteristics for the formation of  vortices in the halo of galaxies in the SIFDM model. However, to fully study the presence and consequences of a rotating halo in those model one needs to perform wave simulations. One study that takes that in consideration is \cite{Hui:2020hbq}.

\paragraph{DM superfluid } \mbox{}\\

%Quantum vortices are a prediction of rotating superfluids and its measurements would be a smoking gun for the superfluid model.  As we saw, superfluids have the property of flowing without friction. When rotated, to velocities above a critical angular velocity, the superfluid circulates around quantized vortex lines that carry quantized orbital angular momentum. This allow the rotation of a superfluid which is always inhomogeneous. 

As we saw above, to calculate the abundance and properties of the vortices, it is necessary to solve the equations of motion of the superfluid coupled to the Poisson equation. For the DM superfluid model, there is still no numerical study of the solution of the equations equivalent to the GP in the presence of gravity in order to understand the formation of vortices in this model.
For this reason, here we present a dimensional analysis and order of magnitude estimation of the presence of vortices in the DM superlfuid context.

Vortices are formed when the angular velocity of the superfluid is larger than the critical velocity: $\Omega_{\mathrm{Sf}} \gg \Omega_{\mathrm{c}}$, with the critical angular velocity given by (\ref{Eq.:critical_angular_velocity}).
For $R \sim 100 \, \mathrm{kpc}$ and $m \sim \mathrm{eV}$, we can  see that $\Omega_{\mathrm{c}}  \sim 10^{-41} \, \mathrm{s}^{-1}$ (neglecting the logarithmic factor). This is much smaller than the rotation velocity of the halo: $\Omega \sim \lambda\sqrt{G \rho} \sim 10^{-18} \mathrm{s}^{-1}$, using a halo mass density $\rho \sim 10^{-25} \mathrm{g/cm}^3$. So, there will be the production of vortices in the halos of the galaxies in the DM superfluid model 
And this production seem to be very numerous. 

We can estimate the number of vortices in the halo $N_v=\Omega/\Omega_{\mathrm{c}} \sim 10^{23}$, with a core radius, given by the healing length, $\xi=1/(m c_{\mathrm{s}}) \sim \mathrm{mm}$ (assumed a MW type galaxy and fiducial values). Although highly numbered, these vortices are small. It is still unclear if it is possible to detect those vortices via, for example, gravitational lensing or any other effect they might have in the galaxy. 

%It worth noticing that this calculation is just a simple estimate of the for the formation of vortices in the DM superfluid model. As we saw on section~\ref{Sec.:vortices}, to get a real estimate of the size, shape and abundance of the vortices in a superfluid model, one has to solve the GP equation for a realistic model of the DM halo. 
%So the estimate above should be taken  with a grain of salt and as only shown that the formation of vortices in the DM superfluid model is expected.
%This topic worth further investigation since the detection of such effect would be an important evidence for the presence of superfluids in galaxies.

\subsubsection{Interference fringes}

Another interesting effect that comes from the wave nature of DM superfluid model is interference. Since the condensate is described by a coherent wave function with the density if give by $|\psi|^2$, in the formation of galaxies, interference patterns are expected to form.

\begin{figure}[htb]
\centering
\includegraphics[scale=0.85]{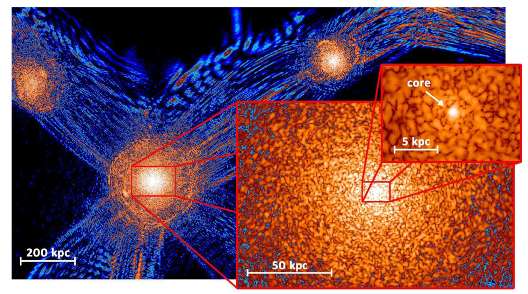}
\includegraphics[scale=0.88]{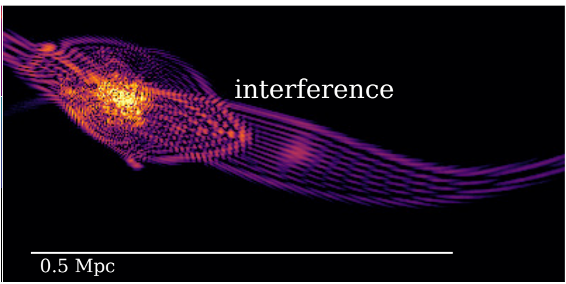}
 \caption{In this figure we show the interference pattern in the DM halos from hydrodynamical simulations of the fuzzy DM model.  \textit{Left panel:} Figure from \cite{numerical_fuzzy_1} showing the density distribution of FDM at redshift $z = 0.1$ at different scales. This simulation uses the \textsc{\tt{GAMER}} code to describe the wave-like FDM using  an adaptive-mesh-refinement (AMR) scheme.
 \textit{Right panel:} Figure from \cite{Mocz} that shows the simulations of the slices of DM through a filament in the DM distribution at redshift $z = 5.5$. This simulation was made by modifying the magneto-hydrodynamics code \textsc{\tt{AREPO}} for FDM model.
 These figures clearly show the interference pattern that the FDM model imprints in the halos which is very distinct from CDM or even other DM models like WDM. These interference fringes  that are present at the scales of the de Broglie wavelength. On top of that, in both stimulations we can see the soliton cores formed in the halos, also characteristic of the FDM model.}
\label{Fig.:fringes}
\end{figure}

As we discussed in the numerical simulations section, the interference patterns can only be seen in wave simulations, since when there is destructive interference, the quantum pressure term is not well defined and this is not present in the fluid simulations. We show in Fig.~\ref{Fig.:fringes}, the result of two wave simulations from different groups that show the interference fringes that appear in the filaments of FDM structure. These patterns appear on scales of order of the de Broglie wavelength. General features like caustics, which are density singularities that usually appear in CDM and even WDM simulations, are regularized in FDM models due  to the uncertainty principle, and do not appear in FDM simulations. 

% Maybe remove this and see it in previous figure. Decide later
%%%%%%%%%%%%%%%%%%%%%%%%%%%%%%%%%%%%%%%%%%
%%%%%%%%%%%%%%%%%%%%%%%%%%%%%%%%%%%%%%%%%

Another effect that could generate interference patterns is the collision of subsonic galaxies. Since those galaxies maintain the coherence of their condensate core described by the coherent wave function, the collision between these cores leaves an interference pattern \citep{Maleki:2019xya}. If this interference fringes could be observed, this would be still another form of probing these models.

Interference patterns are going to be created also  in the merging processes of dark/bright solitons \citep{solitons1,solitons2}. Some authors suggest that this effect could be linked to the shells seen around elliptical galaxies \citep{solitons3}.

%%%%%%%%%%%%%%%%%%%%%%%%%%%%%%%%%
% Conclusions
\section{Summary}

In this review we studied an alternative class of DM model, the ultra-light DM.  These models have been receiving a lot of attention in the literature nowadays given their interesting property of forming a BEC or superfluid on galactic scales. 
In this review we aimed to give a summary of the models of ULDM, suggesting for the first time a classification into three categories according to the their non-relativistic behaviour and the structures they form in galaxies: the fuzzy DM, self-interacting BEC and the DM superfluid. Their different descriptions lead to different phenomenologies and observational effects that can be used to test these DM paradigms. We had also the goal to give a snapshot of the field  as it stands at present. We saw that current observations highly constrain the mass of the FDM and, if these constrains hold, the mass range which shows an interesting phenomenology on small scales and reconciles these scales with the CDM successes on large scales, is challenged for this model. 

We highlight how important the observations on small scales are for helping to determine the nature of DM, and how the observations of the small scales, galaxies and our MW have been advancing very fast in the last few years. With current and new experiments like \textit{Gaia}, LSST, PFS, HERA, SKA, just to cite a few, and the new and exciting probes like stellar streams, 21-cm cosmology, MW observables, BH superradiance, the next few years promise to revolutionize the tests of DM on small scales.

In this review we tried also to give a complete description of the theory behind these models, describing the striking phenomena of BEC and superfluidity.  We stressed the difference between models, showing their different descriptions and phenomenological consequences. Since the understanding of this DM  class of models require knowledge from several fields, including condensed matter physics, we believe that this theoretical summary of these constructions, is very important to  better understand of all the features that these models of DM can present and even help in finding new observables for these models.  It is also important for future progress, since there is still a lot of room for theoretical development of these DM models.

Maybe one of the biggest challenge for these models and testing them against observations is the need for numerical simulations. Those are necessary so we can understand how the formation of structures proceeds in these models. However, performing those simulations in a way that they resolve the small scales in order for us to see  the interesting effects coming from the wave nature of these models, and also that they simulate the structure on cosmological scales is a challenge. Incredible advances in this field have been made in the past few years and there are many groups currently working to improve those simulations. Most of those advance, tough, are only for the FDM model. Simulations for the SIBEC are only a few and there is still no numerical framework to study the DM superfluid.  Thus we should expect continue progress in this field in the future, which will lead to also a better understanding on how to probe those models with observations.

In summary, the study of the ultra-light DM is an active area of research and many challenges are still opened to be addressed theoretically, numerically and observationally. As this field becomes more and more popular we believe this progress will go even faster and we hope this review can help those entering or already in this field supporting their understanding of this fascinating new DM model.

%%%%%%%%%%%%%%%%%%%%%%%%%%%%%%%%%
\begin{acknowledgements}
First, I would like to thank \textit{The Astronomy and Astrophysics Review} for the invitation to write this review. I am very grateful to Eiichiro Komatsu for the help, guidance and encouragement for writing this review. I thank Robert Brandenberger for all the advice and comments that were essential to completion of this work. I am also very grateful to Evan Mcdonough, Jerome Quintin and Guilherme Frazmann for carefully reading and useful comments to the content of the review, and Jerome for the thorough proofreading. Thanks to Justin Khoury for supporting this project, and to both Lasha Berezhiani and Justin for teaching me about this topic so patiently.  I would like to thank Eduardo Toshio Matsushita for teaching so much about BECs and quantum mechanics. I also thank the Referee for the useful comments and suggestions that helped improve the review.
\end{acknowledgements}

%%%%%%%%%%%%%%%%%%%%%%%%%%%%%%%%%
% Bibliography

\bibliographystyle{spbasic}      % basic style, author-year citations
\bibliography{Bibliography.bib}   % name your BibTeX data base

\end{document}